\documentclass[11pt]{article}
\usepackage{fullpage}
\usepackage{graphicx,amsfonts,amsmath,amssymb,epsfig}
\usepackage[usenames,dvipsnames]{xcolor}
\usepackage[ruled,vlined,linesnumbered]{algorithm2e}
\usepackage{algorithmic}

\usepackage{multirow} % Required for the table of results
\usepackage{chngpage} % allows for temporary adjustment of side margins

\usepackage{setspace} % To have the ToC fit on one page

\usepackage[backref, colorlinks,citecolor=blue,bookmarks=true]{hyperref}
\usepackage{aliascnt}

\def\confversion{0}
\def\withcolors{0}
\def\withnotes{0}

\ifnum\withnotes=1
\newcommand{\mnote}[1]{ \marginpar{\tiny\bf

            \begin{minipage}[t]{0.5in}

              \raggedright #1

           \end{minipage}}}
\else
\newcommand{\mnote}[1]{}
\fi

\renewcommand{\paragraph}[1]{{\protect\vspace{8pt}\noindent\sc{#1}}}

\newlength{\saveparindent}

\newlength{\saveparskip}

\ifnum\confversion=0

\fi

\newcommand{\BE}{\begin{enumerate}} \newcommand{\EE}{\end{enumerate}}

\newcommand{\BI}{\begin{itemize}} \newcommand{\EI}{\end{itemize}}

\newcommand{\BDes}{\begin{description}}\newcommand{\EDes}{\end{description}}

\newtheorem{alg}{Algorithm}

\newcommand{\BA}{\begin{alg}} \newcommand{\EA}{\end{alg}}

\newtheorem{thm}{Theorem}
\newcommand{\BT}{\begin{thm}} \newcommand{\ET}{\end{thm}}

\newtheorem{rem}{Remark}

\newcommand{\BR}{\begin{rem}} \newcommand{\ER}{\end{rem}}

\newtheorem{lem}{Lemma}      % A counter for Lemmas
\newcommand{\BL}{\begin{lem}} \newcommand{\EL}{\end{lem}}

\newaliascnt{claim}{lem}
  		\newtheorem{clm}[claim]{Claim}
\aliascntresetthe{claim}
\newcommand{\BCM}{\begin{clm}} \newcommand{\ECM}{\end{clm}}

\newaliascnt{fact}{lem}
  		\newtheorem{fct}[fact]{Fact}
\aliascntresetthe{fact}
\newcommand{\BFCT}{\begin{fct}} \newcommand{\EFCT}{\end{fct}}

\newtheorem{obs}[lem]{Observation}

\newcommand{\BOBS}{\begin{obs}} \newcommand{\EOBS}{\end{obs}}

\newaliascnt{corollary}{thm}
  \newtheorem{techcor}[corollary]{Corollary}
  \newcommand{\BCo}{\begin{techcor}} \newcommand{\ECo}{\end{techcor}}

  \newtheorem{cor}[corollary]{Corollary}      % counter AS FOR Theorems
  \newcommand{\BC}{\begin{cor}} \newcommand{\EC}{\end{cor}}
\aliascntresetthe{corollary}

\newaliascnt{prop}{thm}
  \newtheorem{prop}[prop]{Proposition}     % A counter AS FOR Thms
\aliascntresetthe{prop}

\newcommand{\BP}{\begin{prop}} \newcommand {\EP}{\end{prop}}

\newtheorem{conj} {Conjecture}      % counter AS FOR Theorems

\newcommand{\BCJ}{\begin{conj}} \newcommand{\ECJ}{\end{conj}}

\newtheorem{defn}{Definition}         % A counter for Definition

\newcommand{\BD}{\begin{defn}} \newcommand{\ED}{\end{defn}}

\def\FullBox{\hbox{\vrule width 8pt height 8pt depth 0pt}}

\ifnum\confversion=0
  \newcommand{\qed}{\;\;\;\FullBox}
  \newenvironment{proof}{\noindent{\bf Proof:~~}}{\(\qed\)}
\else
  \renewcommand{\qed}{\;\;\;\FullBox}
  \renewenvironment{proof}{\noindent{\bf Proof:~~}}{\(\qed\)}
\fi

\newcommand{\BPF}{\begin{proof}} \newcommand {\EPF}{\end{proof}}

\newenvironment{proofof}[1]{\noindent{\bf Proof of {#1}:~~}}{\(\qed\)}

\newcommand{\BPFOF}{\begin{proofof}} \newcommand {\EPFOF}{\end{proofof}}

\newcommand{\qedsketch}{\;\;\;\Box}

\newenvironment{smallproof}{\noindent{\bf Proof:~~}}{\(\qedsketch\)}

\newcommand{\bpf}{\begin{smallproof}} \newcommand{\epf}{\end{smallproof}}

\newcommand{\BEQ}{\begin{equation}} \newcommand{\EEQ}{\end{equation}}

\newcommand{\BEQN}{\begin{eqnarray}}\newcommand{\EEQN}{\end{eqnarray}}

 \newcommand{\BEQst}{\begin{equation*}} \newcommand{\EEQst}{\end{equation*}}

 \newcommand{\BEQNst}{\begin{eqnarray*}}\newcommand{\EEQNst}{\end{eqnarray*}}

\newcommand{\eqdef}{\stackrel{\rm def}{=}}

\renewcommand{\Pr}{{\rm Pr}}

\newcommand{\Exp}{{\mbox{\bf\rm E}}}

\newcommand{\no}{{\mbox{\bf\rm No}}}

\newcommand{\poly}{{\rm poly}}

\newcommand{\e}{\epsilon}

\newcommand{\eps}{\epsilon}

\newcommand{\tildeO}{{\tilde{O}}}

\newcommand{\calD}{{\cal D}}

\newcommand{\calP}{{\cal P}}

\newcommand{\calU}{{\cal U}}

\newcommand{\D}{{D}}

\newcommand{\E}{{\rm E}}

\renewcommand{\P}{{\rm P}}

\newcommand{\R}{{\rm R}}

\newcommand{\T}{{\rm T}}

\newcommand{\todo}[1]{\iffalse {#1} \fi }

\newcommand{\setOfSuchThat}[2]{ \left\{\; #1 \;\middle\vert\; #2\; \right\} }           % sets such as "{ elems | condition }"

\newcommand{\indicSet}[1]{\mathbf{1}_{#1}}                                              % indicator function

\newcommand{\indic}[1]{\indicSet{\left\{#1\right\}}}                                             % indicator function

\newcommand{\disjunion}{\amalg}%\coprod, \dotcup...

\newcommand{\totalvardist}[2]{{d_{\rm TV}\!\left({#1, #2}\right)}}%{{\operatorname{d_{\rm TV}}\!\left({#1, #2}\right)}}

\newcommand\restr[2]{{% we make the whole thing an ordinary symbol

  \left.\kern-\nulldelimiterspace % automatically resize the bar with \right

  #1 % the function

  \vphantom{\big|} % pretend it's a little taller at normal size

  \right|_{#2} % this is the delimiter

  }}

\newcommand{\abs}[1]{\left\lvert #1 \right\rvert}

\newcommand{\proba}{\Pr}%{\operatorname{\mathbb{P}}}

\newcommand{\probaOf}[1]{\proba\!\left[\, #1\, \right]}

\newcommand{\probaCond}[2]{\proba\!\left[\, #1 \;\middle\vert\; #2\, \right]} %% Modified (for consistency). I did not change the 3 others by fear of the side effects.

\newcommand{\supp}[1]{\operatorname{supp}\!\left(#1\right)}

\newcommand{\shortexpect}{\mathbb{E}}

\newcommand{\uniform}{\ensuremath{\mathcal{U}}}

\newcommand{\binomial}[2]{\ensuremath{\operatorname{Bin}\!\left( #1, #2 \right)}}

\newcommand{\Algo}{A} % Adversarial algorithm A

\newcommand{\accept}{{\sf ACCEPT}\xspace}

\newcommand{\reject}{{\sf REJECT}\xspace}

\newcommand{\unknown}{{\sf UNKNOWN}\xspace}

\newcommand{\fail}{{\sf FAIL}\xspace}

\newcommand{\half}{\frac{1}{2}}

\newcommand{\ORACLE}{{\sf ORACLE}\xspace}

\newcommand{\ICOND}{{\sf ICOND}\xspace}

\newcommand{\EVAL}{{\sf EVAL}\xspace}

\newcommand{\SAMP}{{\sf SAMP}\xspace}

\newcommand{\COND}{{\sf COND}\xspace}

\newcommand{\PCOND}{{\sf PCOND}\xspace}

\newcommand{\bigO}[1]{{O\!\left({#1}\right)}}

\newcommand{\bigTheta}[1]{{\Theta\!\left({#1}\right)}}

\newcommand{\bigOmega}[1]{{\Omega\!\left({#1}\right)}}

\newcommand{\good}{comparable\xspace}

\newcommand{\constraint}[1]{}

\ifnum\withnotes=1
\newcommand{\modconstraint}[1]{\mnote{{\color{red} \bf \begin{con}{#1}\end{con}}}}
\else
\newcommand{\modconstraint}[1]{}
\fi

\newcommand{\ournote}[1]{}

\ifnum\withcolors=1
  \newcommand{\new}[1]{{\color{red} {#1}}}
  \newcommand{\newer}[1]{{\color{blue} {#1}}}
  \newcommand{\newest}[1]{{\color{magenta} #1}}
  \newcommand{\newerish}[1]{{\color{orange} #1}}

  \newcommand{\red}[1]{{\color{red} #1}}
  \newcommand{\rocco}[1]{{\new{#1}}}
  \newcommand{\dana}[1]{{\newer{#1}}}
  \newcommand{\clement}[1]{{\newerish{#1}}}
\else
  \newcommand{\new}[1]{{ {#1}}}
  
  \newcommand{\newer}[1]{{ {#1}}}
  \newcommand{\newerish}[1]{{#1}}
  \newcommand{\newest}[1]{{#1}}
  
  \newcommand{\red}[1]{{#1}}
  \newcommand{\rocco}[1]{{\new{#1}}}
  \newcommand{\dana}[1]{{\newer{#1}}}
  \newcommand{\clement}[1]{{\newerish{#1}}}
\fi

\ifnum\withnotes=1

\newcommand{\rnote}[1]{\footnote{{\bf [[Rocco: \rocco{#1} ]] }}}

\newcommand{\dnote}[1]{\footnote{{\bf [[Dana: \dana{#1} ]] }}}

\newcommand{\cnote}[1]{\footnote{{\bf [[Clement: \clement{#1} ]] }}}

\else
\newcommand{\rnote}[1]{}

\newcommand{\dnote}[1]{}

\newcommand{\cnote}[1]{}
\fi

\newcommand{\ignore}[1]{}

\newcommand{\N}{\mathbb N}

\renewcommand{\R}{\mathbb{R}}

\newcommand{\dtv}{d_{\rm TV}}

\newcommand{\Bin}{{\mathrm{Bin}}}

\newcommand{\high}{{\textsf{High}}\xspace}

\newcommand{\low}{{\textsf{Low}}\xspace}

\newcommand{\rank}{{\mathrm{rank}}}

\newcommand{\wt}{{\mathrm{sum}}}

\newcommand{\teps}{{\tilde{\eps}}}

\newcommand{\fullOrConfVersion}[2]{
\ifnum\confversion=1
  \xspace#2\xspace
\else%
  \xspace#1\xspace
\fi%
}

\newcommand{\fullOrConfCompactEnums}{%
\ifnum\confversion=1
  \itemsep1pt
  \parskip0pt
  \parsep0pt
\fi
}

\newcommand{\fullOrConfIndentItems}{%
\ifnum\confversion=1
\fi
}

\providecommand{\email}[1]{\href{mailto:#1}{\nolinkurl{#1}\xspace}}

\title{Testing probability distributions\texorpdfstring{\\}{} using conditional samples}

\makeatletter
  \hypersetup{pdftitle=\@title, pdfauthor={Cl\'{e}ment L. Canonne, Dana Ron and Rocco A. Servedio}}
\makeatother

\author{Cl\'{e}ment L. Canonne\thanks{
\email{ccanonne@cs.columbia.edu}, Columbia University. Supported by NSF grants
CCF-1115703 and CCF-1319788.}
\and
Dana Ron\thanks{
\email{danaron@post.tau.ac.il}, Tel Aviv University.  Supported by ISF grants
 246/08 and 671/13.}
\and
Rocco A. Servedio\thanks{
\email{rocco@cs.columbia.edu}, Columbia University.  Supported by NSF grants
CCF-0915929 and CCF-1115703.}
}

\begin{document}

\begin{titlepage}
  \maketitle
  \setcounter{page}{0}
  
\begin{abstract}

We % introduce 
study a new framework for property testing of probability distributions,
by considering distribution testing algorithms that have access to a
\emph{conditional sampling oracle.}\ignore{\footnote{Independently 
from our work, Chakraborty et al.~\cite{CFGM:13} also considered this 
framework. We discuss their work in Subsection~\ref{subsec:CFGM}.}}
  This is an oracle that takes as input
a subset $S \subseteq [N]$ of the domain $[N]$
of the unknown probability distribution $\D$
and returns a draw from the conditional probability distribution $\D$
restricted to $S$.  This new model allows considerable flexibility
in the design of distribution testing algorithms; in particular, testing algorithms in
this model can be adaptive.
% since it makes it possible for such algorithms to be adaptive.

We study a wide range of natural distribution testing problems in this new
framework and some of its variants, giving both upper and lower bounds on query complexity.
These problems include  testing
whether $\D$ is the uniform distribution $\calU$; testing whether $\D
= \D^\ast$ for an explicitly provided $\D^\ast$;
testing whether two unknown distributions $\D_1$ and $\D_2$ are
equivalent; and estimating the variation
distance between $\D$ and the uniform distribution.
At a high level our main finding is that the new conditional sampling
framework we consider is a powerful one:  while all the problems mentioned
above have $\Omega(\sqrt{N})$ sample complexity in the standard model
(and in some cases the complexity must be almost linear in $N$),
we give $\poly(\log N, 1/\eps)$-query algorithms (and in some
cases $\poly(1/\eps)$-query algorithms independent of $N$) for all these
problems in our conditional sampling setting.

\end{abstract}

\end{titlepage}

\setcounter{tocdepth}{2}  % Table of contents: depth down to subsections (included)
\begin{minipage}{\textwidth}
  \begin{spacing}{1.25}
  \tableofcontents
  \end{spacing}
\end{minipage}
% \clearpage
% \listoffigures
% \listoftables
% \listofalgorithms % Uncomment this to add a list of algorithms

% \newpage

\section{Introduction}

\subsection{Background: Distribution testing in the standard model}

One of the most fundamental problem paradigms in statistics is that of
inferring some information about an unknown probability distribution
$\D$ given access to independent samples drawn from it. More than a
decade ago,
% the work of \cite{BFRSW00} introduced problems of this sort
% to the domain of theoretical computer science, by viewing them from the
% perspective of \emph{property testing}.
Batu et al.~\cite{BFRSW00}\footnote{There is a more recent full version of
 this work~\cite{BFRSW:10} and we henceforth reference this recent version.}
initiated the study of problems of this type
from within the framework of \emph{property testing}~\cite{RS,GGR}.
In a property testing problem
there is an unknown ``massive object'' that an algorithm can access
only by making a small number of local inspections of the object,
and the goal is to determine whether % or not
the object has a
particular property.  The algorithm must output $\accept$ if the object
has the desired property and output $\reject$ if the object is far from
every object with the property.
  % \ignore{ (and may output anything if the object is close to having the property)}
  % \dnote{Removed: (and may output anything if the object is close to having the property).
  % We can emphasize this in the formal definition, but I don't think it is needed here.}
  % \ignore{; thus\mnote{Kill last half of this sentence?} property testing algorithms may
  % be viewed as sub-linear time decision algorithms for a particular natural type of ``promise problem''}
(See~\cite{Fischer,Ron:08testlearn,Ron:10FNTTCS,PropertyTestingICS}
\ifnum\confversion=0
for detailed surveys and overviews of the broad field of property testing;
we give precise definitions tailored to our setting in
\autoref{sec:prelim}.)
\else
for detailed surveys and overviews of the broad field of property testing.)
\fi

In distribution property testing the ``massive object'' is an unknown
probability distribution $\D$ over
an $N$-element set, and the algorithm accesses the distribution by drawing
independent samples from it.
A wide range of different properties of
probability distributions have been investigated in this setting,
and upper and lower bounds on the number of samples required have by now
been obtained for many problems.  These include testing whether $\D$ is
uniform~\cite{GRexp:00,BFRSW:10,Paninski:08}, testing whether $\D$ is identical to a given known
distribution $\D^\ast$~\cite{BFFKRW:01}, testing whether two distributions $\D_1$, $\D_2$
(both available via sample access) are identical~\cite{BFRSW00,Valiant:11}, and
testing whether $\D$ has a monotonically increasing probability mass
function~\cite{BFRV11},
  % \dnote{maybe select another problem, e.g., independence~\cite{BFFKRW:01}?}
  % D: I referenced that paper for the \Dstar problem (I put an incorrect ref before), so I'll
  % leave it as is (for independence one needs to define what it means)
 as well as related % algorithmic
problems such as estimating the entropy of $\D$~\cite{BDK+:05,ValiantValiant:11},
and estimating its support size~\cite{RRSS,Valiant:11,ValiantValiant:11}.
  % and estimating the distance
  % between two distributions $\D_1$, $\D_2$.
  % \newer{to the uniform distribution~\cite{ValiantValiant:11,ValiantValiant:10lb}}.
  % \ignore{
  % \dnote{(31.10: I removed the reference to estimating the distance to the uniform distribution at
  % this point. Since it is not a result that is really explicitly stated in that paper as
  % a result, but is somewhat hidden, since at this point we are just giving background,
  % which is not exhaustive, I think it makes more sense to reference it only when we
  % present our result}
  % \dnote{I think the last lower bound only appears in the technical report (in terms of
  % the earth-movers distance, which I am almost sure translates to variation) and not in the
  % proceedings version - should double-check.}\cnote{For the translation EM$\to$TV, I remember the
  % paper mentioning that the latter was upperbounded by the Earth-Mover distance.}
  %}
Similar problems have also been studied by
researchers in other communities, see e.g.,~\cite{Ma:81,Paninski:04,Paninski:08}.

One broad insight that has emerged from this past decade of work is that while sublinear-sample
algorithms do exist for many distribution testing problems, the number of samples required
is in general quite large.  Even the
basic problem of testing whether $\D$ is the uniform distribution $\calU$
over $[N]=\{1,\dots,N\}$ versus $\eps$-far from uniform
requires $\Omega(\sqrt{N})$
\ifnum\confversion=0
samples\footnote{To verify this, consider the
family of all distributions that are uniform over half of the domain, and $0$
elsewhere. Each distribution in this family is $\Theta(1)$-far from the uniform
distribution.
However, it is not possible to distinguish with sufficiently high probability between
the uniform distribution and a distribution selected randomly from this family, given a
   sample of size $\sqrt{N}/c$ (for a sufficiently large constant $c>1$). This is the case
   because for the uniform distribution as well as each
   distribution in this family, the probability of observing the same element more than once
    is very small. Conditioned on such a collision event not occurring, the samples
    are distributed identically.}
\else
samples\footnote{This follows from the difficulty to distinguish between
the uniform distribution and a distribution that is uniform over a random
half of the domain using less than  $\sqrt{N}/c$ samples
(for a sufficiently large constant $c>1$).}
\fi
 for constant $\eps$, and the other problems mentioned above have sample complexities at least this high,
and in some cases \emph{almost linear in $N$}~\cite{RRSS,Valiant:11,ValiantValiant:11}.
%\mnote{check:  entropy?}
Since such sample complexities could be prohibitively high in real-world settings where $N$ can be extremely large,  it is natural to explore problem variants  where it may be possible for algorithms to succeed
using fewer samples.  Indeed, researchers have studied distribution testing in settings where
the unknown distribution is guaranteed to have some special structure, such as being monotone, $k$-modal or a \mbox{``$k$-histogram''} over $[N]$
\cite{BKR:04,DDSVV13,ILR12}, or being monotone over $\{0,1\}^n$
\cite{RubinfeldServedio:09} or over other posets \cite{BFRV11},
and have obtained significantly more sample-efficient algorithms using these additional assumptions.

\subsection{The conditional sampling model}

In this work we pursue a different line of investigation: rather than restricting the class of probability
distributions under consideration, we consider testing algorithms that may use a more powerful form of
access to the unknown distribution $\D$. This is a \emph{conditional sampling oracle}, which
allows the algorithm to obtain a draw from $\D_S$, the conditional distribution of $\D$
restricted to a subset $S$ of the domain (where $S$ is specified by the algorithm).  More
precisely, we have:

\BD \label{def:cond}
Fix a distribution $\D$ over $[N]$.  A \emph{\COND oracle for $\D$}, denoted
$\COND_\D$, is defined as follows:
 The oracle is given as input a \emph{query set}
 $S \subseteq [N]$\newer{, chosen by the algorithm, } that has $\D(S) > 0$. The oracle returns an element $i \in S$, where
 the probability that element $i$ is returned is $\D_S(i) = \D(i)/\D(S),$
 independently of all previous calls to the oracle.\footnote{Note that
as described above the behavior of $\COND_\D(S)$ is undefined if $\D(S)=0$,
i.e., the set $S$ has zero probability under $\D$.  While various definitional choices could be made to deal with this,
%such choices are irrelevant for us because
{we shall assume that in such a case, the oracle (and hence the algorithm)
outputs ``failure'' and terminates. This will not be a problem
for us throughout this paper, as}
(a)  our lower bounds  deal only with distributions that have
$\D(i)>0$ for all $i \in [N]$, and (b) in  our algorithms $\COND_\D(S)$ will only ever be
called on sets $S$ which are ``guaranteed'' to have $\D(S)>0$.  (More precisely, each time
an algorithm calls $\COND_\D(S)$ it will either be on the set $S=[N]$, or
will be on a set $S$ which contains an element $i$ which
has been returned as the output of an earlier call to $\COND_\D$.)}
\ED

\new{As mentioned earlier, a recent work of Chakraborty et al.~\cite{CFGM:13} introduced a very similar conditional model; we discuss their results and how
they relate to our results in \autoref{subsec:CFGM}.}
For compatibility with our $\COND_\D$ notation we will write $\SAMP_\D$ to
denote an oracle that takes no input and, each time it is invoked, returns
an element from $[N]$ drawn according to $\D$ independently from
all previous draws.  This is the sample access to
$\D$ that is used in the standard model of testing distributions, and
this is of course the same as a call to $\COND_\D([N]).$

\medskip
\noindent {\bf Motivation and Discussion.}
\new{One purely theoretical motivation for} the study of the $\COND$ model
is that it may further our understanding regarding what forms of information
(beyond standard sampling) can be helpful for testing properties of
distributions.  In both learning and property testing
it is generally interesting to understand how much power algorithms
can gain by making queries,
%%%%%%%%%%%%%%%
  % \ignore{As in the case of testing properties
  % of other objects, such as graphs or Boolean functions, as well as in the o
  % case of learning, we would like to understand how much power do queries
  % give,}
%%%%%%%%%%%%%%%%
and \COND queries are a natural type of query to investigate
in the context of distributions. As we discuss in more detail below,
in several of our results we actually consider restricted versions of \COND
queries that do not require the full power of obtaining conditional samples
from arbitrary sets.

A second attractive feature of the $\COND$ model is that it
enables a new level of richness for algorithms that deal with
probability distributions.  In the standard model where only access to
$\SAMP_\D$ is provided, all algorithms must necessarily be non-adaptive,
with the same initial step of simply drawing a sample of points
from $\SAMP_\D$, and the difference between two algorithms comes only
from how they process their samples.  In contrast, the
essence of the $\COND$ model is to allow algorithms to \emph{adaptively}
determine later query sets $S$ based on the
outcomes of earlier queries.
\ignore{
% Thus, the \COND model may be viewed as something of a ``bridge''
% between testing probability
% distributions and testing other objects such as graphs or
% Boolean functions, where adaptivity
% often plays an important role.
}

\new{A natural question about the \COND model is its plausibility:  are
there settings in which an investigator could actually make
conditional samples from a distribution of interest?  We feel that
the \COND framework provides a reasonable first approximation for scenarios
that arise in application areas} (e.g., in biology or chemistry)
where the parameters of an experiment can be adjusted so as
to restrict the range of possible outcomes.  For example, a scientist
growing bacteria or yeast cells in a controlled environment may be
able to deliberately introduce environmental factors that allow only
cells with certain desired characteristics to survive,
thus restricting the distribution of all experimental
outcomes to a pre-specified subset.
\new{We further note that techniques which are broadly reminiscent of $\COND$
sampling have long been employed in statistics and polling
design under the name of ``stratified sampling'' (see e.g. \cite{wiki-ss,
Neyman34:ss}).  We thus feel that the study of distribution testing in
the $\COND$ model is well motivated both by theoretical and practical
considerations.}

Given the above motivations, the central question is whether the
\COND model enables
significantly more efficient algorithms than are possible in the
weaker $\SAMP$ model.  Our
results (see \autoref{sec:ourcont}) show that this is indeed the case.

\ignore{
%
%OLD MOTIVATION SECTION STARTS HERE:
%

\medskip
\noindent {\bf Motivation and Discussion.}
The first motivation for studying the $\COND$ model is to capture
scenarios that arise in  application areas (e.g., in biology or chemistry)
where the parameters of some experiment can be adjusted so as
to restrict the range of possible outcomes.
% We briefly discuss three motivations for studying the $\COND$ model.  The first is that the $\COND$ model
% may be a good match for application areas (e.g. in biology or chemistry)
% in which an investigator can adjust the parameters of some experiment
% to restrict the range of possible outcomes.
For example, a scientist growing bacteria or
yeast cells in a controlled environment may be able to deliberately introduce
environmental factors that allow only cells with certain desired characteristics to survive,
corresponding to restricting the distribution of all experimental outcomes to a pre-specified subset.

A second, purely theoretical motivation, is that the study of the $\COND$ model may
further our understanding regarding what forms of information
(beyond standard sampling)
can be helpful for testing properties of distributions.
In both learning and property testing
it is generally interesting to understand how much power algorithms
can gain by making queries,
%%%%%%%%%%%%%%%
  % \ignore{As in the case of testing properties
  % of other objects, such as graphs or Boolean functions, as well as in the o
  % case of learning, we would like to understand how much power do queries give,}
%%%%%%%%%%%%%%%%
and \COND queries are a natural type of query to investigate
in the context of distributions. As we discuss in more detail below,
in several of our results we actually consider restricted versions of \COND
queries that do not require the full power of obtaining conditional samples
from arbitrary sets.
  % \dnote{this new paragraph needs more work. Also, do we need to mention the EVAL oracle
  % somewhere? On one hand, it has been studied. On the other hand, it really just views a
  % distribution as a real-valued function.}
  % \rnote{I made some changes to this paragraph. I think we're OK not
  % mentioning EVAL here -- it does come up later when we mention our
  % approximate eval simulation.}

A {third attractive} feature of the $\COND$ model is that it enables a new level of richness
 for algorithms that deal with probability distributions.
In the standard model where only access to
 $\SAMP_\D$ is provided, all algorithms must necessarily be non-adaptive,
 with the same initial step of simply drawing a sample of points
 from $\SAMP_\D$, and the difference
 between two algorithms comes only from how they process their samples.
 In contrast, the
 essence of the $\COND$ model is to allow algorithms to \emph{adaptively}
 determine later query sets $S$ based on the
 outcomes of earlier queries.
% Thus, the \COND model may be viewed as something of a ``bridge'' between testing probability
% distributions and testing other objects such as graphs or Boolean functions, where adaptivity
% often plays an important role.

% Finally, a third obvious motivation for studying this model is that it
\sloppy
Given the above motivations, the central question is whether the
\COND model enables
significantly more efficient algorithms than are possible in the
weaker $\SAMP$ model.  Our
results (see \autoref{sec:ourcont}) show that this is indeed the case.
\ignore{
\red{With this motivation, the current work focuses exclusively on developing
efficient algorithms for two of the most fundamental distribution testing
problems (testing equality between distributions), as detailed below.}
}

%OLD MOTIVATION SECTION ENDS HERE:
}

%%%%%%%%%%%%%%%%%%%%%%%%%%%%%%%%%%%%%%%%%%%

Before detailing our results, we % briefly
note that several of \new{them}
will in fact deal with a weaker variant of the \COND model,
which we now describe.  In designing \COND-model algorithms
it is obviously desirable to have algorithms that only invoke
the $\COND$ oracle on query sets $S$ which are ``simple'' in some sense.
Of course there are many possible notions of simplicity;
in this work we consider the \ignore{number of elements required to describe}size of a set
as a measure of its simplicity, and consider algorithms which \ignore{minimize
that number}only query small sets.  More precisely, we consider the following
restriction of the general $\COND$ model:
\begin{description}
\item[\PCOND oracle:]
We define a \emph{\PCOND} {(short for ``pair-cond'')
\emph{oracle for $\D$} is} a restricted version
of $\COND_\D$ that only accepts input sets $S$ which are either
$S=[N]$ (thus providing the power of a $\SAMP_\D$ oracle)
or $S=\{i,j\}$ for some $i,j \in [N]$, i.e. sets of size two.  The
\PCOND oracle may be viewed as a minimalist variant of \COND
that essentially permits an algorithm to compare the relative weights of
two items under $\D$ (and to draw random samples from $\D$,
by setting $S=[N]$).
\item[\ICOND oracle:]
We define an \emph{\ICOND} {(short for ``interval-cond'')}
\emph{oracle for $\D$} as a restricted version
of $\COND_\D$ that only accepts input sets $S$ which are intervals
$S=[a,b]=\{a,a+1,\dots,b\}$ for some $a \leq b \in [N]$ (note that
taking $a=1,$ $b=N$ this provides the power of a $\SAMP_{\D}$ oracle).
This is a natural restriction on \COND queries in settings where
the $N$ points are endowed with a total order.
\end{description}

\newer{To motivate the \PCOND model (which essentially gives the ability
to compare two elements), one may consider a setting in which a human domain
expert can provide an estimate of the relative likelihood of two distinct 
outcomes in a limited-information prediction scenario.}

\subsection{Our results} \label{sec:ourcont}

%%%%% Table of results -- requires the {multirow} package
  \begin{table}\renewcommand{\arraystretch}{1.5}\centering
    \begin{adjustwidth}{-.1in}{-.1in}\centering
  \begin{tabular}{@{}|l|lc|c|@{}}\hline
    \multicolumn{1}{|c|}{ \bf Problem } & \multicolumn{2}{|c|}{ \bf Our results } &  \bf Standard model \\ \hline
    \multirow{4}{*}{\parbox{0.30\textwidth}{Is $\D$ uniform? }} & $\COND_\D$ & $\bigOmega{\frac{1}{\eps^2}}$ &  \\ \cline{2-3}
    & $\PCOND_\D$ & $\tildeO\!\left(\frac{1}{\eps^2}\right)$ & \\ \cline{2-3}
    & \multirow{2}{*}{$\ICOND_\D$}& $\tildeO\!\left(\frac{\log^3 N}{\eps^3}\right)$ &
    $\bigTheta{\frac{\sqrt{N}}{\eps^2}}$ \cite{GRexp:00,BFRSW:10,Paninski:08}\\
    & & $\bigOmega{\frac{\log N}{\log\log N}}$ &  \\ \hline
    \multirow{3}{*}{\parbox{0.30\textwidth}{Is $\D = \D^\ast$ for a known $\D^\ast$?}} & $\COND_\D$ &
$\tildeO\!\left(\frac{1}{\eps^4}\right)$ & \\ \cline{2-3}
    & \multirow{2}{*}{$\PCOND_\D$} & $\tildeO\!\left(\frac{\log^4 N}{\eps^4}\right)$ & $\Theta\!\left(\frac{\sqrt{N}}{\eps^{2}}\right)$
\cite{BFFKRW:01,Paninski:08,VV:14} \\
    &  & $\bigOmega{\sqrt{\frac{\log N}{\log\log N}}}$ & \\\hline
%    \multirow{2}{*}{\parbox{0.3\textwidth}{ Are $\D_1,\D_2$ (both unknown) equivalent?}} & $\COND_{\D_1,\D_2}$ & $\tildeO\!\left( \frac{\log^{5} N}{\eps^4}\right)$ & $\tildeO\!\left(\frac{N^{2/3}}{\eps^{8/3}}\right)$ \cite{BFRSW:10} \\  \cline{2-3}
%    & $\PCOND_{\D_1,\D_2}$ & $\tildeO\!\left(\frac{\log^6 N}{\eps^{21}}\!\right)$ & $\bigOmega{N^{2/3}}$ \cite{BFRSW:10,Valiant:11} \\ \hline
    \multirow{2}{*}{\parbox{0.30\textwidth}{ Are $\D_1,\D_2$ (both unknown) equivalent?}} & $\COND_{\D_1,\D_2}$ & $\tildeO\!\left( \frac{\log^{5} N}{\eps^4}\right)$ & \multirow{2}{*}
    {\parbox{0.30\textwidth}{\new{ $\Theta\!\left(\max\left(\frac{N^{2/3}}{\eps^{4/3}}, \frac{\sqrt{N}}{\eps^{2}} \right)\right)$ \cite{BFRSW:10, Valiant:11, CDVV:13} } }} \\  \cline{2-3}
    & $\PCOND_{\D_1,\D_2}$ & $\tildeO\!\left(\frac{\log^6 N}{\eps^{21}}\!\right)$ &  \\ \hline
    \multirow{2}{*}{\parbox{0.30\textwidth}{ How far is $\D$ from uniform? }} & \multirow{2}{*}{$\PCOND_\D$} & \multirow{2}{*}{$\tildeO\!\left(\frac{1}{\eps^{20}}\right)$ } & $O\!\left(\frac{1}{\eps^2}\frac{N}{\log N}\right)$ \cite{ValiantValiant:11,ValiantValiant:10ub} \\
    & & & $\bigOmega{ \frac{N}{\log N} }$ \cite{ValiantValiant:11,ValiantValiant:10lb}\\\hline
    %\multirow{2}{*}{\new{Is $\D$ monotone?}} & \multirow{2}{*}{$\PCOND_\D$} & \multirow{2}{*}{$\tildeO\!\left( \frac{\log N}{\eps^3} \right)$} & $\tildeO\!\left(\frac{\sqrt{N}}{\eps^4}\right)$ \cite{BKR:04} \\
    %& & & $\bigOmega{\sqrt{N}}$ \cite{BKR:04}\\\hline
  \end{tabular}\vspace{-0.25\baselineskip}
  \end{adjustwidth}
\caption{\label{table:summary:results} Comparison between the $\COND$ model
and the standard model on a variety of distribution testing problems over
$[N]$.  The upper bounds for the first three problems
are for testing whether the property holds (i.e. $\dtv = 0$)
versus $\dtv \geq \eps$, and for the last problem
the upper bound is for estimating the distance to uniformity to
within an additive $\pm \eps$.\vspace{-0.8\baselineskip}}
  \end{table}
%%%%% End table

We give a detailed study of a range of
natural distribution testing problems in the $\COND$ model and its
variants described above, establishing both upper and lower bounds on
their query complexity.  Our results show that the ability to do
conditional sampling provides a significant amount of power to property
testers, enabling ${\rm polylog}(N)$-query, or even constant-query, algorithms
for % a wide range of
{problems} whose sample complexities in the standard
model are $N^{\Omega(1)}$; see Table~\ref{table:summary:results}.
While we have considered a % broad range
{variety} of distribution testing problems
in the \COND model, our results are certainly not exhaustive, and many
directions remain to be explored; we discuss some of these
in \autoref{sec:conclusion}.

\ifnum\confversion=0
\subsubsection{Testing distributions over unstructured domains}
\else
\subsubsection{Unstructured domains}
\fi
In this \new{early} work on the
$\COND$ model our main focus has been on the simplest (and, we think,
most fundamental) problems in distribution testing, such as testing
whether $\D$ is the uniform distribution $\calU$; testing whether $\D
= \D^\ast$ for an explicitly provided $\D^\ast$;
testing whether $\D_1 = \D_2$
given $\COND_{\D_1}$ and $\COND_{\D_2}$ oracles;
and estimating the variation
distance between $\D$ and the uniform distribution.
In what follows $\dtv$ denotes the variation distance.

%{\bf Our Results.}  (Start with a little explanation/overview:  clearly
%there is a very wide range of possible directions one could go in with
%a new model like this, essentially any distribution testing problem that's
%been explored in the standard model could be revisited here.
%As described below, in our investigations our initial focus has been on
%the simplest and, we think, most fundamental problems in distribution
%testing:  testing uniformity, identity of two distributions, estimating
%$L_1$ distance between distributions, and the like.  Many of our algorithms
%use only PCOND queries since these are arguably the simplest type of
%COND queries to consider.)

\medskip\noindent
{\bf Testing uniformity.}  We give a $\PCOND_{\D}$ algorithm
that tests whether
{$\D = \calU$} % \mbox{$\dtv(\D,\calU)=0$}
versus  \mbox{$\dtv(\D,\calU)
\geq \eps$} using $\tilde{O}(1/\eps^2)$ calls to $\PCOND_{\D}$,
independent of $N$.  We show that this $\PCOND_{\D}$ algorithm
is nearly optimal by proving that any $\COND_\D$ tester (which may use
arbitrary subsets $S \subseteq [N]$ as its query sets) requires
$\Omega(1/\eps^2)$ queries for this testing problem.

\medskip\noindent
{\bf Testing equivalence to a known distribution.}
As described above, for the simple problem of testing uniformity
we have an essentially optimal $\PCOND$ testing algorithm
and a matching lower bound.  
% Given these results it is natural
% to turn to the more challenging question 
\newer{A more general and challenging problem is that}
of testing whether
$\D$ (accessible via a $\PCOND$ or $\COND$ oracle)
is equivalent to $\D^\ast$, where $\D^\ast$ is an arbitrary
``known'' distribution over $[N]$ that is explicitly
provided to the testing algorithm at no cost (say as a vector
$(\D^\ast(1),\dots,\D^\ast(N))$ of probability values).
For this ``known $\D^\ast$'' problem,
we give a $\PCOND_\D$ algorithm testing whether
% \mbox{$\dtv(\D,\D^\ast)=0$}
{\mbox{$\D = \D^\ast$}}
versus $\dtv(\D,\D^\ast) \geq \eps$ using
$\tilde{O}((\log N)^4/\eps^4)$ queries.   We
further show that the $(\log N)^{\Omega(1)}$ query complexity
of our $\PCOND_\D$ algorithm is inherent in the problem, by proving
that any $\PCOND_\D$ algorithm for this problem must use
$\sqrt{\log(N)/\log\log(N)}$ queries for constant $\eps$.

Given these $(\log N)^{\Theta(1)}$
upper and lower bounds on the query complexity of
$\PCOND_\D$-testing equivalence to a known distribution, it is natural to
ask whether the full $\COND_\D$ oracle provides more power for this problem.
We show that this is indeed the case, by giving a
$\tilde{O}(1/\eps^4)$-query algorithm
(independent of $N$) that uses unrestricted $\COND_\D$ queries.

\medskip\noindent
{\bf Testing equivalence between two unknown distributions.}
We next consider the more challenging problem of testing whether two unknown
distributions $\D_1,\D_2$ over $[N]$ (available via $\COND_{\D_1}$
and $\COND_{\D_2}$ oracles) are identical versus $\eps$-far.
We give two very different algorithms for this problem.  The first uses
$\PCOND$ oracles and has query complexity
$\tilde{O}((\log N)^6/\eps^{21})$, while the second
uses $\COND$ oracles and has query complexity
{$\tilde{O}((\log N)^5/\eps^4)$.}
We % feel
{believe} that the proof technique of the second algorithm is of independent
interest, since it shows how a $\COND_\D$ oracle can efficiently simulate
an ``approximate $\EVAL_\D$ oracle.''  (An $\EVAL_\D$ oracle takes as input a
point $i \in [N]$ and outputs the probability mass $\D(i)$ that $\D$
puts on $i$; {we briefly explain our notion of approximating
such an oracle in \autoref{subsubsec:algs-desc}}.)

\medskip\noindent
{\bf Estimating the distance to uniformity.}  We also consider the problem of
estimating % $\dtv(\D,\calU)$, the total
{the variation distance} between $\D$ and
the uniform distribution $\calU$ over $[N]$, to within an additive error of
$\pm \eps.$  In the standard $\SAMP_\D$ model this is known to be a very
difficult problem, with an $\Omega(N/\log N)$ lower bound established in
{\cite{ValiantValiant:11,ValiantValiant:10lb}}.  In contrast, we give a
$\PCOND_\D$ algorithm that makes only $\tilde{O}(1/\eps^{20})$
queries, independent of $N$.

\ifnum\confversion=0
\subsubsection{Testing distributions over structured domains}
\else
\subsubsection{Structured domains}
\fi

In the final portion of the paper we view the domain $[N]$ as an ordered
set $1 \leq \cdots \leq N$.  (Note that in all the testing problems and results
described previously, the domain could just as well have been viewed
as an unstructured set of abstract points $x_1,\dots,x_N$.)  With this
perspective it is natural to consider 
\newer{an additional oracle.
We define an \emph{\ICOND} {(short for ``interval-cond'')}
\emph{oracle for $\D$} as a restricted version
of $\COND_\D$, which only accepts input sets $S$ that are intervals
$S=[a,b]=\{a,a+1,\dots,b\}$ for some $a \leq b \in [N]$ (note that
taking $a=1,$ $b=N$ this provides the power of a $\SAMP_{\D}$ oracle).
}

We give an $\tilde{O}((\log N)^3/\eps^3)$-query $\ICOND_\D$ algorithm
for testing whether $\D$ is uniform versus $\eps$-far from uniform.
We show that a $(\log N)^{\Omega(1)}$ query complexity is inherent for
uniformity testing using $\ICOND_\D$, by proving
an $\Omega\left(\log N/\log \log N\right)$-query
$\ICOND_\D$ lower bound.

\ignore{
In summary, our results show that the ability %to do
\newer{to perform}
conditional sampling provides a significant amount of power to property
testers, enabling ${\rm polylog}(N)$-query, or even constant-query, algorithms
for these problems, which have sample complexities $N^{\Omega(1)}$ in the standard
model; see Table~\ref{table:summary:results}.}
\new{
Along the way to establishing \newer{our main} testing results
\newer{described above}, we develop several
powerful tools for analyzing distributions in the $\COND$ and $\PCOND$
models, which we believe may be of independent interest and utility
in subsequent work on the $\COND$ and $\PCOND$ models.
  These include as mentioned above a procedure for approximately simulating an ``evaluation
oracle'', as well as a procedure for estimating the weight of the ``neighborhood''
of a given point in the domain of the distribution. (See further discussion of
these tools \newer{in \autoref{subsubsec:algs-desc}}.)
}

\subsubsection{A high-level discussion of our algorithms}\label{subsubsec:algs-desc}
To maintain focus here we describe only the ideas behind our
algorithms; intuition for each of our lower bounds can be found
in an informal discussion preceding the formal
proof, \ifnum\confversion=1
see the beginnings of Sections~{4.2}, {5.2}, and~{8.2} of the full version.
\else
see the beginnings of Sections~\ref{ssec:unif-lb}, \ref{ssec:lb-Pcond-Dstar}, and~\ref{ssec:unif-lb:intcond}.
\fi
As can be seen in the following discussion, our algorithms
share some common themes, though each has its own unique idea/technique,
which we emphasize below.

Our simplest testing algorithm is the algorithm for {\bf testing whether
$\D$ is uniform} over $[N]$ (using $\PCOND_\D$ queries). The algorithm is based on
the observation that if a distribution is $\eps$-far from uniform, then
the total weight (according to $\D$) of points $y\in [N]$ for which
$\D(y)\geq (1+\Omega(\eps))/N$
is $\Omega(\eps)$, and the
fraction of points $x\in [N]$ for which $\D(x) \leq (1-\Omega(\eps))/N$
is $\Omega(\eps)$. If we obtain such a pair of points $(x,y)$,
then we can detect this deviation from uniformity by performing
$\Theta(1/\eps^2)$  $\PCOND_\D$ queries
on the pair. Such a pair can be obtained with high probability by making
$\Theta(1/\eps)$ $\SAMP_\D$ queries (so as to obtain $y$)
as well as selecting $\Theta(1/\eps)$ points uniformly (so as to obtain $x$).
This approach yields an algorithm whose complexity grows like $1/\eps^4$.
To actually get an algorithm with query complexity $\tilde{O}(1/\eps^2)$
(which, as our lower bound shows, is tight),
a slightly more refined approach is applied.

When we take the next step to {\bf testing equality  to an arbitrary (but
fully specified) distribution $\D^\ast$}, the abovementioned observation
generalizes so as to imply that if we sample $\Theta(1/\eps)$ points
from $\D$ and $\Theta(1/\eps)$ from $\D^\ast$, then with high probability
we shall obtain a pair of points $(x,y)$ such that  $\D(x)/\D(y)$
differs by at least $(1\pm \Omega(\eps))$ from $\D^\ast(x)/\D^\ast(y)$.
Unfortunately, this cannot necessarily be detected by a small number
of $\PCOND_\D$ queries since (as opposed to the uniform case), $\D^\ast(x)/\D^\ast(y)$
may be very large or very small. However, we show that by sampling from both
$\D$ and $\D^\ast$ and allowing the number of samples to grow with $\log N$, with
high probability we either obtain a pair of points as described above
for which  $\D^\ast(x)/\D^\ast(y)$ is a constant, or we detect that for
some set of points $B$ we have that  $|\D(B) - \D^\ast(B)|$ is relatively
large.\footnote{Here we use $B$ for ``Bucket'', as we consider a bucketing of the points in
$[N]$ based on their weight according to $\D^\ast$. We note that bucketing has
been used extensively in the context of testing properties of distributions,
see e.g.~\cite{BFRSW:10,BFFKRW:01}.}
% \dnote{should we say something about bucketing and that it was used in the
% past for testing distributions?}

As noted previously, {we prove a lower bound showing that}
a polynomial dependence on $\log N$ is unavoidable if only
$\PCOND_\D$ queries (in addition to standard sampling) are allowed. To obtain
our more efficient $\poly(1/\eps)$-queries algorithm, which uses more general
$\COND_\D$ queries, we extend the observation from
the uniform case in a different
way. Specifically, rather than comparing the relative weight of
pairs of points, we compare the relative weight of pairs in which
{one element is a point and the other is a subset of points.}
Roughly speaking, we show how points can be paired with subsets of points of
comparable weight (according to $\D^\ast$)
 such that the following holds. If $\D$ is far from $\D^\ast$, then
by taking $\tilde{O}(1/\eps)$ samples from $\D$ and selecting subsets of
points in an appropriate manner
(depending on  $\D^\ast$), we can obtain (with high probability) a point $x$
and a subset $Y$ such that $\D(x)/\D(Y)$ differs significantly from
$\D^\ast(x)/\D^\ast(Y)$ \emph{and} $\D^\ast(x)/\D^\ast(Y)$ is a constant.

In our next step, to {\bf testing equality between two unknown distributions
$\D_1$ and $\D_2$},
we need to cope with the fact that we no longer ``have a hold''
on a known distribution. Our $\PCOND$
algorithm can be viewed as creating
such a hold in the following sense. By sampling from $\D_1$ we obtain
(with high probability) a (relatively small) set of points $R$ that
{\em cover\/} the distribution $\D_1$. By ``covering'' we mean that
{except for a subset having small weight according to $\D_1$},
all points $y$ in $[N]$ have a {\em representative\/} $r \in R$,
{i.e. a point $r$}
such that $\D_1(y)$ is close to $\D_1(r)$.
We then show that if $\D_2$ is far from $\D_1$, then one of the following
must hold: (1) There is relatively large weight, either according to
$\D_1$ or according to $\D_2$, on points $y$ such that for
some $r\in R$ we have that $\D_1(y)$ is close to $\D_1(r)$ but $\D_2(y)$
is not sufficiently close to $\D_2(r)$; (2) There exists a point $r\in R$
such that {the} set of points $y$ for which $\D_1(y)$ is close to $\D_1(r)$ has
significantly different  weight according to $\D_2$ as compared to $\D_1$.
  % \dnote{I wrote and then deleted:
  % A central subroutine employed by our algorithm is {\sc Estimate-Neighborhood},
  % which, given a point $x$ and \PCOND access to a distribution $\D$ returns
  % an estimate of the weight of a subset of points whose probability (according
  % to $D$) is similar to that of $x$. The difficulty with
  % performing this task
  % is due to points whose probability is close to the ``similarity threshold''
  % that determines the neighborhood set. }
  % \rnote{I can see where you were coming from writing it, but probably it
  % is too much to get across at this point -- I agree with not including it.}
We note that this algorithm can be viewed as a {variant}
of the \PCOND algorithm
for the case when one of the distributions is known (where the ``buckets'' $B$,
which were defined by $\D^\ast$ in that algorithm (and were disjoint),
are now defined by the points in $R$ (and are not necessarily disjoint)).

As noted previously, our (general) \COND algorithm for testing the equality of
two (unknown) distributions is based on a subroutine that estimates $\D(x)$
(to within $(1\pm O(\eps))$)
for a given point $x$ given access to $\COND_\D$. Obtaining such an
estimate for {\em every\/} $x\in [N]$ cannot be done efficiently for
some distributions.\footnote{As an extreme case consider a distribution $\D$
for which $\D(1) = 1-\phi$ and $\D(2)=\dots=\D(N) = \phi/(N-1)$ for
some very small $\phi$ (which in particular may depend on $N$), and
for which we are interested in estimating $\D(2)$. This requires $\Omega(1/\phi)$
queries.} However, we show that
if we allow the algorithm to output \unknown on some subset of points
with % small total weight,
{total weight $O(\eps)$},
then the relaxed task can be performed using
$\poly(\log N,1/\eps)$ queries, by performing a kind of randomized
binary search
``with exceptions''. This relaxed version, which we refer to as
an {\em approximate \EVAL oracle\/},  suffices for our needs in
distinguishing between the case that $\D_1$ and $\D_2$ are the same
distribution and the case in which they are far from each other.
It is possible that {this procedure} will be useful for other tasks as well.

The algorithm for {\bf estimating the distance to uniformity} (which
uses $\poly(1/\eps)$ $\PCOND_\D$ queries) is based on a subroutine
for finding a {\em reference point\/} $x$ together with an estimate
$\widehat{D}(x)$ of $\D(x)$. A reference point should be such that
$\D(x)$ is relatively close to $1/N$ (if such a point cannot
be found then it is evidence that $\D$ is very far from uniform). Given a reference point $x$
(together with $\widehat{D}(x)$) it is possible to estimate the distance to
uniformity by obtaining (using \PCOND queries)
estimates of the ratio between $\D(x)$ and
$\D(y)$ for $\poly(1/\eps)$ uniformly selected points $y$.
The procedure for finding a reference point $x$ together with $\widehat{D}(x)$ is based
on estimating both the weight and the size of a subset of points $y$ such that
$\D(y)$ is close to $\D(x)$. The procedure shares
a common subroutine, {\sc Estimate-Neighborhood}, with the $\PCOND$ algorithm for testing
equivalence between two unknown distributions.

Finally, the $\ICOND_\D$
algorithm for testing uniformity is based on a version
of the approximate \EVAL oracle mentioned previously,
which on one hand uses only
$\ICOND_\D$ (rather than general $\COND_\D$)
queries, and on the other hand exploits the
fact that we are dealing with the uniform distribution
rather than an arbitrary distribution.

% % \subsection{The independent work~\texorpdfstring{\cite{CFGM:13}}{of Chakraborty et al.}}
% % \label{subsec:CFGM}
% % In what follows we % briefly
% % discuss the work of Chakraborty et al.~\cite{CFGM:13}, which was
% % done independently from our work, and
% % was recently accepted to the ITCS conference
% % (so that we learned about its existence only a few days
% % before the STOC submission deadline).
% % % the discussion that follows might not be complete
% % Chakraborty et al.~\cite{CFGM:13}
% % propose essentially the same \COND model that we
% % propose, differing only in what happens on query sets $S$ such that $D(S)=0$. In
% % our model such a query causes the \COND oracle and algorithm to return \fail,
% % while in their model such a query returns a uniform random $i \in S.$
% % They present the following results.

\subsection{The work of Chakraborty et al.~\cite{CFGM:13}}
\label{subsec:CFGM}
\new{
Chakraborty et al.~\cite{CFGM:13}
proposed essentially the same \COND model that we
study, differing only in what happens on query sets $S$ such that $D(S)=0$. In
our model such a query causes the \COND oracle and algorithm to return \fail,
while in their model such a query returns a uniform random $i \in S.$

Related to testing equality of distributions, \cite{CFGM:13} provides
an (adaptive) algorithm for testing whether $\D$ is equivalent to a specified
distribution $\D^*$ using $\poly(\log^\ast N,1/\eps)$ \COND queries. Recall that
we give an algorithm for this problem that performs $\tilde{O}(1/\eps^4)$
\COND queries.
\cite{CFGM:13} also gives a \emph{non-adaptive}
algorithm for this problem that performs $\poly(\log N,1/\eps)$ \COND queries.\footnote{We
note that it is only possible for them to give a non-adaptive
algorithm because their model is more
permissive than ours (\ignore{that }if a query set $S$ is proposed for which $\D(S)=0$,
their model returns a uniform random element of $S$ while our model
returns \fail).  In our stricter model, any non-adaptive algorithm
which queries a proper subset $S \subsetneq N$ would output $\fail$
on some distribution $\D$.}
Testing equivalence between two unknown distributions is not considered in~\cite{CFGM:13},
and the same is true for testing in the \PCOND model.
}

\cite{CFGM:13} also presents additional results for a range of other problems, \new{which we discuss below:}

\BI
\item An (adaptive) algorithm for testing uniformity that performs $\poly(1/\eps)$
queries.\footnote{The precise polynomial is not specified~--~we believe it is roughly $1/\eps^4$
as it follows from an application of the identity tester of~\cite{BFFKRW:01}
 with distance $\Theta(\eps^2)$ on a domain of size $O(1/\eps)$.} The sets on which the algorithms performs \COND queries are of size linear in $1/\eps$. Recall that our algorithm for this problem performs $\tilde{O}(1/\eps^2)$
\PCOND queries and that we show that every algorithm must perform $\Omega(1/\eps^2)$
queries (when there is no restriction on the types of queries). We note that their
analysis uses the same observation that ours does regarding distributions that
are far from uniform (see the discussion in \autoref{subsubsec:algs-desc}),
but exploits it in a different manner.

They also give a non-adaptive algorithm for this problem that performs
$\poly(\log N,1/\eps)$ \COND queries and show that $\Omega(\log\log N)$ is
a lower bound on the necessary number of queries for
 non-adaptive algorithms.
% We note that their non-adaptive algorithm does not succeed in our stricter
% \COND model, since it relies crucially on the fact that if $D(S)=0$ the oracle
% returns a uniform element $i \in S$ rather than FAIL.
  %   \dnote{
  %   I thought about it some more and I think we shouldn't add that their non-adaptive
  %   algorithm does not succeed in our more strict model unless we can really explain why
  %   our choice is much more ``right'' then theirs. I think this comment might not be
  %   read well, and on the other hand, I am not really worried about people asking why
  %   we didn't consider non-adaptive algorithms. Our focus was on getting the best
  %   algorithms we can, so we used adaptivity. Their study of non-adaptive algorithms is
  %   fine, but it just was't our focus. I think it is problematic to justify the
  %   fact that we didn't go in that direction by saying (implicitly maybe) that non-adaptive algorithms
  %   can't work in our model, since the blame can be put on the model.
  %   }

\item An (adaptive) algorithm for testing whether $\D$ is equivalent to a specified
distribution $\D^*$ using $\poly(\log^\ast N,1/\eps)$ \COND queries. Recall that
we give an algorithm for this problem that performs $\tilde{O}(1/\eps^4)$
\COND queries.

They also give a non-adaptive algorithm for this problem that performs
$\poly(\log N,1/\eps)$ \COND queries.
% As above their non-adaptive algorithm does not succeed in our stricter \COND
% model, for the same reason as above.

\item An (adaptive) algorithm for testing any label-invariant (i.e., invariant
under permutations of the domain) property
that performs $\poly(\log N,1/\eps)$ \COND queries. As noted in~\cite{CFGM:13},
this in particular implies an algorithm with this complexity for estimating the distance to
uniformity. Recall that we give an algorithm for this estimation problem that
performs $\poly(1/\eps)$ \PCOND queries.

The algorithm for testing any label-invariant property is based on learning a
certain approximation of the distribution $\D$ and in this process defining
{some sort of} approximate \EVAL oracle. To the best of our understanding,
our notion of an approximate \EVAL oracle (which is used to obtain one
or our results for testing equivalence between two unknown distributions)
is quite different.

\sloppy
They also show that there exists a label-invariant property
for which any adaptive algorithm must perform $\Omega(\sqrt{\log\log N})$
\COND queries.

\item Finally they show that there exist general properties that require $\Omega(N)$ \COND queries.
\EI

\section{Preliminaries} \label{sec:prelim}

\ifnum\confversion=0
\subsection{Definitions}
\fi

Throughout the paper we shall work with discrete distributions over
an $N$-element set whose elements are denoted $\{1,\dots,N\}$; we write $[N]$
to denote $\{1,\dots,N\}$ and $[a,b]$ to denote $\{a,\dots,b\}.$  For a
distribution $\D$ over $[N]$ we write $\D(i)$ to denote the
probability of $i$ under $\D$, and for $S \subseteq [N]$
we write $\D(S)$ to denote $\sum_{i \in S} \D(i).$
For $S \subseteq [N]$ such that
$\D(S)>0$ we write $\D_S$ to denote the conditional distribution
of $\D$ restricted to $S$, so $\D_S(i) = {\frac {\D(i)}{\D(S)}}$
for $i \in S$ and $\D_S(i)=0$ for $i \notin S.$

As is standard in property testing of distributions, throughout this work we
measure the distance between two distributions $\D_1$ and $\D_2$
using  the {\em total variation distance}:
\ifnum\confversion=1 %%%%%%%%%%%%%%%%%%%%%%%%%%%%%%%%%%%%%%%%%% Conference version
\[
\dtv\left(\D_1,\D_2 \right) \eqdef {\frac 1 2} {\|\D_1 - \D_2\|}_1 = {\frac 1 2} \sum_{i \in
[N]}{|\D_1(i)-\D_2(i)|}.% = \max_{S \subseteq [N]} |\D_1(S)-\D_2(S)|.
\]
{
We use standard definitions from property testing, Chernoff bounds, etc.;
see the full version for details.
\ignore{
We consider the usual definition of property testing, and often resort during our analysis to the standard Chernoff bounds, both additive and multiplicative, and their corollaries (see e.g. Chapter~4 of \cite{MotwaniRaghavan:95} and Exercise~1.1 of \cite{DP09}).
}
}

\else                 %%%%%%%%%%%%%%%%%%%%%%%%%%%%%%%%%%%%%%%%%% Full version
\[
\dtv\left(\D_1,\D_2 \right) \eqdef {\frac 1 2} {\|\D_1 - \D_2\|}_1 = {\frac 1 2} \sum_{i \in
[N]}{|\D_1(i)-\D_2(i)|} = \max_{S \subseteq [N]} |\D_1(S)-\D_2(S)|.
\]

We may view a \emph{property} $\calP$
of distributions over $[N]$ as a subset of all distributions over $[N]$
(consisting of all distributions that have the property).
The distance from $\D$ to a property $\calP$, denoted $\dtv(\D,\calP),$
is defined as $\inf_{\D' \in \calP}\{\dtv(\D,\D')\}.$

We define testing algorithms for properties of distributions over $[N]$ as follows:

\BD \label{def:testing}
Let $\calP$ be a property of distributions over $[N]$.  Let $\ORACLE_\D$
be some type of oracle which provides access to $\D.$
A \emph{$q(\eps,N)$-query $\ORACLE$ testing algorithm for $\calP$}
is an algorithm $T$ which is given $\eps,N$ as input
parameters and oracle access to an $\ORACLE_\D$ oracle.  For any
distribution $\D$ over $[N]$ algorithm
$T$ makes at most $q(\eps,N)$ calls to $\ORACLE_\D$, and:

\begin{itemize}\fullOrConfCompactEnums\fullOrConfIndentItems

\item if $\D \in \calP$ then with probability at least $2/3$ algorithm $T$ outputs $\accept$;

\item if $\dtv(\D,\calP) \geq \eps$ then with probability at least $2/3$ algorithm $T$ outputs $\reject.$

\end{itemize}

\ED

This definition can easily be extended to cover situations in which there are
two ``unknown'' distributions $\D_1,\D_2$ that are accessible via $\ORACLE_{\D_1}$ and $\ORACLE_{\D_2}$ oracles.  In particular we shall consider algorithms for testing whether $\D_1 = \D_2$ versus  $\dtv(\D_1,\D_2)$ in such a setting.
We sometimes write $T^{\ORACLE_\D}$ to indicate that $T$ has access to $\ORACLE_\D$.
%\rnote{This OK?  I thought maybe it's better not to talk extensively about $m$-tuples sine we only ever mention pairs of distributions.}
% \ignore{
% \new{One can generalize this definition to testing algorithms with oracle access to more than one distribution, and testing property over them: namely, if $\calP$ is a property of $m$-tuples of distributions over $[N]$ (e.g, equality), a  \emph{$q(\eps,N)$-query $\ORACLE$ testing algorithm for $\calP$} is an algorithm $T$ which is given $\eps, N$ as input parameters and oracle access to $\ORACLE_{\D_1}$, \dots, $\ORACLE_{\D_m}$ oracles, has correctness as above, and for any $m$-tuple of distributions $\D_1,\dots,\D_m$ makes at most a total of $q(\eps,N)$ calls to the oracles.}
% }

\subsection{Useful tools}

On several occasions we will use the \emph{data processing inequality for variation
distance}.  This fundamental result says that for any two distributions $\D$, $\D'$,
applying any (possibly randomized) function to $\D$ and $\D'$ can never increase their
statistical distance; see e.g. part~(iv) of Lemma~2 of \cite{Rey:11} for a proof of this
lemma.

\BL [Data Processing Inequality for Total Variation Distance]\label{lemma:data:processing:inequality:total:variation}
Let $\D,$ $\D'$ be two distributions over a domain $\Omega$. Fix any randomized function\footnote{Which can be seen as a distribution over functions over $\Omega$.} $F$ on $\Omega$, and let $F(\D)$ be the distribution such that a draw from $F(\D)$ is obtained by drawing independently $x$ from $\D$ and $f$ from $F$ and then outputting $f(x)$ (likewise for $F(\D^\prime)$).
Then we have
\[
\dtv( F(\D),  F(\D'))  \leq \dtv( \D,\D'). \]
%Moreover, we have equality if each realization $f$ of $F$ is one-to-one.
\EL

\ignore{
\BL [Data Processing Inequality for Total Variation Distance]
\label{lemma:data:processing:inequality:total:variation}
Let $\D,$ $\D'$ be two distributions over a domain $\Omega$.
Fix any randomized function $F$ on $\Omega$,
and let $F(\D)$ be the distribution such that a draw from $F(\D)$
is obtained by taking
a draw from $\D$ and applying $F$ (and likewise for $F(\D')$).
Then we have
%such that $F$ is independent of $X$ and $Y$,
\[
\dtv( F(\D),  F(\D'))  \leq \dtv( \D,\D'). \]
%Moreover, we have equality if each realization $f$ of $F$ is one-to-one.
\mnote{this formulation OK?}
\EL
}

% \rnote{Other tools we should put in this section -- do we want explicit
% statements of the additive and/or multiplicative Chernoff bound, or
% anything else?}

We next give several variants of Chernoff bounds (see
e.g. Chapter~4 of \cite{MotwaniRaghavan:95}).
\BT \label{thm:multCB}
Let $Y_1,\dots,Y_m$ be $m$ independent random variables that take on values
in $[0,1]$, where $\Exp[Y_i] = p_i$, and
$\sum_{i=1}^m p_i = P$. For any $\gamma \in (0,1]$ we have
\begin{align}
\label{eq:additive-chernoff}
\text{(additive bound)}& &
\Pr\left[\sum_{i=1}^m Y_i   > P+ \gamma m \right],
\
\Pr\left[\sum_{i=1}^m Y_i  < P - \gamma m\right]
 &\leq \exp(-2 \gamma^2 m)\\
\label{eq:cher-ub}
\text{(multiplicative bound)}&&
\Pr\left[\sum_{i=1}^m Y_i > (1+\gamma)P\right] &< \exp(-\gamma^2 P/3)\\
\text{and}\notag\\
\label{eq:cher-lb}
\text{(multiplicative bound)}&&
\Pr\left[\sum_{i=1}^m Y_i < (1-\gamma)P\right] &< \exp(-\gamma^2 P/2).
\end{align}
The bound in Equation~(\ref{eq:cher-ub}) is derived from the following
more general bound, which holds from any $\gamma > 0$:
\BEQ\label{eq:cher-ub-gen}
\Pr\left[\sum_{i=1}^m Y_i > (1+\gamma)P\right] \leq
     \left(\frac{e^\gamma}{(1+\gamma)^{1+\gamma}}\right)^{P}\;,
\EEQ
and which also implies that for any $B > 2eP$,
\BEQ\label{eq:cher-ub-large}
\Pr\left[\sum_{i=1}^m Y_i > B\right] \leq 2^{-B}\;.
\EEQ
\ET

The following extension of the multiplicative bound is useful when we
only have upper and/or lower bounds on $P$ (see Exercise~1.1 of
\cite{DP09}):
\BC \label{cor:CB-upperlower}
In the setting of \autoref{thm:multCB} suppose that
$P_L \leq P \leq P_H.$ Then for any $\gamma \in (0,1]$, we have
\begin{eqnarray}
\Pr\left[\sum_{i=1}^m Y_i > (1+\gamma)P_H\right] &<& \exp(-\gamma^2 P_H/3)
\label{eq:multCB-upper2}\\
\Pr\left[\sum_{i=1}^m Y_i < (1-\gamma)P_L\right] &<& \exp(-\gamma^2 P_L/2)
\label{eq:multCB-lower}
\end{eqnarray}
\EC
%}

We will also use the following corollary of \autoref{thm:multCB}:
\BC\label{cor:sum-wiXi}
Let $0 \leq w_1,\dots,w_m \in \R$ be such that
$w_i \leq \kappa$ for all $i \in [m]$ where $\kappa \in (0,1]$.
Let $X_1,\dots,X_m$ be i.i.d. Bernoulli random variables
with $\Pr[X_i=1]=1/2$ for all $i$,
and let \mbox{$X = \sum_{i=1}^m w_i X_i$} and
$W = \sum_{i=1}^m w_i$.
For any $\gamma \in (0,1]$,
\[
\Pr\left[X > (1+\gamma)\frac{W}{2}\right]
    < \exp\left(-\gamma^2\frac{W}{6\kappa}\right)
    \;\mbox{ and }\;
\Pr\left[X < (1-\gamma)\frac{W}{2}\right]
    < \exp\left(-\gamma^2\frac{W}{4\kappa}\right)\;,
\]
%\dnew{
and for any $B > e\cdot W$,
\[
\Pr[X > B] < 2^{-B/\kappa}\;.
\]
%}
\EC

\BPF
Let $w_i' = w_i/\kappa$ (so that $w_i' \in [0,1]$), let
$W' = \sum_{i=1}^m w'_i = W/\kappa$, and
for each $i \in [m]$ let $Y_i = w_i' X_i$, so that
$Y_i$ takes on values in $[0,1]$ and $\Exp[Y_i] = w'_i/2$.
Let $X' = \sum_{i=1}^m w'_i X_i = \sum_{i=1}^m Y_i$,
so that $\Exp[X'] = W'/2$.
By the definitions of $W'$ and $X'$ and by Equation~(\ref{eq:cher-ub}),
for any $\gamma \in (0,1]$,
\BEQ
\Pr\left[X > (1+\gamma)\frac{W}{2}\right]
 = \Pr\left[X' > (1+\gamma)\frac{W'}{2}\right]
    < \exp\left(-\gamma^2\frac{W'}{6}\right)
    = \exp\left(-\gamma^2\frac{W}{6\kappa}\right),
\EEQ
and similarly by Equation~(\ref{eq:cher-lb})
\BEQ
\Pr\left[X < (1-\gamma)\frac{W}{2}\right]
    < \exp\left(-\gamma^2\frac{W}{{4}\kappa}\right)\;.
\EEQ
%\dnew{
For $B > e\cdot W = 2e\cdot W/2$
we apply Equation~(\ref{eq:cher-ub-large}) and get
\BEQ
\Pr\left[X > B\right]
 = \Pr\left[X' > B/\kappa\right]
    < 2^{-B/\kappa},
\EEQ
as claimed.%}
\EPF

\fi %%%%%%%%%%%%%%%%%%%%%%%%%%%%%%%%% End conference version

\section{Some useful procedures}
In this section we describe some procedures that will be used by our algorithms. On a first pass the reader may wish to focus on the explanatory prose and performance guarantees of these procedures (i.e., the statements of \autoref{lem:compare} and \autoref{lem:est-n}, as well as \autoref{def:approx-eval} and \autoref{thm:approx-eval-simulator}) and otherwise skip to p.\pageref{sec:uniform}; the internal details of the proofs are not necessary for the subsequent sections that use these procedures.

\subsection{The procedure {\sc Compare}}\label{subsec:compare}
We start by describing a procedure that estimates the ratio between the
weights of two disjoint sets of points by performing \COND queries on
the union of the sets.
More precisely, it estimates the ratio (to within $1\pm\eta$) if the ratio is not
too high and not too low. Otherwise, it may output {\sf high} or {\sf low}, accordingly.
In the special case
when each set is of size one, the queries performed
are \PCOND queries.

\begin{algorithm}[ht!]
\caption{{\sc Compare}}\label{alg:compare}
\ifnum\confversion=1  %%% Conf
  \textbf{Input:} {
\else                 %%% Full
  \SetKwInOut{Input}{Input}
  \Input{
\fi                   %%% End
  \COND query access to a distribution $\D$ over $[N]$,
disjoint subsets $X,Y \subset [N]$, parameters
$\eta\in (0,1]$, $K \geq 1$, and $\delta\in (0,1/2]$.}
\BE\fullOrConfCompactEnums
\item\label{alg:compare-st1}
 Perform $\Theta\!\left(\frac{K\log(1/\delta)}{\eta^2}\right)$
 $\COND_\D$ queries
on the set \mbox{$S = X\cup Y$,} and let $\hat{\mu}$ be the fraction of
times that a point $y \in Y$ is returned.
\item If $\hat{\mu} < \frac{2}{3}\cdot \frac{1}{K+1}$, then return {\sf Low}.
\item Else, if $1-\hat{\mu} < \frac{2}{3}\cdot \frac{1}{K+1}$, then return {\sf High}.
\item Else return $\rho = \frac{\hat{\mu}}{1-\hat{\mu}}$.
\EE
\end{algorithm}

\BL\label{lem:compare}
Given as input
two disjoint subsets of points $X,Y$ together with parameters
$\eta \in (0,1]$, $K \geq 1$, and $\delta\in (0,1/2]$,
as well as \COND query access to a distribution $\D$,
the procedure
{\sc Compare} (Algorithm~\ref{alg:compare})
either outputs a value $\rho > 0$ or
outputs {\sf High} or {\sf Low}, and satisfies the following:
\BE\fullOrConfCompactEnums\fullOrConfIndentItems

% \new{
\item\label{compare:mid}
If $\D(X)/K \leq \D(Y) \leq K \cdot \D(X)$ then with probability
at least $1-\delta$
the procedure outputs a value $\rho \in [1-\eta,1+\eta]\D(Y)/\D(X)$;

\item\label{compare:high}
If $\D(Y) > K \cdot \D(X)$ then with probability at least
$1-\delta$ the procedure outputs
either \high or a value $\rho \in [1-\eta,1+\eta]\D(Y)/\D(X)$;

\item\label{compare:low}
If $\D(Y) < \D(X)/K$ then with probability at least $1-\delta$ the procedure
outputs either \low or a value $\rho \in [1-\eta,1+\eta]\D(Y)/\D(X).$

\EE
The procedure performs $O\!\left(\frac{K\log(1/\delta)}{\eta^2}\right)$
\COND queries on the set $X\cup Y$.
\EL

\BPF
The bound on the number of queries performed by the
algorithm follows directly from the description of the algorithm,
and hence we turn to establish its correctness.

Let $w(X) = \frac{\D(X)}{\D(X) +\D(Y)}$ and let
$w(Y) = \frac{\D(Y)}{\D(X) +\D(Y)}$.
Observe that $\frac{w(Y)}{w(X)} = \frac{D(Y)}{D(X)}$
and that for $\hat{\mu}$ as defined in Line~\ref{alg:compare-st1} of the
algorithm, $\Exp[\hat{\mu}]=w(Y)$ and $\Exp[1-\hat{\mu}] = w(X)$.
Also observe that for any $B \geq 1$,
if $\D(Y) \geq \D(X)/B$,
then $w(Y) \geq \frac{1}{B+1}$ and
if $\D(Y) \leq B\cdot\D(X)$, then
$w(X) \geq \frac{1}{B+1}$.

Let $E_1$ be the event that $\hat{\mu} \in [1-\eta/3,1+\eta/3]w(Y)$
and let $E_2$ be the event that
$(1-\hat{\mu}) \in [1-\eta/3,1+\eta/3]w(X)$.
Given the number of \COND queries
performed on the set $X\cup Y$,
by applying a multiplicative Chernoff bound (see \autoref{thm:multCB}), if
$w(Y) \geq \frac{1}{4K}$ then with probability
at least $1-\delta/2$ the event $E_1$ holds, and if
$w(X) \geq \frac{1}{4K}$, then with probability
at least $1-\delta/2$ the event $E_2$ holds.
We next consider the three cases in the lemma statement.
\BE
\item If $\D(X)/K \leq \D(Y) \leq K \D(X)$, then
by the discussion above, $w(Y) \geq \frac{1}{K+1}$,
$w(X) \geq \frac{1}{K+1}$, and
with probability at least $1-\delta$ we have
that $\hat{\mu} \in [1-\eta/3,1+\eta/3]w(Y)$ and
$(1-\hat{\mu}) \in [1-\eta/3,1+\eta/3]w(X)$.
Conditioned on these bounds holding,
$$\hat{\mu} \geq \frac{1-\eta/3}{K+1} \geq \frac{2}{3}\cdot \frac{1}{K+1}\;
\mbox{ and } \;1-\hat{\mu} \geq \frac{2}{3}\cdot \frac{1}{K+1}\;.$$
It
follows that the procedure outputs a value $\rho = \frac{\hat{\mu}}{1-\hat{\mu}}
\in [1-\eta,1+\eta]\frac{w(Y)}{w(X)}$
as required by Item~\ref{compare:mid}.
\item If $\D(Y) > K\cdot\D(X)$, then we consider two subcases.
  \BE
  \item If $\D(Y) > 3K\cdot\D(X)$, then $w(X) < \frac{1}{3K+1}$,
  so that by a multiplicative Chernoff bound
  (stated in \autoref{cor:CB-upperlower}), with probability
  at least $1-\delta$ we have that
  $$1-\hat{\mu} < \frac{1+\eta/3}{3K+1} \leq \frac{4}{3}\cdot\frac{1}{3K+1}
     \leq \frac{2}{3}\cdot\frac{1}{K+1}, $$
     causing the algorithm to output {\sf High}. Thus Item~\ref{compare:high}
     is established for this subcase.
% (and we do not need to attend to
%     Item~\ref{compare:cor}, since its premise does not hold).
  \item If $K\cdot\D(X) < \D(Y) \leq 3K\cdot\D(X)$, then
  $w(X) \geq \frac{1}{3K + 1}$ and $w(Y) \geq \frac{1}{2}$,
  so that the events $E_1$ and $E_2$ both hold
  with probability at least $1-\delta$. Assume that these events in fact
  hold. This implies that
  $\hat{\mu} \geq \frac{1-\eta/3}{2} \geq\frac{2}{3}\cdot\frac{1}{K+1}$,
  and  the algorithm either outputs {\sf High} or outputs
  $\rho = \frac{\hat{\mu}}{1-\hat{\mu}} \in [1-\eta,1+\eta]\frac{w(Y)}{w(X)}$,
  so Item~\ref{compare:high} is established for this subcase as well.
  \EE
\item If $\D(Y) < \D(X)/K$, so that $\D(X) > K\cdot\D(Y)$, then
the exact same arguments are applied as in the previous case, just
switching the roles of $Y$ and $X$ and the roles of $\hat{\mu}$ and $1-\hat{\mu}$
so as to establish Item~\ref{compare:low}.
% we consider two subcases:
% $\D(X) > 3K D(Y)$ and $K \D(Y) < \D(X) \leq 3K \D(Y)$.
\EE
We have thus established all items in the lemma.
\EPF

\subsection{The procedure {\sc Estimate-Neighborhood}}\label{subsec:est-n}
In this subsection we describe a procedure that, given a point $x$,
provides an estimate of the weight of a set of points $y$ such
that $\D(y)$ is similar to $\D(x)$. In order to specify the behavior
of the procedure more precisely, we introduce the following notation.
For a distribution $\D$ over $[N]$, a point $x \in [N]$ and
a parameter $\gamma \in [0,1]$, let
\BEQ\label{eq:Ugamma}\fullOrConfVersion{}{\hspace{-3mm}}
\fullOrConfVersion{
U^\D_\gamma(x) \eqdef \Big\{y \in [N]:\; \frac{1}{1+\gamma}\D(x)
             \leq \D(y) \leq (1+\gamma)\D(x)\Big\}
}
{
U^\D_\gamma(x) \eqdef \Big\{y \in [N]:\; \frac{\D(x)}{1+\gamma}
             \leq \D(y) \leq (1+\gamma)\D(x)\Big\}
}
\EEQ
denote the set of points whose weight is ``$\gamma$-close'' to
the weight of $x$.
If we take a sample of points distributed according to $\D$, then
the expected fraction of these points that belong to $U^\D_\gamma(x)$
is $\D(U^\D_\gamma(x))$.
% and if we take a uniform sample, then
% the expected fraction is $|U^\D_\gamma(x)|/N$. If both these values are
If this value is
not too small, then the actual fraction in the sample is close to the
expected value. Hence, if we could efficiently determine for
any given point $y$ whether or not it belongs to $U^\D_\gamma(x)$, then
we could obtain a good estimate of $\D(U^\D_\gamma(x))$.
%  and $|U^\D_\gamma(x)|$.
The difficulty is that it is not possible to
perform this task efficiently for {``boundary''} points $y$ such that
$\D(y)$ is very close to $(1+\gamma)\D(x)$ or
to $\frac{1}{1+\gamma}\D(x)$. However, for our purposes, it is not
important that we obtain the weight and size of $U^\D_\gamma(x)$
for a specific $\gamma$, but rather it suffices to do so for $\gamma$
in a given range, as stated in the next lemma.
The parameter $\beta$ in the lemma is the threshold above which we expect the algorithm to provide an estimate of the weight, while $[\kappa,2\kappa)$ is the range in which $\gamma$ is permitted to lie; finally, $\eta$ is the desired (multiplicative) accuracy of the estimate, while $\delta$ is a bound on the probability of error allowed to the subroutine.
\BL\label{lem:est-n}
Given as input a point $x$ together with parameters
$\kappa,\beta,\eta,\delta\in (0,1/2]$ as well as
% \SAMP and
\PCOND query access to a distribution $\D$,
the procedure {\sc Estimate-Neighborhood} (Algorithm~\ref{alg:est-n})
outputs a pair $(\hat{w},\alpha) \in [0,1]\times (\kappa,2\kappa)$
such that $\alpha$ is uniformly distributed in
$\{\kappa+i\theta\}_{i=0}^{\kappa/\theta -1}$
for $\theta = \frac{\kappa\eta\beta\delta}{64}$, and
such that the following holds:
\BE
  \fullOrConfIndentItems\fullOrConfCompactEnums
\item If $\D(U^\D_\alpha(x)) \geq \beta$,
then with probability at least \mbox{$1-\delta$} we have
$\hat{w} \in [1-\eta,1+\eta]\cdot\D(U^\D_\alpha(x))$,
and \mbox{$\D(U^\D_{\alpha+\theta}(x)\setminus U^\D_\alpha(x))
    \leq \eta\beta/16$};
\item If $\D(U^\D_\alpha(x)) < \beta$,
then with probability at least $1-\delta$ we have
$\hat{w}\leq (1+\eta)\cdot\beta$,
%\ignore{\D(U^\D_\alpha(x))$,}
and $\D(U^\D_{\alpha+\theta}(x)\setminus U^\D_\alpha(x))
    \leq \eta\beta/16$.
\EE
The number of \PCOND queries performed by the procedure is
$O\!\left(\frac{\log(1/\delta) \cdot \log(\log(1/\delta)/(\newer{\delta}\beta\eta^2))}
            {\kappa^2\eta^4\beta^3\delta^2}\right)$.
\EL

\begin{algorithm}
\begin{algorithmic}[1]
%\caption{The procedure {\sc Estimate-Neighborhood}}
%\SetKwInOut{Input}{Input}
% \label{alg:est-n}
\REQUIRE{\PCOND query access to a distribution
$\D$ over $[N]$,
a point $x \in [N]$ and parameters
$\kappa,\beta,\eta,\delta\in (0,1/2]$}
\STATE
Set
$\theta = \frac{\kappa\eta\beta\delta}{64}$
and $r = \frac{\kappa}{\theta}=\frac{64}{\eta\beta\delta}$.
\STATE Select a value $\alpha \in \{\kappa+ i\theta\}_{i=0}^{r -1}$
uniformly at random.
% and set $\tau = \alpha + \theta/2$.
\STATE Call the $\SAMP_\D$ oracle $\Theta(\log(1/\delta)/(\beta\eta^2))$
times and let $S$ be the set of points obtained.
% Take a sample $S$ of size $\Theta(\log(1/\delta)/(\beta\eta^2))$
% distributed according to $\D$.
\STATE For each point $y$ in $S$ call
$\mbox{\sc Compare}_\D(\{x\},\{y\},\theta/4,4,\delta/(4|S|))$ (if a point
$y$ appears more than once in $S$, then {\sc Compare} is called only once
on $y$).
\STATE Let $\hat{w}$ be the fraction
of occurrences of points $y$ in $S$ for which
{\sc Compare} returned a value
$\rho(y) \in [1/(1+\alpha+\theta/2),(1+\alpha+\theta/2)]$.
(That is, $S$ is viewed as a multiset.)
% \item Let $T_1(x) = T(x)\cap S_1$ and $T_2(x) = T(x) \cap S_2$, where
% $S_1$ and $S_2$, and hence $T_1(x)$ and $T_2(x)$ are viewed as
% multisets.
\STATE
% If $\hat{w}(x) \geq \beta/2$,
Return $(\hat{w},\alpha)$.
\end{algorithmic}\caption{\label{alg:est-n}{\sc Estimate-Neighborhood}}
\end{algorithm}%\dnote{I left the multiset phrasing because if we consider the points $y_1,\dots,y_s$, then we have to sometimes refer to to points in the corresponding set (when calling Compare) and sometimes to occurrences. I found it easier to leave it as is.}
\BPFOF{\autoref{lem:est-n}}
The number of % \SAMP and
\PCOND queries performed by {\sc Estimate-Neighborhood}
is the size of $S$ times the number of \PCOND queries performed in
each call to {\sc Compare}. By the setting of the parameters
in the calls to {\sc Compare}, the total number of \PCOND queries is
$O\!\left(\frac{(|S|)\cdot \log|S|/\delta)}{\theta^2}\right)
=O\!\left(\frac{\log(1/\delta) \cdot \log(\log(1/\delta)/(\newer{\delta}\beta\eta^2))}{\kappa^2\eta^4\beta^3\delta^2}\right)$.
We now turn to establishing the correctness of the procedure.

Since $\D$ and $x$ are fixed, in what follows we shall use the shorthand $U_\gamma$ for
$U_\gamma^\D(x)$.
% For $i \in [0,\dots,r-1]$ let
% $\gamma(i) = \kappa+i\theta$.
% Consider the corresponding sets $U_{\gamma(i)}(x)$, so that
% $U_{\gamma(0)}(x) = U_\kappa(x)$, $U_{\gamma(r)}=U_{2\kappa}(x)$ and
% $U_{\gamma(i)}(x) \subseteq U_{\gamma(i+1)}(x)$.
% Consider also the {\em difference\/} sets
For \mbox{$\alpha \in \{\kappa+ i\theta \}_{i=0}^{r -1}$}, let $\Delta_\alpha \eqdef U_{\alpha+\theta}\setminus U_{\alpha}$.
% (so that $\tau = \kappa+i(\tau)\theta +\theta/2$).
% $\tau \in\{\kappa+ i\theta + \theta/2\}_{i=1}^{r -1}$ let
% $i(\tau) = \frac{\tau -\kappa-\theta/2}{\theta}$
% (so that $\tau = \kappa+i(\tau)\theta +\theta/2$).
We next define several ``desirable'' events.
In all that follows we view $S$ as a multiset.

\BE
\item Let $E_1$ be the event that $\D(\Delta_{\alpha}) \leq 4/(\delta r)$.
Since there are r disjoint sets $\Delta_\alpha$
for \mbox{$\alpha \in \{\kappa+ i\theta \}_{i=0}^{r -1}$,}
the probability that $E_1$ occurs (taken over the uniform choice of $\alpha$)
is at least $1-\delta/4$.
From this point on we fix $\alpha$ and assume $E_1$ holds.
\item The event $E_2$ is that
$|S\cap \Delta_{\alpha}|/|S| \leq 8/(\delta r)$
(that is, at most
twice the upper bound on the expected value).
By applying the  multiplicative Chernoff bound
% (stated in \autoref{cor:CB-upperlower})
%\rnote{Here we seem to be using the new Corollary of the Chernoff bound}
% D: 1/11/12 moved ref earlier
using the fact that
\mbox{$|S| = \Theta(\log(1/\delta)/(\beta\eta^2)) =
   \Omega(\log(1/\delta) \cdot (\delta r))$,}
we have that $\Pr_S[E_2] \geq 1-\delta/4$.
\item The event $E_3$ is defined as follows:
% (note that $U_{\gamma(i(\tau) )}(x) = U_{\tau-\theta/2}(x)$):
% \dana{D: here there was a central part of the confusion caused by the
% incorrect range of values before for $\tau$}
% \rnote{It seems fine to me now; if we're all happy let's delete these comments}
%
If $\D(U_{\alpha}) \geq \beta$,
%\dnote{it was $\D(U_{\new{\alpha}}\newer{)}(x))$, so that there was some extra parenthesis, but in any case, since $x$ is fixed, I removed it from the notation}
then
$|S\cap U_{\alpha}|/|S| \in
        [1-\eta/2,1+\eta/2]\cdot\D(U_{\alpha})$,
and if
% $\D(U_{\alpha}\newer{)}(x)) < \beta$,
$\D(U_{\alpha}) < \beta$,
then $|S\cap U_{\alpha}|/|S| < (1+\eta/2)\cdot\beta$.
Once again applying the multiplicative Chernoff bound (for both cases)
and using that fact that $|S| = \Theta(\log(1/\delta)/(\beta\eta^2))$,
we have that $\Pr_S[E_3]\geq 1-\delta/4$.
\item Let $E_4$ be the event that all calls to {\sc Compare} return
an output as specified in \autoref{lem:compare}. Given the setting
of the confidence parameter in the calls to {\sc Compare} we have that
$\Pr[E_4] \geq 1-\delta/4$ as well.
\EE
Assume from this point on that events $E_1$ through $E_4$ all hold where
this occurs with probability at least $1-\delta$.
By the definition of $\Delta_\alpha$ and  $E_1$ we have that
% for $\alpha=\tau-\theta/2=\gamma(i(\tau)\new{)}$ and $\theta$
% as output by the algorithm it holds that
$\D(U_{\alpha+\theta}\setminus U_\alpha)
       \leq 4/(\delta r) = \eta\beta/16$,
as required (in both items of the lemma).
Let $T$ be the (multi-)subset of points $y$ in $S$ for which
{\sc Compare} returned a value
$\rho(y) \in [1/(1+\alpha+\theta/2),(1+\alpha+\theta/2)]$
(so that $\hat{w}$, as defined in the algorithm, equals
$|T|/|S|$).
Note first  that conditioned on $E_4$ we have that for every
\mbox{$y \in U_{2\kappa}$} it holds that the output of {\sc Compare}
when called on $\{x\}$ and $\{y\}$,
denoted $\rho(y)$, satisfies
$\rho(y) \in [1-\theta/4,1+\theta/4](\D(y)/\D(x))$,
while for $y \notin U_{2\kappa}$ either {\sc Compare}
outputs {\sf High} or {\sf Low} or it outputs a value
$\rho(y) \in [1-\theta/4,1+\theta/4](\D(y)/\D(x))$.
This implies that if $y \in U_{\alpha}$,
then $\rho(y) \leq (1+\alpha)\cdot(1+\theta/4) \leq 1+\alpha+\theta/2$
and $\rho(y) \geq (1+\alpha)^{-1}\cdot (1-\theta/4)
   \geq (1+\alpha+\theta/2)^{-1}$,
so that $S\cap U_{\alpha} \subseteq T$.
On the other hand, if $y \notin U_{\alpha+\theta}$
%  = U_{\gamma(i(\tau)+1)}$,
then either
$\rho(y) > (1+\alpha+\theta)\cdot(1-\theta/4) \geq 1+\alpha+\theta/2$
or $\rho(y) < (1+\alpha+\theta)^{-1}\cdot(1+\theta/4)
    \leq (1+\alpha+\theta/2)^{-1}$
so that $T \subseteq S\cap U_{\alpha+\theta}$.
Combining the two we have:
\BEQ \label{eq:T-x}
S\cap U_{\alpha} \;\subseteq \;T\;\subseteq\;
              S\cap U_{\alpha+\theta}\;.
\EEQ
Recalling that $\hat{w} = \frac{|T|}{|S|}$,
the left-hand side of Equation~(\ref{eq:T-x}) implies that
\BEQ\label{hat-w-lb}
\hat{w} \geq \frac{|S\cap U_{\alpha}|}{|S|}\;,
\EEQ
and by $E_1$ and $E_2$, the right-hand-side of
Equation~(\ref{eq:T-x}) implies that
\BEQ\label{hat-w-ub}
\hat{w}
 \;\leq\; \frac{|S\cap U_{\alpha)}|}{|S|}
       + \frac{8}{\delta r} \;\leq\;
       \frac{|S\cap U_{\alpha}|}{|S|}
              + \frac{\beta\eta}{8}\;.
\EEQ
We consider the two cases stated in the lemma:
\BE
\item If $\D(U_{\alpha}) \geq \beta$,
%so that $\D(U_{\gamma(i(\tau))}(x)) \geq \beta$ as well,
then by
Equation~(\ref{hat-w-lb}), Equation~(\ref{hat-w-ub}) and (the first
part of) $E_3$, we have that
$\hat{w} \in [1-\eta,1+\eta]\cdot \D(U_{\alpha})$.
% Thus, $\hat{w} \geq \beta/2$ and since
% $\gamma(i(\tau)) = \kappa+ i(\tau)\cdot \theta = \tau - \theta/2$,
% we get that in this case the algorithm outputs a triple $(\hat{w},\alpha,\theta)$
% as required.
\item If $\D(U_\alpha) < \beta$, then
by Equation~(\ref{hat-w-ub}) and (the second
part of) $E_3$, we have that $\hat{w} \leq (1+\eta)\beta$.
% there are two subcases.
% If $\D(U_{\alpha})\geq \beta/4$, then, similarly to
% the previous case (by
% Equation~(\ref{hat-w-lb}), Equation~(\ref{hat-w-ub}) and $E_3$),
% $\hat{w} \in [1-\eta,1+\eta]\cdot \D(U_{\alpha)})$.
% % This implies that either the procedure returns a triple as required,
% % or it returns \low.
% If $\D(U_{\gamma(i(\tau))})< \beta/4$, then by
% Equation~(\ref{hat-w-ub}) and the (second part of) $E_3$,
% we have that $\hat{w} \leq (1+\eta)\beta$.
\EE
The lemma is thus established.
\EPFOF

\subsection{The procedure {\sc Approx-Eval-Simulator}}\label{subsec:approx-eval}

\subsubsection{Approximate \EVAL oracles.}

We begin by defining the notion of an ``approximate $\EVAL$ oracle''
that we will use.  Intuitively this is an oracle which gives
a multiplicatively $(1 \pm  \eps)$-accurate estimate of
the value of $\D(i)$ for
all $i$ in a fixed set of probability weight at least $1-\eps$
under $\D$.
More precisely, we have the following definition:

\BD \label{def:approx-eval}
Let $\D$ be a distribution over $[N]$.
An \emph{$(\eps,\delta)$-approximate $\EVAL_\D$ simulator} is a randomized
procedure $\ORACLE$ with the following property:
For each $0 < \eps < 1$, there is a fixed set $S^{(\eps,\D)} \subsetneq [N]$
with $\D(S^{(\eps,\D)}) < \eps$ for which the following holds.
Given as input an element $i^\ast \in [N]$, the procedure
$\ORACLE$ either
outputs a value $\alpha \in [0,1]$ or outputs \unknown{} \newer{or \fail}.
The following holds for all $i^\ast \in [N]$:

%Let $\ORACLE$ be a randomized procedure which operates as follows:  It takes
%as input an element $i \in [N]$ and
%returns either a real number or \unknown.
%Each time it is
%invoked (say on an element $i \in [N]$),
%$\ORACLE$ independently generates a
%draw $\alpha_i$ from a distribution $\D^{(i)}$ over $[0,1] \cup
%\{\unknown\}$ and returns $\alpha_i.$
%Let $\D$ be a distribution over $[N]$.
%We say that a procedure $\ORACLE$ as described above
%For all $i \in [N]$, each time $\ORACLE$ is invoked on input $i$,
%its output independently with probability at least $1 - \delta$
%satisfies the following conditions:

\BI

\item [(i)] If $i^\ast \notin S^{(\eps,\D)}$ then
with probability at least $1 - \delta$
%(over the internal randomness of $\ORACLE$,
%i.e. the probability is taken w.r.t. $\D^{(i^\ast)}$)
the output of $\ORACLE$ on input $i^\ast$ is a value
$\alpha \in [0,1]$ such that $\alpha \in [1-\eps,1+\eps]\D(i^\ast)$;

\item [(i)] If $i^\ast \in S^{(\eps,\D)}$ then with probability at least
$1-\delta$ the procedure either outputs \unknown or outputs a value
$\alpha \in [0,1]$ such that $\alpha \in [1-\eps,1+\eps]\D(i^\ast)$.

\EI

\ED
We note that according to the above definition, it may be the case that
different calls to $\ORACLE$ on the same input element $i^\ast \in [N]$
may return different values.  However, the ``low-weight'' set
$S^{(\eps,\D)}$ is an
\emph{a priori} fixed set that does not depend in any way on the input
point $i^\ast$ given to the algorithm.  The key
property of an
$(\eps,\delta)$-approximate \EVAL$_\D$ oracle is that it
reliably gives a multiplicatively $(1 \pm  \eps)$-accurate estimate of
the value of $\D(i)$ for all $i$ in some
fixed set of probability weight at least $1-\eps$ under $\D$.

\subsubsection{Constructing an approximate \texorpdfstring{$\EVAL_\D$}{\EVAL} simulator using \texorpdfstring{$\COND_\D$}{\COND}}
\label{sec:approx-eval-simulator}

In this subsection we show that a $\COND_\D$ oracle can be used to
obtain an approximate $\EVAL$ simulator:

\BT \label{thm:approx-eval-simulator}
Let $\D$ be any distribution over $[N]$ and let
$0 < \eps,\delta < 1.$ The algorithm
{\sc Approx-Eval-Simulator} has the following properties:  It
uses
\[
\tilde{O}\left(
{\frac {(\log N)^5 \cdot (\log (1/\delta))^2}{\eps^3}}
\right)
\]
calls to
$\COND_\D$ and it is an $(\eps,\delta)$-approximate $\EVAL_\D$ simulator.
\ET

A few notes:  First, in the proof we give below
of \autoref{thm:approx-eval-simulator} we assume throughout that
$0 < \eps \leq \newer{1/40}$. %, where $c$ is a small absolute constant.
This incurs no loss of generality
because if the desired $\eps$ parameter is in \newer{$(1/40,1)$} then the parameter
can simply be set to \newer{$1/40$}.  We further
note that in keeping with our requirement on a $\COND_\D$ algorithm,
the algorithm {\sc Approx-Eval-Simulator} only ever calls the $\COND_\D$
oracle on sets $S$ which are either $S=[N]$ or else
contain at least one element $i$ that has been returned as the output
of an earlier call to $\COND_\D.$
To see this, note that Line~\ref{st:cond-queries-on-Sj-1}
is the only line when $\COND_\D$ queries are performed.
In the first execution of the outer ``For'' loop clearly
all $\COND$ queries are on set $S_{0}=[N].$  In subsequent
stages the only way a set $S_j$ is formed is if
either (i) $S_j$ is set to $\{i^\ast\}$ in Line~\ref{st:setDhat1}, in which
case clearly $i^\ast$ was previously received as the response of a $\COND_\D(
S_{j-1})$ query, or else (ii) a nonzero fraction of elements $i_1,\dots,i_m$
received as responses to $\COND_\D(S_{j-1})$ queries belong to
$S_j$ (see Line~\ref{st:unknown2}).

\medskip

{
\noindent {\bf A preliminary simplification.}
Fix a distribution $\D$ over $[N]$.  Let $Z$ denote $\supp{\D}$,
i.e. $Z = \{i \in [N]: \D(i) > 0\}.$
We first claim that in proving \autoref{thm:approx-eval-simulator}
we may assume without loss of generality
that no two distinct elements $i, j \in Z$ have $\D(i)=\D(j)$ -- in other
words, we shall prove the theorem under this assumption on $\D$, and
we claim that this implies the general result.
To see this, observe that if $Z$ contains elements $i \neq j$
with $\D(i)=\D(j)$,
then for any arbitrarily
small $\xi>0$ and any arbitrarily large $M$  we can perturb the weights
of elements in $Z$ to obtain a distribution $\D'$ supported on $Z$
such that (i) no two elements of $Z$
have the same probability under $\D'$, and (ii)
for every $S \subseteq [N],$ $S \cap Z \neq \emptyset$
we have $\dtv(\D_S,\D'_{S}) \leq \xi/M.$
%(We will see that our simulation of
%an approximate $\EVAL$ oracle indeed only calls $\COND_\D$ on
%sets $S$ that are guaranteed to have $S \cap Z \neq \emptyset$.)
Since the variation distance between $\D'_S$ and $\D_S$ is
at most $\xi/M$ for an arbitrarily small $\xi$,
the variation distance between (the execution of any $M$-query $\COND$
algorithm run on $\D$) and (the execution of any $M$-query $\COND$
algorithm run on $\D'$) will be at most $\xi.$  Since $\xi$ can be
made arbitrarily small this means that indeed without loss of generality
we may work with $\D'$ in what follows.%\rnote{Add more explanation later, perhaps in line with the 10-08-2012 email, if we think it's necessary.}

Thus, we henceforth assume that the distribution $\D$ has no two elements
in $\supp{\D}$ with the same weight.  For such a distribution
we can explicitly describe the set $S^{(\eps,\D)}$ from
\autoref{def:approx-eval} that our analysis will deal with.
Let $\pi: \{1,\dots,|Z|\} \to Z$ be the bijection such
that $\D(\pi(1)) > \cdots > \D(\pi(|Z|))$ (note that
the bijection $\pi$ is uniquely defined by the assumption that $\D(i)
\neq \D(j)$ for all distinct $i,j \in Z$).
Given a value $0 < \tau < 1$
%and a nonempty subset $S \subseteq [N],$
we define the set $L_{\tau,D}$ to be $
([N] \setminus Z) \cup \{\pi(s),\dots,\pi(|Z|)\}$ where $s$ is
the smallest index in $\{1,\dots,|Z|\}$
such that $\sum_{j=s}^{|Z|} \D(\pi(j)) < \tau$
(if $\D(\pi(|Z|))$ itself is at least $\tau$ then we define
$L_{\tau,\D} = [N] \setminus Z$).
\ignore{%(Note
%that specifying $\D$ completely specifies $L_{\tau,\D}$ for
%every $0 < \tau < 1.$)
}
Thus intuitively $L_{\tau,\D}$ contains the $\tau$
fraction (w.r.t. $\D$) of $[N]$ consisting of the lightest elements.
The desired set $S^{(\eps,\D)}$ is precisely $L_{\eps,\D}.$
}

\medskip

\noindent
{\bf Intuition for the algorithm.}
The high-level idea of the $\EVAL_D$ simulation is the
following:  Let $i^\ast \in [N]$ be the input element given to the
$\EVAL_\D$ simulator.   The algorithm works in a sequence of
stages.  Before performing
the $j$-th stage it maintains a set $S_{j-1}$ that contains $i^\ast$,
and it has a high-accuracy estimate $\hat{\D}(S_{j-1})$ of
the value of $\D(S_{j-1})$.  (The initial set $S_0$ is simply $[N]$
and the initial estimate $\hat{\D}(S_0)$ is of course 1.)
In the $j$-th stage the algorithm attempts to construct a subset
$S_j$ of $S_{j-1}$ in such a way that (i) $i^\ast \in S_j$, and
(ii) it is possible to obtain a high-accuracy estimate of
$\D(S_{j})/\D(S_{j-1})$ (and thus a high-accuracy
estimate of $\D(S_{j})$).
If the algorithm cannot construct such a
set $S_j$ then it outputs \unknown; otherwise, after at most (essentially)
$O(\log N)$ stages, it reaches a situation where $S_j=\{i^\ast\}$ and
so the high-accuracy estimate of $\D(S_{j}) = \D(i^\ast)$ is the desired value.
%\mnote{fix explan b/c of early stopping}

A natural first idea towards implementing this high-level plan is
simply to split $S_{j-1}$ randomly into two pieces and use one of them
as $S_j$.  However this simple approach may not work; for example,
if $S_{j-1}$ has one or more elements which are very heavy compared to
$i^\ast$, then with a random split it may not be possible to
efficiently estimate $D(S_j)/\D(S_{j-1})$ as required in (ii) above.
Thus we follow a more careful approach which first identifies and removes
``heavy'' elements from $S_{j-1}$ in each stage.

In more detail, during the $j$-th stage, the algorithm first performs
$\COND_\D$ queries on the set $S_{j-1}$ to identify a set $H_{j}
\subseteq S_{j-1}$ of ``heavy'' elements; this set essentially consists
of all elements which individually each
contribute at least a $\kappa$ fraction of the total mass
$\D(S_{j-1})$.  (Here $\kappa$ is a ``not-too-small'' quantity
but it is significantly less than $\eps.$)
Next, the algorithm performs additional $\COND_\D$ queries
to estimate $\D(i^\ast)/\D(S_{j-1}).$  If this fraction exceeds
$\kappa/20$ then it is straightforward to estimate
%If $i^\ast \in H_{j}$
%then it is straightforward to estimate
$\D(i^\ast)/\D(S_{j-1})$ to high accuracy, so using $\hat{D}(S_{j-1})$ it is
possible to obtain a high-quality estimate of $\D(i^\ast)$
and the algorithm can conclude.  However, the typical case is
that $\D(i^\ast)/\D(S_{j-1}) < \kappa/20.$
In this case, the algorithm next estimates $\D(H_{j})/\D(S_{j-1})$.
If this is larger than $1-\eps/10$ then
the algorithm outputs \unknown (see below for more discussion of this).
%(assuming all estimates
%are reasonably accurate) $i^\ast$ must not belong to $H_{1-\eps,\D}$,
%so the algorithm can output \unknown.
%$i^\ast$ must not belong to $H_{1-\eps,\D}$, so
%\rnote{The logic:  We are in a situation where
%all elements of $H_{j}$ have weight at least $\kappa$, and
%those guys alone weigh at least $1-\eps/10$.  The point $i^\ast$
%weighs less than $\kappa/2$ and so it must come later in the list so
%it's not in $H_{1-\eps,\D}$.}
If %the estimate of
$\D(H_{j})/\D(S_{j-1})$ is
less than $1-\eps/10$ then $\D(S_{j-1} \setminus H_{j})/\D(S_{j-1})
\geq \eps/10$ (and so $\D(S_{j-1} \setminus H_{j})/
\D(S_{j-1})$ can be efficiently estimated to high accuracy),
 but each element $k$ of $S_{j-1} \setminus H_{j}$ has
$\D(k)/\D(S_{j-1}) \leq \kappa \ll \eps/10 \leq \D(S_{j-1} \setminus H_{j})
/\D(S_{j-1}).$
Thus it must be the case that the weight under $\D$ of
$S_{j-1} \setminus H_{j}$ is ``spread out'' over many ``light'' elements.

Given that this is the situation, the algorithm next chooses
$S'_j$ to be a random subset of
$S_{j-1} \setminus (H_{j} \cup \{i^\ast\})$, and sets
$S_j$ to be $S'_j \cup \{i^\ast\}.$
%Using a structural lemma from recent work in property
%testing of Boolean functions,
It can be shown that
with high probability (over the random choice of $S_j$) it will be the case
that $\D(S_j) \geq {\frac 1 3} \D(S_{j-1} \setminus H_{j})$
(this relies crucially on the fact that
the weight under $\D$ of
$S_{j-1} \setminus H_{j}$ is ``spread out'' over many ``light'' elements).
This makes
it possible to efficiently estimate $\D(S_j)/\D(S_{j-1}
\setminus H_{j})$; together with the high-accuracy estimate of
$\D(S_{j-1} \setminus H_{j})/\D(S_{j-1})$ noted above, and
the high-accuracy estimate $\hat{D}(S_{j-1})$ of $D(S_{j-1})$,
this means it is
possible to efficiently estimate $\D(S_j)$ to high accuracy
as required for the next stage.
(We note that after defining $S_j$ but before proceeding to the next stage,
the algorithm actually checks to be sure that  $S_j$ contains at least one
point that was returned from the $\COND_{\D}(S_{j-1})$ calls made in the
past stage.  This check ensures that whenever the algorithm calls
$\COND_{\D}(S)$ on a set $S$, it is guaranteed that $\D(S)>0$
as required by our $\COND_\D$ model.  Our analysis shows that doing this
check does not affect correctness of the algorithm since with high
probability the check always passes.)

\medskip

\noindent {\bf Intuition for the analysis.}
We require some definitions to give the intuition for the analysis
establishing correctness.  Fix a nonempty subset $S \subseteq [N]$.
Let $\pi_S$ be the bijection mapping $\{1,\dots,|S|\}$ to
$S$ in such a way that
$\D_S(\pi_S(1)) > \cdots > \D_S(\pi_S(|S|))$, i.e.
$\pi_S(1),\dots,\pi_S(|S|)$ is a listing of the elements of $S$
in order from heaviest under $\D_S$ to lightest under $\D_S.$
Given $j \in S$, we define the \emph{$S$-rank of $j$}, denoted
$\rank_S(j)$, to be
the value $\sum_{i: \D_S(\pi(i)) \leq \D_S(j)} \D_S(\pi(i))$,
i.e. $\rank_S(j)$ is the sum of the weights (under $\D_S$) of
all the elements in $S$ that are no heavier than $j$ under $\D_S.$
Note that having $i^\ast \notin L_{\eps,N}$ implies that
$\rank_{[N]}(i^\ast) \geq \eps$.

We first sketch the argument for correctness.
(It is easy to show that the algorithm only outputs \fail with very small
probability so we ignore this possibility below.)
Suppose first that $i^\ast \notin L_{\eps,\D}$.  A key lemma shows that if
$i^\ast \notin L_{\eps,\D}$ (and hence
$\rank_{[N]}(i^\ast) \geq \eps$), then with high
probability every set $S_{j-1}$ constructed by the algorithm
is such that $\rank_{S_{j-1}}(i^\ast) \geq \eps/2$.  (In other words,
if $i^\ast$ is not initially among the $\eps$-fraction (under $\D$)
of lightest elements, then it never ``falls too far''
to become part of the $\eps/2$-fraction (under $\D_{S_{j-1}}$)
of lightest elements for $S_{j-1}$, for any $j$.).  Given that
(\newer{with high probability}) $i^\ast$ always has $\rank_{S_{j-1}}(i^\ast) \geq \eps/2$, though,
then it must be the case that (\newer{with high probability}) the procedure does not output \unknown
(and hence it must \newer{with high probability} output a numerical value).
This is because there are only two places where the procedure can
output \unknown, in Lines~\ref{st:unknown} and~\ref{st:unknown2};
we consider both cases below.

\BE

\item
In order for the procedure to output \unknown in Line~\ref{st:unknown},
it must be the case that the elements of $H_{j}$ -- each of
which individually has weight at least $\kappa/2$ under $\D_{S_{j-1}}$ --
collectively have weight at least $1-{3\eps/20}$ under $\D_{S_{j-1}}$
by Line~\ref{st:how-big-is-heavy}.
But $i^\ast$ has weight at most {$3\kappa/40$} under
$\D_{S_{j-1}}$ (because the procedure did not go to Line~\ref{st:startfor} in
Line~\ref{st:setDhat1}),
and thus $i^\ast$ would need to be in the bottom {$3\eps/20$} of
the lightest elements, i.e. it would need to have $\rank_{S_{j-1}}(i^\ast)
\leq {3\eps/20}$; but this contradicts
$\rank_{S_{j-1}}(i^\ast) \geq \eps/2$.

\item
In order for the procedure to output \unknown in Line~\ref{st:unknown2},
it must be the case that all elements $i_1,\dots,i_m$ drawn in
Line~\ref{st:cond-queries-on-Sj-1} are not chosen for inclusion in $S_j.$
In order for the algorithm to reach Line~\ref{st:unknown2}, though,
it must be the case that at least $(\eps/10 - \kappa/20)m$
of these draws do not belong to $H_{j} \cup \{i^\ast\}$;
since these draws do not belong to $H_{j}$
each one occurs only a small number
of times among the $m$ draws, so there must be many distinct values,
and hence the probability that none of these distinct values is chosen
for inclusion in $S'_j$ is very low.

\EE

Thus we have seen that if $i^\ast \notin L_{\eps,\D}$, then \newer{with high probability}
the procedure outputs a numerical value;
%\rnote{The logic:  We are in a situation where
%all elements of $H_{j}$ have weight at least $\kappa$, and
%those guys alone weigh at least $1-\eps/2$.  The point $i^\ast$
%weighs less than $\kappa/2$ and so it must come later in the list so
%it's not in $H_{1-\eps,\D}$.}
it remains to show that \newer{with high probability} this value is a high-accuracy
estimate of $\D(i^\ast).$
%Assuming estimates are all accurate as required, the only way the
%algorithm can output \unknown is if $i^\ast \notin H_{1-\eps,\D}$,
%so indeed if $i \in H_{1-\eps,\D}$ the algorithm outputs a numerical value
%rather than \unknown.  To show that the definition of an
%approximate $\EVAL_\D$ simulator is satisfied, we must argue that
%(again assuming all estimates are accurate as required)
%whenever the algorithm outputs a value, that value is multiplicatively
%$(1 \pm \eps)$-accurate for $\D(i^\ast)$.
However, this follows easily from the
fact that we inductively maintain a high-quality estimate of
$\D(S_{j-1})$ and the fact that the algorithm ultimately constructs
its estimate of $\hat{D}(i^\ast)$ only when it additionally
has a high-quality estimate of
$\D(i^\ast)/\D(S_{j-1}).$
This fact also handles the case in which
$i^\ast \in L_{\eps,\D}$ -- in such a case it is allowable
for the algorithm to output \unknown, so since the algorithm
\newer{with high probability} outputs a high-accuracy estimate when it outputs a numerical value,
this means the algorithm performs as required in Case~(ii) of
\autoref{def:approx-eval}.

We now sketch the argument for query complexity.
We will show that the heavy elements can be identified in each stage
using $\poly(\log N,1/\eps)$ queries.
Since the algorithm constructs $S_j$ by taking a random subset of
$S_{j-1}$ (together with $i^\ast$) at each stage, the number of stages
is easily bounded by (essentially) $O(\log N).$  Since the
final probability estimate for $\D(i^\ast)$ is a product of
$O(\log N)$ conditional probabilities, it suffices to estimate
each of these conditional probabilities to within a multiplicative
factor of $(1 \pm O\left({\frac {\eps}{\log N}}\right)).$
We show that each conditional
probability estimate can be carried out to this required precision
using only $\poly(\log N,1/\eps)$ calls to $\COND_\D$;
given this, the overall $\poly(\log N,1/\eps)$ query bound
follows straightforwardly.

%\rnote{In what follows ``$K$'' is a constant (the \# of bad events that we eventually take a union bound over in the proof of \autoref{thm:approx-eval-simulator}).  I had thought it was 8 but now it seems to be 9.}

\begin{algorithm}\begin{algorithmic}[1]
\REQUIRE access to $\COND_\D$; parameters $0 < \eps,\delta <1$;
input element $i^\ast \in [N]$
\STATE \label{st:initialize}
Set $S_0 = [N]$ and $\hat{D}(S_0)=1.$% {Set $K=9.$}
Set $M = \log N + \log(\newer{9}/\delta) +1.$
Set $\kappa = \Theta(\eps/(M^2 \log({M}/\delta))).$
\FOR{ $j=1$ to $M$} \label{st:startfor}
  \IF {$|S_{j-1}|=1$}
  \STATE {return $\hat{D}(S_{j-1})$ (and exit)} \label{st:just-one}
  \ENDIF
  \STATE Perform {$m = \Theta( \max\{M^2 \log(M / \delta)/(\eps^2 \kappa),
   \log(M/(\delta \kappa))/\kappa^2
%\ignore{,\log(MN/\delta)/\kappa^2}
 \} )$} $\COND_\D$ queries on $S_{j-1}$ to
obtain points $i_1,\dots,i_m \in S_{j-1}$. \label{st:cond-queries-on-Sj-1}
  \STATE Let $H_{j} = \{k \in [N]: k$ appears at least ${\frac 3 4} \kappa m$
times in the list $i_1,\dots,i_m\}$ \label{st:call-id-heavy}
  \STATE {Let $\hat{D}_{S_{j-1}}(i^\ast)$ denote the fraction of times that
$i^\ast$ appears in $i_1,\dots,i_m$}
  \IF {$\hat{D}_{S_{j-1}}(i^\ast) \geq {\frac \kappa {20}}$}
\label{st:hatD-iast-big}
  \STATE Set $S_{j} = \{i^\ast\}$, set
$\hat{D}(S_j)=\hat{D}_{S_{j-1}}(i^\ast) \cdot \hat{D}(S_{j-1})$,
increment $j$, and go to Line~\ref{st:startfor}. \label{st:setDhat1}
  \ENDIF
  \STATE { Let
$\hat{D}_{S_{j-1}}(H_{j})$ denote the fraction of elements among
$i_1,\dots,i_m$ that belong to $H_{j}$. }
\label{st:hatD-Hj-accurate}
  \IF {$\hat{D}_{S_{j-1}}(H_{j} ) > 1 - \eps/10$} \label{st:how-big-is-heavy}
  \STATE return \unknown (and exit) \label{st:unknown}
  \ENDIF
  \STATE Set $S'_{j}$ to be a uniform random subset
of $S_{j-1} \setminus (H_{j} \cup \{i^\ast\})$
and set $S_j$ to be $S'_j \cup \{i^\ast\}$. \label{st:random-subset}
  \STATE { Let
$\hat{D}_{S_{j-1}}(S_j)$ denote the fraction of elements among
$i_1,\dots,i_m$ that belong to $S_j$}
\label{st:hatD-Sj-large}
  \IF {$\hat{D}_{S_{j-1}}(S_j)$ = 0} \label{st:none}
  \STATE return \unknown (and exit) \label{st:unknown2}
  \ENDIF
  \STATE {Set $\hat{D}(S_j) =
\hat{D}_{S_{j-1}}(S_j) \cdot \hat{D}(S_{j-1})$} \label{st:setDhat2}
  \ENDFOR  \label{st:endfor}
  \STATE Output \fail.  \label{st:fail}
\end{algorithmic}\caption{\label{algo:approx-eval-simulator}\sc
Approx-Eval-Simulator}
\end{algorithm}

Now we enter into the actual proof.
We begin our analysis with a simple but useful lemma about the ``heavy''
elements identified in Line~\ref{st:call-id-heavy}.

\BL \label{lem:Identify-Heavy}
With probability at least $1-\delta/\newer{9}$, every set $H_{j}$ that is
ever constructed in Line~\ref{st:call-id-heavy} satisfies the following
for all $\ell \in S_{j-1}$:
\BI
\item [(i)] If $\D(\ell)/\D(S{_{j-1}}) > \kappa$, then $\ell \in H{_j}$;
\item [(ii)] If $\D(\ell)/\D(S{_{j-1}}) < \kappa/2$
then $\ell \notin H{_j}.$
\EI
\EL

{
\BPF
Fix an iteration $j$. By Line~\ref{st:call-id-heavy} in the algorithm,
a point $\ell$ is included in $H_j$ if it appears at least
${\frac 3 4} \kappa m$ times among $i_1,\dots,i_{m}$ (which are
the output of $\COND_\D$ queries on $S_{j-1}$).
For the first item, fix an element $\ell$ such that
$\D(\ell)/\D(S_{j-1}) > \kappa$.
Recall that
$m = \Omega(M^2 \log( M / \delta)/(\eps^2 \kappa)) =
   \Omega(\log(M N/\delta)/\kappa)$ (since $M = \Omega(\log(N))$).
By a multiplicative Chernoff bound,
the probability (over the choice of $i_1,\dots,i_m$ in
$S_{j-1}$) that $\ell$ appears less than
${\frac 3 4} \kappa m$ times among $i_1,\dots,i_{m}$
(that is, less than $3/4$ times the lower bound on the expected value) is
at most $\delta/(\newer{9}MN)$ (for an appropriate constant in the
setting of $m$). %{\rnote{Here in both applications of the mult CB it seems to me we need the new corollary version, that lets us plug in bounds on the expected value}}
On the other hand, for each fixed $\ell$ such that
$\D(\ell)/\D(S_{j-1}) < \kappa/2$, the probability that
$\ell$ appears at least ${\frac 3 4} \kappa m$ times
(that is, at least $3/2$ times the upper bound on the expected value)
is at most $\delta/(\newer{9}MN)$ as well.
The lemma follows by taking a union bound over all (at most $N$)
points considered above and over all $M$ settings of $j$.
\EPF
}

Next we show that with high probability Algorithm~{\sc Approx-Eval-Simulator}
returns either \unknown or a numerical value (as opposed to outputting
\fail in Line~\ref{st:fail}):

\BL \label{lem:givesvalue}
For any $\D$, $\eps, \delta$ and $i^\ast$,
Algorithm~{\sc Approx-Eval-Simulator} outputs \fail with probability
at most $\delta/\newer{9}.$
\EL

\BPF
Fix any element $i \neq i^\ast$.  The probability (taken only
over the choice of the random subset in each execution of
Line~\ref{st:random-subset}) that $i$ is placed in $S'_j$ in each of
the first $\log N + \log(\newer{9}/\delta)$ executions of Line~\ref{st:random-subset}
is at most
${\frac \delta {\newer{9}N}}.$  Taking a union bound over all $N-1$ points
$i \neq i^\ast$, the probability that any point other than $i^\ast$
remains in $S_{j-1}$ through all of the first $ \log N + \log(\newer{9}/\delta)$
executions of the outer ``for'' loop is at most
$\frac{\delta}{\newer{9}}.$  Assuming that this holds, then in the execution of the
outer ``for'' loop when $j=\log N + \log(\newer{9}/\delta) + 1$, the algorithm
will return $\hat{D}(S_{j-1}) = \hat{D}(i^\ast)$ in Line~\ref{st:just-one}.
\EPF

For the rest of the analysis
it will be helpful for us to define several ``desirable'' events
and show that they all hold with high probability:

\BE

\item Let $E_1$ denote the event that every set $H_{j}$ that is
ever constructed in Line~\ref{st:call-id-heavy}
satisfies both properties (i) and (ii)
stated in \autoref{lem:Identify-Heavy}.  By \autoref{lem:Identify-Heavy}
the event $E_1$ holds with probability at least $1 - \delta/\newer{9}$.

\item Let $E_2$ denote the event that in every execution of
Line~\ref{st:hatD-iast-big}, the estimate $\hat{D}_{S_{j-1}}(i^\ast)$
is within an additive $\pm {\frac \kappa {40}}$ of the true value of
$\D(i^\ast)/\D(S_{j-1})$.  By the choice of $m$ in
Line~\ref{st:cond-queries-on-Sj-1}
{(i.e., using $m =\Omega(\log(M/\delta)/\kappa^2)$)},
an additive Chernoff bound, {and a union bound over all iterations},
the event $E_2$ holds with probability at least $1 - \delta/\newer{9}$.

\item Let $E_3$ denote the event that if Line~\ref{st:setDhat1} is executed,
the resulting value
$\hat{D}_{S_{j-1}}(i^\ast)$ lies in \mbox{$[1 - {\frac \eps {2M}},
1+{\frac \eps {2M}}]\D(i^\ast)/\D(S_{j-1})$.} Assuming that event $E_2$ holds, if Line~\ref{st:setDhat1} is reached then
the true value of $\D(i^\ast)/\D(S_{j-1})$ must be at least $\kappa/40$,
and consequently
a multiplicative Chernoff bound and the choice of $m$
{(i.e. using $m = \Omega(M^2\log(M/\delta)/(\eps^2\kappa))$)}
together imply that
$\hat{D}_{S_{j-1}}(i^\ast)$ lies in $[1 - {\frac \eps {2M}},
1+{\frac \eps {2M}}]\D(i^\ast)/\D(S_{j-1})$
except with failure probability at most $\delta/\newer{9}$.

\item Let $E_4$ denote the event that
in every execution of
Line~\ref{st:hatD-Hj-accurate}, the estimate $\hat{D}_{S_{j-1}}(H_{j})$
is within an additive error of $\pm {\frac \eps {20} }$ from the true value of
$\D(H_{j})/\D(S_{j-1})$.  By the choice of $m$ in
Line~\ref{st:cond-queries-on-Sj-1}
{(i.e., using  $m = \Omega(\log(M/\delta)/\eps^2)$)}
and an additive Chernoff bound,
the event $E_4$ holds with probability at least $1 - \delta/\newer{9}$.

\EE

The above arguments show that $E_1,E_2,E_3$ and $E_4$ all hold with probability
at least $1 - 4\delta/\newer{9}$.

Let $E_5$ denote the event that in every execution of
Line~\ref{st:random-subset}, the set $S'_j$ which is drawn satisfies
\mbox{$\D(S'_j)/\D(S_{j-1} \setminus (H_{j} \cup \{i^\ast\})) \geq 1/3$}.
The following lemma says that conditioned
on $E_1$ through $E_4$  all holding, event $E_5$
holds with high probability:

\BL \label{lem:S'j-heavy}
Conditioned on $E_1$ through $E_4$ the probability that
$E_5$ holds is at least
$1 - \delta/\newer{9}$.
\EL

\BPF
Fix a value of $j$ and consider the $j$-th iteration of
Line~\ref{st:random-subset}. Since events $E_2$ and $E_4$ hold,
it must be the case that
$\D(S_{j-1} \setminus (H_{j} \cup \{i^\ast\}))/\D(S_{j-1}) \geq \eps/40.$
Since event $E_1$ holds, it must be the case that every
$i \in (S_{j-1} \setminus (H_{j} \cup \{i^\ast\}))$ has
$\D(i)/\D(S_{j-1}) \leq \kappa.$
Now since $S'_{j}$ is chosen by independently including each
element of $S_{j-1} \setminus (H_{j} \cup \{i^\ast\})$
with probability $1/2$,
{we can apply the first part of \autoref{cor:sum-wiXi}
and get
\[
\Pr\left[\D(S'_j) <
{\frac 1 3}\D(S_{j-1} \setminus (H_{j} \cup \{i^\ast\}))\right]
\leq \e^{-4\eps/(40\cdot 9\cdot 4 \kappa)} < \frac{\delta}{\newer{9}M}\;,
\]
where the last inequality follows by the setting of
$\kappa = \newer{\Omega}(\eps/(M^2 \log(1/\delta)))$.
}
\EPF

Thus we have established that $E_1$ through $E_5$ all hold with probability
at least $1 - 5\delta/\newer{9}$.

Next, let $E_{6}$ denote the event that the algorithm never returns
\unknown and exits in Line~\ref{st:unknown2}.  Our next lemma shows
that conditioned on events $E_1$ through $E_5$, the probability
of $E_{6}$ is at least $1-\delta/\newer{9}$:

\BL \label{lem:no-exit-unknown2}
Conditioned on $E_1$ through $E_5$ the probability that $E_{6}$ holds is at
least $1-\delta/\newer{9}$.
\EL
\BPF
Fix any iteration $j$ of the outer ``For'' loop.  In order
for the algorithm to reach Line~\ref{st:none} in this iteration,
it must be the case (by Lines~\ref{st:hatD-iast-big}
and~\ref{st:how-big-is-heavy}) that at least $(\eps/10 - \kappa/20)m
> (\eps/20)m$ points in $i_1,\dots,i_m$ do not belong to $H_{j} \cup
\{i^\ast\}.$  Since each point not in $H_{j}$ appears at most ${\frac 3 4}
\kappa m$ times in the list $i_1,\dots,i_m$, there must
be at least ${\frac \eps {15 \kappa}}$ distinct such values.
Hence the probability that none of these values is selected to
belong to $S'_j$ is at most $1/2^{\eps/(15 \kappa)} <
\delta/(\newer{9}M).$ A union bound over all (at most $M$) values of $j$
gives that the probability the algorithm ever returns \unknown and
exits in Line~\ref{st:unknown2} is at most $\newer{\delta/9}$,
so the lemma is proved.
\EPF

Now let $E_7$ denote the event that
in every execution of Line~\ref{st:hatD-Sj-large}, the estimate
$\hat{D}_{S_{j-1}}(S_j)$ lies in
$[ 1 - {\frac \eps {2M}}, 1 + {\frac \eps {2M}}] \D(S_j)/\D(S_{j-1}).$
The following lemma says that conditioned on $E_1$ through $E_5$,
event $E_7$ holds with probability at least $1-\delta/\newer{9}$:

\BL \label{lem:Sj-accurate}
Conditioned on $E_1$ through $E_5$, the probability that
$E_7$ holds is at least $1 - \delta/\newer{9}$.
\EL

\BPF
Fix a value of $j$ and consider the $j$-th iteration of
Line~\newer{\ref{st:hatD-Sj-large}}.%\ref{st:random-subset}.
The expected value of $\hat{D}_{S_{j-1}}(S_j)$ is precisely
\begin{equation} \label{eq:exp}
{\frac {\D(S_{j})} {\D(S_{j-1})}} =
{\frac {\D(S_{j})} {\D(S_{j-1} \setminus (H_{j} \cup \{i^\ast\}))}} \cdot
{\frac {\D(S_{j-1} \setminus (H_{j} \cup \{i^\ast\}))}{\D(S_{j-1})}}.
\end{equation}
Since events $E_2$ and $E_4$ hold we have that
${\frac {\D(S_{j-1} \setminus (H_{j} \cup \{i^\ast\}))}{\D(S_{j-1})}} \geq
\eps/40$, and
since event $E_5$ holds we have that
${\frac {\D(S_{j})} {\D(S_{j-1} \setminus (H_{j} \cup \{i^\ast\}))}} \geq 1/3$
(note that $\D(S_j) \geq \D(S'_j)$).
Thus we have that (\ref{eq:exp}) is at least $\eps/120.$
Recalling the value of $m$ (i.e., using $m=\Omega(M^2\log(M/\delta)/\eps^2\kappa) = \Omega(M^2\log(\newer{M}/\delta)/\eps^3)$)
a multiplicative Chernoff bound gives that indeed
$\hat{D}_{S_{j-1}}(S_j) \in [1 - {\frac \eps {2M}},
1 + {\frac \eps {2M}}] \D(S_j)/\D(S_{j-1})$
with failure probability at most $\delta/(\newer{9}M).$  A
union bound over all $M$ possible values
of $j$ finishes the proof.
\EPF

At this point we have established that events $E_1$ through $E_7$
all hold with probability at least $1-7\delta/\newer{9}$.

We can now argue that each estimate $\hat{D}(S_j)$ is indeed a high-accuracy
estimate of the true value $\D(S_j)$:

\BL \label{lem:hatD-Sj-good}
With probability at least $1 - 7\delta/\newer{9}$ each estimate
$\hat{D}(S_j)$ constructed by {\sc Approx-Eval-Simulator}
lies in $[(1 - {\frac {\eps}{2M}})^{j},
(1+{\frac \eps {2M}})^{j}] \D(S_j)$.
\EL
\BPF
We prove the lemma by showing that if all events $E_1$ through $E_7$ hold, then
the following claim (denoted (*)) holds:  each estimate
$\hat{D}(S_j)$ constructed by {\sc Approx-Eval-Simulator}
lies in $[(1 - {\frac \eps {2M}})^{j},
(1+ {\frac \eps {2M}})^{j}] \D(S_j)$.  Thus for the rest of the proof we assume
that indeed all events $E_1$ through $E_7$ hold.

The claim (*) is clearly true for $j=0$.  We prove (*) by induction on $j$ assuming
it holds for $j-1.$
The only places in the algorithm where
$\hat{D}(S_j)$ may be set are Lines~\ref{st:setDhat1} and~\ref{st:setDhat2}.
If $\hat{D}(S_j)$  is set in Line~\ref{st:setDhat2} then
(*) follows from the inductive claim for $j-1$ and \autoref{lem:Sj-accurate}.  If
$\hat{D}(S_j)$  is set in Line~\ref{st:setDhat1}, then (*) follows from
the inductive claim for $j-1$ and the fact that event~$E_{3}$ holds.  This concludes the
proof of the lemma.
\EPF

%We require one more crucial lemma to prove \autoref{thm:approx-eval-simulator}.  To state this
%lemma we need some new notation.  Given a subset $S \subseteq [N]$ and a parameter
%$0 < \tau < 1$, we define
%$H_{\tau,\D,S}$ to be $H_{\tau,\D_S,S}$.  In other words, $H_{\tau,\D,S}$ is the subset
%consisting of the heaviest elements of $S$, taken from heaviest to lightest, until the total
%weight of the subset is at least a $\tau$ fraction.

Finally, we require the following crucial lemma which establishes that
if $i^\ast \notin L_{\eps,N}$ (and hence the initial rank
$\rank_{[N]}$ of $i^\ast$ is at least $\eps$),
then with very high probability
the rank of $i^\ast$ never becomes too low
during the execution of the algorithm:

\BL \label{lem:rankOK}
Suppose $i^\ast \notin L_{\eps,N}.$  Then with probability at least
$1 - \delta/\newer{9}$, every set $S_{j-1}$ constructed by the
algorithm has $\rank_{S_{j-1}}(i^\ast) \geq \eps/2.$
\EL
We prove \autoref{lem:rankOK} in Section~\ref{sec:proof-of-rankOK} below.

With these pieces in place we are ready to prove \autoref{thm:approx-eval-simulator}.

\noindent {\bf Proof of \autoref{thm:approx-eval-simulator}:}
It is straightforward to verify that algorithm
{\sc Approx-Eval-Simulator} has the claimed query complexity.  We now argue that {\sc Approx-Eval-Simulator}
meets the two requirements (i) and (ii) of \autoref{def:approx-eval}.  Throughout the
discussion below we assume that all the ``favorable events'' in the
above analysis (i.e. events $E_1$ through $E_7$, \autoref{lem:givesvalue},
and \autoref{lem:rankOK}) indeed hold as desired (incurring
an overall failure probability of at most $\delta$).
%\ignore{\rnote{This is where ``$K=9$'' comes from.}}

Suppose first that $i^\ast \notin L_{\eps,\D}$.  
\newer{
We claim that by \autoref{lem:rankOK}
it must be the case that the algorithm does not return \unknown
in Line~\ref{st:unknown}.
To verify this, observe that in order to reach Line~\ref{st:unknown}
it would need to be the case that
$\D(i^\ast)/\D(S_{j-1}) \leq 3 \kappa/40$ (so the algorithm does
not instead go to Line~\ref{st:endfor}
in Line~\ref{st:setDhat1}). Since
by \autoref{lem:Identify-Heavy} every element $k$ in $H_{j}$
satisfies $\D(k)/\D(S_{j-1}) \geq \kappa/2$, this means
that $i^\ast$ does not belong to $H_{j}.$
In order to reach Line~\ref{st:unknown}, by event $E_4$ we must have
$\D(H_{j})/\D(S_{j-1}) \geq 1 - 3 \eps/20$.  Since
every element of $H_{j}$ has more mass under $\D$ (at least $\kappa/2$)
than $i^\ast$ (which has at most $3 \kappa/40$),
this would imply that $\rank_{S_{j-1}}(i^\ast) {\leq } 3 \eps/20$,
contradicting \autoref{lem:rankOK}.
Furthermore, by \autoref{lem:no-exit-unknown2} it must
be the case that the algorithm does not return \unknown
in Line~\ref{st:unknown2}.}
Thus the algorithm terminates by returning an estimate
$\hat{D}(S_j) = \hat{D}(i^\ast)$ which, by \autoref{lem:hatD-Sj-good}, lies
in $[(1 -  {\frac \eps {2M}})^{j},
(1+ {\frac \eps {2M}})^{j}] \D(i^\ast)$.  Since $j \leq M$ this estimate lies in $[1-\eps,1+\eps]\D(i^\ast)$
as required.

Now suppose that $i^\ast \in L_{\eps,\D}.$  By \autoref{lem:givesvalue}
we may assume that the algorithm either outputs \unknown or a numerical value.
As above, \autoref{lem:hatD-Sj-good} implies that if the algorithm
outputs a numerical value then the value lies in $[1-\eps,1+\eps]\D(i^\ast)$
as desired. This concludes the proof of
\autoref{thm:approx-eval-simulator}.
\qed

\subsubsection{Proof of \autoref{lem:rankOK}.}
\label{sec:proof-of-rankOK}

%Our goal is to prove (a variant of) Lemma~6 of the d1d2.tex file.  Recall that
%that lemma says the following:
%
%\BL \label{lem:rankOK}
%Suppose $i^\ast \notin L_{\eps,N}.$  Then with probability at least
%$1 - \delta/K$, every set $S_{j-1}$ constructed by the
%algorithm has $\rank_{S_{j-1}}(i^\ast) \geq \eps/2.$
%\EL
%
%\ignore{
%We will modify the {\sc Approx-Eval-Simulator} procedure so that it only
%enters the FOR loop at a given step $j$ if
%$|S_{j-1}| \geq n(N,\eps,\kappa)= \poly(\log(N),1/\eps)$.
%Thus in proving \autoref{lem:rankOK} we can
%assume throughout that $S_{j}$ is ``relatively large''; this will be
%helpful.
%}

The key to proving \autoref{lem:rankOK}
will be proving the next lemma.
(In the following, for $S$ a set of real numbers we write
$\wt(S)$ to denote $\sum_{\alpha \in S}\alpha.$)

\BL \label{lem:tech}
Fix $0 < \eps \leq \newer{1/40}$. %\newer{(recall that we assumed for the proof of \autoref{thm:approx-eval-simulator} that $c\in(0,1)$ was some absolute constant)}.
Set $\kappa = \newer{\Omega}(\eps/(M^2 \log(1/\delta))).$
Let $T=\{\alpha_1,\dots,\alpha_n\}$ be a set of\ignore{ $n \geq n_0(N,\eps,\kappa)$}
values $\alpha_1 < \cdots < \alpha_n$ such that
$\wt(T)=1.$
Fix $\ell \in [N]$ and let
$T_L = \{\alpha_1,\dots,\alpha_\ell\}$ and let
$T_R = \{\alpha_{\ell+1},\dots,\alpha_n\}$, so $T_L \cup T_R = T.$
Assume that $\wt(T_L) \geq \eps/2$
and that $\alpha_\ell \leq \kappa/10.$

Fix $H$ to be any subset of $T$ satisfying the following two properties:
(i) $H$ includes every $\alpha_j$ such that $\alpha_j \geq \kappa$; and
(ii) $H$ includes no $\alpha_j$ such that $\alpha_j < \kappa/2.$
(Note that consequently $H$ does not intersect $T_L.$)

Let $T'$ be a subset of $(T \setminus (H \cup \{\alpha_\ell\})$
selected uniformly at random.
Let $T'_L = T' \cap T_L$ and let $T'_R = T' \cap T_R$.

Then we have the following:

\BE

\item If $\wt(T_L) > 20 \eps$,  then with probability at least
$1 - \delta{/M}$
(over the random choice of $T'$) it holds that
\[
{\frac {\wt(T'_L \cup \{\alpha_\ell\})}
       {\wt(T' \cup \{\alpha_\ell\})}} \geq 9 \eps;
\]

\item If $\eps/2 \leq \wt(T_L) < 20 \eps$, then with probability at
least $1 - \delta{/M}$ (over the random choice of $T'$) it holds that
\[
{\frac {\wt(T'_L \cup \{\alpha_\ell\})}
       {\wt(T' \cup \{\alpha_\ell\})}} \geq
\wt(T_L) \left( 1 - \rho \right),
\]
where $\rho = {\frac {\ln 2}{M}}.$
\EE
\EL

\noindent {\bf Proof of \autoref{lem:rankOK} using \autoref{lem:tech}:}
We apply \autoref{lem:tech} repeatedly at each iteration $j$ of the
outer ``For'' loop.
The set $H$ of \autoref{lem:tech}
corresponds to the set $H_{j}$ of ``heavy''
elements that are removed at a given iteration, the set of values
$T$ corresponds to the values $\D(i)/\D(S_{j-1})$ for $i \in S_{j-1}$,
and the element $\alpha_\ell$ of \autoref{lem:tech} corresponds
to $\D(i^\ast)/\D(S_{j-1}).$  The value $\wt(T_L)$ corresponds
to $\rank_{S_{j-1}}(i^\ast)$
and the value
\[
{\frac {\wt(T'_L \cup \{\alpha_\ell\})}
       {\wt(T' \cup \{\alpha_\ell\})}}
\]
corresponds to $\rank_{S_j}(i^\ast).$
Observe that since $i^\ast \notin L_{\eps,N}$ we know that
initially $\rank_{[N]}(i^\ast) \geq \eps$,
which means that the first time we apply \autoref{lem:tech}
(with $T=\{\D(i): i \in [N]\}$) we have $\wt(T_L) \geq \eps$.

By \autoref{lem:tech} the probability of failure in any of the (at most $M$)
iterations is at most $\delta/\newer{9}$, so we assume that there is never a
failure.  Consequently for all $j$ we have that if
$\rank_{S_{j-1}}(i^\ast) \geq 20 \eps$
then
$\rank_{S_{j}}(i^\ast) \geq {9\eps}$, and if
$\eps/2 \leq \rank_{S_{j-1}}(i^\ast) < 20 \eps$
then
$\rank_{S_{j}}(i^\ast) \geq \rank_{S_{j}}(i^\ast) \cdot \left(1-\rho\right).$
Since $\rank_{S_{0}}(i^\ast) \geq \eps$,
it follows that for all $j \leq M$ we have $\rank_{S_{j}}(i^\ast)
\geq \eps \cdot (1 - \rho)^M > \eps/2.$
\qed

\noindent {\bf Proof of \autoref{lem:tech}.}
We begin with the following claim:

\BCM \label{claim:TL}
With probability at least $1-\delta/{(2M)}$ (over the random choice of $T'$)
it holds that
$\wt(T'_L) \geq {\frac 1 2} \cdot \wt(T_L) \cdot (1-\rho/2).$
\ECM

\BPF
Recall from the setup that every element $\alpha_i \in T_L$ satisfies $\alpha_i \leq \kappa/10$,
and $\wt(T_L) \geq \eps/2$. Also recall that $\kappa = \newer{\Omega}(\eps/(M^2 \log(1/\delta)))$ and that
$\rho = {\frac {\ln 2}{M}}$, so that $\rho^2\eps/(6\kappa) \geq \ln (2M/\delta).$
The claim follows by applying the first part of \autoref{cor:sum-wiXi} (with $\gamma = \rho/2$).
\EPF

Part~(1) of \autoref{lem:tech} is an immediate
consequence of \autoref{claim:TL}, since in part (1) we have
\[
{\frac {\wt(T'_L \cup \{\alpha_\ell\})}
       {\wt(T' \cup \{\alpha_\ell\})}} %\ignore{ \geq{\frac {\wt(T'_L)} {\wt(T' \cup \{\alpha_\ell\})}}}
\geq \wt(T'_L) \geq
{\frac 1 2} \cdot \wt(T_L) \cdot \left(1-\frac{\rho}{2}\right) \geq
{\frac 1 2} \cdot 20 \eps \cdot \left(1 - {\frac \rho 2}\right) \geq 9 \eps.
\]

It remains to prove Part~(2) of the lemma.  We will do this using the
following claim:

\BCM \label{claim:TR}
Suppose $\eps/2 \leq \wt(T_L) \leq 20 \eps.$  Then with probability
at least $1-\delta/(2M)$ (over the random choice of $T'$)
it holds that
$\wt(T'_R) \leq {\frac 1 2} \wt(T_R) \cdot (1 + \rho/2).$
\ECM

\BPF
Observe first that $\alpha_i < \kappa$ for each $\alpha_i \in T_R \setminus H$.
We consider two cases.

% NB: the "4\eps" threshold below is used to get an upperbound of \exp(-\rho^2 \eps/(6 \kappa)) and use Claim 24 to upper bound further.
If $\wt(T_R \setminus H) \geq 4 \eps$, then we apply the first part of
\autoref{cor:sum-wiXi} to the $\alpha_i$'s in $T_R \setminus H$
% (and recalling that $T'_R = T' \cap T_R$ where $T'$ is a subset of
% $T \setminus (H \cup \{\alpha_\ell\})$
% selected uniformly at random),
and get that
\begin{eqnarray}
\Pr\left[\wt(T'_R) > {\frac 1 2} \wt(T_R) \cdot (1 + \rho/2)\right]
 &\leq& \Pr\left[\wt(T'_R) > {\frac 1 2} \wt(T_R \setminus H)
    \cdot (1 + \rho/2)\right] \nonumber \\
    &<& \exp(-\rho^2 \wt(T_R \setminus H)/24\kappa)\\
    &\leq& \exp(-\rho^2 \eps/(6 \kappa)) \;\leq\; \frac{\delta}{2M}\;
\end{eqnarray}
(recall from the proof of \autoref{claim:TL} that $\rho^2 \eps / (6 \kappa) \geq \ln(2M/\delta)$).

If $\wt(T_R \setminus H) < 4 \eps$,
(so that the expected value of $\wt(T'_R)$ is less than
{$2\eps$}) then we can apply the second part
of \autoref{cor:sum-wiXi} as we explain next.
Observe that by the premise of the lemma,
$\wt(T_R) \geq 1 - 20 \eps$ which is at least
$1/2$ (recalling that $\eps$ is at most \newer{1/40}). %a small absolute constant $c$).
Consequently, the event ``$\wt(T'_R) \geq {\frac 1 2} \cdot \wt(T_R)\cdot
(1 + \rho/2)$'' implies the event
``$\wt(T'_R) \geq {\frac 1 4}$'', and by applying the second part
of \autoref{cor:sum-wiXi} we get
\BEQ
\Pr\left[\wt(T'_R) > {\frac 1 2} \wt(T_R) \cdot (1 + \rho/2)\right]
 \leq \Pr\left[\wt(T'_R) > \frac{1}{4}\right] < 2^{-1/4\kappa}
   < \frac{\delta}{2\newer{M}}\;,
\EEQ
as required.
\EPF

%%%%%%%%%%%%%%%%%%%%%%%%%%%%%%%%%%%%%%%%%%%%

Now we can prove \autoref{lem:tech}.  Using Claims~\ref{claim:TL}
and~\ref{claim:TR} we have that with probability at least
{ $1-\delta/M$,}
\[
\wt(T'_L) \geq {\frac 1 2} \cdot \wt(T_L) \cdot (1-\rho/2)
\quad \quad \text{and} \quad \quad
\wt(T'_R) \leq {\frac 1 2} \wt(T_R) \cdot (1 + \rho/2);
\]
we assume that both these inequalities hold going forth.  Since
\[
{\frac {\wt(T'_L \cup \{\alpha_\ell\})}
       {\wt(T' \cup \{\alpha_\ell\})}}
=
{\frac {\wt(T'_L) + \alpha_\ell}
       {\wt(T') + \alpha_\ell}}
>
{\frac {\wt(T'_L)}
       {\wt(T')}},
\]
it is sufficient to show that
$ {\frac {\wt(T'_L)} {\wt(T')}} \geq \wt(T_L) (1-\rho)$; we now show this.
As \mbox{$\wt(T')=\wt(T'_L)+\wt(T'_R)$},
\begin{eqnarray*}
{\frac {\wt(T'_L)} {\wt(T')}} &=&
{\frac {\wt(T'_L)} {\wt(T'_L) + \wt(T'_R)}} =
{\frac 1 {1 + {\frac {\wt(T'_R)}{\wt(T'_L)}}}}\\
&\geq&
{\frac 1 {1 + {\frac {(1/2)\cdot \wt(T_R) \cdot (1 + \rho/2)}
{(1/2) \cdot \wt(T_L) \cdot (1-\rho/2)}}}}\\
&=&
{\frac {\wt(T_L) \cdot (1 - \rho/2)}
 {\wt(T_L)\cdot (1-\rho/2) + \wt(T_R) \cdot (1 + \rho/2)}}\\
&\geq&
{\frac {\wt(T_L) \cdot (1 - \rho/2)}
 {\wt(T_L)\cdot (1+\rho/2) + \wt(T_R) \cdot (1 + \rho/2)}}\\
&=&
\wt(T_L) \cdot {\frac {1-\rho/2}{1+\rho/2}} > \wt(T_L) \cdot (1-\rho).
\end{eqnarray*}
This concludes the proof of \autoref{lem:tech}.
\qed

\section{Algorithms and lower bounds for testing uniformity}
\label{sec:uniform}

\subsection{A \texorpdfstring{$\tilde{O}(1/\eps^2)$}{O(1/eps2)}-query \PCOND algorithm for testing uniformity} \label{ssec:unif-alg}

In this subsection we present an algorithm {\sc $\PCOND_\D$-Test-Uniform}
and prove the following theorem:

\BT \label{thm:paircond-test-uniform}
{\sc $\PCOND_\D$-Test-Uniform} is a $\tilde{O}(1/\eps^2)$-query $\PCOND_\D$
testing algorithm for uniformity, i.e. it outputs $\accept$ with probability
at least $2/3$ if $\D = \calU$ and outputs $\reject$
with probability at least $2/3$ if $\dtv(\D,\calU) \geq \eps.$
\ET

\ignore{

%\BR[Extension to almost-uniformity]
%This can be generalized to $\eps$-testing against a known, fully-specified
%distribution $\D^\ast$ over $N$ elements $\{1,\dots,N\}$ provided
%that $\D^\ast$ is \emph{almost uniform} (in the sense that for some
%constant $c$, we have $\D^\ast(k) \leq c \D^\ast(\ell)$ for
%every $k,\ell
%\in \{1,\dots,n\}$): this is done in
%\autoref{section:testing:almost:uniformity}.
%\ER
}

\noindent {\bf Intuition.}  For the sake of intuition we first
describe a simpler approach that yields a $\tilde{O}(1/\eps^4)$-query
algorithm, and then build on those ideas to obtain our real algorithm with
its improved $\tilde{O}(1/\eps^2)$ bound.
Fix $\D$ to be a distribution over $[N]$
that is $\eps$-far from uniform.  Let
\[
H = \setOfSuchThat{ h  \in [N]}{ \D(h) \geq \frac{1}{N} }
\text{~and~}
L = \setOfSuchThat{ \ell  \in [N]}{ \D(\ell) < \frac{1}{N} }.
\]
It is easy to see that since $\D$ is $\eps$-far from uniform, we have
\begin{equation}\label{eq:eps-far1}
\sum_{h\in H} \left( \D(h) - \frac{1}{N} \right) =
\sum_{\ell\in L} \left( \frac{1}{N} - \D(\ell) \right)
\geq \frac{\eps}{2}\;.
\end{equation}
From this it is not hard to show that

\BI

\item [(i)] many elements of $[N]$ must be ``significantly light'' in the
following sense:  Define $L' \subseteq L$ to be $L' = \setOfSuchThat{ \ell
\in [N]}{\D(\ell) < {\frac 1 N} - {\frac \eps {4N}}}$.  Then
it must be the case that $|L'| \geq (\eps/4)N.$

\item [(ii)] $\D$ places significant weight on elements that are
``significantly heavy'' in the following sense:  Define $H' \subseteq H$
to be  $H' = \setOfSuchThat{ h
\in [N]}{\D(h) \geq {\frac 1 N} + {\frac \eps {4N}}}$.  Then
it must be the case that $\D(H') \geq (\eps/4).$

\EI

Using (i) and (ii) it is fairly straightforward to give a
$O(1/\eps^4)$-query $\PCOND_\D$ testing algorithm as follows:
we can get a point in $L'$ with high probability by
randomly sampling $O(1/\eps)$ points uniformly at random from $[N]$,
and we can get a point in $H'$ with high probability by drawing
$O(1/\eps)$ points from $\SAMP_\D$.
Then at least one of the $O(1/\eps^2)$ pairs that have
one point from the first sample and one point from the second will have
a multiplicative factor difference of $1 + \Omega(\eps)$ between the
weight under $\D$ of the two points, and this can be detected
by calling the procedure {\sc Compare} (see \autoref{subsec:compare}).
Since there are $O(1/\eps^2)$ pairs and for each one the invocation of
{\sc Compare} uses $\tilde{O}(1/\eps^2)$ queries,
the overall sample complexity of this simple approach is $
\tilde{O}(1/\eps^4).$

Our actual algorithm {\sc $\PCOND_\D$-Test-Uniform} for testing uniformity extends
the above ideas to get a $\tilde{O}(1/\eps^2)$-query algorithm.
More precisely, the algorithm works as follows:
it first draws a ``reference sample'' of $O(1)$ points uniformly
from $[N]$.  Next, repeatedly for $\bigO{\log \frac{1}{\eps}}$ iterations,
the algorithm draws two other samples,
one uniformly from $[N]$ and the other from $\SAMP_\D$.
(These samples have different sizes at different iterations;
intuitively, each iteration is meant to deal with a different
``scale'' of probability mass that points could have under $\D$.)
At each iteration it then uses {\sc Compare} to
do comparisons between
pairs of elements, one from the reference sample and the other from one of the
two other samples.  If $\D$ is $\eps$-far from uniform, then
with high probability at some iteration the algorithm will either
draw a point from $\SAMP_\D$ that has ``very big'' mass under $\D$, or
draw a point from the uniform distribution over $[N]$ that has
``very small'' mass under $\D,$ and this will be detected by the
comparisons to the reference points.  Choosing the sample sizes
and parameters for the {\sc Compare} calls
carefully at each iteration yields the improved query bound.
%\rnote{More on the idea of where the savings comes from?}
%\dnote{I think it is fine as is.}

\begin{algorithm}
\begin{algorithmic}[1]
\REQUIRE error parameter $\eps > 0$; query access to $\PCOND_\D$ oracle
\STATE Set $t = \log(\frac{4}{\eps})+ 1$.
\STATE\label{st:unif} Select $q=\bigTheta{1}$ points $i_1,\dots,i_q$
independently and uniformly from $[N]$.
\FOR{ $j = 1$ to $t$ }
   \STATE \label{st:D}
Call the $\SAMP_\D$ oracle $s_j = \bigTheta{2^j\cdot t}$
times to obtain points $h_1,\dots,h_{s_j}$ distributed according
to $\D$.
%Select $s_j = \bigTheta{2^j\cdot t}$ points
%       $h_1,\dots,h_{s_j}$ according to $\D$.
   \STATE \label{st:unif2} Select $s_j $ points
       $\ell_1,\dots,\ell_{s_j}$ independently and uniformly from $[N].$
   \FORALL{ pairs $(x,y)=(i_r,h_{r'})$ and $(x,y)=(i_r,\ell_{r'})$ (where $1\leq r \leq q$, $1\leq r' \leq s_j$) }\label{st:COND}
  \STATE{Call {\sc Compare}$_\D(\{x\},\{y\},\Theta(\eps 2^j), 2,
\newer{\exp(-\Theta(t))})$.} \label{st:call-compare:unif}
  \IF{the {\sc Compare} call does not return a value in $[1 - 2^{j-5} {\frac \eps 4}, 1 + 2^{j-5} {\frac \eps 4}]$ }
      \STATE output \reject (and exit). \label{st:rej}
    \ENDIF
    \ENDFOR
   \ENDFOR
\STATE Output \accept
\end{algorithmic}\caption{\label{algo:testing-uniformity}\sc $\PCOND_\D$-Test-Uniform}
\end{algorithm}

% \rnote{Note:  we should fix notation of how we specify the distribution
% that {\sc Compare} is being called with.  We've been leaving it implicit
% so far but maybe later there will be {\sc Compare} calls in settings where
% there are both $\D_1$ and $\D_2$ in the picture so perhaps it's better to
% always spell this out?  I did this in the algorithm above,
% writing {\sc Compare}$_{\D}$.}
% \dnote{I tend to say that when there is only a single distribution
% $\D$, then we don't need to state it explicitly, but I don't mind very
% much - I do prefer writing {\sc Compare}$^{\D}$ (as an oracle notation).}

% \rnote{I changed the parameters of the algorithm slightly; it had been using $m_j = \bigTheta{ \frac{\log t}{\eps^2  2^{2j}} }$ $\PCOND$ calls to do the comparison, but it seems to me we need $t$ rather than $\log t$ in the numerator for the union bound in the completeness case, since we have $\poly(1/\eps)$ pairs that are tested. I changed the next line giving the query complexity bound accordingly -- it just replaces a $\log\log(1/\eps)$ by $\log(1/\eps)$.}
% \dnote{You are right.}

% Let $m_j = \bigTheta{ \frac{t}{\eps^2  2^{2j}} }$ denote the number
Let $m_j$ denote the number of $\PCOND_\D$ queries used to run {\sc Compare$_\D$} in a given execution
of Line~\ref{st:call-compare:unif} during the $j$-th iteration of the outer
loop. By the setting of the parameters in each such
call and \autoref{lem:compare},
$m_j = \bigO{ \frac{t}{\eps^2  2^{2j}} }$.
It is easy to see that the algorithm only performs $\PCOND_\D$ queries
and that
the total number of queries that the algorithm performs is
\[
  \bigO{ \sum_{j=1}^t q\cdot s_j \cdot m_j }
      = \bigO{ \sum_{j=1}^t 2^j\log\left(\frac{1}{\eps}\right) \cdot \frac{\log(\frac{1}{\eps})}{\eps^2  2^{2j}} }
      = \bigO{ \frac{(\log(\frac{1}{\eps}))^2}{\eps^2} }\;.
\]
We prove \autoref{thm:paircond-test-uniform} by arguing completeness and
soundness below.

%%%%%%%%%%%%%%%%%%%%%%%%%%%%%%%%%%%%%%%%%%%%%%%%%%%%%%%%%%%%%%%%%%%%%%%%%%%%%%%%%%%%%%%%%%%%%%%%%%%%%%%%%%%%%%%%%%%%%%%%%%%%%%%%%%%%%%%%%%%%%%%%%%%%%%%%%

\noindent {\bf Completeness:}  Suppose that $\D$ is the uniform
distribution. Then for any fixed pair of points $(x,y)$,
\autoref{lem:compare} implies that the call to {\sc Compare}
on $\{x\},\{y\}$ in Line~\ref{st:call-compare:unif} causes the algorithm
to output \reject in Line~\ref{st:rej} with probability
at most $e^{-\Theta(t)}=
\poly(\eps).$
By taking a union bound over all $\poly(1/\eps)$ pairs of points considered
by the algorithm, the algorithm will accept with probability at least
$2/3$, as required.

\medskip

\noindent {\bf Soundness:}
Now suppose that $\D$ is $\eps$-far from uniform (we assume
throughout the analysis that $\eps=1/2^k$ for some integer $k$,
which is clearly without loss of generality).
We define $H,L$ as above and further
partition $H$ and $L$ into ``buckets'' as follows:
for $j = 1,\dots, t-1 = \log(\frac{4}{\eps})$, let
\[
  H_j \eqdef \setOfSuchThat{ h }{ \left(1+ 2^{j-1}\cdot \frac{\eps}{4}\right)\cdot \frac{1}{N} \leq  \D(h) < \left(1+ 2^{j}\cdot \frac{\eps}{4}\right)\cdot \frac{1}{N}  }\;,
\]
and \newer{for $j=1,\dots,t-2$ let
\[
  L_j \eqdef \setOfSuchThat{ \ell }{ \left(1-2^{j}\cdot \frac{\eps}{4}\right)\cdot \frac{1}{N} <  \D(\ell) \leq \left(1- 2^{j-1}\cdot \frac{\eps}{4}\right)\cdot \frac{1}{N}  }\;.
\]
}
Also define
\[
  H_0 \eqdef \setOfSuchThat{ h }{ \frac{1}{N} \leq \D(h) < \left(1+ \frac{\eps}{4}\right)\cdot \frac{1}{N}  }\;,\;\;
  L_0 \eqdef \setOfSuchThat{ \ell }{ \left(1- \frac{\eps}{4}\right)\cdot \frac{1}{N} < \D(\ell) < \frac{1}{N} }\;,
\]
and
\[
  H_t \eqdef \setOfSuchThat{ h }{ \D(h) \geq \frac{2}{N}  }\;,\;\;
  L_{\newer{t-1}} \eqdef \setOfSuchThat{ \ell }{ \D(\ell) \leq \frac{1}{2N} }\;.
\]
First observe that by the definition of $H_0$ and $L_0$, we have
\[
  \sum_{h\in H_0} \left( \D(h) - \frac{1}{N} \right) \leq \frac{\eps}{4}
  \;\text{ and }\;
  \sum_{\ell\in L_0} \left( \frac{1}{N} - \D(\ell) \right) \leq \frac{\eps}{4}\;.
\]
Therefore (by Equation~(\ref{eq:eps-far1})) we have
\[
  \sum_{j=1}^t \sum_{h\in H_j} \left( \D(h) - \frac{1}{N} \right) \geq \frac{\eps}{4}
  \;\text{ and }\;
  \sum_{j=1}^{\newer{t-1}}\sum_{\ell\in L_j} \left( \frac{1}{N} - \D(\ell) \right) \geq \frac{\eps}{4}\;.
\]
This implies that for some $1\leq j(H) \leq t$,
and some $1 \leq j(L) \leq \newer{t-1}$, we have
\begin{equation}\label{eq:j-H}
  \sum_{h\in H_{j(H)}} \left( \D(h) - \frac{1}{N} \right) \geq \frac{\eps}{4t}
  \;\text{ and }\;
  \sum_{\ell\in L_{j(L)}} \left(\frac{1}{N} - \D(\ell) \right) \geq \frac{\eps}{4t}\;.
\end{equation}

The rest of the analysis is divided into two cases depending on
whether $\abs{L} \geq \frac{N}{2}$ or $\abs{H} > \frac{N}{2}$.

\smallskip\noindent
{\bf Case 1: $\abs{L} \geq \frac{N}{2}$.}
In this case, with probability at least $99/100$,
in Line~\ref{st:unif} the algorithm will select at least one point $i_r \in L$.
We consider two subcases: $j(H) = t$, and $j(H) \leq t-1$.
\BI
  \item $j(H)=t$:  In this subcase, by Equation~(\ref{eq:j-H}) we have that
$\sum_{h\in H_{j(H)}} \D(h) \geq \frac{\eps}{4t}$. This implies that
when $j = j(H) = t = \log(\frac{4}{\eps}) + 1$,
so that $s_j = s_t = \bigTheta{\frac{t}{\eps}}$, with
probability at least $99/100$ the algorithm selects
 a point $h_{r'} \in H_t$ in Line~\ref{st:D}.
Assume that indeed such a point $h_{r'}$ is selected.
Since $\D(h_{r'}) \geq \frac{2}{N}$,
while $\D(i_{r}) < \frac{1}{N}$,
\autoref{lem:compare} implies that with probability at least
$1 - \poly(\eps)$ the {\sc Compare} call in Line~\ref{st:call-compare:unif}
outputs either \high or a value that is at least $
{\frac 7 {12}} = {\frac 1 2} + {\frac 1 {12}}.$
Since ${\frac 7 {12}} > {\frac 1 2} + 2^{j-5} {\frac \eps 4}$ for
$j=t$, the algorithm will output \reject in Line~\ref{st:rej}.

\item $j(H) < t$:  By Equation~(\ref{eq:j-H}) and the definition of the
buckets, we have
\[ \sum_{h\in H_{j(H)}} \left(\left(1+2^{j(H)}
\frac{\eps}{4}\right)\frac{1}{N} - \frac{1}{N}\right) \geq \frac{\eps}{4t}\;,
\]
implying that
$\abs{ H_{j(H)}} \geq \frac{N}{2^{j(H)} t}$ so that $\D(H_{j(H)})
\geq \frac{1}{2^{j(H)} t}$. Therefore, when $j = j(H)$ so that \mbox{$s_j =
\bigTheta{2^{j(H)} t}$}, with probability at least $99/100$
the algorithm will
get a point $h_{r'} \in H_{j(H)}$ in Line~\ref{st:D}.
Assume that indeed such a point $h_{r'}$ is selected.
Since \mbox{$\D(h_{r'}) \geq \left(1 + 2^{j(H)-1}\frac{\eps}{4}\right)
\frac{1}{N}$ }, while $\D(i_{r}) \leq \frac{1}{N}$, for $\alpha_{j(H)}
= 2^{j(H)-1}\frac{\eps}{4}$, we have
\[
  \frac{\D(h_{r'})}{\D(i_{r})} \geq 1 +\alpha_{j(H)}\;.
\]
Since {\sc Compare} is called in Line~\ref{st:call-compare:unif}
on the pair $\{i_r\},\{h_{r'}\}$
with the ``$\delta$'' parameter set to $\Theta(\eps 2^j)$,
with probability $1 - \poly(\eps)$ the algorithm outputs \reject
as a result of this {\sc Compare} call.

\EI

\noindent
{\bf Case 2: $\abs{H} > \frac{N}{2}$.}  This proceeds similarly to Case~1.
In this case we have that with high constant probability the algorithm
selects a point $i_r \in H$ in Line~\ref{st:unif}.
Here we consider the subcases $j(L) = \newer{t-1}$ and $j(L) \leq
\newer{t-2}$.
In the first subcase we have that
$\sum_{\ell \in L_t} \frac{1}{N} \geq \frac{\eps}{4t}$, so that $\abs{ L_t} \geq (\frac{\eps}{4t})N$,
and in the second case we have that
$\sum_{\ell \in L_{j(L)}} (2^{j(L)}\frac{\eps}{4})\frac{1}{N} \geq \frac{\eps}{4t}$,
so that $\abs{ L_{j(L)}} \geq \frac{N}{2^{j(L)} t}$.
The analysis of each subcase is similar to Case~1. %\rnote{should we write it out or no?}\dnote{maybe we should put this under a lower priority right now}
This concludes the proof of \autoref{thm:paircond-test-uniform}. \qed

%%%%%%%%%%%%%%%%%%%%%%%%%%%%%%%%%%%%%%%%%%%%%%%%%%%%%%%%%%%%%%%%%%%%%%%%%%%%%%%%%%%%%%%%%%%%%%%%%%%%%%%%%%%%%%%%%%%%%%%%%%%%%%%%%%%%%%%%%%%%%%%%%%%%%%%%%

\subsection{An \texorpdfstring{$\Omega(1/\eps^2)$}{Omega(1/eps2)} lower bound for \texorpdfstring{$\COND_\D$}{COND} algorithms that test uniformity} \label{ssec:unif-lb}

In this subsection we give a lower bound showing that
the query complexity of the $\PCOND_\D$ algorithm
of the previous subsection is essentially optimal, even for
algorithms that may make general $\COND_\D$ queries:

\BT \label{thm:unif-lb}
Any $\COND_\D$ algorithm for testing whether $\D = \calU$
versus $\dtv(\D,\calU) \geq \eps$ must make $\Omega(1/\eps^2)$
queries.
\ET

The high-level idea behind \autoref{thm:unif-lb} is to
reduce it to the well-known fact that distinguishing a fair coin
from a $(\half + 4\eps)$-biased coin requires $\bigOmega{\frac{1}{\eps^2}}$
coin tosses.  We show that any $q$-query algorithm $\COND_\D$ testing
algorithm $A$ can be transformed into an algorithm $A'$ that successfully
distinguishes $q$ tosses of a fair coin from $q$ tosses of a
$(\half + 4 \eps)$-biased coin.

\medskip

\noindent {\bf Proof of \autoref{thm:unif-lb}:}
First note that
we may assume without loss of generality that $0 < \eps \leq 1/8.$
Let $A$ be any $q$-query algorithm that makes $\COND_\D$ queries
and tests whether $\D = \calU$ versus $\dtv(\D,\calU) \geq \eps.$
We may assume without loss of generality that in every possible
execution algorithm $A$ makes precisely $q$ queries (this will be
convenient later).

Let $\D_{\text \no}$ be the distribution that has $\D_{\text \no}(i)=
\frac{1+2\eps}{N}$ for each $i \in \left[1,\frac{N}{2}\right]$ and
has $\D_{\text \no}(i) = \frac{1-2\eps}{N}$ for
each $i \in \left[\frac{N}{2}+1, N\right].$
(This is the ``no''-distribution for our lower bound; it is
$\eps$-far in variation
distance from the uniform distribution $\calU.$)
By \autoref{def:testing}, it must be the case
that
\[
Z := \left|
\Pr\left[A^{\COND_{\D_{\text \no}}} \text{~outputs~}\accept\right] -
\Pr\left[A^{\COND_\calU} \text{~outputs~}\accept\right]
\right| \geq 1/3.
\]

The proof works by showing that given $A$ as described above, there
must exist an algorithm $A'$ with the following properties:
$A'$ is given as input a $q$-bit string $(b_1,\dots,b_q) \in \{0,1\}^q$.
Let $\D_0$ denote the uniform distribution over $\{0,1\}^q$
and let $\D_{4\eps}$ denote the distribution over
$\{0,1\}^q$ in which each coordinate is independently set to 1
with probability $1/2 + 4 \eps$.
Then algorithm $A'$ has
\BEQ \label{eq:fair-vs-bias}
\left|
\Pr_{b \sim \D_0}[A'(b) \text{~outputs~}\accept] -
|\Pr_{b \sim \D_{4 \eps}}[A'(b) \text{~outputs~}\accept]
\right|
= Z.
\EEQ
Given (\ref{eq:fair-vs-bias}), by the data processing inequality for total
variation distance
(\autoref{lemma:data:processing:inequality:total:variation})
we have that $Z \leq \dtv(\D_0,\D_{4 \eps}).$
It is easy to see that $\dtv(\D_0,\D_{4\eps})$ is precisely equal to
the variation distance $\dtv(\Bin(q,1/2),\Bin(q,1/2 + 4 \eps)).$
However, in order for the variation distance
between these two binomial distributions to be as large as $1/3$
it must be the case that $q \geq \Omega(1/\eps^2)$:

\BFCT[Distinguishing Fair from Biased Coin]\label{fact:favourite:fact:fair:biased:coin}
Suppose $m\leq\frac{c}{\eps^2}$, with $c$ a sufficiently small constant
and $\eps \leq 1/8.$ Then,
\[ \totalvardist{ \binomial{m}{\half} }{ \binomial{m}{\half+4\eps} } %ROCCO:  was \half+2\eps
\leq \frac{1}{3}. \]
%In particular, distinguish between a fair and a $2\eps$-biased coins
%requires $\bigOmega{\frac{1}{\eps^2}}$ samples.
\EFCT
(\autoref{fact:favourite:fact:fair:biased:coin} is well known; it follows,
for example, as an immediate consequence of Equations~(2.15) and~(2.16)
of~\cite{AdellJodra:06}.)
Thus to prove \autoref{thm:unif-lb} it remains only to describe
algorithm~$A'$ and prove Equation~(\ref{eq:fair-vs-bias}).

As suggested above, algorithm~$A'$ uses algorithm~$A$; in order to do this,
it must perfectly
simulate the $\COND_\D$ oracle that $A$ requires, both in the
case when $\D = \calU$ and in the case when
$\D = \D_{\text \no}.$  We show below that when its input $b$
is drawn from $\D_0$ then $A'$ can perfectly simulate the
execution of $A$ when it is run on the $\COND_\calU$ oracle,
and when $b$ is drawn from $\D_{4 \eps}$ then
$A'$ can perfectly simulate the execution of $A$ when it is run on the
$\COND_{\D_{\text \no}}$ oracle.

Fix any step $1 \leq t \leq q$.  We now describe how $A'$ perfectly
simulates the $t$-th step of the execution of $A$ (i.e. the $t$-th call
to $\COND_\D$ that $A$ makes, and the response of $\COND_\D$).
We may inductively assume that $A'$ has perfectly simulated the
first $t-1$ steps of the execution of $A$.

For each possible prefix of $t-1$ query-response pairs to $\COND_\D$
\[
\mathit{PREFIX} = ((S_1,s_1),...,(S_{t-1},s_{t-1}))
\]
(where each $S_i \subseteq [N]$ and each $s_i \in S_i$),
there is some distribution $\P_{A,\mathit{PREFIX}}$ over possible
$t$-th query sets $S_t$ that $A$ would make given that its first $t-1$
query-response pairs were $\mathit{PREFIX}.$
So for a set $\newer{S_t} \subseteq [N]$ and a possible prefix
$\mathit{PREFIX},$ the value $P_{A,\mathit{PREFIX}}(\newer{S_t})$ is the probability that algorithm $A,$
having had the transcript of its execution thus far be $\mathit{PREFIX},$
generates set $S_t$ as its $t$-th query set.
For any query set {$S \subseteq [N]$}, let us write $S$ as a disjoint union
$S = S_{0} \disjunion S_{1}$, where $S_{0}=S \cap
\left[1,\frac{N}{2}\right]$ and $S_{1}=S \cap [\frac{N}{2}+1,N]$.
We may assume that every query $S$ ever used by $A$ has
$\abs{ S_0 },\abs{ S_1 } \geq 1$ (for otherwise $A$ could perfectly
simulate the response of $\COND_\D(S)$ whether $\D$ were
$\calU$ or $\D_{\text \no}$ by simply choosing a uniform point
from $S$, so there would be no need to call $\COND_\D$ on such an $S$).
Thus we may assume that $P_{A,\mathit{PREFIX}}(S)$ is nonzero only for sets $S$
that have $|S_0|,|S_1| \geq 1.$

Consider the bit $b_t \in \{0,1\}$. As noted above, we inductively have
that (whether $\D$ is $\calU$ or $\D_{\text \no}$) the algorithm
$A'$ has perfectly simulated the execution of $A$ for its first
$t-1$ query-response pairs; in this simulation some prefix
$\mathit{PREFIX}=((S_1,s_1),\dots,(S_{t-1},s_{t-1}))$ of query-response
pairs has been constructed.  If $b=(b_1,\dots,b_q)$ is distributed
according to $\D_0$ then $\mathit{PREFIX}$ is distributed exactly according to
the distribution of $A$'s prefixes of length $t-1$ when $A$
is run with $\COND_\calU$, and if $b=(b_1,\dots,b_q)$ is distributed
according to $\D_{4\eps}$ then the distribution of $\mathit{PREFIX}$ is
exactly the distribution of $A$'s prefixes of length $t-1$ when $A$
is run with $\COND_{\D_{\text \no}}$.

\noindent Algorithm $A'$ simulates the $t$-th stage of the execution of $A$
as follows:
\begin{enumerate}
\item Randomly choose a set {$S\subseteq [N]$} according to the
distribution $P_{A,\mathit{PREFIX}}$; let $S=S_0 \disjunion S_1$
be the set that is selected.  Let us write $\alpha(S)$ to denote
\newer{$|S_1|/|S_0|$} (so $\alpha(S) \in [2/N,N/2]$).

\item If $b_t = 1$ then set the bit $\sigma \in \{0,1\}$ to
be 1 with probability $u_t$ and to be 0 with
probability $1-u_t$. If $b_t= 0$ then set $\sigma$ to be 1
with probability $v_t$ and to be 0 with probability $1-v_t$.
(We specify the exact values of $u_t,v_t$ below.)

\item Set $s$ to be a uniform random element of $S_{\sigma}.$
Output the query-response pair $(S_t,s_t)=(S,s).$

\end{enumerate}

It is clear that Step~1 above perfectly simulates the $t$-th query that
algorithm~$A$ would make (no matter what is the distribution $\D$).  To
show that the $t$-th response is simulated perfectly, we must show that

\begin{itemize}

\item [(i)] if $b_t$ is uniform random over $\{0,1\}$ then
$s$ is distributed exactly as it would be distributed if $A$
were being run on $\COND_\calU$ and had just proposed $S$ as a query to
$\COND_\calU$; i.e. we must show that $s$ is a uniform random element
of $S_1$ with probability $p(\alpha) \eqdef {\frac {\alpha}{\alpha + 1}}$
and is a uniform random element of $S_0$ with probability $1-p(\alpha).$

\item [(ii)] if $b_t \in \{0,1\}$ has $\Pr[b_t=1]=1/2 + 4 \eps$,
then $s$ is distributed exactly as it would be distributed if $A$
were being run on $\COND_{\D_{\text \no}}$
and had just proposed $S$ as a query to
$\COND_\calU$; i.e. we must show that $s$ is a uniform random element
of $S_1$ with probability $q(\alpha) \eqdef {\frac {\alpha}{\alpha +
(1+2\eps)/(1-2\eps)}}$
and is a uniform random element of $S_0$ with probability $1-q(\alpha).$

\end{itemize}

By (i), we require that
\BEQ \label{eq:p}
{\frac {u_t} 2} + {\frac {v_t} 2} = p(\alpha) = {\frac \alpha {\alpha+1}},
\EEQ
and by (ii) we require that
\BEQ \label{eq:q}
\left({\frac 1 2} + 4\eps\right)u_t +
\left({\frac 1 2} - 4\eps\right)v_t = q(\alpha) =
{\frac \alpha {\alpha + {\frac {1+2\eps}{1-2\eps}}}}
\EEQ
It is straightforward to check that
\ignore{
% START OLD EQUATION SOLUTION, NOW BELIEVED TO BE WRONG -- ROCCO
\[
u_t = {\frac {\alpha + 2 \alpha^2 + 4 \alpha \eps - 2 \alpha^2 \eps}
       {2 + 4 \alpha + 2 \alpha^2 + 4 \eps - 2 \alpha^2 \eps}},
\quad \quad
v_t = {\frac {3 \alpha + 2 \alpha^2 + 4 \alpha \eps - 2 \alpha^2 \eps}
       {2 + 4 \alpha + 2 \alpha^2 + 4 \eps - 2 \alpha^2 \eps}},
\]
% END OLD EQUATION SOLUTION, NOW BELIEVED TO BE WRONG -- ROCCO
}
\newer{
\[
u_t = {\frac \alpha {\alpha+1}} \left(
1 - {\frac 1 {2((1-2\eps)\alpha + 1 + 2 \eps)}}
\right),
\quad \quad
v_t = {\frac \alpha {\alpha+1}} \left(
1 + {\frac 1 {2((1-2\eps)\alpha + 1 + 2 \eps)}}
\right)
\]
}
satisfy the above equations, and that for $0 < \alpha$, $0
< \eps \leq 1/8$ we have $0 \leq \newer{u_t,v_t} \leq 1$.
So indeed $A'$ perfectly simulates the execution of $A$ in all
stages $t=1,\dots,q$. Finally, after simulating the $t$-th stage
algorithm $A'$ outputs whatever is output by its simulation of $A$,
so Equation~\eqref{eq:fair-vs-bias} indeed holds.
This concludes the proof of \autoref{thm:unif-lb}. \qed

%        * both numerators are positive (the only negative part in either
%          one is -2alpha^2eps which is dominated by 2alpha^2)

%        * both denominators are positive (same reason, note both denoms
%          are same)

%        * (denom) > (numerator of v) > (numerator of u) (second inequality
%          is trivial, first seems to boil down to " 2 alpha eps <= 2 +
%          alpha + 2 eps" which is OK because we can assume eps<1/2 or 1/4
%          or whatever).

\section{Testing equivalence to a known distribution \texorpdfstring{$\D^\ast$}{D*}}
\label{sec:knownDstar}
\ifnum\confversion=0 %%%%%%%%%%%%%%%%%%%%%%%%%%%%%%%%%%%%%%%%%% Full version
\subsection{\texorpdfstring{A $\poly(\log n, 1/\eps)$-query $\PCOND_\D$ algorithm}{A poly(log n, 1/eps)-query \PCOND algorithm}}

In this subsection we present an algorithm {\sc $\PCOND$-Test-Known}
and prove the following theorem:

\BT \label{thm:paircond-test-known}
{\sc $\PCOND$-Test-Known} is a $\tilde{O}((\log N)^4/\eps^4)$-query $\PCOND_\D$
testing algorithm for testing equivalence to a known distribution $\D^\ast$.
{That is,}
for every pair of distributions $\D,\D^\ast$ over $[N]$
{(such that $\D^\ast$ is fully specified and there is $\PCOND$ query access to $\D$)}
the algorithm outputs \accept with probability at least $2/3$ if $\D =
\D^\ast$ and outputs \reject with probability at least $2/3$ if
\mbox{$\dtv(\D,\D^\ast) \geq \eps.$}
\ET

\ignore{
Recall that in this setting $\D^\ast$ is a fully specified distribution on
$[N]$, which is ``known'' to the algorithm, while $\D$ is unknown
and can only be accessed through queries to the \COND oracle.
}

\noindent
{\bf Intuition.}
Let $\D^\ast$ be a fully specified distribution,
and let $\D$ be a distribution
that may be accessed via a $\PCOND_\D$ oracle.
The high-level idea of the {\sc $\PCOND$-Test-Known} algorithm is the
following:  As in the case of
testing uniformity, we shall try to ``catch'' a pair of points $x,y$
such that\ignore{$\frac{\D(x)}{\D(x)+\D(y)}$}
{$\frac{\D(x)}{\D(y)}$} differs significantly
from
\ignore{$\frac{\D^\ast(x)}{\D^\ast(x)+\D^\ast(y)}$}
{$\frac{\D^\ast(x)}{\D^\ast(y)}$}
(so that
\ignore{querying $\COND_{\D}(\{x,y\})$ repeatedly}
{calling {\sc Compare}$_{\D}$ on $\{x\},\{y\}$}
will reveal this difference).
In the uniformity case,
where $\D^\ast(z) = 1/N$ for every $z$ (so that
$\frac{\D^\ast(x)}{\D^\ast(x)+\D^\ast(y)} = 1/2$), to get a
$\poly(1/\eps)$-query algorithm it was sufficient to show
that sampling
$\Theta(1/\epsilon)$ points uniformly (i.e., according to
$\D^\ast$)
with high probability
yields a point $x$ for which
$\D(x) < \D^\ast(x) - \Omega(\epsilon/N)$, and that sampling
$\Theta(1/\epsilon)$ points from $\SAMP_\D$
with high probability yields a point $y$ for
which $\D(x) > \D^\ast(y) + \Omega(\epsilon/N)$.
However, for general $\D^\ast$ it is not sufficient to get such a pair
because it is possible that $\D^\ast(y)$ could be much larger than
$\D^\ast(x)$.  If this were the case then it might happen that
both\ignore{$\frac{\D^\ast(x)}{\D^\ast(x)+\D^\ast(y)}$}
{$\frac{\D^\ast(x)}{\D^\ast(y)}$}
and\ignore{$\frac{\D(x)}{\D(x)+\D(y)}$}
{$\frac{\D(x)}{\D(y)}$} are very small, so\ignore{ very many queries
would be required to $\COND_{\D}(\{x,y\})$ to determine that}
{calling {\sc Compare}$_{\D}$ on $\{x\},\{y\}$ cannot efficiently
demonstrate that
$\frac{\D^\ast(x)}{\D^\ast(y)}$ differs from
$\frac{\D(x)}{\D(y)}$.}
\ignore{
$\frac{\D^\ast(x)}{\D^\ast(x)+\D^\ast(y)}$ differs from
$\frac{\D(x)}{\D(x)+\D(y)}$.
}

To address this issue
we partition the points into $O(\log N/\epsilon)$
``buckets'' so that within each bucket
all points have similar probability according to $\D^\ast$. We
show that if $\D$ is $\epsilon$-far from $\D^\ast$, then either
the probability weight of one of these buckets according to $\D$
differs significantly from what it is according to $\D^\ast$
(which can be observed by sampling from $\D$), or we can get
a pair $\{x,y\}$ {\em that belong to the same bucket\/} and for
which $\D(x)$ is sufficiently smaller than $\D^\ast(x)$ and
$\D(y)$ is sufficiently larger than $\D^\ast(y)$.  {For such a pair
{\sc Compare} will efficiently give evidence that
$\D$ differs from $\D^\ast$.}
\ignore{ with only a moderate
number of calls to {\sc Compare} $\COND_\D$ queries.
}

\medskip

\noindent {\bf The algorithm and its analysis.}
We define some quantities that are used in the algorithm and its analysis.
Let ${\eta} \eqdef \epsilon/c$ for some sufficiently large constant $c$
that will be determined later.
As described above we partition the domain elements $[N]$ into
``buckets'' according
to their probability weight in $\D^\ast$.
Specifically, for $j = 1,\dots, \lceil \log(N/\eta)+1\rceil$, we let
\begin{equation} \label{eq:Bj-def}
B_j \eqdef \{x \in [N]: 2^{j-1}\cdot \eta/N \leq \D^\ast(x) < 2^j\cdot \eta/N\}
\end{equation}
and we let $B_0 \eqdef \{x \in [N]: \D^\ast(x) < \eta/N\}$.
Let $b \eqdef \lceil \log(N/\eta)+1\rceil +1$ denote the number of buckets.

We further define $J^h \eqdef \{j: \D^\ast(B_j) \geq \eta/b\}$ to denote the set of indices
of ``heavy'' buckets, and let
$J^\ell \eqdef\{j: \D^\ast(B_j)< \eta/b\}$ denote the set of indices
of ``light'' buckets.  Note that we have
\begin{equation} \label{eq:Jell-light}
\sum_{j\in J^\ell\cup \{0\}}\D^\ast(B_j) < 2\eta.
\end{equation}

\begin{algorithm}
\begin{algorithmic}[1]
\REQUIRE error parameter $\eps > 0$; query access to $\PCOND_\D$ oracle;
explicit description $(\D^\ast(1),\dots,\D^\ast(N))$ of distribution $\D^\ast$
\STATE \label{st:init-sample}
Call the $\SAMP_\D$ oracle $m = \Theta(b^2 (\log b)/\eta^2)$
times to obtain points $h_1,\dots,h_{m}$ distributed according
to $\D$.
\FOR{ $j=0$ to $b$ }\STATE \label{st:est-hatDBj}Let
${\widehat{\D}}(B_j)$ be the fraction of points $h_1,\dots,
h_m$ that lie in $B_j$ \newer{(where the buckets $B_j$ are as defined in Equation~(\ref{eq:Bj-def}))}.
\IF {some $j$ has $|\D^\ast(B_j) - \widehat{\D}(B_j)| > \eta/b$}
\STATE \label{st:rej1} {output \reject and exit}
\ENDIF
\ENDFOR
\STATE \label{st:Dstarsamp}
Select $s=\Theta(b/\eps)$ points $x_1,\dots,x_s$ independently from $\D^\ast$.
\STATE \label{st:Dsamp}Call the $\SAMP_\D$ oracle $s=\Theta(b/\eps)$ times to obtain points $y_1,\dots,y_s$
distributed according to $\D.$
\FORALL{ pairs $(x_i,y_j)$ (where $1 \leq i,j \leq s$)
{ such that $\frac{D^\ast(x)}{D^\ast(y)} \in [1/2,2]$} } \label{st:loop}
\STATE \label{st:many}
{ Call {\sc Compare}$(\{x\},\{y\},\eta/(4b),2,1/(10s^2))$}
% perform $q=\Theta(b^2 (\log s)/\eta^2)$ $\PCOND_\D$ queries
% on the pair.
\IF
{{ {\sc Compare} returns \low or a value smaller than
$(1-\eta/(2b))\cdot \frac{D^\ast(x)}{D^\ast(y)}$}}
\label{st:checkfreq}
% {in some pair $(x_i,y_j)$ that satisfies
% $\frac{\D^\ast(x_i)}{\D^\ast(x_i) + \D^\ast(y_{j})}
% \in [1/3,2/3]$ the point $x_i$ is received with frequency less
% than$(1 - \eta/8b) \cdot
% \frac{\D^\ast(x_i)}{\D^\ast(x_i) + \D^\ast(y_{j})}$} \label{st:checkfreq}
\STATE output \reject (and exit) \label{st:r2}
\ENDIF
\ENDFOR
\STATE output \accept
\end{algorithmic}\caption{\label{algo:pcond-test-known}\sc $\PCOND_\D$-Test-Known}
\end{algorithm}
% \rnote{Things to check:
% \begin{itemize}
%   \item I changed the setting of $m=\#$ samples in Step~1;   it had been $\Theta(b^3/\delta^2)$ but it seems to me for the additive Chernoff bound all we need it to be is $\Theta(b^2 (\log b)/\delta^2).$
%   \item Some constant factors of $\delta$ and $\eps$ changed in various places -- they are currently colored red to be more  easily findable
% \end{itemize}}
% \dnote{The changes look fine to me. In addition, I changed $\delta$ to $\eta$ (since we usually use $\delta$ as a confidence parameter (here we don't need that) and we use $\eta$ as an estimation parameter.}

The query complexity of the algorithm is dominated by the number of $\PCOND_\D$ queries performed in the executions of {\sc Compare},
which by \autoref{lem:compare} is upper bounded by
\[
{O}(s^2 \cdot b^2 \cdot (\log s) / \eta^2) =
{O}\left(
{\frac {(\log {\frac N \eps})^4 \cdot \log \left( (\log {\frac N \eps})/\eps\right)}
{\eps^4}}
\right)\;.
\]
% calls to $\PCOND_\D.$
We argue completeness and soundness below.

\medskip

\noindent {\bf Completeness:}  Suppose that $\D = \D^\ast.$
Since the expected value of $\widehat{D}(B_j)$
(defined in Line~\ref{st:est-hatDBj}) is precisely
$\D^\ast(B_j)$, for any fixed value of
$j \in \{0,\dots,\lceil \log(N/\eta)+1\rceil\}$ an additive Chernoff
bound implies that $\left|\D^\ast(B_j) - \widehat{D}(B_j)\right| {>} \eta/b$
with failure probability at most $1/(10b)$.
By a union bound over all $b$ values of $j$, the algorithm outputs
\reject in Line~\ref{st:rej1} with probability at most $1/10$.
Later in the algorithm,
since $\D = \D^\ast$, no matter what points $x_i,y_j$
are sampled from $\D^\ast$ and $\D$ respectively,
{the following holds for each pair $(x_i,y_j)$
such that $\D^\ast(x)/\D^\ast(y) \in [1/2,2]$.
By \autoref{lem:compare} (and the setting of the parameters
in the calls to {\sc Compare}), the probability that {\sc Compare}
returns \low or a value smaller than
$(1-\delta/(2b))\cdot (\D^\ast(x)/\D^\ast(y))$, is at most
$1/(10s^2)$.}
% the expected frequency with which $x_i$ is received from a call to $\PCOND_\D(\{x_i,y_j\})$ is precisely $\frac{\D^\ast(x_i)}{\D^\ast(x_i) + \D^\ast(y_{j})}$.
% By the choice of $q$, a multiplicative Chernoff bound implies that the observed frequency for any fixed $(x_i,y_j)$ pair for which $\frac{\D^\ast(x_i)}{\D^\ast(x_i) + \D^\ast(y_{j})} \in [1/3,2/3]$
% is at least $(1 - \eta/8b) \frac{\D^\ast(x_i)}{\D^\ast(x_i) + \D^\ast(y_{j})}$
% with probability at least $1/(10s^2).$
A union bound over all (at most $s^2)$ pairs $(x_i,y_j)$
{for which $\D^\ast(x)/\D^\ast(y) \in [1/2,2]$},
% that have $\frac{\D^\ast(x_i)}{\D^\ast(x_i) + \D^\ast(y_{j})} \in [1/3,2/3]$
gives that the probability
of outputting \reject in Line~\ref{st:r2} is at most $1/10$.
Thus with overall probability
at least $8/10$ the algorithm outputs \accept.

\bigskip

\noindent {\bf Soundness:}  Now suppose that $\dtv(\D,\D^\ast) \geq \eps$; our goal is to
show that the algorithm rejects with probability at least $2/3.$
Since the algorithm rejects if any estimate
$\widehat{\D}(B_j)$ obtained in Line~\ref{st:est-hatDBj} deviates from $\D^\ast(B_j)$ by
more than $\pm \eta/b$, we may assume that all these estimates are indeed $\pm \eta/b$-close
to the values $\D^\ast(B_j)$ as required.
Moreover, by an additive Chernoff bound (as in the completeness analysis), we have that
with overall failure probability at most $1/10$, each $j$ has
$|\widehat{\D}(B_j) - \D(B_j)| \leq \eta/b$; we condition on this event going forth.
Thus, for every $0 \leq j \leq b$,
\begin{equation}
\D^\ast(B_j) - 2\eta/b \leq \D(B_j) \leq \D^\ast(B_j) + 2\eta/b\;.
\label{est-b.eq}
\end{equation}
Recalling the definition of $J^\ell$ and Equation~(\ref{eq:Jell-light}), we see that

% By our assumption on the estimates, we have that for every
% $j \in J^h$,
% \begin{equation}
% (1-\eta)\D^\ast(B_j) \leq \D(B_j) \leq (1+\eta)\D^\ast(B_j)\;.
% \label{heavy-b.eq}
% \end{equation}
% On then other hand
\begin{equation}
%\sum_{j\in J^\ell\cup \{0\}}\D^\ast(B_j) < 2\eta\;\;\mbox{ and }\;\;
\sum_{j\in J^\ell \cup \{0\}}\D(B_j) < 4\eta\;.
\label{light-b.eq}
\end{equation}
Let
\BEQ
d_j \eqdef \sum_{x\in B_j} |\D^\ast(x) - \D(x)|\;,
\EEQ
so that $\|\D^\ast - \D\|_1 = \sum_j d_j$.
By Equations~(\ref{eq:Jell-light}) and~(\ref{light-b.eq}), we have
\begin{equation}
\sum_{j\in J^\ell\cup\{0\}} d_j \leq
\sum_{j\in J^\ell\cup\{0\}} \left(\D^\ast(B_j) + \D(B_j)\right)
\leq 6\eta\;.
\end{equation}
Since we have (by assumption) that $\|\D^\ast - \D\|_1 = 2 \dtv(\D^\ast,\D) \geq 2\epsilon,$
we get that
\begin{equation}
\sum_{j \in J^h\setminus \{0\}} d_j > 2\epsilon-6\eta\;.
\label{sum-high.eq}
\end{equation}

Let ${N_j} \eqdef |B_j|$ and observe that
$N_j \leq \D^\ast(B_j)/p_j\leq 1/p_j$, where
$p_j \eqdef 2^{j-1}\cdot\eta/N$ is the lower bound on
the probability (under $\D^\ast$) of all elements in $B_j$.
For each $B_j$ such that $j \in J^h\setminus \{0\}$, let
$H_j \eqdef \{x \in B_j: \D(x) > \D^\ast(x)\}$
and $L_j \eqdef \{x\in B_j: \D(x)< \D^\ast(x)\}$.
Similarly to the ``testing uniformity'' analysis, we have that
\begin{equation}
\sum_{x\in L_j} (\D^\ast(x) - \D(x))
+ \sum_{x\in H_j} (\D(x) - \D^\ast(x)) = d_j\;.
\end{equation}
Equation~(\ref{est-b.eq}) may be rewritten as
\begin{equation}
\left| \sum_{x\in L_j} (\D^\ast(x) - \D(x))
- \sum_{x\in H_j} (\D(x) - \D^\ast(x)) \right|
\leq 2\eta/b\;,
%    \leq \eta \D^\ast(B_j)\;,
\end{equation}
and so we have both
\begin{equation}
\sum_{x\in L_j} (\D^\ast(x) - \D(x)) \geq d_j/2 - \eta/b
\;\;
\mbox{ and }
\;\;
\sum_{x\in H_j} (\D(x) - \D^\ast(x)) \geq d_j/2 - \eta/b\;.
\label{Lj-Hj.eq}
\end{equation}
Also similarly to what we had before, let
$H'_j \eqdef \{x \in B_j: \D(x) > \D^\ast(x)+ \eta/(b N_j)\}$,
and $L'_j  \eqdef \{x\in B_j: \D(x)< \D^\ast(x)-\eta/(b N_j)\}$
(recall that $N_j = |B_j|$); these are the elements of $B_j$ that are
``significantly heavier'' (lighter, respectively) under $\D$ than
under $\D^\ast$.  We have
\begin{equation}
\sum_{x\in L_j\setminus L'_j} (\D^\ast(x) - \D(x)) \leq \eta/b
\;\;
\mbox{ and }
\;\;
\sum_{x\in H_j\setminus H'_j} ({\D(x) - \D^\ast(x)}) \leq \eta/b\;.
\label{jj.eq}
\end{equation}

By Equation~(\ref{sum-high.eq}), there exists
$j^\ast \in J^h\setminus \{0\}$
for which $d_{j^\ast} \geq (2\epsilon-6\eta)/b$. For this
index, applying Equations~(\ref{Lj-Hj.eq}) and~(\ref{jj.eq}), we get that
\begin{equation}
\sum_{x\in L'_{j^\ast}} \D^\ast(x) \geq
\sum_{x\in L'_{j^\ast}} (\D^\ast(x) - \D(x))
\geq (\epsilon - 5\eta)/b \;,
\end{equation}
and similarly,
\begin{equation}
\sum_{x\in H'_{j^\ast}} \D(x) \geq
\sum_{x\in H'_{j^\ast}} (\D(x) - \D^\ast(x))
\geq (\epsilon - 5\eta)/b\;.
\end{equation}

\newer{Recalling that $\eta = \eps/c$ and setting the
constant $c$ to $6$}, we have that
$(\epsilon - 5\eta)/b = \epsilon/6b$.  Since $s=\Theta(b/\eps)$,
with probability at least 9/10 it is the case both that some $x_i$
drawn in Line~\ref{st:Dstarsamp} belongs to
$L'_{j^\ast}$ and that some $y_{i'}$ drawn in Line~\ref{st:Dsamp}
belongs to $H'_{j^\ast}$.
By the definitions of $L'_{j^\ast}$ and $H'_{j^\ast}$
and the fact for each $j >0$ it holds that
$N_j \leq 1/p_j$
and $p_j \leq \D^\ast(x) < 2 p_j$ for each $x_{\newer{i}} \in B_j$, we
have that
\begin{equation}
\D(x_i) < \D^\ast(x_i) - \eta/(b N_{j^\ast})
       \leq \D^\ast(x_i) - (\eta/b)p_{j^\ast}
       \leq (1-\eta/(2b))\D^{\ast}(x_i)\;
\end{equation}
and
\begin{equation}
\D(y_{i'}) > \D^\ast(y_{i'}) + \eta/(b N_{j^\ast})
     \geq \D^\ast(y_{i'})  + (\eta/b)p_j
     \geq (1+\eta/(2b)) \D^\ast(y_{i'})\;.
\end{equation}
Therefore,
{
\BEQ
\frac{\D(x_i)}{\D(y_{i'})}
   < \frac{1-\eta/(2b)}{1+\eta/(2b)}
   \cdot \frac{\D^\ast(x_i)}{\D^\ast(y_{i'})}
    < \left(1-\frac{3\eta}{4b}\right) \cdot
       \frac{\D^\ast(x_i)}{\D^\ast(y_{i'})}\;.
\EEQ
By \autoref{lem:compare}, with probability at least $1-1/(10s^2)$,
the output of {\sc Compare} is either \low or is at most
$\left(1-\frac{3\eta}{4b}\right)\cdot\left(1+\frac{\eta}{4b}\right)
 < \left(1-\frac{\eta}{2b}\right)$, causing the algorithm to reject.
}
%%%%%%%%%%%%%%
  % \ignore{
  % \begin{eqnarray}
  % \frac{\D(x_i)}{\D(x_i) + \D(y_{i'})} &=&
  % \frac{1}{1+ \D(y_{i'})/\D(x_i)} \\
  % &\leq&  \frac{1}{1+ ((1+\eta/(2b))\D^\ast(y_{i'}))/(1-\eta/(2b))\D^\ast(x_i))} \\
  % &=&\frac{(1-\eta/(2b))\D^\ast(x_i)}{\D^\ast(x_i)+\D^\ast(y_{i'})
  %     -(\eta/(2b))(\D^\ast(x_i)-\D^\ast(y_{i'}))}\;.\\
  % \end{eqnarray}
  % Since $p_j \leq \D^\ast(x_i),\D^\ast(y_{i'}) \leq 2p_j$ we have that
  % $\D^\ast(x_i) - \D^\ast(y_{i'}) \leq (\D^\ast(x_i)+\D^\ast(y_{i'}))/2$, and so
  % \begin{equation}
  % \frac{\D(x_i)}{\D(x_i) + \D(y_{i'})} \leq {\frac {1-\eta/(2b)}{1-\eta/(4b)}} \cdot
  % \frac{\D^\ast(x_i)}{\D^\ast(x_i) + \D^\ast(y_{i'})}
  % \leq
  % (1-\eta/(4b))\cdot\frac{\D^\ast(x_i)}{\D^\ast(x_i) + \D^\ast(y_{i'})}.
  % \end{equation}
  % We have that
  % $\frac{\D^\ast(x_i)}{\D^\ast(x_i) + \D^\ast(y_{i'})} \in [1/3,2/3].$
  % Consequently, by the choice of $q$, a multiplicative Chernoff bound implies
  % that if the algorithm reaches Step~\ref{st:checkfreq},
  % the pair $(x_i,y_{i'})$ will cause it
  % to output \reject with probability at least $1-1/(10s^2).$
  % }
%%%%%%%%%%%%%%%%
Thus the overall probability
that the algorithm outputs \reject is at least $8/10 - 1/(10s^2) > 2/3$,
and the theorem is proved.
\qed

\subsection{A \texorpdfstring{$(\log N)^{\Omega(1)}$}{(log N)\^{}Omega(1)}
   lower bound for \texorpdfstring{$\PCOND_\D$}{\PCOND}}
\label{ssec:lb-Pcond-Dstar}

In this subsection we prove that any $\PCOND_\D$ algorithm
for testing equivalence to a known distribution must have
query complexity at least $(\log N)^{\Omega(1)}$:

\BT \label{thm:pcond-ddstar-lb}
Fix $\eps = 1/2.$ There is a distribution $\D^\ast$ over $[N]$
(described below), which is such that
any $\PCOND_\D$ algorithm for testing whether $\D = \D^\ast$
versus $\dtv(\D,\D^\ast) \geq \eps$ must make $\Omega\left(
\sqrt{{\frac {\log N}{\log \log N}}}
\right)$ queries.
\ET

\noindent
{\bf The distribution $\D^\ast$.}
Fix parameters $r = \bigTheta{\frac{\log N}{\log\log N}}$ and $K=\Theta(\log N).$
We partition $[N]$ from left ($1$) to right ($N$)
into $2r$ consecutive intervals $B_1,\dots,B_{2r}$, which we henceforth refer to as ``buckets.''  The $i$-th bucket has $|B_i|=K^i$ (we may assume without loss of generality that
$N$ is of the form $\sum_{i=1}^{\newer{2r}} K^i$).
The distribution $\D^\ast$ assigns equal probability weight to each bucket, so
$\D^\ast(B_i)=1/(2r)$ for all $1 \leq i \leq 2r.$  Moreover $\D^\ast$ is uniform within each
bucket, so for all $j \in B_i$ we have $\D^\ast(j) = 1/(2r K^i).$  This completes
the specification of $\D^\ast$.

To prove the lower bound we construct
a probability distribution $\calP_{\text \no}$ over possible
``No''-distributions.  To define the distribution $\calP_{\text \no}$
it will be useful to have
the notion of a ``bucket-pair.''  A bucket-pair $U_i$ is $U_i = B_{2i-1} \cup B_{2i}$, i.e. the
union of the $i$-th pair of consecutive buckets.

A distribution $\D$ drawn from $\calP_{\text \no}$ is obtained by
selecting a string $\pi=(\pi_1,\dots,\pi_r)$ uniformly at random
from $\{ \downarrow\uparrow, \uparrow\downarrow \}^r$
and setting $\D$ to be $D_\pi$, which we now define.
The distribution $\D_\pi$ is obtained by perturbing $\D^\ast$ in the following
way:
for each bucket-pair $U_i=(B_{2i-1},B_{2i})$,

\begin{itemize}

\item If $\pi_i = \uparrow \downarrow$ then the weight of $B_{2i-1}$ is
uniformly ``scaled up'' from $1/(2r)$ to $3/(4r)$ (keeping the distribution
uniform within $B_{2i-1}$) and the weight of $B_{2i}$ is uniformly
``scaled down'' from $1/(2r)$ to $1/(4r)$ (likewise keeping the distribution
uniform within $B_{2i}$).

\item If $\pi_i = \downarrow \uparrow$ then the weight of $B_{2i-1}$ is uniformly ``scaled down'' from $1/(2r)$ to $1/(4r)$ and the weight of $B_{2i}$ is uniformly ``scaled up''
from $1/(2r)$ to $3/(4r)$.

\end{itemize}

Note that for any
distribution $\D$ in the support of $\calP_{\text \no}$
and any $1 \leq i \leq r$
we have that $\D(U_i)=\D^\ast(U_i)=1/r.$

Every distribution $\D$ in the support of $\calP_{\text \no}$
has $\dtv(\D^\ast,\D) = 1/2.$  Thus
\autoref{thm:pcond-ddstar-lb} follows immediately from the following:

\BT \label{thm:ddstar-hard-to-distinguish}
Let $A$ be any (possibly adaptive) algorithm. which makes at most
$q \leq {\frac 1 3} \cdot  \sqrt{r}
$
calls to $\PCOND_\D$.
Then
\BEQ \label{eq:ddstar}
\left|
\Pr_{D \leftarrow \calP_{\text \no}}
   \left[A^{\PCOND_\D} \text{~outputs~} \mbox{\accept}\right] -
\Pr\left[A^{\PCOND_{\D^\ast}} \text{~outputs~} \mbox{\accept}\right]
\right|
\leq 1/5.
\EEQ
\ET
Note that in the first probability of
Equation~(\ref{eq:ddstar}) the randomness is over
the draw of $\D$ from $\calP_{\text \no}$,
the internal randomness of $A$ in selecting its query sets, and the randomness
of the responses to the $\PCOND_\D$ queries.
In the second probability the randomness is
just over the internal coin tosses of $A$ and the randomness of the responses
to the $\PCOND_\D$ queries.

\medskip\noindent
{\bf Intuition for \autoref{thm:ddstar-hard-to-distinguish}.}
A very high-level intuition for the lower bound is that $\PCOND_\D$
queries are only useful for ``comparing'' points whose probabilities
are within a reasonable multiplicative ratio of each other.
But $\D^\ast$ and every distribution $\D$ in the support of
\newer{$\calP^{\text \no}$} are such that every two points either have the
same probability mass under all of these distributions (so
a $\PCOND_\D$ query is not informative), or else the ratio of their
probabilities is so skewed that a small number of $\PCOND_\D$ queries
is not useful for comparing them.

In more detail, we may suppose without loss of generality that in every possible
execution, algorithm $A$ first makes $q$ calls to $\SAMP_\D$ and
then makes $q$ (possibly adaptive) calls to $\PCOND_\D.$
The more detailed intuition for the lower bound is as follows:  First consider
the $\SAMP_\D$ calls.  Since every
possible $\D$ (whether $\D^\ast$ or a distribution drawn from
$\calP_{\text \no}$) puts weight $1/r$ on each bucket-pair
$U_1,\dots,U_r$, a birthday paradox argument implies that in
both scenarios, with probability
at least $9/10$ (over the randomness in the responses to the $\SAMP_\D$
queries) no two of the $q \leq {\frac 1 3} \sqrt{r}$
calls to $\SAMP_\D$ return points from the
same bucket-pair.
Conditioned on this, the distribution of responses to the $\SAMP_\D$ queries
is exactly the same under $\D^\ast$ and under $\D$ where $\D$ is
drawn randomly from $\calP_{\text \no}.$

For the pair queries, the intuition is that in either setting
(whether the distribution $\D$ is $\D^\ast$ or a randomly
chosen distribution from $\calP_{\text \no}$),
making $q$ pair queries
will with $1-o(1)$ probability provide no information
that the tester could not simulate for itself.  This is because
any pair query $\PCOND_{\D}(\{x,y\})$ either has $x,y$ in the same
bucket $B_i$ or in different buckets $B_i \neq B_j$ with $i < j.$
If $x,y$ are both in the same bucket $B_i$ then in either setting
$\PCOND_{\D}(\{x,y\})$ is equally likely to return $x$ or $y$.
If they belong to buckets $B_i,B_j$ with $i<j$ then in
either setting $\PCOND_\D(\{x,y\})$ will return the
one that belongs to $P_i$ with probability
$1 - 1/\Theta(K^{j-i}) \geq 1 - 1/\Omega(K).$

\ignore{
For pair queries, if the pair is within the same bucket, then of course we see exactly the same conditional probability distribution. If they belong to neighboring buckets, then in less than $\frac{\log N}{c}$ samples (for some constant $c$), we'll just get the heavier point all the time (and if they belong to non-neighboring buckets, then it will be even more extreme).

For sampling, in less than $(\log N)^{\frac{1}{2}}$ samples we won't hit the same bucket-pair twice, the probability that we get any bucket-pair is the same, and within the pair, each bucket has equal probability (and each point in the bucket has equal probability).
}

\noindent
{\bf Proof of \autoref{thm:ddstar-hard-to-distinguish}:}
As described above, we may fix $A$ to be any $\PCOND_\D$
algorithm that makes exactly $q$ calls to $\SAMP_\D$
%(i.e. to $\PCOND_\D([N])$),
% \dnote{removed: (i.e. to $\PCOND_\D([N])$)}
followed by
exactly $q$ adaptive calls to $\PCOND_\D.$

A \emph{transcript} for $A$ is a full specification of the sequence
of interactions that $A$ has with the
$\PCOND_\D$ oracle in a given execution. More precisely,
it is a pair $(Y,Z)$ where $Y=(s_1,\dots,s_q) \in [N]^q$ and
$Z=((\{x_1,y_1\},p_1),\dots,(\{x_q,y_q\},p_q))$, where $p_i \in \{x_i,y_i\}$
and $x_i,y_i \in [N].$
The idea is that $Y$ is a possible
sequence of responses that $A$ might receive
to the initial $q$ $\SAMP_\D$ queries, $\{x_i,y_i\}$ is a possible
pair that could be the input to an $i$-th $\PCOND_\D$ query,
and $p_i$ is a possible response that could be received from that query.

We say that a \emph{length-$i$ transcript prefix} is a
pair $(Y,Z^i)$ where $Y$ is as above and $Z^i=((\{x_1,y_1\},p_1),\dots,
(\{x_i,y_i\},p_i))$.
A $\PCOND$ algorithm $A$ may be viewed as a collection of distributions
over pairs $\{x,y\}$ in the following way:  for each
length-$i$ transcript-prefix $(Y,Z^i)$ ($0 \leq i \leq q-1$),
there is a distribution over pairs
$\{x_{i+1},y_{i+1}\}$ that $A$ would use to select the
$(i+1)$-st query pair for $\PCOND_\D$ given that the
length-$i$ transcript prefix of $A$'s execution thus far was $(Y,Z^i)$.
We write $\T_{(Y,Z^i)}$ to denote this distribution over pairs.

Let $\P^\ast$ denote the distribution over
transcripts induced by running $A$ with oracle
$\PCOND_{D^\ast}$.
Let $\P^{\text \no}$ denote the distribution
over transcripts induced by first (i) drawing $\D$
from $\calP_{\text \no}$, and then (ii) running $A$ with
oracle $\PCOND_{\D}.$
To prove \autoref{thm:ddstar-hard-to-distinguish} it is sufficient
to prove that the distribution  over transcripts of $A$ is statistically close
whether the oracle is $\D^\ast$ or is a random $\D$ drawn from
\newer{$\calP^{\text \no}$}, i.e. it is sufficient to prove that
\BEQ \label{eq:close}
\dtv(\P^\ast,\P^{\text \no}) \leq 1/5.
\EEQ

For our analysis we will need to consider variants of algorithm $A$
that, rather than making $q$ calls to $\PCOND_\D$, instead ``fake''
the final $q-k$ of these $\PCOND_\D$ queries as described below.
For $0 \leq k \leq q$ we define $A^{(k)}$ to be the algorithm
that works as follows:

\begin{enumerate}

\item $A^{(k)}$ exactly simulates the execution of $A$ in making
an initial $q$ $\SAMP_\D$ calls and making the first $k$ $\PCOND_\D$
queries precisely like $A$.  Let $(Y,Z^{k})$ be the length-$k$ transcript
prefix of $A$'s execution thus obtained.

\item Exactly like $A$, algorithm $A^{(k)}$ draws a pair
$\{x_{k+1},y_{k+1}\}$ from $\T_{(Y,Z^k)}$.
However, instead of calling $\PCOND_\D(\{x_{k+1},y_{k+1}\})$
to obtain $p_{k+1}$, algorithm $A^{(k)}$ generates $p_{k+1}$
in the following manner:

\begin{itemize}

\item [(i)] If $x_{k+1}$ and $y_{k+1}$ both belong to the same bucket $B_\ell$
then $p_{k+1}$ is chosen uniformly from $\{x_{k+1},y_{k+1}\}.$

\item [(ii)] If one of $\{x_{k+1},y_{k+1}\}$ belongs to $B_\ell$
and the other belongs to $B_{\ell'}$ for some $\ell < \ell'$,
then $p_{k+1}$ is set to be the element of
$\{x_{k+1},y_{k+1}\}$ that belongs to $B_\ell$.

\end{itemize}

Let $(Y,Z^{k+1})$ be the length-$(k+1)$ transcript prefix
obtained by appending $(\{x_{k+1},y_{k+1}\},p_{k+1})$ to $Z^k$.
Algorithm \newer{$A^{(k)}$} continues in this way for a total of $q-k$ stages; i.e.
it next draws $\{x_{k+2},y_{k+2}\}$ from $\T_{(Y,Z^{k+1})}$ and
generates $p_{k+2}$ as described above; then $(Y,Z^{k+2})$ is
the length-$(k+2)$ transcript prefix obtained by
appending $(\{x_{k+2},y_{k+2}\},p_{k+2})$ to $Z^{k+1}$;
and so on.  At the end of the process a transcript
$(Y,Z^q)$ has been constructed.

\end{enumerate}

Let $\P^{\ast,(k)}$ denote the distribution over final
transcripts $(Y,Z^q)$ that are obtained by running $A^{(k)}$ on a
$\PCOND_{\D^\ast}$ oracle.
Let $\P^{\text \no,(k)}$ denote the distribution over final
transcripts $(Y,Z^q)$ that are obtained by (i) first drawing
$\D$ from \newer{$\calP^{\text \no}$}, and then (ii) running $A^{(k)}$ on a
$\PCOND_{\D}$ oracle.
Note that $\P^{\ast,(q)}$ is identical to $\P^\ast$
and $\P^{\text \no, (q)}$ is identical to $\P^{\text \no}$
(since algorithm $A^{(q)}$, which does not fake any queries,
is identical to algorithm $A$).

Recall that our goal is to prove Equation~(\ref{eq:close}).
Since
$\P^{\ast,(q)} = \P^\ast$
and $\P^{\text \no, (q)} = \P^{\text \no}$,
Equation~(\ref{eq:close}) is an immediate consequence (using the triangle
inequality for total variation distance) of the following
two lemmas, which we prove below:

\BL \label{lem:init-close}
$\dtv(\P^{\ast, (0)},\P^{\text \no , (0)}) \leq 1/10.$
\EL

\BL \label{lem:small-change}
For all $0 \leq k < q$, we have
$\dtv(\P^{\ast,(k)},\P^{\ast, (k+1)}) \leq 1/(20q)$ and
$\dtv(\P^{\text \no, (k)},\P^{\text \no , (k+1) }) \leq 1/(20q)$.
\EL

\noindent {\bf Proof of \autoref{lem:init-close}:}
Define
$\P^\ast_0$ to be the distribution over outcomes of the $q$ calls
to $\SAMP_\D$ (i.e. over length-0 transcript prefixes) when $\D=\D^\ast.$
Define $\P^{\text \no}$
to be the distribution over outcomes of the $q$ calls
to $\SAMP_\D$
when $\D$ is drawn from $\calP_{\text \no}.$
We begin by noting that by the data processing inequality
for total variation distance \newer{(\autoref{lemma:data:processing:inequality:total:variation})}, we have
$\dtv(\P^{\ast,(0)},\P^{\text \no,(0)}) \leq
\dtv(\P^\ast_0, \P^{\text \no}_0)$ {(indeed, after the calls to respectively $\SAMP_\D$
and $\SAMP_{\D^\ast}$, the same randomized function $F$ -- which fakes all remaining oracle calls -- is applied to the two resulting distributions over length-0 transcript prefixes $\P^\ast_0$ and $\P^{\text \no}_0$)}. In the rest of the proof we show that $\dtv(\P^\ast_0, \P^{\text \no}_0) \leq 1/10.$
% \rnote{Is this clear or does it need additional justification?  It's
% a consequence of the DPI for TVD because after the calls to $\SAMP_\D$
% versus $\SAMP_{\D^\ast}$,
% the same (randomized) procedure (``fake all remaining oracle calls'')
% is followed in both cases, i.e. the same function $F$
% is applied to the two distributions over length-0
% transcript prefixes, in the terminology of
% \autoref{lemma:data:processing:inequality:total:variation}).}
% \dnote{Maybe we need to explain the application once and then
% we don't need to do this anymore? (possibly can reference to first
% explanation)? On one hand it is simple, on the other hand, if
% someone is not used to this, then it might help to spill it out
% the first time.}\cnote{Added a sentence in that direction, adapted from Rocco's note; and commented the next $\backslash$rnote regarding the same issue, on next page.}

Let $E$ denote the event that the $q$ calls to $\SAMP_\D$
yield points $s_1,\dots,s_q$ such that no bucket-pair $U_i$
contains more than one of these points.
Since $\D^\ast(U_i)=1/r$ for all $i$,
% it is easy to see that
\BEQ \label{eq:P0E}
\P^\ast_0(E) = \prod_{j=0}^{q-1} \left(1 - {\frac j r}\right) \geq 9/10\;,
\EEQ
where Equation~(\ref{eq:P0E}) follows from a standard birthday paradox
analysis and the fact that $q \leq {\frac 1 3} \sqrt{r}.$
Since for each possible outcome of $\D$ drawn
from $\calP_{\text \no}$ we have $\D(U_i)=1/r$ for all $i$, we
further have
that also
\BEQ \label{eq:PNOE}
\P^{\text \no}_0(E)= \prod_{j=0}^{q-1} \left(1 - {\frac j r}\right) .
\EEQ
We moreover claim that
the two conditional distributions
$(\P^\ast_0|E)$ and $(\P^{\text \no}_0|E)$ are identical,
i.e.
\BEQ \label{eq:cond-identical}
(\P^\ast_0|E) = (\P^{\text \no}_0|E).
\EEQ
To see this, fix any sequence $(\ell_1,\dots,\ell_q) \in [r]^q$
such that $\ell_i \neq \ell_j$ for all $i \neq j$.
Let \mbox{$(s_1,\dots,s_q) \in [N]^q$} denote a draw from $(\P^\ast_0|E)$.
The probability that ($s_i \in U_{\ell_i}$ for all $1 \leq i \leq q$)
is precisely $1/r^q$.
Now given that $s_i \in U_{\ell_i}$ for all $i$, it is clear that
$s_i$ is equally likely to lie in $B_{2\ell_i-1}$ and in $B_{2 \ell_i}$,
and given that it lies in a particular
one of the two buckets, it is equally likely
to be any element in that bucket.  This is true independently for all
$1 \leq i \leq q.$

Now let $(s_1,\dots,s_q) \in [N]^q$ denote
a draw from $(\P^{\text \no}_0|E).$
Since each distribution $\D$ in the support of $\calP_{\text \no}$
has $\D(U_i)=1/r$ for all $i$, we likewise have
that the probability that ($s_i \in U_{\ell_i}$ for all $1 \leq i \leq q$)
is precisely $1/r^q$.
Now given that $s_i \in U_{\ell_i}$ for all $i$, we have that
$s_i$ is equally likely to lie in $B_{2\ell_i-1}$ and in $B_{2 \ell_i}$;
this is because $\pi_i$ (recall that $\pi$ determines $\D=\D_\pi$)
is equally likely to be $\uparrow \downarrow$ (in which case
$\D(B_{2 \ell_i-1})= 3/(4r)$ and $\D(B_{2 \ell_i})=1/(4r)$)
as it is to be $\downarrow \uparrow$ (in which case
$\D(B_{2 \ell_i-1})=1/(4r)$ and $\D(B_{2\ell_i})=3/(4r)$).
Additionally,
 given that $s_i$ lies in a particular one of the two buckets, it is equally
likely to be any element in that bucket.  This is true
independently for all $1 \leq i \leq q$ (because conditioning on $E$
ensures that no two elements of $s_1,\dots,s_q$
lie in the same bucket-pair, so there is
``fresh randomness for each $i$''),
and so indeed the two conditional distributions
$(\P^\ast_0|E)$ and $(\P^{\text \no}_0|E)$ are identical.

Finally, the claimed bound
$\dtv(\P^\ast_0, \P^{\text \no}_0) \leq 1/10$
follows directly from Equations~(\ref{eq:P0E}),
~(\ref{eq:PNOE}) and~(\ref{eq:cond-identical}).
\qed

\noindent {\bf Proof of \autoref{lem:small-change}:}
Consider first the claim that $\dtv(\P^{\ast,(k)},\P^{\ast, (k+1)})
\leq 1/(20q)$.  Fix any \mbox{$0 \leq k < q$}.
The data processing inequality for total variation distance implies that
$\dtv(\P^{\ast,(k)},\P^{\ast, (k+1)})$
is at most the variation distance between random variables $X$ and \newer{$X'$},
where
\begin{itemize}

\item $X$ is the random variable obtained by running $A$ on $\COND_{\D^\ast}$
to obtain a length-$k$ transcript prefix
$(Y,Z^k)$, then drawing $\{x_{k+1},y_{k+1}\}$
from $\T_{(Y,Z^k)}$, then setting $p_{k+1}$ to be the output
of $\PCOND_{\D^\ast}(\{x_{k+1},y_{k+1}\newer{\}})$; and

\item \newer{$X'$} is the random variable obtained by running $A$ on $\COND_{\D^\ast}$
to obtain a length-$k$ transcript prefix
$(Y,Z^k)$, then drawing $\{x_{k+1},y_{k+1}\}$
from $\T_{(Y,Z^k)}$, then setting $p_{k+1}$ according to the
\newer{aforementioned} rules 2(i) and 2(ii).
\end{itemize}
%\rnote{Is this clear or does it need additional justification?  It's a consequence of the DPI for TVD because after this one step, the same (randomized) procedure is followed (i.e. the same function $F$ is applied to random variables $X$ and $Y$, in the terminology of \autoref{lemma:data:processing:inequality:total:variation}) to generate the final draws from $\P^{\ast,(k)}$ and from $\P^{\ast, (k+1)}$.}

Consider any fixed outcome of $(Y,Z^k)$ and $\{x_{k+1},y_{k+1}\}.$
If rule 2(i) is applied ($x_{k+1}$ and $y_{k+1}$ are in the same bucket),
then % it is easy to see that
there is
zero contribution to the variation distance between $X$ and \newer{$X'$},
because choosing a uniform element of $\{x_{k+1},y_{k+1}\}$ is a
perfect simulation of $\PCOND_\D(\{x_{k+1},y_{k+1}\}).$
If rule 2(ii) is applied, then the contribution is upper bounded by $O(1/K)
< 1/20q$, because
$\PCOND_{\D^\ast}(\{x_{k+1}y_{k+1}\})$ would return a different
outcome from rule 2(ii) with probability $1/\Theta(K^{\ell'-\ell}) =
O(1/K).$  Averaging over all possible outcomes
of $(Y,Z^k)$ and $\{x_{k+1},y_{k+1}\}$ we get that the variation distance
between $X$ and \newer{$X'$} is at most $1/20q$ as claimed.

An identical argument shows that similarly
$\dtv(\P^{\text \no,(k)},\P^{\text \no, (k+1)}) \leq 1/(20q)$.
The key observation is that for any distribution $\D$
in the support of \newer{$\calP^{\text \no}$},
as with $\D^\ast$ it is the case that points in the same
bucket have equal probability under $\D$ and 
% a point $y$ that is
% $\ell' - \ell$ buckets lower than $x$ has probability
% only $1/\Theta(K^{\ell'-\ell})$ of being returned by a
% call to $\PCOND_\D(\{x,y\}).$
\newer{for a pair of points $\{x,y\}$ such that $x\in B_\ell$
and $y \in B_{\ell'}$ for $\ell' > \ell$, the probability that 
 a call to $\PCOND_\D(\{x,y\})$ returns $y$ is
 only $1/\Theta(K^{\ell'-\ell})$.}
This concludes the proof of \autoref{lem:small-change} and of
\autoref{thm:pcond-ddstar-lb}.
\qed

\subsection{\texorpdfstring{A $\poly(1/\eps)$-query $\COND_\D$ algorithm}{A poly(1/eps)-query \COND algorithm}}
\fi %%%%%%%%%%%%%%%%%%%%%%%%%%%%%%%%%%%%%%%%%% End of "only Full version"

In this
\ifnum\confversion=0
subsection
we present an algorithm {\sc $\COND$-Test-Known}
and prove the following theorem:
\else
{section}
{we describe the idea behind an algorithm {\sc $\COND$-Test-Known},
whose properties are stated in the following theorem.}
\fi

\BT \label{thm:cond-test-known}
{\sc $\COND$-Test-Known} is a $\tilde{O}(1/\eps^4)$-query $\COND_\D$
testing algorithm for testing equivalence to a known distribution $\D^\ast$.
{That is,} for every pair of distributions $\D,\D^\ast$ over $[N]$
{(such that $\D^\ast$ is fully specified and there is $\COND$ query access to $\D$),}
the algorithm outputs $\accept$ with probability at least $2/3$ if $\D =
\D^\ast$ and outputs $\reject$ with probability at least $2/3$ if
\mbox{$\dtv(\D,\D^\ast) \geq \eps.$}
\ET

This constant-query testing algorithm stands in interesting contrast
to the $(\log N)^{\Omega(1)}$-query lower bound for
$\PCOND_\D$ algorithms for this problem.

\medskip

\noindent {\bf High-level overview of the % argument:}
{algorithm and its analysis:}}
First, we note that by reordering elements of $[N]$ we may assume
without loss of
generality that $\D^\ast(1) \leq \cdots \leq \D^\ast(N)$; this
will be convenient for us.
% Throughout the rest of this subsection we assume that
% $\D^\ast(1) \leq \cdots \leq \D^\ast(N)$.

Our $(\log N)^{\Omega(1)}$ query
lower bound for $\PCOND_\D$ algorithms exploited the
intuition that comparing two points using the $\PCOND_\D$ oracle
might not provide much information (e.g. if one of the two
points was a priori ``known'' to be much heavier than the other).
In contrast, with a general $\COND_\D$ oracle at our disposal,
we can compare a given point $j \in [N]$ with
\emph{any subset} of $[N] \setminus \{j\}.$  Thus the following definition will be useful:

\BD [\good points]
Fix $0 < \lambda \leq 1$. A point $j\in\supp{D^\ast}$ is said to be \emph{$\lambda$-\good}
%(or just \emph{\good}if there is no ambiguity),
if there exists a set $S \subseteq
([N] \setminus \{j\})$ such that
\[ D^\ast(j) \in [ \lambda D^\ast(S), D^\ast(S)/\lambda ]. \]
Such a set $S$ is then said to be a \emph{$\lambda$-\good-witness for $j$} (according to $D^\ast$), which is denoted $
S\cong^\ast j$.
We say that a set $T \subseteq [N]$ is $\lambda$-\good if
every $i \in T$ is $\lambda$-\good.
\ED
\ifnum\confversion=0
We stress that the notion of being $\lambda$-\good deals only with the
known distribution $\D^\ast$; this will be important later.
\else
We stress that the notion of being $\lambda$-\good deals only with the
known distribution $\D^\ast$.
\fi
%$D^\ast$ is $\lambda$-\good if all points from its support are.

%\clement{ Would ``matchable'' be a suitable term for this concept? ``Good'' is nice, but not very %self-explanatory\dots}

%\BR
%$\lambda$ will be taken to be some ${\eps_1}=\bigTheta{\eps}$.
%\ER

Fix ${\eps_1} = \Theta(\eps)$\fullOrConfVersion{(we specify ${\eps_1}$ precisely in Equation~(\ref{eq:choice:epsilons}) below)}{{(see full version for its exact value)}}.
Our analysis and algorithm consider two possible cases for the distribution $\D^\ast$
(where it is not hard to verify, and we provide an explanation
\ifnum\confversion=0
subsequently,
\else
{in the full version,}
\fi
that one of the two cases must hold):

\BE
  \fullOrConfCompactEnums\fullOrConfIndentItems

\item The first case is that for some $i^\ast \in [N]$ we have
\begin{equation} \label{eq:gap}
\D^\ast(\{1,\dots,i^\ast\}) > 2 {\eps_1} \fullOrConfVersion{\quad}{\;\,} \text{but} \fullOrConfVersion{\quad}{\;\,} \D^\ast(\{1,\dots,i^\ast-1\}) \leq
{\eps_1}.
\end{equation}
In this case % it is easy to see that
$1-{\eps_1}$ of the total probability mass of $\D^\ast$
must lie on a set of at most $1/{\eps_1}$ elements, and in such a situation it is
easy to efficiently test whether $\D = \D^\ast$
using $\poly(1/\eps)$ queries (see Algorithm~{\sc $\COND_\D$-Test-Known-Heavy} and\fullOrConfVersion{\autoref{lem:test-heavy}}{{its analysis in the full version}}).

\item The second case is that there exists an element $k^\ast \in [N]$ such that
\begin{equation} \label{eq:nogap}
{\eps_1} < D^\ast(\{1,\dots,k^\ast\}) \leq 2{\eps_1} <
D^\ast(\{1,\dots,k^\ast + 1\}).
\end{equation}
This is the more challenging (and typical) case.
In this case, it can be shown
%(see \autoref{claim:good:after:trimming})
that every element $j > k^\ast$ has at least one ${\eps_1}$-\good-witness
within $\{1,\dots,j\}.$
In fact,\fullOrConfVersion{we show (see \autoref{claim:partition})}{{one can show}} that
either (a) $\{1,\dots,j-1\}$ is an ${\eps_1}$-\good witness for $j$, or (b) the
set $\{1,\dots,j-1\}$ can be partitioned
into disjoint sets\footnote{In fact the sets are intervals (under the assumption
\mbox{$\D^\ast(1)\leq\dots\leq\D^\ast(n)$}),
but that is not really important for our arguments.}
$S_1,\dots,S_t$
such that \emph{each} $S_i$, $1 \leq i \leq t$, is a ${\frac 1 2}$-\good-witness for $j$.
Case (a) is relatively easy to handle so we focus on (b) in our informal description below.

\EE

The partition $S_1,\dots,S_t$ is useful to us for the following reason:  Suppose that
$\dtv(\D,\D^\ast) \geq \eps.$  It is not difficult to show\fullOrConfVersion{(see \autoref{claim:finding:heavy:guy}) that}{that}
unless $\D(\{1,\dots,k^\ast\}) > 3 {\eps_1}$ (which can be easily detected and provides evidence that the tester should reject), a random sample of {$\Theta(1/\eps)$} draws from $\D$ will with high probability contain a ``heavy'' point $j > k^\ast$, that is, a point $j > k^\ast$ such that $\D(j) \geq (1+{\eps_2})\D^\ast(j)$ (where ${\eps_2} = \Theta(\eps)$).  Given such a point $j$, there are two possibilities:

\BE
  \fullOrConfCompactEnums\fullOrConfIndentItems

\item The first possibility is that a significant fraction of the
sets $S_1,\dots,S_t$ have $\D(j)/\D(S_i)$ ``noticeably
different'' from $\D^\ast(j)/\D^\ast(S_i).$
(Observe that since each set $S_i$ is a ${\frac 1 2}$-\good
witness for $j$, it is possible to efficiently check whether this is the
case.)  If this is the case then our tester should reject since
this is evidence that $\D \neq \D^\ast.$

\item The second possibility is that almost every
$S_i$ has $\D(j)/\D(S_i)$ very close to $\D^\ast(j)/\D^\ast(S_i)$.
If this is the case, though, then since
$\D(j) \geq (1+{\eps_2})\D^\ast(j)$ and the union of
$S_1,\dots,S_t$ is $\{1,\dots,j-1\}$, it must be the case that
$\D(\{1,\dots,j\})$ is ``significantly larger'' than $\D^\ast(\{1,\dots,j\}).$
This will be revealed by random sampling from $\D$ and thus our
testing algorithm can reject in this case as well.
%\ignore{Observe that since $j > k^\ast$, the value $\D(\{1,\dots,j\})$ must be at least ${\eps_1}$, and thus by random sampling it is easy to efficiently estimate $\D(\{1,\dots,j\})$ and compare it with the known value $\D^\ast(\{1,\dots,j\})$.  So our testing algorithm can detect this possibility and reject in this case as well. }
\EE

%%%------ Ignore
\ignore{
For such a point, it must either be the case that (i) some subset $S_i$ has $\D(S_i)$ noticeably larger than $\D^\ast(S_i)$, or else if (i) does not hold, then it must be the case that (ii)
}
\ignore{
%Observe that $D^\ast$ induces a partition of $[N]$ into \good and \bad points: namely, $[N]=U\disjunion M$. %The first step (\textsf{trimming}) will then allow us to get rid of $U$, that is to  assume at little cost %that all points are \good. We can also suppose (up to a relabeling of the points) that $D^\ast$ is %non-decreasing.
}
\ignore{
One high-level idea underlying our algorithm is that if a significant amount of probability mass under $\D^\ast$ is on points $i$ which are ``incomparable'' with any subset of $[N] \setminus \{i\}$, then $\D^\ast$ must have a very special structure (almost all of its weight is on a set of $O(1/\eps)$ points, see {XXX} below).  This special structure makes it possible to test whether $\D = \D^\ast$ versus $\dtv(\D,\D^\ast) \geq \eps$
using only $\poly(1/\eps)$ queries, see {YYY}.

The other possibility is that almost all the weight of $\D^\ast$ is on elements $i$ which indeed are ``comparable'' to some
subset of $[N] \setminus \{i\}$; roughly speaking we refer to such a distribution $\D^\ast$ as a ``comparable'' distribution.  If $\D^\ast$ is comparable, then in order to have $\dtv(\D,\D^\ast) \geq \eps$ it must be the case that there is significant mass under {ZZZ} on points which are significantly heavier/lighter under $\D$ than under $\D^\ast.$ {[[[XXX finish sketch of the argument ]]]}
}
%%%------ End Ignore

\ifnum\confversion=1 %%%%%%%%%%%%%%%%%%%%%%%%%%%%%%%%%%%%%%%%%% Conf version
{The detailed proof {of \autoref{thm:cond-test-known}}, including a complete description and analysis
of the algorithm, is given in the full version.}
\else                %%%%%%%%%%%%%%%%%%%%%%%%%%%%%%%%%%%%%%%%%% Only Full version

\noindent {\bf Key quantities and useful claims.}
We define some quantities that are used in the algorithm and its analysis.
Let
\BEQ
\label{eq:choice:epsilons}
{\eps_1} \eqdef {\frac \eps {10}}; \quad
{\eps_2} \eqdef {\frac \eps {2}}; \quad
{\eps_3} \eqdef {\frac \eps {48}}; \quad
{{\eps_4}}  \eqdef {\frac \eps {6}}.
\EEQ

\ignore{
  % %%%%%%%
  % OLD
  % %%%%%%%
  \begin{align}\label{eq:choice:epsilons}
  \eps' &= {\frac \eps 2} \\
  {\eps_1} &= \frac{\eps}{10} \leq {\frac 1 {10}}\\
  \eps_1 &= {\eps_1} \\
  \eps_1 &= 3\eps_2 \\
  \eps_2 &> 21Z3.\\
  {\eps' = 6 Z3}
  \end{align}
  \rnote{Was $\eps_2 > 7 Z3$, but our ``$Z3$'' factor is 3 times the old one, because  we only know that  $\D(\{1,\dots,i_j\}) \in  [1-3Z3,1+3Z3]\D^\ast(\{1,\dots,i_j\}).$}
  % %%%%%%%
  % END OLD
  % %%%%%%%
}

\BCM \label{claim:partition}
Suppose there exists an element $k^\ast \in [N]$
that satisfies Equation~(\ref{eq:nogap}).
%$ {\eps_1} < D^\ast(\{1,\dots,k^\ast\}) \leq 2{\eps_1}.$
Fix any $j > k^\ast.$  Then

\BE

\item If $\D^\ast(j) \geq {\eps_1}$, then $S_1 \eqdef \{1,\dots,j-1\}$ is an
${\eps_1}$-\good witness for $j$;

\item If $\D^\ast(j) < {\eps_1}$ then the set $\{1,\dots,j-1\}$ can be partitioned
into disjoint sets $S_1,\dots,S_t$ such that each
$S_i$, $1 \leq i \leq t$, is a ${\frac 1 2}$-\good-witness for $j$.

\EE

\ECM
\BPF
First consider the case that $\D^\ast(j)\geq {\eps_1}.$  In this case $S_1=\{1,\dots,j-1\}$  is an ${\eps_1}$-\good witness
for $j$ because $\D^\ast(j) \geq {\eps_1} \geq {\eps_1} \D^\ast(\{1,\dots,j-1\})$
and $\D^\ast(j) \leq 1 \leq {\frac 1 {{\eps_1}}} \D^\ast(\{1,\dots,k^\ast\}) \leq {\frac 1 {{\eps_1}}} \D^\ast(\{1,\dots,j-1\})$,
where the last inequality holds since $k^\ast \leq j-1$.

Next, consider the case that $\D^\ast(j) < {\eps_1}.$  In this case we build our
intervals iteratively from right to left, as follows.  Let $j_1=j-1$
and let $j_2$ be the minimum index in $\{0,\dots,j_1-1\}$
such that
\[ \D^\ast(\{j_2+1,\dots,j_1\}) \leq \D^\ast(j). \]
(Observe that we must have $j_2 \geq 1$, because $\D^\ast(\{1,\dots,k^\ast\})
> {\eps_1} > \D^\ast(j).$)  Since \mbox{$\D^\ast(\{j_2,\dots,j_1\}) > \D^\ast(j)$}
and the function $\D^\ast(\cdot)$ is monotonically increasing, it must be the
case that
\[
\frac{1}{2}\D^\ast(j) \leq
\D^\ast(\{j_2+1,\dots,j_1\}) \leq
\D^\ast(j).
\]
Thus the interval $S_1 \eqdef \{j_2+1,\dots,j_1\}$ is
a ${\frac 1 2}$-\good witness for $j$ as desired.

We continue in this fashion from right to left; i.e. if
we have defined $j_2,\dots,j_t$ as above and there is an index $j'
\in \{0,\dots,j_t-1\}$ such that
$
\D^\ast(\{j'+1,\dots,j_t\}) > \D^\ast(j),
$
then we define $j_{t+1}$ to be the minimum index in $\{0,\dots,j_{t}-1\}$
such that
\[
\D^\ast(\{j_{t+1}+1,\dots,j_t\}) \leq \D^\ast(j),
\]
and we define
$S_{t}$ to be the interval $\{j_{t+1}+1,\dots,j_t\}$.
The argument of the previous paragraph tells us that
\begin{equation} \label{eq:comp}
\frac{1}{2}\D^\ast(j) \leq
\D^\ast(\{j_{t+1}+1,\dots,j_t\}) \leq
\D^\ast(j)
\end{equation}
and hence $S_{t}$ is an ${\frac 1 2}$-\good witness for $j$.

At some point, after intervals $S_1 = \{j_2+1,\dots,j_1\},\dots,S_t
= \{j_{t+1}+1,\dots,j_t\}$ have been defined in this way, it will be the
case that there is no index $j' \in \{0,\dots,j_t-1\}$ such that
\mbox{$\D^\ast(\{j' + 1,\dots,j_t\}) > \D^\ast(j).$}
At this point there are two possibilities:  first, if $j_{t+1}+1=1$,
then $S_1,\dots,S_t$ give the desired partition of $\{1,\dots,j-1\}$.
If $j_{t+1}+1>1$ then it must be the case that
$\D^\ast(\{1,\dots,j_{t+1}\}) \leq \D^\ast(j).$
In this case we simply add the elements $\{1,\dots,j_{t+1}\}$
to $S_{t}$, i.e. we redefine $S_t$ to be
$\{1,\dots,j_t\}$.  By Equation~(\ref{eq:comp}) we have that
\[
\frac{1}{2}\D^\ast(j) \leq
\D^\ast(S_t) \leq 2\D^\ast(j)
\]
and thus $S_t$ is an ${\frac 1 2}$-\good witness for $j$ as desired. This
concludes the proof.
\EPF

\BD[Heavy points]
A point $j\in\supp{D^\ast}$ is said to be \emph{$\eta$-heavy}
%or just \emph{heavy},
if \mbox{$D(j) \geq (1+\eta)D^\ast(j)$}.
%Similarly, it is \emph{$\eta$-light}
%if \mbox{$D(j) \leq (1-\eps)D^\ast(j)$}.
\ED

\BCM\label{claim:finding:heavy:guy}
\sloppy
Suppose that $\totalvardist{D}{D^\ast} \geq \eps$ and
Equation~(\ref{eq:nogap}) holds.
Suppose moreover that $\D(\{1,\dots,k^\ast\}) \leq 4 {\eps_1}.$
Let $i_1,\dots,i_\ell$ be i.i.d. points drawn from $\D$.
Then for $\ell = \Theta(1/\eps)$, with probability at least $99/100$
(over the i.i.d. draws of $i_1,\dots,i_\ell \sim \D$)
there is some point $i_j \in \{i_1,\dots,i_\ell\}$ such that
$i_j > k^\ast$ and $i_j$ is ${\eps_2}$-heavy.
% and $\eps^\prime < \frac{\eps}{2}$. Then, by
%drawing $\frac{1}{\eps^\prime}$ samples from $D$, one gets with
%constant probability an $\eps^\prime$-heavy sample.
\ECM
\BPF
Define $H_1$ to be the set of all ${\eps_2}$-heavy points and $H_2$ to be the
set of all ``slightly lighter'' points as follows:
\begin{align*}
H_1 &= \setOfSuchThat{ i\in [N] }{ D(i) \geq (1+{\eps_2})D^\ast(i) } \\
H_2 &= \setOfSuchThat{ i\in [N] }{ (1+{\eps_2})D^\ast(i) > D(i) \geq D^\ast(i)  }
\end{align*}
By definition of the total variation distance, we have
\begin{eqnarray*}
\eps \leq \totalvardist{D}{D^\ast}
&=& \sum_{i : D(i) \geq D^\ast(i)} \left( D(i)
- D^\ast(i) \right) =  \left( D(H_1)-D^\ast(H_1) \right)
+ \left( D(H_2)-D^\ast(H_2) \right) \\
&\leq& D(H_1) + \left( (1+{\eps_2})D^\ast (H_2)-D^\ast(H_2) \right) \\
&=& \D(H_1) + {\eps_2} D^\ast(H_2)
< \D(H_1) + {\eps_2} = \D(H_1) + {\frac \eps 2}.
\end{eqnarray*}
So it must be the case that $\D(H_1) \geq \eps/2 = 5 {\eps_1}.$  Since by assumption we have $\D(\{1,\dots,k^\ast\}) \leq 4 {\eps_1}$, it must be the case that $\D(H_1 \setminus \{1,\dots,k^\ast\}) \geq  {\eps_1}.$ The claim follows from the definition of $H_1$ and the size, $\ell$, of the sample.
\EPF

\begin{algorithm}
\begin{algorithmic}[1]
\REQUIRE error parameter $\eps > 0$; query access to $\COND_\D$ oracle;
explicit description $(\D^\ast(1),\dots,\D^\ast(N))$ of distribution $\D^\ast$
satisfying $\D^\ast(1) \leq \cdots \leq \D^\ast(N)$
\STATE Let $i^\ast$ be the minimum index $i \in [N]$ such that
$\D^\ast(\{1,\dots,i\}) > 2 {\eps_1}$.
\IF{ $D^\ast(\{1,\dots,i^\ast-1\})\leq {\eps_1}$} \label{st:istar-exists}
  \STATE Call algorithm $\COND_\D$-Test-Known-Heavy$(\eps,
   \COND_\D, \D^\ast, i^\ast)$ (and exit)
\ELSE
  \STATE Call algorithm $\COND_\D$-Test-Known-Main$(\eps,
  \COND_\D, \D^\ast,i^\ast-1)$ (and exit).
\ENDIF
\end{algorithmic}\caption{\label{algo:cond-test-known}\sc $\COND_\D$-Test-Known}
\end{algorithm}

\begin{algorithm}
\begin{algorithmic}[1]
\REQUIRE error parameter $\eps > 0$; query access to $\COND_\D$ oracle;
explicit description $(\D^\ast(1),\dots,\D^\ast(N))$ of distribution $\D^\ast$
satisfying $\D^\ast(1) \leq \cdots \leq \D^\ast(N)$;
value $i^\ast \in [N]$ satisfying
$\D^\ast(\{1,\dots,i^\ast-1\}) \leq {\eps_1}$, $\D^\ast(\{1,\dots,i^\ast\}) >
2 {\eps_1}$
\STATE \label{st:cond:heavy:sampling} Call the $\SAMP_\D$ oracle $m = \Theta((\log(1/\eps))/\eps^4)$
times.
For {each} ${i} \in [i^\ast,N]$ let $\widehat{\D}(\newer{i})$ be the fraction of the
$m$ calls to $\SAMP_\D$ that returned $i.$  {Let $\widehat{D}'= 1 - \sum_{i \in [i^\ast,N]}
\widehat{D}(i)$ be the fraction of the $m$ calls that returned values in $\{1,\dots,i^\ast-1\}.$}
\IF {{either (any $i \in [i^\ast,N]$ has $|\widehat{\D}(i) - \D^\ast(i)| > {\eps_1}^2$) or
($\widehat{D}' - \D^\ast(\{1,\dots,i^\ast-1\}) > {\eps_1}$)}}
    \STATE output \reject (and exit)
\ENDIF
\STATE Output \accept
\end{algorithmic}\caption{\label{algo:cond-test-known-heavy}\sc
$\COND_\D$-Test-Known-Heavy}
\end{algorithm}

\begin{algorithm}
\begin{algorithmic}[1]
\REQUIRE error parameter $\eps > 0$; query access to $\COND_\D$ oracle;
explicit description $(\D^\ast(1),\dots,\D^\ast(N))$ of distribution $\D^\ast$
satisfying $\D^\ast(1) \leq \cdots \leq \D^\ast(N)$;
value $k^\ast \in [N]$ satisfying
${\eps_1} < \D^\ast(\{1,\dots,k^\ast\}) \leq 2 {\eps_1} < \D^\ast(\{1,\dots,k^\ast+1\})$

\STATE Call the $\SAMP_\D$ oracle $\Theta(1/\eps^2)$ times and let $\widehat{\D}(\{1,\dots,k^\ast\})$ denote the
fraction of responses that lie in $\{1,\dots,k^\ast\}$.  If
$\widehat{\D}(\{1,\dots,k^\ast\}) \notin [{\frac {{\eps_1}} 2}, {\frac {5 {\eps_1}} 2}]$ then output \reject (and exit).
\label{st:check-prefix-not-too-heavy}

\STATE Call the $\SAMP_\D$ oracle $\ell = \Theta(1/\eps)$ times to obtain points
 $i_1,\dots,i_\ell.$
 \label{st:draw-l-samples}

\FOR {all $j\in \{1,\dots,\ell\}$ such that $i_j > k^\ast$}

\STATE Call the $\SAMP_\D$ oracle $m = \Theta(\log(1/\eps)/\eps^{\newer{2}})$ times and let $\widehat{D}(\{1,\dots,i_j\})$ be the
fraction of responses that lie in $\{1,\dots,i_j\}$.  If
$\widehat{D}(\{1,\dots,i_j\}) \notin [1-{\eps_3},1+{\eps_3}]\D^\ast(\{1,\dots,i_j\})$ then output \reject
(and exit). \label{st:weigh-prefix}

\IF {$\D^\ast(i_j) \geq {\eps_1}$} \label{eq:Dast-big}

\STATE Run
{\sc Compare}$(\{i_j\},\{1,\dots,i_j-1\},{\frac {{\eps_2}}{16}},
{\frac 2 {{\eps_1}}}, {\frac 1 {10\ell}})$ and let
$v$ denote its output.  If $v \notin [1-{\frac {{\eps_2}}{8}},1+{\frac
{{\eps_2}}{8}}]{\frac {\D^\ast(\{1,\dots,i_j-1\})}{\D^\ast(\{i_j\})}}$
  then output \reject (and exit). \label{st:lastcheck-Dast-big}

\ELSE

\STATE Let $S_1,\dots,S_t$ be the partition of $\{1,\dots,i_j-1\}$ such that
each $S_i$ is an ${\eps_1}$-\good witness for $i_j$, which is provided by \autoref{claim:partition}.
\label{st:make-partition}

\STATE Select a list of {$h=\Theta(1/\eps)$} %$n=\Theta(1/\eps_4)$
elements $S_{a_1},\dots,S_{a_h}$ independently
and uniformly from $\{S_1,\dots,S_j\}$. \label{st:randomsets}

\STATE For each $S_{a_r}$, $1 \leq r \leq h$, run
{\sc Compare}$(\{i_j\},S_{a_r},{\frac {{\eps_4}} 8},4, {\frac 1 {10\ell h}})$ and let
$v$ denote its output.  If $v \notin [1-{\frac {{\eps_4}} 4},1+{\frac {{\eps_4}} 4}]{\frac {\D^\ast(S_{a_r})}{\D^\ast(\{i_j\})}}$ then output \reject (and exit). \label{st:lastcheck}

\ENDIF

\ignore{

\STATE {make $\bigTheta{\frac{1}{\eps^2}}$ \COND queries on $S\cup\{j\}$ to estimate $\frac{D(j)}{D(S)}$}

\IF {(things are out of line)}
\STATE {output \reject (and exit)}
\ENDIF

}
\ENDFOR

\STATE Output \accept.

\end{algorithmic}\caption{\label{algo:cond-test-known-main}\sc
$\COND_\D$-Test-Known-Main}
\end{algorithm}

\subsubsection{Proof of \newer{\autoref{thm:cond-test-known}}}

It is straightforward to verify that
the query complexity of $\COND_\D$-Test-Known-Heavy is
$\tilde{O}(1/\eps^4)$ and the query complexity of $\COND_\D$-Test-Known-Main
is also $\tilde{O}(1/\eps^4)$, so the overall query complexity of
{\sc $\COND$-Test-Known} is as claimed.

{By the definition of $i^\ast$ (in the first line of the algorithm), either
Equation~(\ref{eq:gap}) holds for this setting of $i^\ast$, or
Equation~(\ref{eq:nogap}) holds for $k^\ast = i^\ast -1$.}
To prove correctness of the algorithm, we first deal with the simpler case, which is
that Equation~(\ref{eq:gap}) holds:

\BL \label{lem:test-heavy}
Suppose that $\D^\ast$ is such that $\D^\ast(\{1,\dots,i^\ast\}) >
2 {\eps_1}$ but $\D^\ast(\{1,\dots,i^\ast-1\}) \leq {\eps_1}.$
Then {\sc $\COND_\D$-Test-Known-Heavy$(\eps,\COND_D,\D^\ast,i^\ast)$}
returns \accept with probability at least $2/3$ if $\D = \D^\ast$
and returns \reject with probability at least $2/3$ if \mbox{$\dtv(\D,\D^\ast)\geq \eps.$}
\EL

\BPF
The conditions of \autoref{lem:test-heavy}, together with the fact that
$\D^\ast(\cdot)$ is monotone non-decreasing, imply
that each $i \geq i^\ast$ has $\D^\ast(i) \geq {\eps_1}$.  Thus there can
be at most $1/{\eps_1}$ many values $i \in \{i^\ast,\dots,N\},$
i.e. it must be the case that $i^\ast \geq N - 1/{\eps_1} + 1.$
Since the expected value of $\widehat{D}(i)$ (defined in Line~\ref{st:cond:heavy:sampling} of
{\sc $\COND_\D$-Test-Known-Heavy}) is precisely $\D(i)$,
for any fixed value of $i \in \{i^\ast,\dots,n\}$
an additive Chernoff bound implies that $|\D(i) - \widehat{D}(i)|
\leq ({\eps_1})^2$
with failure probability at most ${\frac 1 {10
\left(1 + {\frac 1 {{\eps_1}}}\right)}}.$
Similarly $|\widehat{D}' - \D(\{1,\dots,i^\ast-1\})| \leq {\eps_1}$
with failure probability at most ${\frac 1 {10
\left(1 + {\frac 1 {{\eps_1}}}\right)}}.$
A union bound over all failure events  gives that with probability at least
$9/10$ each value $i \in \{i^\ast,\dots,N\}$ has $|\D(i) - \widehat{D}(i)|
\leq {\eps_1}^2$ and additionally $|\widehat{D}' - \D(\{1,\dots,i^\ast-1\})|
\leq {\eps_1}$; we refer to this compound event as (*).

If $\D^\ast = \D$, by (*) the algorithm
outputs \accept with probability at least $9/10.$

Now suppose that $\dtv(\D,\D^\ast) \geq \eps.$  With probability at least 9/10 we have (*) so we suppose that indeed (*) holds.  In this case we have
\begin{eqnarray*}
\eps \leq \dtv(\D,\D^\ast) &=& \sum_{i < i^\ast} |\D(i)-\D^\ast(i)| +
\sum_{i \geq i^\ast} |\D(i) - \D^\ast(i)|\\
&\leq& \sum_{i < i^\ast} \left(\D(i) + \D^\ast(i) \right) +   \sum_{i \geq i^\ast} |\D(i) - \D^\ast(i)|\\
&\leq& \D(\{1,\dots,i^\ast-1\}) + {\eps_1}  + \sum_{i \geq i^\ast} \left(|\widehat{\D}(i) - \D^\ast(i)|
+ {\eps_1}^2\right)\\
&\leq& \widehat{\D}' + {\eps_1} + 2{\eps_1}  + \sum_{i \geq i^\ast} \left(|\widehat{D}(i) - \D^\ast(i)|\right)\\
\end{eqnarray*}
where the first inequality is by the triangle inequality, the second is by (*) and the fact that
$\D^\ast(\{1,\dots,i^\ast-1\}) \leq {\eps_1}$, and the third inequality is by (*) and the
fact that there are at most $1/{\eps_1}$ elements in $\{i^\ast,\dots,N\}.$  Since
% \mnote{{uses $\eps_1=\eps/10$}}
${\eps_1} = \eps/10,$ the above inequality implies that
\[
{\frac 7 {10}} \eps \leq \widehat{\D}' +
\sum_{i \geq i^\ast} \left(|\widehat{\D}(i) - \D^\ast(i)|\right).
\]
If any $i \in \{i^\ast,\dots,N\}$ has $|\widehat{\D}(i) - \D^\ast(i)| > ({\eps_1})^2$ then the
algorithm outputs \reject so we may assume that $|\widehat{D}(i) - \D^\ast(i)| \leq {\eps_1}^2$ for all $i$.
This implies that
\[
6 {\eps_1} = {\frac 6 {10}} \eps \leq \widehat{\D}'
\]
but since $\D^\ast(\{1,\dots,i^\ast-1\}) \leq {\eps_1}$ the algorithm must \reject.
\EPF

Now we turn to the more difficult (and typical) case,
that Equation~(\ref{eq:nogap}) holds (for $k^\ast = i^\ast -1$), i.e.
\[
{\eps_1} < D^\ast(\{1,\dots,k^\ast\}) \leq 2{\eps_1} < D^\ast(\{1,\dots,k^\ast + 1\}).
\]

With the claims we have already established it is straightforward to argue completeness:

\BL \label{lem:CKTM-complete}
Suppose that $\D = \D^\ast$ and Equation~(\ref{eq:nogap}) holds.  Then with probability
at least $2/3$ algorithm {\sc $\COND_\D$-Test-Known-Main} outputs \accept.
\EL

\BPF
We first observe that the expected value of the quantity $\widehat{\D}(\{1,\dots,k^\ast\})$ defined in Line~\ref{st:check-prefix-not-too-heavy} is precisely $\D(\{1,\dots,k^\ast\}) =
\D^\ast(\{1,\dots,k^\ast\})$ and hence lies in $[{\eps_1},2{\eps_1}]$
by Equation~(\ref{eq:nogap}).
% A standard
The additive Chernoff bound implies that the probability the algorithm outputs \reject
in Line~\ref{st:check-prefix-not-too-heavy} is at most $1/10.$  Thus we may assume the algorithm continues to Line~\ref{st:draw-l-samples}.

\sloppy
In any given execution of Line~\ref{st:weigh-prefix}, since the expected value
of $\widehat{\D}(\{1,\dots,i_j\})$ is precisely $\D(\{1,\dots,i_j\})=\D^\ast(\{1,\dots,i_j\}) > {\eps_1}$,
a % standard
multiplicative Chernoff bound
\ignore{\rnote{Here I am using the Chernoff bound
\[
\Pr[X \geq (1 + \tau)\E[X]] \leq \exp(-\tau^2 \E[X]/3)
\]
and the lower-tail counterpart.}}%\dnote{Have we decided about giving references to the statements in prelim?}
gives that the algorithm outputs \reject with
probability at most $1/(10 \ell).$  Thus the probability that the algorithm outputs
\reject in any execution of Line~\ref{st:weigh-prefix} is at most $1/10.$
We henceforth assume that the algorithm never outputs \reject in this step.

Fix a setting of $j \in \{1,\dots,\ell\}$ such that $i_j > k^\ast.$
Consider first the case
that $\D^\ast(i_j) \geq {\eps_1}$ so the algorithm enters Line~\ref{st:lastcheck-Dast-big}.
By item (1) of \autoref{claim:partition} and item (1) of
\autoref{lem:compare}, we have that
with probability at least $1 - {\frac 1 {10 \ell}}$
{\sc Compare} outputs a value $v$ in the range $[1-{\frac {{\eps_2}}{16}},
1+{\frac {{\eps_2}}{16}}]{\frac {\D^\ast(\{1,\dots,i_j-1\})}
{\D^\ast(\{i_j\})}}$ (recall that $\D = \D^\ast$),
so the algorithm does not output \reject in Line~\ref{st:lastcheck-Dast-big}.
Now suppose that $D^\ast(i_j) < {\eps_1}$ so the algorithm enters Line~\ref{st:make-partition}.
Fix a value $1 \leq r \leq {h}$ in Line~\ref{st:lastcheck}.  By \autoref{claim:partition} we have
that $S_{a_r}$ is a ${\frac 1 2}$-\good witness for $i_j.$  By item (1) of \autoref{lem:compare}, we have that with probability at least $1 - {\frac 1 {10 \ell h}}$
{\sc Compare} outputs a value $v$ in the range $[1-{\frac {{\eps_4}} 4},1+{\frac {{\eps_4}} 4}]{\frac {\D^\ast(S_{a_r})}
{\D^\ast(\{i_j\})}}$ (recall that $\D = \D^\ast$).
 A union bound over all $h$ values of
$r$ gives that the algorithm
outputs \reject in Line~\ref{st:lastcheck} with probability at most $1/(10 \ell).$
So in either case, for this setting of $j$, the algorithm outputs \reject on that iteration
of the outer loop with probability at most $1/(10\ell).$  A union bound over all
$\ell$ iterations of the outer loop gives that the algorithm outputs \reject at any
execution of Line~\ref{st:lastcheck-Dast-big} or Line~\ref{st:lastcheck} is at most $1/10.$

Thus the overall probability that the algorithm outputs \reject is at most $3/10$, and the lemma is proved.
\EPF

Next we argue soundness:

\BL \label{lem:CKTM-sound}
Suppose that $\dtv(\D,\D^\ast) \geq \eps$ and Equation~(\ref{eq:nogap}) holds.  Then with probability
at least $2/3$ algorithm {\sc $\COND_\D$-Test-Known-Main} outputs \reject.
\EL

\BPF
If $\D(\{1,\dots,k^\ast\}) \notin [{\eps_1}, 3 {\eps_1}]$ then a standard additive Chernoff bound implies that the algorithm
outputs \reject in Line~\ref{st:check-prefix-not-too-heavy} with probability at least $9/10.$
Thus we may assume going forward in the argument that $\D(\{1,\dots,k^\ast\}) \in [{\eps_1}, 3 {\eps_1}]$.
As a result we may apply \autoref{claim:finding:heavy:guy}, and we have that with probability
at least $99/100$ there is an element $i_j \in \{i_1,\dots,i_\ell\}$ such that
$i_j > k^\ast$ and $i_j$ is ${\eps_2}$-heavy,
i.e. $\D(i_j) \geq (1+{\eps_2})\D^\ast(i_j).$  We condition on this event going forward (the rest of our
analysis will deal with this specific element $i_j$).

We now consider two cases:

\noindent {\bf Case 1:}  Distribution $\D$ has $\D(\{1,\dots,i_j\}) \notin
[1-3{\eps_3},1+3{\eps_3}]\D^\ast(\{1,\dots,i_j\}).$  Since the quantity $\widehat{D}(\{1,\dots,i_j\})$ obtained in Line~\ref{st:weigh-prefix}
has expected value $\D(\{1,\dots,i_j\}) \geq \D(\{1,\dots,k^\ast\}) \geq {\eps_1}$,
% a standard
{applying the}
multiplicative Chernoff bound implies that
$\widehat{D}(\{1,\dots,i_j\}) \in [1-{\eps_3},1+{\eps_3}]\D(\{1,\dots,i_j\})$
except with failure probability at most $\eps/10 \leq 1/10$.  If this failure event does not occur
then since $\D(\{1,\dots,i_j\}) \notin
[1-3{\eps_3},1+3{\eps_3}]\D^\ast(\{1,\dots,i_j\})$ it must hold that $
\widehat{D}(\{1,\dots,i_j\}) \notin [1-{\eps_3},1+{\eps_3}]\D^\ast(\{1,\dots,i_j\})$ and consequently
the algorithm outputs \reject.  Thus in Case~1 the algorithm outputs \reject with overall failure
probability at least $89/100.$

\noindent {\bf Case 2:} Distribution $\D$ has $\D(\{1,\dots,i_j\}) \in
[1-3{\eps_3},1+3{\eps_3}]\D^\ast(\{1,\dots,i_j\}).$  This case is divided into two sub-cases depending
on the value of $\D^\ast(i_j)$.

\noindent {\bf Case 2(a):}  $\D^\ast(i_j) \geq {\eps_1}$.
In this case the algorithm reaches Line~\ref{st:lastcheck-Dast-big}.
\ignore{
moreover, by item~1 of \autoref{claim:partition}, we have
that $\{1,\dots,i_j-1\}$ is an ${\eps_1}$-\good witness for $i_j.$  }We
use the following {claim}:

\BCM \label{clm:truth-will-out-1}
In Case~2(a), suppose that
$i_j > k^\ast$ is such that\ignore{{$\D^\ast(i_j) \geq {\eps_1}$,}}
$\D(i_j) \geq (1+{\eps_2})\D^\ast(i_j)$, and
$\D(\{1,\dots,i_j\}) \in
[1-3{\eps_3},1+3{\eps_3}]\D^\ast(\{1,\dots,i_j\})$.
Then
\[
{\frac {\D(\{1,\dots,i_j-1\})}{\D(i_j)}} \leq
\left(
1 - {\frac {{\eps_2}} 4}
\right)
\cdot
{\frac {\D^\ast(\{1,\dots,i_j-1\})}{\D^\ast(i_j)}}.
\]
\ECM

\BPF %\ignore{As noted above the set $\{1,\dots,i_j-1\}$ is an
% ${\eps_1}$-\good witness for $i_j$, so  we have
% ${\eps_1} \D^\ast(i_j) \leq \D^\ast(\{1,\dots,i_j-1\}) \leq {\frac 1 {{\eps_1}}}
% \D^\ast(i_j).$}
To simplify notation we write
\[
a \eqdef \D(i_j); \quad b \eqdef \D^\ast(i_j); \quad
c \eqdef \D(\{1,\dots,i_j-1\}); \quad d \eqdef \D^\ast(\{1,\dots,i_j-1\}).
\]
We have that
\begin{equation}
\label{eq:abcd}
a \geq (1+{\eps_2})b \quad \text{and} \quad a+c \leq (1 + 3{\eps_3}) (b+d).
\end{equation}
This gives
\begin{equation} \label{eq:z1}
c \leq (1+3 {\eps_3}) (b+d) - (1+{\eps_2})b = (1+3{\eps_3})d + (3{\eps_3} - {\eps_2})b \;< (1+3{\eps_3})d\;,
\end{equation}
where in the last inequality we used $\eps_2 > 3\eps_3$.
Recalling that $a \geq (1+{\eps_2})b$ and using $\eps_3 = \eps_2/24$ we get
\BEQ
\frac{c}{a} < \frac{(1+3{\eps_3})d}{(1+{\eps_2})b} = \frac{d}{b}\cdot \frac{1+\eps_2/8}{1+\eps_2}
          < \frac{d}{b}\cdot \left(1-\frac{\eps_2}{4}\right)\;.
\EEQ
%%%%%%%%%% OLD TEXT (NEW IS ESSENTIALLY THE SAME JUST A BIT SHORTER SINCE GOT RID OF TERM
% Define $r \eqdef d/b$\ignore{, so we have $r \in [{\eps_1},{\frac 1 {{\eps_1}}}].$}.
% Equation~(\ref{eq:z1}) then becomes
% \[
% c \leq ((1+ 3 {\eps_3})r + 3 {\eps_3} - {\eps_2})b,
% \]
% which, recalling that $a \geq (1+{\eps_2})b$, gives
% \BEQN
% {\frac c a} &\leq& {\frac {(1+ 3 {\eps_3})r + 3 {\eps_3} - {\eps_2}}{1+{\eps_2}}} \;=\;
% r \cdot \left( {\frac {1 + 3 {\eps_3} +
%         {\frac {3 {\eps_3} - {\eps_2}} r}} {1 + {\eps_2}} } \right)
% \;=\; {\frac d b} \cdot \left(
% {\frac
% {1 + {\frac {{\eps_2}} 8}   - {\frac {{3 \eps_3 - \eps_2} } {r}}}
% {1 + {\eps_2}} } \right)\nonumber\\
% &<& {\frac d b} \cdot \left( {\frac
%            {1 + {\frac {{\eps_2}} 8}} {1 + {\eps_2}} } \right)
% \;<\; {\frac d b} \cdot \left( 1 - {\frac {{\eps_2}} 4} \right),
% \EEQN
% where for the penultimate equality we used the fact that
% ${\eps_2} = 24 {\eps_3}.$ \mnote{{uses $\eps_2 = 24 \eps_3$}}
%%%%%%%%%%%%%%%%5
This proves the claim.
% \rnote{It seems to me this proof nowhere uses the fact that
% $\{1,\dots,i_j-1\}$ is an ${\eps_1}$-\good witness for $i_j$.
% This surprised me when I realized it, but we do use this in
% the completeness argument, so maybe it is OK.}
% \dnote{see the response in my email (Oct. 21)}
\EPF

Applying \autoref{clm:truth-will-out-1}, we get that
in Line~\ref{st:lastcheck-Dast-big} we have
\BEQ
{\frac {\D(\{1,\dots,i_j-1\})}{\D(i_j)}} \leq
  \left(1 - {\frac {{\eps_2}} 4} \right) \cdot
   {\frac {\D^\ast(\{1,\dots,i_j-1\})}{\D^\ast(i_j)}}\;.
\EEQ
Recalling that by the premise of this case $\D^\ast(i_j) \geq {\eps_1}$,
by applying \autoref{claim:partition} we have that
\mbox{$\{1,\dots,i_j-1\}$} is an ${\eps_1}$-\good witness for $i_j$.
Therefore, by \autoref{lem:compare}, with probability at least $1 - {\frac 1 {10\ell}}$
the call to {\sc Compare}$(\{i_j\},\{1,\dots,i_j-1\},{\frac {{\eps_2}}{16}},
\frac{2}{\eps_1}, {\frac 1 {10 \ell}})$ in Line~\ref{st:lastcheck-Dast-big}
either outputs an element of $\{\high,\low\}$ or outputs a value
$v \leq (1 - {\frac {{\eps_2}} 4}) (1 + {\frac {{\eps_2}}{16}})
{\frac {\D^\ast(\{1,\dots,i_j-1\})}{\D^\ast(i_j)}}
<
(1 - {\frac {{\eps_2}}{8}})
{\frac {\D^\ast(\{1,\dots,i_j-1\})}{\D^\ast(i_j)}}.
$
In either case the algorithm outputs \reject in
Line~\ref{st:lastcheck-Dast-big}, so we are done with Case~2(a).

\medskip

\noindent {\bf Case 2(b):}  $\D^\ast(i_j) < {\eps_1}$.
In this case the algorithm reaches Line~\ref{st:lastcheck}, and by
item~2 of \autoref{claim:partition}, we have
that $S_1,\dots,S_t$ is a partition of
$\{1,\dots,i_j-1\}$ and
each set $S_1,\dots,S_t$ is a ${\frac 1 2}$-\good
witness for $i_j,$ i.e.,
\begin{equation}
\label{eq:construction}
\text{for all~}i \in \{1,\dots,t\}, \quad
{\frac 1 2} \D^\ast(\newer{i_j}) \leq \D^\ast(S_i) \leq 2 \D^\ast(\newer{i_j}).
\end{equation}
We use the following \newer{claim}:

\BCM \label{clm:truth-will-out}
In Case~2(b) suppose $i_j > k^\ast$ is such that
$\D(i_j) \geq (1+{\eps_2})\D^\ast(i_j)$ and
$\D(\{1,\dots,i_j\}) \in
[1-3{\eps_3},1+3{\eps_3}]\D^\ast(\{1,\dots,i_j\})$.
Then at least %a $1-\rho = {\frac {{\eps_4}} 8}$
$(\eps_4/8)$-fraction of the sets $S_1,\dots,S_t$
are such that
\[
\D(S_i) \leq (1 + {{\eps_4}}) \D^\ast(S_i).
\ignore{
{\frac {\D(S_i)}{\D(i_j)}} \leq
\left(
1 - \gamma
\right)
\cdot
{\frac {\D^\ast(S_{a_r})}{\D^\ast(i_j)}}.
}
\]
\ECM

\BPF
The proof is by contradiction.
Let $\rho = 1- \eps_4/8$ and suppose that there are $w$ sets (without loss of generality we
call them $S_1,\dots,S_w$) that satisfy $\D(S_i) > (1+{{\eps_4}})\D^\ast(S_i)$, where $\rho' = {\frac w t} > \rho.$
We first observe that the weight of the $w$ subsets $S_1,\dots,S_w$ under $\D^\ast$,
as a fraction of $\D^\ast(\{1,\dots,i_j-1\})$, is at least
\[
{\frac {\D^\ast(S_1 \cup \dots \cup S_w)}
{\D^\ast(S_1 \cup \dots \cup S_w) + (t-w) \cdot 2 \D^\ast(\newer{i_j})}} \geq
{\frac {w {\frac {\D^\ast(i_{j})} 2}}
{w {\frac {\D^\ast(i_j)} 2} + (t-w) \cdot 2 \D^\ast(\newer{i_j})}}
= {\frac w {4t - 3w}} = {\frac {\rho'}{4-3\rho'}},
\]
where we used the right inequality in Equation~(\ref{eq:construction}) on
$S_{w+1},\dots,S_t$ to obtain the leftmost expression above, and the left inequality
in Equation~(\ref{eq:construction}) (together with the fact that ${\frac x {x+c}}$ is an increasing function of
$x$ for all $c>0$) to obtain the inequality above.
This implies that
\begin{eqnarray}
\D(\{1,\dots,i_j-1\}) &=&
\sum_{i=1}^w \D(S_i) +
\sum_{i=w+1}^t \D(S_i) \geq
(1+{{\eps_4}}) \sum_{i=1}^w \D^\ast(S_i) + \sum_{i=w+1}^t \D(S_i) \nonumber \\
&\geq&
(1+{{\eps_4}}) {\frac {\rho'}{4-3\rho'}} \D^\ast(\{1,\dots,i_j-1\}) \nonumber \\
&\geq&
(1+{{\eps_4}}) {\frac {\rho}{4-3\rho}} \D^\ast(\{1,\dots,i_j-1\})\;. \label{eq:ratio}
\end{eqnarray}
From Equation~(\ref{eq:ratio}) we have
\begin{eqnarray*}
\D(\{1,\dots,i_j\}) &\geq&
(1+{{\eps_4}}) {\frac {\rho}{4-3\rho}} \D^\ast(\{1,\dots,i_j-1\}) + (1 + {\eps_2}) \D^\ast(i_j)\\
&\geq& \left(1 + {\frac {3\eps_4} 8}\right)
\D^\ast(\{1,\dots,i_j-1\}) + (1 + {\eps_2}) \D^\ast(i_j)
\end{eqnarray*}
where for the first inequality above we used $\D(i_j) \geq (1+{\eps_2})\D^\ast(i_j)$ and for the second inequality we used $(1+{{\eps_4}}) {\frac {\rho}{4-3\rho}}  \geq 1 + {\frac {3\eps_4} 8}$. This implies that
\begin{eqnarray*}
\D(\{1,\dots,i_j\}) &\newer{>}&
\left(
1 + {\frac {3 {{\eps_4}}} 8}
\right)
\D^\ast(\{1,\dots,i_j-1\}) + \left(1 + {\frac {3 {{\eps_4}}} 8}\right) \D^\ast(i_j) =
\left( 1 + {\frac {3\eps_4} 8} \right)
\D^\ast(\{1,\dots,i_j\})
\\
&
\end{eqnarray*}
where the inequality follows from \newer{${\eps_2} > {\frac {3 {{\eps_4}}} 8}$}.
Since \newer{${\frac {3 {{\eps_4}}} 8} = 3{\eps_3}$,} though, this is a contradiction and the claim is proved.
%\mnote{{Uses ${\eps_2} \geq {\frac {3 {{\eps_4}}} 8}$ and ${\frac {3 {{\eps_4}}} 8} > 3{\eps_3}$}}
\EPF

Applying \autoref{clm:truth-will-out}, and recalling that
{$h=\Theta(1/\eps) = \Theta(1/{{\eps_4}})$} sets are chosen
randomly in Line~\ref{st:randomsets}, we have that with probability at
least $9/10$ there is some $r \in \{1,\dots,h\}$ such that
\mbox{$\D(S_{a_r}) \leq (1+{{\eps_4}})\D^\ast(S_{a_r})$}.
Combining this with $\D(i_j) \geq (1+{\eps_2})\D^\ast(i_j)$, we get that
\[
{\frac {\D(S_{a_r})}{\D(i_j)}} \leq
{\frac {1 + {{\eps_4}}}{1 + {\eps_2}}} \cdot
{\frac {\D^\ast(S_{a_r})}{\D^\ast(i_j)}}
\leq \left(1 -  {\frac {{\eps_4}} 2}\right) \cdot
{\frac {\D^\ast(S_{a_r})}{\D^\ast(i_j)}}.
\]
By \autoref{lem:compare}, with probability at least
$1 - {\frac 1 {10\ell h}}$ the call to
{\sc Compare}$(\{i_j\},S_{a_r},{\frac {{\eps_4}} 8},
4, {\frac 1 {10 \ell n}})$ in Line~\ref{st:lastcheck}
either outputs an element of $\{\high,\low\}$ or outputs a value
$v \leq (1 - {\frac {{\eps_4}} 2}) (1 + {\frac {{{\eps_4}}}{8}})
{\frac {\D^\ast(S_{a_r})}{\D^\ast(i_j)}}
<
(1 - {\frac {{{\eps_4}}}{4}})
{\frac {\D^\ast(S_{a_r})}{\D^\ast(i_j)}}.
$
In either case the algorithm outputs \reject in
Line~\ref{st:lastcheck}, so we are done in Case~2(b).
This concludes the proof of soundness and the proof of \autoref{thm:paircond-test-known}.
\EPF

\fi %%%%%%%%%%%%%%%%%%%%%%%%%%%%%%%%%%%%%%%%%% Enf of full version only

\ifnum\confversion=0
\def\fulldetails{1}
\else
\def\fulldetails{0}
\fi

\section{Testing equality between two unknown distributions}
\label{sec:d1d2}
\ifnum\confversion=0 %%%%%%%%%%%%%%%%%%%%%%%%%%%%%%%%%%%%%%%%%% Full version
\subsection{An approach based on \PCOND queries}
\fi

In this
\ifnum\confversion=0
subsection
\else
section
\fi
we consider the problem of testing whether two unknown
distributions $\D_1,\D_2$ are identical versus $\eps$-far, given
% sample
\PCOND
access to these distributions. Although this is known to require
$\bigOmega{N^{2/3}}$ many samples in the standard
model~\cite{BFRSW:10,Valiant:11}, we are able to give
a $\poly(\log N, 1/\eps)$-query algorithm using
$\PCOND$ queries,
% \cnote{Remove last part of this sentence?}
% \dnote{I think it is Ok.}
by taking advantage of comparisons to perform some
sort of \emph{clustering} of the domain.

On a high level the algorithm works as follows. First it obtains
(with high probability) a small set of points $R$ such that
almost every element in $[N]$, except possibly for some negligible
subset according to $\D_1$, has probability weight (under $\D_1$)
close to some ``representative'' in $R$.
% Then, it draws a
% sufficiently large sample according to both distributions,
% and considers their union $S$: for each of those representatives $r$,
% the set $S$ is used to estimate
%  \begin{enumerate}
%    \item the mass, both under $\D_1$ and $\D_2$, of the points
%    associated with $r$ (i.e., its ``neighborhood under $\D_1$'');
%    \item the fraction of points from this neighborhood that $\D_2$
%    classifies as far from $r$ (and therefore on which $\D_1$ and $\D_2$
%    differ).
% \end{enumerate}
Next, for each representative $r$ in $R$
 it obtains an estimate of the weight, according
to $\D_1$, of a set of points \newer{$U(r)$ such that $\D_1(u)$
is close to $\D_1(r)$ for each $u$ in $U(r)$}
(i.e., $r$'s ``neighborhood under $\D_1$'').
This is done using the procedure
{\sc Estimate-Neighborhood}\fullOrConfVersion{from \autoref{subsec:est-n}}{{(Algorithm~\ref{alg:est-n})}}.
Note that these neighborhoods can be interpreted roughly % like
as a succinct \emph{cover} of the support of $\D_1$ into
(not necessarily disjoint) sets of
points, where within each set the points have similar
weight (according to $\D_1$). % with similar mass.
Our algorithm is based on the observation that,
if $\D_1$ and $\D_2$ are far from each other, it must be the case
that one of these sets, \newer{denoted $U(r^\ast)$, reflects it
in one of the following ways: (1) $\D_2(U(r^\ast))$ differs
significantly from $\D_1(U(r^\ast))$; (2) $U(r^\ast)$ contains
a subset of points $V(r^\ast)$ such that $\D_2(v)$ differs
significantly from $\D_2(r^\ast)$ for each $v$ in $V(r^\ast)$, and
either $\D_1(V(r^\ast))$ is relatively large or
$\D_2(V(r^\ast))$ is relatively large.
}
(This structural result is made precise in
\autoref{lem:bad-j}). We thus take additional samples, both
from $\D_1$ and from $\D_2$, and compare the weight
(according to both distributions)
of each point in these samples to the representatives in $R$
(using the procedure
{\sc Compare}\fullOrConfVersion{from \autoref{subsec:compare})}{{(Algorithm~\ref{alg:compare})}}.
In this manner we detect (with high probability) that either
(1) or (2) holds.
% In this case, provided we obtained
% sufficiently accurate estimates, this difference will be detected;
% and the algorithm will reject.

\noindent We begin by formalizing the notion of a cover discussed above:

\BD[Weight-Cover]
Given a distribution $D$ on $[N]$ and a parameter $\eps_1>0$,
we say that a point $i \in [N]$ is
{\em $\eps_1$-covered\/} by a set $R = \{r_1,\dots,r_t\}\subseteq [N]$ if
there exists a point $r_j \in R$ such that
$\D(i) \in [1/(1+\eps_1), 1+\eps_1]\D(r_j)$.
Let the set of points in $[N]$ that are $\eps_1$-covered
by $R$ be denoted by $\newer{U}^\D_{\eps_1}(R)$.
We say that $R$ is an
\mbox{\emph{$(\eps_1,\eps_2)$-cover for $\D$}}
if $\D([N]\setminus \newer{U}_{\eps_1}^\D(R)) \leq \eps_2$.
\ED

\newer{For a singleton set $R=\{r\}$ we slightly abuse notation and write
$U^\D_\eps(r)$ to denote $U^\D_\eps(R)$; note that this aligns
with the notation established in (\ref{eq:Ugamma}).}

The following lemma says that a small sample of points % Removed "The following [straightforward] lemma"
drawn from $\D$ gives a cover with high probability:
\BL\label{lem:cover}
Let $\D$ be any distribution over $[N]$.
Given any fixed $c>0$, there exists a constant $c'>0$ such that
with probability at least $99/100$, a sample $R$ of size
$m=c' {\frac {\log(N/\eps)}{\eps^2}} \cdot \log \left(
{\frac {\log(N/\eps)}{\eps}}\right)$
drawn according to
distribution $\D$ is an $(\eps/c,\eps/c)$-cover for $\D$.
\EL

\BPF
Let $t$ denote $\lceil\ln(2cN/\eps) \cdot {\frac c \eps}\rceil$.
We define $t$ ``buckets'' of points with similar weight under $D$ as follows:
for \mbox{$i=0,1,\dots,t-1$,} define $B_i \subseteq [N]$
to be
\[
B_i \eqdef \left\{x \in [N] : {\frac 1 {(1+\eps/c)^{i+1}}}
< \D(x) \leq {\frac 1 {(1+\eps/c)^i}} \right\}.
\]
Let $L$ be the set of points $x$ which are not in any of $B_0,\dots,B_{t-1}$
(because $\D(x)$ is too small); since every point in $L$ has
$\D(x) < {\frac \eps {2c N}}$, one can see that $D(L) \leq {\frac \eps
{2c}}.$

It is easy to see that if the sample $R$ contains a point from a bucket
$B_j$ then every point $y \in B_j$ is ${\frac \eps c}$-covered by $R$.
We say that bucket $B_i$ is \emph{insignificant} if $\D(B_i) \leq
{\frac \eps {2ct}}$; otherwise bucket $B_i$ is \emph{significant}.
It is clear that the total weight under $D$ of all insignificant buckets is
at most $\eps/2c$.  Thus if we can show that for the claimed sample size,
with probability at least $99/100$ every significant
bucket has at least one of its points in $R$, we will have established the
lemma.

This is a simple probabilistic calculation:  fix any significant bucket
$B_j$.  The probability that $m$ random draws from $\D$ all miss $B_j$
is at most $(1- {\frac \eps {2ct}})^m$, which is at most ${\frac 1 {100 t}}$
for a suitable (absolute constant) choice of $c'$.
Thus a union bound over all (at most $t$) significant buckets gives that
with probability at least $99/100$, no significant bucket is missed by
$R$.
\EPF

\BL\label{lem:bad-j}
  Suppose $\totalvardist{\D_1}{\D_2}\geq \eps$, and let \mbox{$R=\{r_1,\dots,r_t\}$}
  be an $(\teps,\teps)$-cover for $\D_1$ where $\teps \leq \eps/100$.
  Then,
  there exists $j\in[t]$ such that at least one of the
  following conditions holds for every $\alpha \in [\teps,2\teps]$:
    \begin{enumerate} \fullOrConfIndentItems%[(i), noitemsep]
      \item  \label{lem:bad-j-a1}
      $\D_1( U^{\D_1}_{\alpha}(r_j) ) \geq \frac{\teps}{t}$ and
      $\D_2( U^{\D_1}_{\alpha}(r_j) )
           \notin [1-\teps, 1+\teps]\D_1( U^{\D_1}_{\alpha}(r_j) )$,
           or
      $\D_1( U^{\D_1}_{\alpha}(r_j) ) < \frac{\teps}{t}$ and
      \mbox{$\D_2(U^{\D_1}_{\alpha}(r_j) ) > \frac{2\teps}{t}$;}
       \item \label{lem:bad-j-a2}\sloppy
      $\D_1( U^{\D_1}_{\alpha}(r_j) ) \geq \frac{\teps}{t}$,
      and at least a $\teps$-fraction  of the points $i$
      in $U^{\D_1}_{\alpha}(r_j)$ satisfy
      \mbox{$\frac{\D_2(i)}{\D_2(r_j)}
        \notin[1/(1+\alpha+\teps), 1+\alpha+\teps]$};
      \item \label{lem:bad-j-a3}\sloppy
      $\D_1( U^{\D_1}_{\alpha}(r_j) ) \geq \frac{\teps}{t}$,
      and the total weight according to $\D_2$ of the points
      $i$ in $U^{\D_1}_{\alpha}(r_j)$ for which
      \mbox{$\frac{\D_2(i)}{\D_2(r_j)}
        \notin\left[1/(1+\alpha+\teps), 1+\alpha+\teps\right]$}
      is at least $\frac {\teps^2}{t}$;
    \end{enumerate}
\EL

\BPF
Without loss of generality, we can assume that \mbox{$\eps\leq 1/4$}.
Suppose, contrary to the claim, that
for each $r_j$ there exists $\alpha_j \in [\teps,2\teps]$ such
that if we let $U_j \eqdef U^{\D_1}_{\alpha_j}(r_j)$, then the following holds:
\BE
\item  If $\D_1(U_j) < \frac{\teps}{t}$, then
      $\D_2(U_j) \leq \frac{2\teps}{t}$;
\item If $\D_1(U_j) \geq \frac{\teps}{t}$, then:
  \BE
  \item $\D_2(U_j)
           \in [1-\teps,1+\teps]\D_1(U_j)$;
 \item  \sloppy Less than an $\teps$-fraction of the points $y$
      in $U_j$ satisfy
      \mbox{$\frac{\D_2(y)}{\D_2(r_j)}
        \notin\left[1/(1+\alpha_j+\teps), 1+\alpha_j+\teps\right]$};
 \item\sloppy
      The total weight according to $\D_2$ of the points $y$
      in $U_j$ for which
      \mbox{$\frac{\D_2(y)}{\D_2(r_j)}
        \notin\left[1/(1+\alpha_j+\teps), 1+\alpha_j+\teps\right]$}
      is at most $\frac{\teps^2}{t}$;
 \EE
\EE
We show that in such a case $\dtv(\D_1,D_2) < \eps$, contrary to
the premise of the claim.

\sloppy
Consider each point $r_j \in R$ such that
\mbox{$\D_1( U_j) \geq \frac{\teps}{t}$}.
By the foregoing discussion (point 2(a)), \mbox{$\D_2( U_j) \in [1-\teps,1+\teps]\D_1( U_j)$}.
By the definition of  $U_j$ (and since
$\alpha_j \leq 2\teps$),
\BEQ
\D_1(r_j) \in \left[1/(1+2\teps),1+2\teps\right]
     \frac{\D_1(U_j )}
     {\left|U_j\right|}\;.
\EEQ
% \cnote{I removed above the last inequality, which wasn't used for
% the next bounds.}
Turning to bound $\D_2(r_j)$,
%%%%%%%%%%%%%%%%%%%%%%%%%%%%%%%%%
\ifnum\fulldetails=0
by 2(b),
\BEQ
\D_2(r_j) \leq
    \frac{(1+3\teps)\D_2( U_j )}
     {(1-\teps)\left|U_j\right|}
     \leq (1+6\teps)\frac{\D_1( U_j )}
     {\left|U_j\right|}\;,
\EEQ
and by 2(c),
\BEQN
\D_2(r_j) &\geq&
    \frac{\D_2( U_j) -\teps^2/t}
     {(1+3\teps)\left|U_j\right|} % \nonumber \\
   % &\geq&
   \geq
    \frac{(1-\teps)\D_1( U_j)
           -\teps\D_1(U_j)}
     {(1+3\teps)\left|U_j\right|} \nonumber \\
     &\geq& (1-5\teps)\frac{\D_1( U_j )}
     {\left|U_j\right|}\;.
\EEQN
%%%%%%%%%%%%%%%%%%%%%%%%%%%%%%
\else
on one hand (by 2(b))
\BEQ
\D_2(U_j) = \sum_{y\in U_j} \D_2(y) \geq \teps|U_j|\cdot 0
  + (1-\teps)|U_j|\cdot \frac{\D_2(r_j)}{1+3\teps}\;,
\EEQ
and so
\BEQ
\D_2(r_j) \leq
    \frac{(1+3\teps)\D_2( U_j )}
     {(1-\teps)\left|U_j\right|}
     \leq (1+6\teps)\frac{\D_1( U_j )}
     {\left|U_j\right|}\;.
\EEQ
On the other hand (by 2(c)),
\BEQ
\D_2(U_j) = \sum_{y\in U_j} \D_2(y) \leq \frac{\teps^2}{t}
  + |U_j|\cdot (1+3\teps) \D_2(r_j)\;,
\EEQ
and so
\BEQ
\D_2(r_j) \geq
    \frac{\D_2( U_j) -\teps^2/t}
     {(1+3\teps)\left|U_j\right|}
   \geq
    \frac{(1-\teps)\D_1( U_j)
           -\teps\D_1(U_j)}
     {(1+3\teps)\left|U_j\right|}
     \geq (1-5\teps)\frac{\D_1( U_j )}
     {\left|U_j\right|}\;.
\EEQ
\fi
%%%%%%%%%%%%%%%%%%%%%%%%%
Therefore, for each such $r_j$ we have
\BEQ\label{eq:D2-rj-D1-rj}
\D_2(r_j) \in [1-8\teps,1+10\teps]\D_1(r_j)\;.
\EEQ
Let $C \eqdef \bigcup_{j=1}^t U_j$.
We next partition the points in $C$ so that each
point $i \in C$ is assigned to some $r_{j(i)}$
such that $i \in U_{j(i)}$. We define the following
``bad'' subsets of points in $[N]$:
\BE
  \fullOrConfCompactEnums\fullOrConfIndentItems
\item $B_1 \eqdef [N]\setminus C$,
so that $\D_1(B_1) \leq \teps$ (we later bound $\D_2(B_1)$);
\item $B_2 \eqdef
   \left\{i\in C:\D_1(U_{j(i)}) < \teps/t\right\}$, so that
     $\D_1(B_2) \leq \teps$
     and $\D_2(B_2) \leq 2\teps$;
\item $B_3\eqdef \left\{i\in C\setminus B_2:
         \D_2(i) \notin
          [1/(1+3\teps),1+3\teps]\D_2(r_{j(i)})\right\}$, so that
% \cnote{Since we deal with a $\teps$ fraction of the points of each bucket,
% and not of the mass, I think we can only prove a $(25/16)\teps\leq 2\teps$
% upper bound (it can be the case that all points are amongst the heaviest
% of their bucket).}
          $\D_1(B_3) \leq 2\teps$ and \mbox{$\D_2(B_3)\leq \teps^2$.}
\EE
Let $B \eqdef B_1\cup B_2\cup B_3$.
Observe that for each $i \in [N]\setminus B$
we have that
\fullOrConfVersion{
\BEQ\label{eq:D2-i-D1-i}
\D_2(i) \in [1/(1+3\teps),1+3\teps]\D_2(r_{j(i)})
  \subset [1-{15}\teps,1+{15}\teps]\D_1(r_{j(i)})   %  \in [1-15\teps,1+15\teps]\D_1(r_{j(i)})
  \subset [1-{23}\teps,1+{23}\teps]\D_1(i)\;,        %%% Full/Conf        %  \in [1-20\teps,1+20\teps]\D_1(i)\;,
\EEQ
}{
\BEQN\label{eq:D2-i-D1-i}
\D_2(i) &\in& [1/(1+3\teps),1+3\teps]\D_2(r_{j(i)}) \nonumber \\
  & \subset& [1-{15}\teps,1+{15}\teps]\D_1(r_{j(i)}) \nonumber \\
  & \subset& [1-{23}\teps,1+{23}\teps]\D_1(i)\;,
\EEQN
}
where the first containment follows from the fact that
$i \notin B$, the second follows
from Equation~(\ref{eq:D2-rj-D1-rj}), and the third from
the fact that $i\in U_{j(i)}$.
In order to complete the proof we need a bound on $\D_2(B_1)$,
which we obtain next.
\fullOrConfVersion{       %%% Full/Conf
\BEQN
\D_2(B_1) &=& 1-\D_2([N]\setminus B_1) \;\leq\; 1-\D_2([N]\setminus B)
 \;\leq\; 1 - (1-{23}\teps)\D_1([N]\setminus B)\nonumber \\
 & \leq& 1- (1-{23}\teps)(1-{4}\teps) \; \leq\; {27}\teps\;.
\EEQN
}{
\BEQN
\D_2(B_1) &=& 1-\D_2([N]\setminus B_1) \;\leq\; 1-\D_2([N]\setminus B) \nonumber \\
& \leq& 1 - (1-{23}\teps)\D_1([N]\setminus B)\nonumber \\
 & \leq& 1- (1-{23}\teps)(1-{4}\teps) \; \leq\; {27}\teps\;.
\EEQN
}                        %%% End Full/Conf
Therefore,
\fullOrConfVersion{       %%% Full/Conf
\BEQN
\dtv(\D_1,\D_2) &=& \frac{1}{2}\sum_{i=1}^N \left|\D_1(i)-\D_2(i)\right|
          \nonumber \\
&\leq& \frac{1}{2}
  \Big(\D_1(B) + \D_2(B)
  + \sum_{i\notin B} {23}\teps\D_1(i)\Big)\;\; \;<\; \;\eps\;,
\EEQN
}{
\BEQN
\dtv(\D_1,\D_2) &=& \frac{1}{2}\sum_{i=1}^N \left|\D_1(i)-\D_2(i)\right|
          \nonumber \\
&\leq& \frac{1}{2}
  \Big(\D_1(B) + \D_2(B)
  + \sum_{i\notin B} {23}\teps\D_1(i)\Big)\;\; \nonumber \\
  & <& \;\eps\;,
\EEQN
}                         %%% End Full/Conf
and we have reached a contradiction.
\EPF

%%%%%%%%%%%%%%% References of the form 1-a, 1-b... (instead of 1a, 1b...)
\makeatletter
 \renewcommand{\p@enumii}{\theenumi-}
\makeatother
%%%%%%%%%%%%%%%
\begin{algorithm}[ht!]
\label{alg:test-equality}
\caption{Algorithm {\sc $\PCOND_{\D_1,\D_2}$-Test-Equality-Unknown}}
\ifnum\confversion=1  %%% Conf
  \textbf{Input:} {
\else                 %%% Full
  \SetKwInOut{Input}{Input}
  \Input{
\fi                   %%% End
\PCOND % and \SAMP
query access to distributions $\D_1$
and $\D_2$ and a parameter $\eps$.}
\BE
  \fullOrConfCompactEnums %% Conf version

\item Set $\teps = \eps/100$.
\item Draw a sample $R$ of size
$t= \tilde{\Theta}\!\left(\frac{\log N}{\eps^2}\right)$
from $\D_1$.
% \item In what follows we use the shorthand {\sc EN} for the procedure
% {\sc Estimate-Neighborhood}.
\item For each $r_j \in R$:
  \BE
    \fullOrConfCompactEnums\fullOrConfIndentItems %% Conf version

    \item \label{st:call-est-n}
     Call $\mbox{\sc Estimate-Neighborhood}_{\D_1}$ on $r_j$ with
    $\kappa=\teps$, $\eta=\frac{\teps}{8}$,
    $\beta=\frac{\teps}{2t}$, $\delta=\frac{1}{100t}$ and
    let the output be denoted by $(\hat{w}^{(1)}_j,\alpha_j)$.
    \item Set $\theta = \kappa\eta\beta\delta/64 = \tilde{\Theta}(\eps^7/\log^2 N)$.
     \item Draw a sample $S_1$ from $\D_1$,
    of size $s_1= \Theta\!\left(\frac{t}{\eps^2}\right)
                  = \tilde{\Theta}\!\left(\frac{\log N}{\eps^4}\right)$.
    % \Theta\!\left(\frac{\log N}{\eps^4}\right)$.
    \item Draw a sample $S_2$ from $\D_2$,
    of size $s_2= \Theta\!\left(\frac{t\log t}{\eps^3}\right)
                   = \tilde{\Theta}\!\left(\frac{\log N}{\eps^5}\right)$.
          % \Theta\!\left(\frac{\log N}{\eps^4}\right)$.
    \item \label{st:call-compare}
     For each point $i \in S_1\cup S_2$
     call $\mbox{\sc Compare}_{\D_1}(\{r_j\},\{i\},\theta/4,4,1/(200t(s_1+s_2)))$
     and $\mbox{\sc Compare}_{\D_2}(\{r_j\},\{i\},\theta/4,4,1/(200t(s_1+s_2)))$,
       and let the outputs be denoted $\rho^{(1)}_{r_j}(i)$ and
       $\rho^{(2)}_{r_j}(i)$, respectively (where in particular these
       outputs may be \high or \low).
  \item Let $\hat{w}^{(2)}_j$ be the fraction of occurrences of $i\in S_2$
      such that $\rho^{(1)}_{r_j}(i) \in [1/(1+\alpha_j+\theta/2),1+\alpha_j+\theta/2]$.
  \item\label{st:comp-w} If
  ( $\hat{w}^{(1)}_j\leq \frac{3}{4}\frac{\teps}{t}$ and
  $\hat{w}^{(2)}_j > \frac{3}{2}\frac{\teps}{t}$ ) or
   ( $\hat{w}^{(1)}_j >\frac{3}{4}\frac{\teps}{t}$ and
  $\hat{w}^{(2)}_j/\hat{w}^{(1)}_j \notin [1-\teps/2,1+\teps/2]$ ),
   then output \reject.
  \item \label{st:comp-i-rj}If there exists $i \in S_1\cup S_2$ such that
  $\rho^{(1)}_{r_j}(i) \in [1/(\alpha_j+\teps/2),1+\alpha_j+\teps/2]$
  and
  $\rho^{(2)}_{r_j}(i) \notin [1/(\alpha_j+3\teps/2),1+\alpha_j+3\teps/2]$,
  then output \reject.
  \EE
\item Output \accept.
\EE
\end{algorithm}
%%%%%%%%%%%%%%%
\makeatletter
 \renewcommand{\p@enumii}{\theenumi}
\makeatother
%%%%%%%%%%%%%%%

\BT\ignore{\mnote{$>\leadsto\geq$}}
If $\D_1 = \D_2$ then with probability at least $2/3$
Algorithm {\sc \PCOND-Test-Equality-Unknown} returns
\accept, and if {$\dtv(\D_1,\D_2) \geq \eps$}, then
with probability at least $2/3$
Algorithm {\sc \PCOND-Test-Equality-Unknown} returns
\reject.
The number of % \SAMP and
\PCOND queries performed by the
algorithm is
$\tildeO\big(\frac{\log^6 N}{\eps^{21}}\big)$.
\ET

\BPF
The number of queries performed by the algorithm is the
sum of: (1) $t$ times the number of queries performed in each
execution of {\sc Estimate-Neighborhood} (in
Line~\ref{st:call-est-n}) and (2) $t\cdot(s_1+s_2)=O(t \cdot s_2)$
times the number of queries performed in each
execution of {\sc Compare} (in
Line~\ref{st:call-compare}). By \autoref{lem:est-n}
(and the settings of the parameters in the calls to
{\sc Estimate-Neighborhood}),
 the first term is
 $O\!\left(t\cdot \frac{\log(1/\delta)
        \cdot \log(\log(1/\delta)/(\beta\eta^2))}
            {\kappa^2\eta^4\beta^3\delta^2}\right)
 = \tilde{O}\!\left(\frac{\log^6 N}{\eps^{{19}}}\right)$,
 and by \autoref{lem:compare}
(and the settings of the parameters in the calls to
{\sc Compare}), the second term is
$O\!\left(t \cdot s_2\cdot \frac{\log(t\cdot s_2)}{\theta^2}\right)
= \tilde{O}\!\left(\frac{\log^6 N}{\eps^{21}}\right)$,
so that we get the bound stated in the theorem.

We now turn to establishing the correctness of the algorithm.
We shall use the shorthand $U_j$ for $U^{\D_1}_{\alpha_j}(r_j)$,
and $U'_j$ for $U^{\D_1}_{\alpha_j+\theta}(r_j)$.
We consider the following ``desirable'' events.
\BE
\item The event $E_1$ is that the sample $R$ is a
$(\teps,\teps)$-weight-cover for $\D_1$ (for
$\teps = \eps/100$). By
\autoref{lem:cover} (and an appropriate
constant in the $\Theta(\cdot)$ notation for the
size of $R$), the probability that $E_1$ holds is
at least $99/100$.
\item The event $E_2$ is that all calls to the procedure
{\sc Estimate-Neighborhood} are as specified by \autoref{lem:est-n}.
By the setting of the confidence parameter in the calls to the procedure,
the event $E_2$ holds with probability at least $99/100$.
\item The event $E_3$ is that all calls to the procedure
{\sc Compare} are as specified by \autoref{lem:compare}.
By the setting of the confidence parameter in the calls to the procedure,
the event $E_3$ holds with probability at least $99/100$.
\item The event $E_4$ is that
 $\D_2(U'_j \setminus U_j) \leq \eta\beta/16 = \teps^2/({256} t)$
 for each $j$. If $\D_2 = \D_1$ then this event follows from
 $E_2$. Otherwise,  it holds with probability at least $99/100$
 by the setting
% \cnote{Not really easy to see, unless familiar with the proof of \autoref{lem:est-n}
%(for \textsf{Compare}).}
of $\theta$ and the choice of $\alpha_j$
 (as shown in the proof of \autoref{lem:est-n} in the analysis
 of the event $E_1$ there.%\fullOrConfVersion{}{(see full version of this paper)}).
\item The event $E_5$ is defined as follows. For each $j$,
  if $\D_2(U_j) \geq \teps/(4t)$, then
  $|S_2 \cap U_j|/|S_2| \in [1-\teps/10,1+\teps/10]\D_2(U_j)$,
   and if
  $\D_2(U_j) < \teps/(4t)$ then
  $|S_2 \cap U_j|/|S_2| < (1+\teps/10)\teps/(4t)$.  This event holds with
  probability at least $99/100$ by applying a multiplicative
  Chernoff bound in the first case, and\fullOrConfVersion{\autoref{cor:CB-upperlower}}{the corollary from Exercise~1.1 of \cite{DP09})} in the second.
\item The event $E_6$ is that for each $j$ we have
  $|S_2 \cap (U'_j\setminus U_j)|/|S_2| \leq \teps^2/({128} t)$.
  Conditioned on $E_4$, the event $E_6$ holds
  with probability at least $99/100$ by applying\fullOrConfVersion{\autoref{cor:CB-upperlower}}{the abovementioned corollary}.
\EE
From this point on we assume that events $E_1-E_6$ all hold.
Note that in particular this implies the following:
% \mnote{D: Broke into itemization for easier reading}
\BE
\item By $E_2$, for every $j$:
  \BI\fullOrConfIndentItems\fullOrConfCompactEnums
  \item If $\D_1(U_j) \geq \beta = \teps/(2t)$,
  then
  $ \hat{w}_j^{(1)} \in [1-\eta,1+\eta]\D_1(U_j) =
     [1-\teps/8,1+\teps/8]\D_1(U_j)$. %\D_2(U_j)$,
  \item If $\D_1(U_j) < \teps/(2t)$, then
  $\hat{w}_j^{(1)} \leq (1+\teps/8)(\teps/(2t))$.
  \EI
\item By $E_3$, for every $j$ and
for each point $i \in S_1\cup S_2$:
  \BI\fullOrConfIndentItems\fullOrConfCompactEnums
  \item If $i \in U_j$, {then} \ifnum\confversion=1 \\ \fi
  \mbox{$\rho_{r_j}^{(1)}(i) \in [1/(1+\alpha_j+\frac{\theta}{2}),1+\alpha_j+\frac{\theta}{2}]$}.
  \item If
  $i \notin U'_j$, then
  $\rho_{r_j}^{(1)}(i) \notin [1/(1+\alpha_j+\frac{\theta}{2}),1+\alpha_j+\frac{\theta}{2}]$.
  \EI
\item By the previous item and $E_4$--$E_6$:
 \BI\fullOrConfIndentItems\fullOrConfCompactEnums
 \item If
$\D_2(U_j) \geq \teps/(4t)$, then $\hat{w}^{(2)}_j\geq (1-\teps/10)\D_2(U_j)$
 and $\hat{w}^{(2)}_j \leq (1+\teps/10)\D_2(U_j) + \teps^2/({128} t)
      \leq (1+\teps/8)\D_2(U_j)$.
 \item If $\D_2(U_j) < \teps/(4t)$ then
  $\hat{w}^{(2)}_j \leq (1+\teps/10)\teps/(4t) + \teps^2/({128} t)
      \leq (1+\teps/4)(\teps/(4t))$.
  \EI
\EE

\noindent{\bf Completeness.}~
% To establish the completeness of the algorithm,
\sloppy
Assume
$\D_1$ and $\D_2$ are the same distribution $\D$.
For each $j$, if $\D(U_j) \geq \teps/t$, then
by the foregoing discussion,
$\hat{w}^{(1)}_j \geq (1-\teps/8)\D(U_j) > 3\teps/(4t)$
and
$\hat{w}^{(2)}_j/\hat{w}^{(1)}_j \in [(1-\teps/8)^2,(1+\teps/8)^2]
 \subset [1-\teps/2,1+\teps/2]$,
so that the algorithm does not reject in Line~\ref{st:comp-w}.
%%%%%%
\ignore{
If $\D(U_j) < \teps/t$, then there are two cases.
If $\D(U_j) \leq \teps/(2t)$, then $\hat{w}^{(1)}_j \leq 3\teps/(4t)$
and $\hat{w}^{(2)}_j \leq 3\teps/(2t)$, while if
$\D(U_j) > \teps/(2t)$, then
$\hat{w}^{(2)}_j/\hat{w}^{(1)}_j \in
   [(1-\teps/8)^2,(1+\teps/8)^2] \subset [1-\teps/2,1+\teps/2]$,
and the algorithm does not reject in this case either
in Line~\ref{st:comp-w}. By $E_3$, the algorithm does not
reject in Line~\ref{st:comp-i-rj} either.
\dnote{This needs doublechecking and maybe a bit more elaboration}}
%%%%%%%
Otherwise (i.e., $\D(U_j) < \teps/t$), we consider two subcases.
Either \mbox{$\D(U_j) \leq \teps/(2t)$}, in which case
$\hat{w}^{(1)}_j \leq 3\teps/(4t)$, or \mbox{$\teps/(2t) < \D(U_j)< \teps/t$},
and then % $w^{(2)}_j/\hat{w}^{(1)}_j
$\hat{w}^{(1)}_j\in [1-\teps/8,1+\teps/8]\D_1(U_j)$.
Since in both cases
$\hat{w}^{(2)}_j \leq (1+\teps/8)\D(U_j)\leq 3\teps/(2t)$,
the algorithm does not reject in Line~\ref{st:comp-w}.
By $E_3$, the algorithm does not reject in Line~\ref{st:comp-i-rj} either.
We next turn to establish soundness.

% We now turn to establishing soundness.
\medskip\noindent
{\bf Soundness.}~
Assume {$\dtv(\D_1,\D_2) \geq \eps$}.\ignore{\mnote{$>\leadsto\geq$}}
By applying \autoref{lem:bad-j}
on $R$ (and using $E_1$),
there exists an index $j$ for which one of the items in the
lemma holds. We denote this index by $j^\ast$, and consider
the three items in the lemma.

\BE
\item If Item~\ref{lem:bad-j-a1} holds, then
% by the previous discussion
% regarding the quality of
% the estimates $\hat{w}^{(1)}_{j^\ast}$
% and $\hat{w}^{(2)}_{j^\ast}$, the algorithm
% rejects in Line~\ref{st:comp-w}
we consider its two cases:
  \BE
  \item In the first case, $\D_1(U_{j^\ast}) \geq \teps/t$ and
  $\D_2(U_{j^\ast}) \notin [1-\teps,1+\teps]\D_1(U_{j^\ast})$. Due to the lower
  bound on $\D_1(U_{j^\ast})$ we have that
  $\hat{w}^{(1)}_{j^\ast} \in [1-\teps/8,1+\teps/8]\D_1(U_{j^\ast})$,
  so that in particular $\hat{w}^{(1)}_{j^\ast} > 3\teps/(4t)$.
  As for $\hat{w}^{(2)}_{j^\ast}$,
  either $\hat{w}^{(2)}_{j^\ast} < (1-\teps)(1+\teps/8)\D_1(U_{j^\ast})$
  (this holds both when $\D_2(U_{j^\ast})\geq \teps/(4t)$ and when
  $\D_2(U_{j^\ast})< \teps/(4t)$)
  or $\hat{w}^{(2)}_{j^\ast} >(1+\teps)(1-\teps/10)\D_1(U_{j^\ast})$.
  In either (sub)case $\hat{w}^{(2)}_{j^\ast}/\hat{w}^{(1)}_{j^\ast}\notin [1-\teps/2,1+\teps/2]$,
  causing the algorithm to
  reject in (the second part of ) Line~\ref{st:comp-w}.
  \item In the second case, \mbox{$\D_1(U_{j^\ast}) < \teps/t$} and
  $\D_2(U_{j^\ast}) > 2\teps/t$. Due to the lower bound on
  $\D_2(U_{j^\ast})$ we have that
  $\hat{w}^{(2)}_{j^\ast} \geq (1-\teps/10)\D_2(U_{j^\ast}) >(1-\teps/10)(2\teps/t)$,
  so that in particular $\hat{w}^{(2)}_{j^\ast} > (3\teps/(2t))$.
  As for $\hat{w}^{(1)}_{j^\ast}$, if $\D_1(U_{j^\ast}) \leq \teps/(2t)$, then
   $\hat{w}^{(1)}_{j^\ast} \leq 3\teps/(4t)$,  causing the algorithm
   to reject in (the first part of) Line~\ref{st:comp-w}.
   If $\teps/(2t)< \D_1(U_{j^\ast}) \leq \teps/t $, then
   $\hat{w}^{(1)}_{j^\ast} \in [1-\teps/8,1+\teps/8]\D_1(U_{j^\ast})
      \leq (1+\teps/8)(\teps/t)$, so that
   $\hat{w}^{(2)}_{j^\ast} /\hat{w}^{(1)}_{j^\ast}
       \geq \frac{(1-\teps/10)(2\teps/t)}{(1+\teps/8)\teps/t}>(1+\teps/2)$,
   causing the algorithm
   to reject in (either the first or second part
   of) Line~\ref{st:comp-w}.
  \EE
% \cnote{I'm not quite sure here: if, in the first item, we get that
% ``$D_1(U_{j*}) < \teps/t$ and $D_2(U_{j*}) > 2\teps/t$'',
% the guarantee on $\hat{w}_j^{(1)}$ does not imply that
% it will be at most $(3/4)\teps/t$ -- it still can be slightly
% more, eg $(3/2)\teps/t$ -- can't it?}.
% \dnote{Is it Ok? or did I miss something?}
\item If Item~\ref{lem:bad-j-a2} holds,
then, by the choice of the size of $S_1$,
which is $\Theta(t/\teps^2)$, \newer{and since all points in $U_{j^\ast}$
have approximately the same weight according to $D_1$},
with probability
at least $99/100$, the sample $S_1$ will contain a point $i$ for
which \mbox{$\frac{\D_2(i)}{\D_2(r_{j^\ast})}
        \notin[1/(1+\alpha_{j^\ast}+\teps), 1+\alpha_{j^\ast}+\teps]$},
        and by $E_3$ this will be detected in Line~\ref{st:comp-i-rj}.
\item Similarly, if Item~\ref{lem:bad-j-a3} holds,
then by the choice of the size of $S_2$, with probability
at least $99/100$, the sample $S_2$ will contain a point $i$ for
which \mbox{$\frac{\D_2(i)}{\D_2(r_{j^\ast})}
        \notin[1/(1+\alpha_{j^\ast}+\teps), 1+\alpha_{j^\ast}+\teps]$},
        and by $E_3$ this will be detected in Line~\ref{st:comp-i-rj}.
\EE
The theorem is thus established.
% \dnote{All the settings of constants need doublechecking here as well,
% and maybe also some more elaboration.}
\EPF

% \newpage

%\section{Testing Equality between Two Unknown Distributions}
%\label{sec:d1d2}

\subsection{An approach based on simulating \EVAL}
\label{ssec:sim-eval}

\ignore{

%In the previous subsection we gave a $\poly(\log N, 1/\eps)$-query
%$\PCOND$ algorithm for testing whether two unknown distributions
%$\D_1$, $\D_2$ are identical versus $\eps$-far.  In this section we
%present a very different algorithm for this problem, using a
%$\COND$ oracle.  While the query complexity of our new algorithm
%is similar ($\poly(\log N, 1/\eps)$) to that of the previous
%algorithm, and the $\COND$ oracle is stronger than the
%$\PCOND$ oracle, we feel that the algorithm described in this
%section is interesting because of the technique employed.

}

% \rnote{Add in prose connecting this to previous subsection,
% where we will have given a $\PCOND$ algorithm}

In this subsection we present an alternate approach for testing
whether two unknown distributions $\D_1,\D_2$ are identical versus
$\eps$-far.  We prove the following theorem:

\BT \label{thm:cond-test-d1d2}
{\sc $\COND$-{Test-Equality-Unknown}} is a
\[
\tilde{O}\!\left( {\frac {(\log N)^5}{\eps^4}} \right)
\]
-query
algorithm with the following properties: given $\COND_{\D_1}$,
$\COND_{\D_2}$ oracles for any two distributions
$\D_1,\D_2$ over $[N]$,
it outputs $\accept$ with probability at least $2/3$ if $\D_1 =
\D_2$ and outputs $\reject$ with probability at least $2/3$ if
$\dtv(\D_1,\D_2) \geq \eps.$
\ET

At the heart of this result is our efficient simulation of an ``approximate
$\EVAL_\D$ oracle'' using a $\COND_\D$ oracle.  (Recall
that an $\EVAL_\D$ oracle is an oracle which, given as input
an element $i \in [N]$, outputs the numerical value $\D(i).$)
We feel that this efficient simulation of an approximate $\EVAL$ oracle
using a $\COND$ oracle is of independent interest since it sheds
light on the relative power of the $\COND$ and $\EVAL$ models.

In more detail, the starting point of our approach to prove
\autoref{thm:cond-test-d1d2} is a simple algorithm from
\cite{RubinfeldServedio:09} that uses an $\EVAL_\D$
oracle to test equality between $\D$ and a known distribution $\D^\ast.$
We first show
(see \autoref{thm:d1d2-using-approx-eval})
that a modified version of the algorithm, which uses
a $\SAMP$ oracle and an ``approximate'' $\EVAL$ oracle,
can be used to efficiently test equality between two unknown
distributions $\D_1$ and $\D_2$.  As we show (in \autoref{sec:approx-eval-simulator}) % how
the required ``approximate'' $\EVAL$ oracle can be efficiently implemented using
a $\COND$ oracle, \newer{and so} \autoref{thm:cond-test-d1d2} follows straightforwardly by combining Theorems~\ref{thm:d1d2-using-approx-eval} and~\ref{thm:approx-eval-simulator}.

\subsubsection{Testing equality between \texorpdfstring{$\D_1$ and $\D_2$}{\D1 and \D2} using an approximate \EVAL oracle.}
\label{sec:test-d1d2-approx-eval}

We now show how an approximate $\EVAL_{\D_1}$
oracle, an approximate $\EVAL_{\D_2}$ oracle, and a $\SAMP_{\D_1}$
oracle can be used together to test whether
$\D_1 = \D_2$ versus $\dtv(\D_1,\D_2) \geq \eps.$
As mentioned earlier, the approach is a simple extension of the $\EVAL$
algorithm
given in Observation~24 of \cite{RubinfeldServedio:09}.

\BT \label{thm:d1d2-using-approx-eval}
Let $\ORACLE_1$ be an $(\eps/100,\eps/100)$-approximate $\EVAL_{\D_1}$
simulator and let $\ORACLE_2$ be an $(\eps/100,\eps/100)$-approximate
$\EVAL_{\D_2}$ simulator.  There is an algorithm
{\sc Test-Equality-Unknown} with the following properties:  for
any distributions $\D_1,\D_2$ over $[N]$, algorithm
{\sc Test-Equality-Unknown} makes $O(1/\eps)$ queries to
$\ORACLE_1$, $\ORACLE_2$, $\SAMP_{\D_1}$, \newer{$\SAMP_{\D_2}$}, and it outputs
\accept with probability at least $7/10$ if $\D_1 = \D_2$ and outputs
\reject with probability at least $7/10$ if $\dtv(\D_1,\D_2) \geq \eps.$
\ET

\begin{algorithm}\begin{algorithmic}[1]
\REQUIRE query access to $\ORACLE_1$, to $\ORACLE_2$, and access to
$\SAMP_{\D_1}$\newer{, $\SAMP_{\D_2}$ oracles}
\STATE \label{st:init-sample1}
Call the $\SAMP_{\D_1}$ oracle $m = 5/\eps$
times to obtain points $h_1,\dots,h_{m}$ distributed according
to ${\D_1}$.
\STATE \label{st:init-sample2}
Call the $\SAMP_{\D_2}$ oracle $m = 5/\eps$
times to obtain points $h_{m+1},\dots,h_{2m}$ distributed according
to ${\D_2}$.
\FOR{ $j=1$ to $2m$ }
    \STATE \label{st:call-eval-d1}  Call $\ORACLE_1(h_j)$.  If it
returns \unknown then output \reject, otherwise let $v_{1,i} \in [0,1]$
be the value it outputs.
    \STATE \label{st:call-eval-d2}  Call $\ORACLE_2(h_j)$.  If it
returns \unknown then output \reject, otherwise let $v_{2,i} \in [0,1]$
be the value it outputs.
\IF {$v_{1,j} \notin [1 - \eps/8,1+\eps/8]v_{2,j}$}
\label{st:check-close}
\STATE \label{st:reject-because-far} {output \reject and exit}
\ENDIF\ENDFOR
\STATE output \accept
\end{algorithmic}\caption{\label{algo:test-equality-unknown}\sc Test-Equality-Unknown}
\end{algorithm}

It is clear that {\sc Test-Equality-Unknown} makes $O(1/\eps)$ queries
as claimed.  To prove
\autoref{thm:d1d2-using-approx-eval} we argue completeness and
soundness below.

\medskip

\noindent {\bf Completeness:}
Suppose that $\D_1 = \D_2.$  Since $\ORACLE_1$ is an $(\eps/100,
\eps/100)$-approximate $\EVAL_{\D_1}$ simulator, the probability
that any of the $2m=10/\eps$ points $h_1,\dots,h_{2m}$ drawn
in Lines~1 and~2 lies in $S^{(\eps/100,\D_1)}$ is at most $1/10.$
Going forth, let us assume that all points $h_i$ indeed
lie outside $S^{(\eps/100,\D_1)}$.
Then for each execution of Line~4 we have that with probability
at least $1-\eps/100$ the call to $\ORACLE(h_i)$ yields a value
$v_{1,i}$ satisfying $v_{1,i} \in [1-{\frac \eps {100}},
1+{\frac \eps {100}}] \D_1(i)$.  The same holds for each execution of
Line~5.  Since there are $20/\eps$ total executions of
Lines~4 and~5, with overall probability at least 7/10 we have that
each $1 \leq j \leq m$ has
$v_{1,j},v_{2,j} \in [1-{\frac \eps {100}}, 1+{\frac \eps {100}}] \D_1(i)$.
If this is the case then $v_{1,j},v_{2,j}$ pass the check in
Line~\ref{st:check-close},
and thus the algorithm outputs \accept with overall probability at least
$7/10.$

\medskip

\noindent {\bf Soundness:}
Now suppose that $\dtv(\D_1,\D_2) \geq \eps$.
Let us say that $i \in [N]$ is \emph{good} if
$\D_1(i) \in [1 - \eps/5,1+\eps/5]\D_2(i)$.  Let $\textrm{BAD} \subseteq [N]$
denote the set of all $i \in [N]$ that are not good.
We have
\[
2\dtv(\D_1,\D_2) =
\sum_{i \text{~is good}}|\D_1(i)-\D_2(i)|
+ \sum_{i \text{~is bad}}|\D_1(i)-\D_2(i)| \geq 2 \eps.
\]
Since
\[
\sum_{i \text{~is good}}|\D_1(i)-\D_2(i)| \leq
\sum_{i \text{~is good}} {\frac \eps 5} |\D_2(i)| \leq
{\frac \eps 5},
\]
we have
\[
\sum_{i \text{~is bad}}\left(|\D_1(i)| + |\D_2(i)|\right) \geq
\sum_{i \text{~is bad}}|\D_1(i)-\D_2(i)|
\geq {\frac 9 5}\eps.
\]
Consequently it must be the case that either
$
\D_1(\textrm{BAD}) \geq {\frac 9 {10}}\eps$ or
$
\D_2(\textrm{BAD}) \geq {\frac 9 {10}}\eps.$  For the rest
of the argument we suppose that
$\D_1(\textrm{BAD}) \geq {\frac 9 {10}}\eps$
(by the symmetry of the algorithm, an identical argument to the
one we give below but with the roles of $D_1$ and $D_2$ flipped throughout
handles the other case).

Since $\D_1(\textrm{BAD}) \geq {\frac 9 {10}}\eps$, a simple calculation shows that
with probability at least $98/100$ at least one of the $5/\eps$ points
$h_1,\dots,h_{m}$ drawn in Line~1 belongs to $\textrm{BAD}$.
For the rest
of the argument we suppose that indeed (at least) one of these points
is in $\textrm{BAD}$; let $h_{i^\ast}$ be such a point.
Now consider the execution of Line~4 when $\ORACLE_1$ is called
on $h_{i^\ast}.$  By \autoref{def:approx-eval}, whether
or not $i^\ast$ belongs to $S^{(\eps/100,\D_1)}$, with probability
at least $1 - \eps/100$ the call to $\ORACLE_1$ either causes
{\sc Test-Equality-Unknown} to \reject in Line~4 (because $\ORACLE_1$
returns \unknown) or it returns a value $v_{1,i^\ast} \in
[1-{\frac \eps {100}},1+ {\frac \eps {100}}]\D_1(i^\ast)$.   We
may suppose that it returns a value $v_{1,i^\ast} \in
[1-{\frac \eps {100}},1+ {\frac \eps {100}}]\D_1(i^\ast)$.
Similarly, in the execution of Line~5 when $\ORACLE_2$ is called on
$h_{i^\ast}$, whether or not $i^\ast$ belongs to $S^{(\eps/100,\D_2)}$,
with probability at least $1-\eps/100$ the call to $\ORACLE_2$ either
causes {\sc Test-Equality-Unknown} to reject in Line~5 or it returns
a value $v_{2,i^\ast} \in [1-{\frac \eps {100}},1+{\frac \eps {100}}]
\D_2(i^\ast).$  We may suppose that it returns a
value $v_{2,i^\ast} \in
[1-{\frac \eps {100}},1+ {\frac \eps {100}}]\D_2(i^\ast)$.
%Putting the pieces together, we have that
%$v_{1,i^\ast} \geq (1 + {\frac {18\eps}{100}})\D_2(i^\ast)$
%and $v_{2,i^\ast} \leq (1 + {\frac \eps {100}}) \D_2(i^\ast).$
But recalling that $i^\ast \in \textrm{BAD}$, an easy calculation shows that
the values $v_{1,i^\ast}$ and $v_{2,i^\ast}$ must be
multiplicatively far enough from each other that
the algorithm will output \reject in Line~7.  Thus with
overall probability at least $96/100$ the algorithm outputs
\reject.
\qed

\section{An algorithm for estimating the distance to uniformity}
In this section we describe an algorithm that estimates the distance
between a distribution $\D$ and the uniform distribution $\uniform$ by performing
$\poly(1/\eps)$ \PCOND (and \SAMP) queries. We start by giving a high
level description of the algorithm.

By the definition of the variation distance (and the uniform distribution),
\BEQ
\dtv(\D,\uniform) \;=\; \sum_{i:D(i) < 1/N}\left(\frac{1}{N} - \D(i)\right)\;.
\EEQ
We define the following function over $[N]$:
\BEQ\label{eq:alpha-def}
\psi^\D(i) = (1-N\cdot\D(i)) \;\mbox{ for }\; \D(i) < \frac{1}{N}\;,\;
\mbox{ and } \psi^\D(i) = 0 \;\mbox{ for }\; \D(i) \geq \frac{1}{N}\;.
\EEQ
Observe that $\psi^\D(i) \in [0,1]$ for every $i \in [N]$ and
\BEQ\label{eq:dtv-alpha-Dist}
\dtv(\D,\uniform) \;=\; \frac{1}{N}\sum_{i=1}^{N} \psi^\D(i)\;.
\EEQ
Thus $\dtv(\D,\uniform)$ can be viewed as an average value of
a function whose range is in $[0,1]$. Since $\D$ is fixed throughout
this subsection, we shall use the shorthand $\psi(i)$ instead
of $\psi^\D(i)$.
Suppose we were able to compute $\psi(i)$ exactly for any $i$ of our
choice. Then we could obtain an estimate $\hat{d}$ of
$\dtv(\D,\uniform)$ to within an additive error of $\eps/2$ by simply
selecting $\Theta(1/\eps^2)$ points in $[N]$ uniformly at random
and setting $\hat{d}$ to be the average value of $\psi(\cdot)$ on
the sampled points. By an additive Chernoff bound (for an appropriate
constant in the $\Theta(\cdot)$ notation), with high constant
probability the estimate $\hat{d}$ would deviate by at most $\eps/2$
from $\dtv(\D,\uniform)$.

Suppose next that instead of being able to compute $\psi(i)$
exactly, we were able to compute an estimate $\hat{\psi}(i)$
such that
$|\hat{\psi}(i) -\psi(i)| \leq \eps/2$.
By using $\hat{\psi}(i)$ instead of $\psi(i)$ for
each of the $\Theta(1/\eps^2)$ sampled points we would incur
an additional additive error of at most $\eps/2$.
Observe first that for $i$ such that $\D(i) \leq \eps/(2N)$
we have that $\psi(i) \geq 1-\eps/2$, so the
estimate $\hat{\psi}(i)=1$ meets our requirements.
Similarly, for $i$ such that $\D(i) \geq 1/N$, any
estimate $\hat{\psi}(i)\in [0,\eps/2]$ can be used.
Finally, for $i$ such that $\D(i) \in [\eps/(2N),1/N]$,
if we can obtain an estimate $\widehat{\D}(i)$ such that
$\widehat{\D}(i) \in [1-\eps/2,1+\eps/2]\D(i)$, then we can use
$\hat{\psi}(i) = 1 - N\cdot \widehat{\D}(i)$.

In order to obtain such estimates $\hat{\psi}(i)$, we
shall be interested in finding a {\em reference point\/} $x$.
Namely, we shall be interested in finding a pair $(x,\widehat{\D}(x))$
such that $\widehat{\D}(x) \in [1-\eps/c,1+\eps/c]\D(x)$ for
some sufficiently large constant $c$, and such that
$\D(x) = \Omega(\eps/N)$ and $\D(x) = O(1/(\eps N))$.
In \autoref{subsec:find-ref} we describe a procedure for finding
such a reference point. More precisely, the procedure is required to
find such a reference point (with high constant probability) only
under a certain condition on $\D$. It is not hard to verify (and we show
this subsequently), that if this condition is not met, then
$\dtv(\D,\uniform)$ is very close to $1$.
In order to
state the lemma we introduce the following notation. For
$\gamma\in[0,1]$, let
\BEQ\label{eq:H-def}
H^\D_\gamma \eqdef \left\{i:\; \D(i) \geq \frac{1}{\gamma N}\right\}\;.
\EEQ
\BL\label{lem:find-ref}
Given an input parameter $\kappa \in (0,1/4]$ as
well as \SAMP and \PCOND query access to a distribution $\D$,
the procedure {\sc Find-Reference} (Algorithm~\ref{alg:find-ref})
 either returns a pair
$(x,\widehat{\D}(x))$ where $x\in [N]$ and $\widehat{\D}(x) \in [0,1]$ or
returns {\sf No-Pair}. The procedure satisfies the following:
% for some fixed constants $c_1,c_2 > 1$.
\BE
\item If $\D(H^\D_\kappa) \leq 1-\kappa$,
then with probability at least $9/10$, the
procedure returns a pair $(x,\widehat{\D}(x))$ such that
$\widehat{\D}(x) \in [1-2\kappa,1+3\kappa]\D(x)$ and
$\D(x) \in \left[\frac{\kappa}{8},\frac{4}{\kappa}\right]\cdot \frac{1}{N}$.
\item If $\D(H^\D_\kappa) > 1-\kappa$, 
then with probability at least $9/10$, the
procedure either returns {\sf No-Pair} or it returns
a pair $(x,\widehat{\D}(x))$ such that
$\widehat{\D}(x) \in [1-2\kappa,1+3\kappa]\D(x)$ and
$\D(x) \in \left[\frac{\kappa}{8},\frac{4}{\kappa}\right]\cdot \frac{1}{N}$.
\EE
The procedure performs ${\tildeO(1/\kappa^{20})}$ \PCOND and \SAMP queries.
\EL
Once we have a reference point $x$ we can use it to obtain an
estimate $\hat{\psi}(i)$ for any $i$ of our choice, using
the procedure {\sc Compare}, whose properties are stated in
\autoref{lem:compare} (see \autoref{subsec:compare}).

\begin{algorithm}[ht!]
% \begin{algorithm}{\sc (Estimating the Distance to Uniformity)}
\caption{Estimating the Distance to Uniformity}
\SetKwInOut{Input}{Input}
\label{alg:est-dist-unif}
\Input{\PCOND and \SAMP query access to a distribution $\D$ and
a parameter $\eps \in [0,1]$.}
\BE
\item Call the procedure {\sc Find-Reference} (Algorithm~\ref{alg:find-ref})
with $\kappa$ set to $\eps/8$.
If it returns {\sf No-Pair}, then output $\hat{d} = 1$ as the estimate for the
distance to uniformity. Otherwise, let $(x,\widehat{\D}(x))$ be its
output.
\item Select a sample $S$ of $\Theta(1/\eps^2)$ points uniformly.
\item Let
$K = \max\left\{\frac{2/N}{\widehat{\D}(x)},
           \frac{\widehat{\D}(x)}{\eps/(4N)}\right\}$.
\item For each point $y \in S$:
 \BE
 \item Call {\sc Compare}$\left(\{x\},\{y\},\kappa,K,\frac{1}{10|S|}\right)$.
 \item If {\sc Compare} returns \high or it returns a
 value $\rho(y)$ such that $\rho(y)\cdot\widehat{\D}(x)\geq 1/N$, then set
 $\hat{\psi}(y) = 0$;
 \item Else, if {\sc Compare} returns \low or it returns a value
 $\rho(y)$ such that $\rho(y)\cdot \widehat{\D}(x)\leq \eps/4N$,
 then set $\hat{\psi}(y) = 1$;
 \item Else set $\hat{\psi}(y) = \newer{1-}N\cdot \rho(y)\cdot \widehat{\D}(x)$.
 \EE
\item Output $\hat{d} = \frac{1}{|S|}\sum_{y\in S}\hat{\psi}(y)$.
\EE
\end{algorithm}

\BT\label{thm:est-dist-unif}
With probability at least $2/3$,  the
estimate $\hat{d}$ returned by Algorithm~\ref{alg:est-dist-unif}
satisfies: $\hat{d}=\dtv(\D,\uniform) \pm O(\newer{\eps})$.
The number of queries performed by the algorithm
is $\tilde{O}(1/\eps^{20})$.
\ET

\BPF
In what follows we shall use the shorthand $H_\gamma$ instead of
$H^\D_\gamma$.
 Let $E_0$ denote the event that
the procedure {\sc Find-Reference} (Algorithm~\ref{alg:find-ref}) obeys the
requirements in \autoref{lem:find-ref}, where
by  \autoref{lem:find-ref} the event $E_0$ holds
with  probability at least $9/10$. Conditioned
on $E_0$, the algorithm outputs $\hat{d} = 1$
right after calling the procedure (because the procedure returns
{\sf No-Pair}) only when $\D(H_\kappa) > 1-\kappa = 1-\eps/8$.
We claim that in this case $\dtv(\D,\uniform) \geq 1-2\eps/8=1-\eps/4$.
To verify this, observe that
\BEQ
\dtv(\D,\uniform) = \sum_{i:D(i)>1/N}\left(D(i) - \frac{1}{N}\right)
\geq \sum_{i \in H_\kappa} \left(D(i) - \frac{1}{N}\right) =
\D(H_\kappa) - \frac{|H_\kappa|}{N}
    \geq \D(H_\kappa) - \kappa\;.
\EEQ
Thus, in this case the estimate $\hat{d}$ is as required.

We turn to the case in which\ignore{$\D(H_\kappa) \leq 1-\kappa$ and} the
procedure {\sc Find-Reference} returns a pair $(x,\widehat{\D}(x))$
such that $\widehat{\D}(x) \in [1-2\kappa,1+3\kappa]\D(x)$ and
$\D(x) \in \left[\frac{\kappa}{8},\frac{4}{\kappa}\right]\cdot \frac{1}{N}$.

We start by defining two more ``desirable'' events, which
hold (simultaneously) with high constant probability, and
then show that conditioned on these events holding (as well as $E_0$),
the output of the algorithm is as required.
Let $E_1$ be the event that the sample $S$ satisfies
\BEQ\label{eq:samp-S-eps}
\left|\frac{1}{|S|}\sum_{y\in S} \psi(y)- \dtv(\D,\uniform)\right|
\leq \eps/2\;.
\EEQ
By an additive Chernoff bound, the event $E_1$ holds with probability
at least $9/10$.

%\dana{
\ignore{
  Details: Let the points in $S$ be denoted
  $y_1,\dots,y_s$ for $s=|S|$. We define $s$
  random variables $\chi_1,\dots,\chi_s$,
  where $\chi_j = \psi(y_j)$. We have that
  $\chi_j \in [0,1]$ and  $\Exp[\chi_j] = \dtv(\D,\uniform)$
  for each $j$.
  By applying the additive Chernoff bound:
  \BEQ
  \Pr\left[\left|\frac{1}{s}\sum_{j=1}^s \chi_j - \dtv(\D,\uniform)\right| > \eps/2\right]
   \leq 2\exp(-2s(\eps/2)^2)\;,
   \EEQ
   which is at most $1/10$ for $s \geq 8/\eps^2$.
}
% }

 Next, let $E_2$ be the event that all calls to the procedure
 {\sc Compare} return answers as specified in \autoref{lem:compare}.
 Since {\sc Compare} is called $|S|$ times, and for each call the probability
 that it does not return an answer as specified in the lemma is
 at most $1/(10|S|)$, by the union bound
 the probability that $E_2$ holds is at least
 $9/10$.

From this point on assume events $E_0$, $E_1$ and $E_2$ all occur, which
holds with probability at least $1- 3/10\geq 2/3$.
Since $E_2$ holds, we get the following.
\BE
\item When {\sc Compare} returns \high for $y\in S$
(so that $\hat{\psi}(y)$ is set to $0$) we have that
\BEQ
\D(y) > K \cdot \D(x) \geq \frac{2/N}{\widehat{\D}(x)}\cdot\D(x)
            > \frac{1}{N}\;,
\EEQ
implying that $\hat{\psi}(y) =\psi(y)$.
\item When {\sc Compare} returns \low for $y\in S$
(so that $\hat{\psi}(y)$ is set to $1$) we have that
\BEQ
\D(y) <  \frac{\D(x)}{K} \leq  \frac{\D(x)}{\widehat{\D}(x)/(\eps/4N)}
            \leq \frac{\eps}{2N}\;,
\EEQ
implying that $\hat{\psi}(y) \leq \psi(y) + \eps/2$
(and clearly $\psi(y) \leq \hat{\psi}(y)$).
\item When {\sc Compare} returns a value $\rho(y)$
it holds that $\rho(y) \in [1-\kappa,1+\kappa](\D(y)/\D(x))$,
so that $\rho(y)\cdot \widehat{\D}(x) \in [(1-\kappa)(1-2\kappa),(1+\kappa)(1+3\kappa)]\D(y)$. %%\cnote{I think this is the bound we have, instead of $[(1-\kappa)^2,(1+\kappa)^2]$. The statement of the rest of the paragraph appear to hold even like this, however. }
Since $\kappa = \eps/8$, if $\rho(y)\cdot\widehat{\D}(x)\geq 1/N$
(so that $\hat{\psi}(y)$ is set
to $0$), then $\psi(y) < \eps/2$,
if $\rho(y)\cdot\widehat{\D}(x) \leq \eps/4N$
(so that $\hat{\psi}(y)$ is set to $1$), then
$\psi(y) \geq 1- \eps/2$, and otherwise
$|\hat{\psi}(y) - \psi(y)|\leq \eps/2$.\ignore{\mnote{D: maybe give 
more details.}}
%%%%
% \item If {\sc Compare} returns \high or it returns a
% value $\rho(y)$ such that $\rho(y)\hat{D}(x)\geq 1/N$, then set $\hat{\psi}(y):=0$;
% \item Else, if {\sc Compare} returns \low or it returns a value
% $\rho(y)$ such that $\rho(y)\cdot \hat{D}(x)\leq \eps/4N$,
% then set $\hat{\psi}(y):=1$;
% \item Else set $\hat{\psi}(y):= N\cdot \rho(y)\cdot \hat{D}(x)$.
%%%
\EE
It follows that
\BEQ
\hat{d} = \frac{1}{|S|}\sum_{y\in S}\hat{\psi}(y)
   \in \left[\frac{1}{|S|}\sum_{y\in S} \psi(y) -\eps/2,
 \frac{1}{|S|}\sum_{y\in S} \psi(y) +\eps/2\right]
 \subseteq [\dtv(\D,\uniform)-\eps,\dtv(\D,\uniform)+\eps]\;
\EEQ
as required.

The number of queries performed by the algorithm is
the number of queries performed by the procedure
{\sc Find-Reference}, which is {$\tildeO(1/\eps^{20})$}, plus
$\Theta(1/\eps^2)$ times the number of queries performed in each
call to {\sc Compare}. The procedure  {\sc Compare} is
called with the parameter $K$, which is
bounded by $O(1/\eps^2)$, the parameter $\eta$, which is
$\Omega(\eps)$, and $\delta$, which is $\Omega(1/\eps^2)$.
By \autoref{lem:compare}, the number of queries performed
in each call to {\sc Compare} is
$O(\log(1/\eps)/\eps^4)$.
The total number of queries performed is hence {$\tildeO(1/\eps^{20})$}.
\EPF

\subsection{Finding a reference point}\label{subsec:find-ref}
In this subsection we prove \autoref{lem:find-ref}.
We start by giving the high-level idea behind the procedure.
For a point $x\in [N]$ and $\gamma \in [0,1]$, let
$U^\D_\gamma(x)$ be as defined in Equation~(\ref{eq:Ugamma}).
Since $\D$ is fixed throughout this subsection, we shall use
the shorthand $U_\gamma(x)$ instead of $U^\D_\gamma(x)$.
Recall that $\kappa$ is a parameter given to the procedure.
Assume we had a point $x$ for which $\D(U_\kappa(x)) \geq \kappa^{d_1}$
and  $|U_\kappa(x)| \geq \kappa^{d_2}N $
for some constants $d_1$ and $d_2$
(so that necessarily $\D(x) = \Omega(\kappa^{d_1}/N)$ and
$\D(x) = O(1/(\kappa^{d_2}N)$).
It is not hard to verify (and we show this in detail subsequently),
that if  $\D(H) \leq 1-\kappa$, then a sample of size
$\Theta(1/\poly(\kappa))$ distributed according to $\D$
will contain such a point $x$
with high constant probability. Now suppose that we could obtain an estimate
$\hat{w}$ of $\D(U_\kappa(x))$ such that
$\hat{w} \in [1-\kappa,1+\kappa]\D(U_\kappa(x))$ and an estimate
$\hat{u}$ of $|U_\kappa(x)|$ such that
$\hat{u}\in [1-\kappa,1+\kappa]|U_\kappa(x)|$. By the definition of
$U_\kappa(x)$ we have that
$(\hat{w}/\hat{u})\in  [1-O(\kappa),1+O(\kappa)]\D(x)$.

Obtaining good estimates
of $\D(U_\kappa(x))$  and $|U_\kappa(x)|$ (for $x$ such that
both $|U_\kappa(x)|$ and $\D(U_\kappa(x))$ are sufficiently large)
might be infeasible. This is
due to the possible existence
of many points $y$ for which $\D(y)$ is very close
to $(1+\kappa)\D(x)$ or $\D(x)/(1+\kappa)$
which define the boundaries of the set $U_\kappa(x)$.
For such points it is not possible to efficiently
distinguish between those among them that belong
to $U_\kappa(x)$ (so that they are within the borders
of the set) and those that do not belong to
$U_\kappa(x)$ (so that they are just outside the borders
of the set).
%%%%%%%%%%%%%%% VERSION USING RANDOM THRESHOLD (ESTIMATE-NEIGHBORHOOD PROC) %%%
However, for our purposes it suffices to estimate
the weight and size  of {\em some\/} set $U_\alpha(x)$
such that $\alpha \geq \kappa $ (so that $U_\kappa(x) \subseteq U_\alpha(x)$)
and $\alpha$ is not much larger than $\kappa $ (e.g., $\alpha \leq 2\kappa)$).
To this end we can apply Procedure {\sc Estimate-Neighborhood}
(see \autoref{subsec:est-n}), which (conditioned
on $\D(U_\kappa(x))$ being above a certain threshold),
returns a pair $(\hat{w}(x),\alpha)$
such that $\hat{w}(x)$ is a good estimate of $\D(U_\alpha(x))$.
Furthermore, $\alpha$ is such that for $\alpha'$ slightly larger
than $\alpha$, the weight of $U_{\alpha'}(x)\setminus U_\alpha(x)$
is small, allowing us to obtain also a good estimate $\hat{\mu}(x)$
of  $|U_\alpha(x)|/N$.

\begin{algorithm}[ht!]
\label{alg:find-ref}
\caption{Procedure \sc Find-Reference}
\SetKwInOut{Input}{Input}
\Input{\PCOND and \SAMP query access to a distribution $\D$ and
a parameter $\kappa \in (0,1/4]$}
\BE
\item Select a sample $X$ of
$\Theta(\log(1/\kappa)/\kappa^2)$ points distributed according to $\D$.
\item\label{st:x-S} For each $x \in X$ do the following:
 \BE
 \item Call  {\sc Estimate-Neighborhood}  with the parameters $\kappa$
 as in the
 input to {\sc Find-Reference}, $\beta = \kappa^2/(40\log(1/\kappa))$,
 $ \eta = \kappa$, and $\delta = 1/(40|X|)$. Let
 $\theta = \kappa\eta\beta\delta/64 = \Theta(\kappa^6/\log^2(1/\kappa))$
 (as in {\sc Find-Reference}).
 \item If {\sc Estimate-Neighborhood} returns
 % \low or
 {a pair $(\hat{w}(x),\alpha(x))$ such that
 $\hat{w}(x) < \kappa^2/20\log(1/\kappa)$,} then go to Line~\ref{st:x-S}
 and continue with next $x \in X$.
 \item Select a sample $Y_x$ of size $\Theta(\log^2(1/\kappa)/\kappa^5)$
  distributed uniformly.
 \item For each  $y \in Y_x$
      call {\sc Compare}$(\{x\},\{y\},\theta/4,4,1/40|X||Y_x|)$,
       and let the output be denoted $\rho_x(y)$.
  \item Let $\hat{\mu}(x)$ be the fraction of occurrences of
$y \in Y_x$ such that $\rho_x(y) \in [1/(1+\alpha+\theta/2),1+\alpha+\theta/2]$.
\item Set $\widehat{\D}(x) = \hat{w}(x)/(\hat{\mu}(x) N)$.
\EE
\item\label{st:find-ref-out}
If for some point $x\in X$ we have
$\hat{w}(x) \geq \kappa^2/20\log(1/\kappa)$,
$\hat{\mu}(x) \geq \kappa^3/20\log(1/\kappa)$, and
$\kappa/4N \leq \widehat{\D}(x) \leq 2/(\kappa N)$,
then return $(x,\widehat{\D}(x))$. Otherwise return {\sf No-Pair}.
\EE
\end{algorithm}

\BPFOF{\autoref{lem:find-ref}}
We first introduce the following notation.
\BEQ
L \eqdef \left\{i :\; \D(i) < \frac{\kappa}{2N}\right\},\;
M \eqdef \left\{i:\; \frac{\kappa}{2 N}
           \leq \D(i) < \frac{1}{\kappa N}\right\}\;.
% L \eqdef \left\{i :\; \D(i) < \frac{\kappa^2}{c(\log(1/\kappa)n}\right\},\;
% M \eqdef \left\{i:\; \frac{\kappa^2}{c(\log(1/\kappa)n}
%           \leq \D(i) < \frac{c\log(1/\kappa)}{\kappa^3 n}\right\}\;.
\EEQ
Let $H=H^\D_\kappa$ where $H^\D_\kappa$
is as defined in Equation~(\ref{eq:H-def}).
Observe that $\D(L) < \kappa/2$, so that if $\D(H) \leq 1-\kappa$,
then $\D(M) \geq \kappa/2$.
Consider further partitioning the set $M$ of ``medium weight'' points
into buckets $M_1,\dots,M_r$ where
% $r = \log_{1+\kappa}(c^2\log^2(1/\kappa)/\kappa^5) = \Theta(\log(1/\kappa)/\kappa)$
$r = \log_{1+\kappa}(2/\kappa^2) = \Theta(\log(1/\kappa)/\kappa)$
and the bucket $M_j$ is defined as follows:
\BEQ
M_j \eqdef \left\{i:\; (1+\kappa)^{j-1}\cdot \frac{\kappa}{2N}
     \leq \D(i) < (1+\kappa)^j \cdot \frac{\kappa}{2 N} \right\}\;.
\EEQ
We consider the following ``desirable'' events.
\BE
\item Let $E_1$ be the event that conditioned on the existence of
a bucket $M_j$ such that
$\D(M_j) \geq \kappa/2r = \Omega(\kappa^2/\log(1/\kappa))$, there
exists a point $x^\ast \in X$ that belongs to $M_j$. By the setting
of the size of the sample $X$, the (conditional) event $E_1$ holds
with probability at least $1-1/40$.
\item Let $E_2$ be the event that all calls to {\sc Estimate-Neighborhood}
return an output as specified by \autoref{lem:est-n}.
By \autoref{lem:est-n},
the setting of the confidence parameter
$\delta$ in each call and a union bound over all $|X|$
calls, $E_2$ holds with probability at least $1-1/40$.
\item Let $E_3$ be the event that for each $x\in X$
% such that {\sc Estimate-Neighborhood} returned a pair $(\hat{w}(x),\alpha(x))$
we have the following.
  \BE
   \item If $\frac{|U_{\alpha(x)}(x)|}{N}\geq
       \frac{\kappa^3}{40\log(1/\kappa)}$,
   then
   $\frac{|Y_x\cap U_{\alpha(x)}(x)|}{|Y_x|}
     \in [1-\eta/2,1+\eta/2]\frac{|U_{\alpha(x)}(x)|}{N}$; \\
   If $\frac{|U_{\alpha(x)}(x)|}{N}< \frac{\kappa^3}{40\log(1/\kappa)}$,
   then $\frac{|Y_x\cap U_{\alpha(x)}(x)|}{|Y_x|}
        < \frac{\kappa^3}{30\log(1/\kappa)}$;
    \item Let
  $\Delta_{\alpha(x),\theta}(x) \eqdef
     U_{\alpha(x)+\theta}(x)\setminus U_{\alpha(x)}(x)$ (where
     $\theta$ is as specified by the algorithm). \\
  If $\frac{|\Delta_{\alpha(x),\theta}(x)|}{N}
      \geq \frac{\kappa^4}{240\log(1/\kappa)}$, then
   $\frac{|Y_x\cap \Delta_{\alpha(x),\theta}(x)|}{|Y_x|}
     \leq 2\cdot \frac{|\Delta_{\alpha(x),\theta}(x)|}{N}$;\\
   If $\frac{|\Delta_{\alpha(x),\theta}(x)|}{N}
        < \frac{\kappa^4}{240\log(1/\kappa)}$, then
   $\frac{|Y_x\cap \Delta_{\alpha(x),\theta}(x)|}{|Y_x|}
        < \frac{\kappa^4}{120\log(1/\kappa)}$.
   \EE
  By the size of each set $Y_x$ and a union bound over all $x\in X$,
  the event $E_3$ holds with probability at least $1-1/40$.
\item Let $E_4$ be the event that all calls to {\sc Compare}
return an output as specified by \autoref{lem:compare}.
By \autoref{lem:compare}, the setting of
the confidence parameter $\delta$ in each call and a union bound
over all (at most) $|X|\cdot |Y|$ calls, $E_3$ holds with probability
at least $1-1/40$.
\EE
Assuming events $E_1$--$E_4$ all hold (which occurs with
probability at least $9/10$) we have the following.
\BE
\item By $E_2$, for each $x \in X$
% such that {\sc Estimate-Neighborhood} returns
% a pair $(\hat{w}(x),\alpha(x))$
% we have that
% $\hat{w}(x) \in [1-\eta,1+\eta]\D(U_{\alpha(x)}(x))$.
% In particular this implies that for each $x\in X$
such that
$\hat{w}(x) \geq \kappa^2/20\log(1/\kappa)$
(so that $x$ may be
selected for the output of the procedure) we have that
$\D(U_{\alpha(x)}(x)) \geq \kappa^2/40\log(1/\kappa)$.

The event $E_2$ also implies that for each % such
$x \in X$ we have that
$\D(\Delta_{\alpha(x),\theta}(x))\leq \eta\beta/16
    \leq (\eta/16)\cdot \D(U_{\alpha(x)}(x))$, so that
\BEQ\label{eq:Delta-size}
\frac{|\Delta_{\alpha(x),\theta}(x)|}{N}
\leq \frac{\eta(1+\alpha(x))(1+\alpha(x)+\theta)}
     {16}\cdot \frac{|U_{\alpha(x)}(x)|}{N}
\leq \frac{\eta}{6}\cdot \frac{|U_{\alpha(x)}(x)|}{N}\;.
\EEQ

\item Consider any $x\in X$ such that
$\hat{w}(x) \geq \kappa^2/20\log(1/\kappa)$.
Let $T_x \eqdef \{\newer{y\in Y_x} : \rho_x(y) \in [1/(1+\alpha+\theta/2),(1+\alpha+\theta/2]\}$,
so that $\hat{\mu}(x) = |T_x|/|Y_x|$. By $E_4$,
for each $y \in Y_x\cap U_{\alpha(x)}(x)$ we have that
$\rho_x(y) \leq (1+\alpha)(1+\theta/4) \leq (1+\alpha+\theta/2)$
and $\rho_x(y) \geq (1+\alpha)^{-1}(1-\theta/4) \geq (1+\alpha+\theta/2)^{-1}$,
so that $y \in T_x$. On the other hand, for each
$y \notin Y_x \cap U_{\alpha(x)+\theta}(x)$ we have that
$\rho_x(y) > (1+\alpha+\theta)(1-\theta/4) \geq 1+\alpha+\theta/2$
or $\rho_x(y) < (1+\alpha+\theta)^{-1}(1-\theta/4) < (1+\alpha+\theta/2)^{-1}$,
so that $y \notin T_x$. It follows that
\BEQ\label{eq:T-x-2}
  Y_x \cap U_{\alpha(x)}(x) \;\subseteq\; T_x \;\subseteq\;
      Y_x \cap (U_{\alpha(x)}(x) \cup \Delta_{\alpha(x),\theta}(x))\;.
\EEQ
By $E_3$, when $\hat{\mu}(x) = |T_x|/|Y_x| \geq \kappa^3/20\log(1/\kappa)$,
then necessarily $\hat{\mu}(x) \in [1-\eta,1+\eta]|U_{\alpha(x)}(x)|/N$.
To verify this consider the following cases.
  \BE
  \item If $\frac{|U_{\alpha(x)}(x)|}{N}\geq \frac{\kappa^3}{40\log(1/\kappa)}$,
  then (by the left-hand-side of
  Equation~(\ref{eq:T-x-2}) and the definition of $E_3$) we get that
  $\hat{\mu}(x) \geq (1-\eta/2)\frac{|U_{\alpha(x)}(x)|}{N}$, and
  (by the right-hand-side of Equation~(\ref{eq:T-x-2}),
  Equation~(\ref{eq:Delta-size}) and $E_3$) we get that
  $\hat{\mu}(x) \leq (1+ \eta/2)\frac{|U_{\alpha(x)}(x)|}{N}
    + 2(\eta/6)\frac{|U_{\alpha(x)}(x)|}{N} < (1+\eta)\frac{|U_{\alpha(x)}(x)|}{N}$.
  \item If $\frac{|U_{\alpha(x)}(x)|}{N}< \frac{\kappa^3}{40\log(1/\kappa)}$,
  then (by the right-hand-side of Equation~(\ref{eq:T-x-2}),
  Equation~(\ref{eq:Delta-size}) and $E_3$)
  we get that
  $\hat{\mu}(x) < \frac{\kappa^3}{30\log(1/\kappa)}
      + \frac{\kappa^4}{120\log(1/\kappa)} <  \kappa^3/20\log(1/\kappa)$.
  \EE
\item If $\D(H) \leq 1-\kappa$, so that $\D(M) \geq \kappa/2$,
then there exists at least one bucket $M_j$ such that
$\D(M_j) \geq \kappa/2r = \Omega(\kappa^2/\log(1/\kappa))$.
By $E_1$, the sample $X$ contains a point $x^\ast\in M_j$.
By the definition of the buckets, for this point $x^\ast$ we have that
$\D(U_\kappa(x^\ast)) \geq \kappa/2r \geq \kappa^2/(10\log(1/\kappa))$ and
$|U_\kappa(x^\ast)| \geq (\kappa^2/2r)N \geq \kappa^3/(10\log(1/\kappa))$.
\EE
By the first two items above and the setting $\eta = \kappa$ we have that
for each $x$ such that $\hat{w}(x) \geq \kappa^2/20\log(1/\kappa)$ and
$\hat{\mu}(x) \geq \kappa^3/20\log(1/\kappa)$,
$$\widehat{\D}(x) \in
  \left[\frac{1-\kappa}{1+\kappa},\frac{1+\kappa}{1-\kappa}\right]\D(x)
     \subset [1-2\kappa,1+3\kappa]\D(x)\;.$$
Thus, if the algorithm outputs a pair $(x,\widehat{\D}(x))$
then it satisfies the condition stated in both items of the lemma.
This establishes the second item in the lemma.
By combining all three items we get that if $\D(H) \geq 1-\kappa$
then the algorithm outputs a pair $(x,\widehat{\D}(x))$
(where possibly, but not necessarily, $x = x^\ast$), and the first
item is established as well.

Turning to the query complexity, the total number of \PCOND queries performed
in the \mbox{$|X| = O(\log(1/\kappa)/\kappa^2)$}
calls to {\sc Estimate-Neighborhood}  is
$O\!\left(\frac{|X|\log(1/\newer{\delta})^2\log(1/(\beta\eta))}
            {\kappa^2\eta^4 \beta^3\newer{\delta}^2}\right)
            = \tildeO( 1/\kappa^{18} )$, and
the total number of \PCOND queries performed in the
calls to {\sc Compare} (for at most all pairs $x\in X$ and $y \in Y_x$)
is $\tildeO( 1/\kappa^{20} )$. 
\EPFOF

\section{A \texorpdfstring{$\tildeO\!\left(\frac{\log^3 N}{\eps^3}\right)$}{\~O(log\^{}3 N/eps\^{}3)}-query \texorpdfstring{$\ICOND_\D$}{\ICOND}
           algorithm for testing uniformity}
\label{ssec:unif-alg:icond}

In this \red{and the next} section we consider $\ICOND$ algorithms for testing
whether an unknown distribution $\D$ over $[N]$ is the uniform distribution
versus $\eps$-far from uniform.
Our results show that $\ICOND$ algorithms are not as powerful as
$\PCOND$ algorithms for this basic testing problem;
\red{in this section}
we give a $\poly(\log N, 1/\eps)$-query $\ICOND_\D$ algorithm, and
\red{in the next section} we
prove that any $\ICOND_\D$ algorithm must make
$\tilde{\Omega}(\log N)$ queries.

\red{In more detail, in this section} we describe an algorithm {\sc
$\ICOND_\D$-Test-Uniform} and prove the following theorem:

\BT \label{thm:intcond-test-uniform} {\sc $\ICOND_\D$-Test-Uniform} is a
$\tildeO(\frac{\log^3 N}{\eps^3})$-query $\ICOND_\D$ testing algorithm
for uniformity, i.e. it outputs $\accept$ with probability at least
$2/3$ if $\D = \calU$ and outputs $\reject$ with probability at
least $2/3$ if $\dtv(\D,\calU) \geq \eps.$ \ET

\noindent {\bf Intuition.}  Recall that, as mentioned in
\autoref{ssec:unif-alg}, any distribution $\D$ which is $\eps$-far
from uniform must put $\Omega(\eps)$ probability mass on ``significantly
heavy'' elements (that is, if we define \mbox{$H' = \setOfSuchThat{ h
\in [N]}{\D(h) \geq {\frac 1 N} + {\frac \eps {4N}}}$}, it must hold
that $\D(H') \geq \eps/4$).
Consequently a sample of $O(1/\eps)$ points drawn from $\D$
will contain such a point with high
probability. Thus, a natural approach to testing whether $\D$ is uniform
is to devise a procedure that, given an input point $y$, can
distinguish between the case that $y \in H^\prime$ and the case that
$\D(y)=1/N$ (as it is when $\D= \calU$).

\noindent We give such a procedure, which uses
the $\ICOND_\D$ oracle to perform a sort of
binary search over intervals.   The procedure successively
``weighs'' narrower and narrower intervals until it converges
on the single point $y$.
In more detail, we
consider the \emph{interval tree} whose root is
the whole domain $[N]$, with  two children
$\{1,\dots,N/2\}$ and $\{N/2+1,\dots, N\}$, and so on, with a single
point at each of the $N$ leaves.
Our algorithm starts at the root of the tree and
goes down the path that corresponds to $y$; at each child node
it \newer{uses} {\sc Compare} to compare the weight of the current node to the
weight of its sibling under $\D$.
If at any point the estimate deviates significantly from the value it
should have if $\D$ were uniform (namely the weights should be essentially
equal, with slight deviations because of even/odd issues),
then the algorithm \newer{rejects.}
Assuming the algorithm does not reject, it provides a
$(1\pm O(\eps))$-accurate
multiplicative estimate of $\D(y)$, and the algorithm checks
whether this estimate is sufficiently close to $1/N$ (rejecting if this
is not the case). If no point in a sample of $\Theta(1/\eps)$ points (drawn according to $\D$) causes rejection, then the algorithm accepts.

\begin{algorithm}
  \begin{algorithmic}[1]
    \REQUIRE parameter $\eps > 0$; integers $1\leq a \leq b \leq N$; $y\in[a,b]$; query access to $\ICOND_\D$ oracle
%     \IF{ $y\notin \{a,\dots,b\}$}
%       \RETURN \fail
%     \ELSIF{$a=b$}
    \IF{$a=b$}
      \RETURN 1
    \ENDIF
    \STATE Let $c = \left\lfloor\frac{a+b}{2}\right\rfloor$; $\Delta = (b-a+1)/2$.
    \IF{ $y\leq c$ }
      \STATE Define $I_y=\{a,\dots,c\}$, $I_{\bar{y}}=\{c+1,\dots,b\}$ and $\rho = \lceil\Delta\rceil/\lfloor\Delta\rfloor$
    \ELSE
      \STATE Define $I_{\bar{y}}=\{a,\dots,c\}$, $I_y=\{c+1,\dots,b\}$ and $\rho = \lfloor\Delta\rfloor/\lceil\Delta\rceil$
    \ENDIF
    \STATE Call \textsc{Compare} on $I_y$, $I_{\bar{y}}$ with parameters $
\eta=\frac{\eps}{48\log N}$, $K=2$, $\delta=\frac{\eps}{100(1 + \log N)}$ to get an estimate $\hat{\rho}$ of $\D(I_y)/\D(I_{\bar{y}})$
    \IF{ $\hat{\rho}\notin[1-\frac{\eps}{48\log N}, 1+\frac{\eps}{48\log N}]\cdot\rho$ (this includes the
    case that $\hat{\rho}$ is \high or \low) }
      \RETURN \reject   \label{algo:intcond:uniformity:ub:step:binary:descent:reject}
    \ENDIF
    \STATE Call recursively \textsc{Binary-Descent} on input
     ($\eps$, the endpoints of $I_y$, $y$);
\label{st:do-compare}
    \IF{ \textsc{Binary-Descent} returns a value $\nu$}
      \RETURN $\frac{\hat{\rho}}{1+\hat{\rho}}\cdot\nu$
    \ELSE % (\textsc{Binary-Descent} returns \reject)
      \RETURN \reject
    \ENDIF
    \end{algorithmic}\caption{\label{algo:testing-uniformity:icond:binary-descent}\sc Binary-Descent}
\end{algorithm}

\begin{algorithm}
  \begin{algorithmic}[1]
    \REQUIRE error parameter $\eps > 0$; query access to $\ICOND_\D$ oracle
    \STATE Draw $t=\frac{20}{\eps}$ points $y_1,\dots,y_t$ from $\SAMP_\D$. \label{algo:intcond:uniformity:ub:sampling}
    \FOR{ $j = 1$ to $t$ }
       \STATE Call
{\sc Binary-Descent}$(\eps,1,N,y_j)$ and return \reject if it rejects,
otherwise let $\hat{d}_j$ be the value it returns as its estimate of $\D(y_j)$
\ignore{
       \IF{ \textsc{Binary-Descent} rejects }
            \RETURN \reject \label{algo:intcond:uniformity:ub:reject:1}
}
       \IF{ $\hat{d}_j\notin[1-\frac{\eps}{12}, 1+\frac{\eps}{12}] \cdot
{\frac 1 N}$ }
            \RETURN \reject \label{algo:intcond:uniformity:ub:reject:2}
       \ENDIF
    \ENDFOR
    \RETURN \accept
  \end{algorithmic}\caption{\label{algo:testing-uniformity:icond}\sc $\ICOND_\D$-Test-Uniform}
\end{algorithm}

The algorithm we use to perform the ``binary search'' described above is
Algorithm~\ref{algo:testing-uniformity:icond:binary-descent},
{\sc Binary-Descent}.  We begin by proving correctness for it:

\begin{lem}\label{lem:intcond-test-uniform:binary-search} Suppose the
algorithm \textsc{Binary-Descent} is run with inputs $\eps\in(0,1]$, $a=1$, $b= N$, and $y\in[N]$, and is provided
\ICOND oracle access to distribution $\D$ over $[N]$. It
performs $\tildeO(\log^3 N/\eps^2)$ queries and either outputs a value
$\hat{\D}(y)$ or \reject, where the following holds:

\BE

    \item if $\D(y) \geq\frac{1}{N} + \frac{\eps}{4N}$, then with probability at least $1-\frac{\eps}{100}$ the procedure either outputs a value $\hat{\D}(y)\in[1-\eps/12, 1+\eps/12]\D(y)$ or \reject;

    \item if $\D=\calU$, then with probability at least $1-\frac{\eps}{100}$ the procedure outputs a value
$\hat{\D}(y)\in[1-\eps/12, 1+\eps/12] \cdot {\frac 1 N}$.

\EE
\end{lem}

\begin{proofof}{\autoref{lem:intcond-test-uniform:binary-search}}
The claimed query bound is easily verified, since the
recursion depth is at most $1 + \log N$
and the only queries made  are  during calls to
\textsc{Compare}, each of which performs
$O(\log(1/\delta)/\gamma^2)=\tildeO(\log^2 N/\eps^2)$ queries.

  \noindent Let $E_0$ be the event that all calls to \textsc{Compare}
satisfy the conditions in \autoref{lem:compare}; since each of them
succeeds with probability at least $1-\delta=1-{\frac {\eps}{100 (1 + \log N)}}$, a union
bound shows that $E_0$ holds with probability at least $1-\eps/100$.
We hereafter condition on $E_0$.

% \rnote{I rephrased things a bit --- previously we
% were ``globally'' asserting that it must be the case that
% \mbox{$\hat{\rho}_{j}\in[1-\frac{\eps}{48\log N}, 1+\frac{\eps}{48\log
% N}]\cdot\D(I^{(j)}_y)/\D(I^{(j)}_{\bar{y}})$}, but this seemed to ignore the
% possibility that {\sc Compare} outputs \high or \low.  This is only
% a possibility in case (i) so it seemed to me better to discuss the
% cases separately.}
% \dnote{Looks right to me.}

We first prove the second part of the lemma where $\D = \calU$.
Fix any specific recursive call, say the $j$-th, during the
execution of the procedure.
The intervals $I^{(j)}_y,I^{(j)}_{\bar{y}}$ used in that execution of the
algorithm are easily seen to satisfy $\D(I_y)/\D(I_{\bar{y}})\in[1/K,K]$
(for $K=2$), so by event $E_0$ it must be the case that {\sc Compare} returns an estimate
\mbox{$\hat{\rho}_{j}\in[1-\frac{\eps}{48\log N}, 1+\frac{\eps}{48\log
N}]\cdot\D(I^{(j)}_y)/\D(I^{(j)}_{\bar{y}})$}.
Since
$\D = U$, we have that $\D(I^{(j)}_y)/\D(I^{(j)}_{\bar{y}}) = \rho^{(j)}$,
so the overall procedure returns a numerical value
rather than \reject.

Let $M = \lceil \log N \rceil$
be the number of recursive calls (i.e., the number of executions
of Line~\ref{st:do-compare}).
Note that we can write
$\D(y)$ as a product
\BEQ \label{eq:Dy}
\D(y) = \prod_{j=1}^{M} \frac{ \D(I^{(j)}_y) }{ \D(I^{(j)}_y) + \D(I^{(j)}_{\bar{y}}) }
      = \prod_{j=1}^{M} \frac{\D(I^{(j)}_y)/\D(I^{(j)}_{\bar{y}})}{\D(I^{(j)}_y)/\D(I^{(j)}_{\bar{y}})+1}\;.
\EEQ
We next observe that for any
$0 \leq \eps' < 1$ and $\rho,d>0$, if
$\hat{\rho} \in [1-\eps',1+\eps']d$ then we have
${\frac {\hat{\rho}}{\hat{\rho}+1}} \in [1-{\frac {\eps'}2},1+\eps']
{\frac {d}{d+ 1}}$
(by straightforward algebra).
Applying this $M$ times, we get

\begin{eqnarray*}
\prod_{j=1}^{M} \frac{\hat{\rho}_{j}}{\hat{\rho}_{j}+1}
&\in&\left[\left(1-\frac{\eps}{96\log N}\right)^M,
\left(1+\frac{\eps}{48\log N}\right)^M\right]\cdot
\prod_{j=1}^{M}
\frac{\D(I^{(j)}_y)/\D(I^{(j)}_{\bar{y}})}{\D(I^{(j)}_y)/\D(I^{(j)}_{\bar{y}})+1}\\
&\in&
\left[\left(1-\frac{\eps}{96\log N}\right)^M,
\left(1+\frac{\eps}{48\log N}\right)^M\right]\cdot
\D(y)\\
&\in&
\left[1 - {\frac \eps {12}}, 1 + {\frac \eps {12}} \right]
\D(y).
\end{eqnarray*}
\ignore{
where we used the fact that $M=\lceil\log N\rceil$, and the
inequalities $\frac{(1+\alpha)x}{1+(1-\alpha)x}\leq (1+3\alpha)x$ and
\mbox{$\frac{(1-\alpha)x}{1+(1+\alpha)x}\geq (1-3\alpha)x$} for
$x\in(0,1/3)$.
}
Since
$\prod_{j=1}^{M} \frac{\hat{\rho}_{j}}{\hat{\rho}_{j}+1}$ is the value that
the procedure outputs, the second part of the lemma is proved.

The proof of the first part of the lemma is virtually identical.  The only difference is that now
it is possible that {\sc Compare} outputs \high or \low at some call (since $\D$ is not uniform it need not be the case that
$\D(I^{(j)}_y)/\D(I^{(j)}_{\bar{y}}) = \rho^{(j)}$), but this is not a problem for (i) since in that case
{\sc Binary-Descent} would output \reject.
\end{proofof}

See Algorithm~\ref{algo:testing-uniformity:icond:binary-descent}
for a description of the
testing algorithm {\sc $\ICOND_\D$-Test-Uniform}.
We now prove \autoref{thm:intcond-test-uniform}:

\begin{proofof}{\autoref{thm:intcond-test-uniform}}
Define $E_1$ to be the event that all calls to \textsc{Binary-Descent} satisfy the conclusions of \autoref{lem:intcond-test-uniform:binary-search}. With a union bound over all these $t=20/\eps$ calls, we have $\Pr[E_1] \geq 8/10$.
\begin{description}

  \item[Completeness:] Suppose $\D=\calU$, and condition again on $E_1$.
Since this implies that \textsc{Binary-Descent} will always return a
value, the only case \textsc{$\ICOND_\D$-Test-Uniform} might reject is
by reaching Line~\ref{algo:intcond:uniformity:ub:reject:2}. However,
since it is the case that every value $\hat{d}_j$ returned by the
procedure satisfies \mbox{$\hat{\D}(y)\in[1-\eps/12, 1+\eps/12]\cdot
{\frac 1 N}$},
this can never happen.

  \item[Soundness:] Suppose $\totalvardist{\D}{\calU} \geq \eps$.
Let $E_2$ be the event that at least one of the $y_i$'s drawn in
Line~\ref{algo:intcond:uniformity:ub:sampling} belongs to $H^\prime$.
As $\D(H') \geq \eps/4$, we have
$\Pr[E_2] \geq 1-(1 - \eps/4)^{20/\eps} \geq 9/10$.
Conditioning on both $E_1$ and $E_2$, for such a $y_j$, one of two cases below holds:
  \begin{itemize}

    \item either the call to \textsc{Binary-Descent} outputs \reject and \textsc{$\ICOND_\D$-Test-Uniform}  outputs \reject;

   \item or a value $\hat{d}_j$ is returned, for which
      $\hat{d}_j \geq (1-\frac{\eps}{12})(1+\frac{\eps}{4})\cdot {\frac 1 N} > (1+\eps/12)/N$
    (where we used the fact that $E_1$ holds); and \textsc{$\ICOND_\D$-Test-Uniform} reaches Line~\ref{algo:intcond:uniformity:ub:reject:2} and rejects.

  \end{itemize}

\end{description}

Since $\Pr[E_1\cup E_2] \geq 7/10$, \textsc{$\ICOND_\D$-Test-Uniform} is
correct with probability at least $2/3$. Finally, the claimed query
complexity directly follows from the $t=\Theta(1/\eps)$ calls to
\textsc{Binary-Descent}, each of which makes $\tildeO(\log^3
N/\eps^2)$ queries to $\ICOND_\D$.
% \end{proofof}
\qed

% %%%%%%%%%%%%%%%%%%%%%%%%%%%%%%%%%%%%%%%%%%%%%%%%%%%%%%%%%%%%%%%%
% \begin{figure}[ht]\centering
%   \begin{tikzpicture}[every node/.style={rectangle, draw=black, solid,thin, minimum size = 0.5cm},scale=0.8]\scriptsize
%   \tikzstyle{level 1}=[sibling distance=35mm]
%   \node  (root){$[N]$}
%     child { node  (l1) {$\intinterv{1}{\frac{N}{2}}$}
%           child { node  (l2) {$\intinterv{1}{\frac{N}{4}}$} }
%           child { node  (r2) {$\intinterv{\frac{N}{4}+1}{\frac{N}{2}}$}
%               child[dotted] { node  (r3) {$\intinterv{i-2}{i-1}$} }
%               child[dotted] { node  (r3) {$\intinterv{i}{i+1}$}
%                 child[solid] { node (l4) {\color{black}$\{i\}$} }
%                 child[solid] { node  (r4) {$\{i+1\}$} }
%             }
%            }
%      }
%     child { node  (r1) {$\intinterv{\frac{N}{2}+1}{N}$}
%     };
%   \end{tikzpicture}
%   \caption{\label{fig:binary:search:tree} Interval tree.}
% \end{figure}
% %%%%%%%%%%%%%%%%%%%%%%%%%%%%%%%%%%%%%%%%%%%%%%%%%%%%%%%%%%%%%%%%

%%%%%%%%%%%%%%%%%%%%%%%%%%%%%%%%%%%%%%%%%%%%%%%%%%%%%%%%%%%%%%%%%%%%%%%%%%%%%%%%%%%%%%%%%%%%%%%%%%%%%%%%%%%%%%%%%%%%%%%%%%%%%%%%%%%%%%%%%%%%%%%%%%%%%%%%%

\section{An \texorpdfstring{$\Omega(\log N/\log \log N)$}{Omega(log N/ log log N)} lower bound for
\texorpdfstring{$\ICOND_\D$}{\ICOND} algorithms that test uniformity} \label{ssec:unif-lb:intcond}

\red{In this section we} prove that any $\ICOND_\D$ algorithm that $\eps$-tests
uniformity even for constant $\eps$
must have query complexity $\tilde{\Omega}(\log N).$
This shows that our algorithm in the previous subsection is not too
far from optimal, and sheds light on a key difference between $\ICOND$ and
$\PCOND$ oracles.

\begin{thm}\label{thm:intcond-test-uniform:lb}
\sloppy
Fix $\eps=1/3$. Any $\ICOND_\D$ algorithm for testing whether $\D = \calU$
versus \mbox{$\dtv(\D,\D^\ast) \geq \eps$} must make $\Omega\!\left(\frac{\log N}{\log \log N}\right)$ queries.
\end{thm}

To prove this lower bound we define a probability distribution
$\calP_{\text \no}$ over possible ``No''-distributions (i.e.
distributions that have variation distance at least $1/3$ from $\calU$).
A distribution drawn from $\calP_{\text \no}$ is constructed
as follows:  first \newer{(assuming without loss of generality that $N$ is a power of $2$)}, we partition $[N]$ into $b=2^X$ consecutive
intervals of the same size $\Delta=\frac{N}{2^X}$, which we refer to as
``blocks'', where $X$ is a random variable distributed uniformly on the
set \mbox{$\{\frac{1}{3}\log N, \frac{1}{3}\log N + 1, \dots,
\frac{2}{3}\log N\}$}. Once the block size $\Delta$ is determined, a random
offset $y$ is drawn uniformly at random in $[N]$, and all block
endpoints are shifted by $y$ modulo $[N]$ (intuitively, this prevents
the testing algorithm from ``knowing'' a priori that
specific points are endpoints of blocks).
Finally, independently for each block, a fair coin is thrown to determine its
\emph{profile}: with probability $1/2$, each point in the first half
of the block
will have probability weight $\frac{1-2\eps}{N}$ and each
point in the second half will have probability
$\frac{1+2\eps}{N}$ (such a block is said to be
a \emph{low-high} block, with profile
$\downarrow\uparrow$).  With probability $1/2$ the reverse is true:
each point in the first half has probability ${\frac {1+2\eps} N}$
and each point in the second half has probability ${\frac {1-2\eps} N}$
(a \emph{high-low} block $\uparrow\downarrow$).
It is clear that each distribution $\D$ in the support of
$\calP_{\text \no}$ defined in this way indeed has
$\dtv(\D,\calU) = \eps.$

To summarize, each ``No''-distribution $\D$ in the support of
$\calP_{\text \no}$ is parameterized
by $(b+2)$ parameters: its block size $\Delta$, offset $y$, and profile
$\vartheta\in\{\downarrow\uparrow,\uparrow\downarrow\}^b$. Note
that regardless of the profile vector,
each block always has weight exactly $\Delta/N$.
% % \ignore{\cnote{I removed the part mentioning that
% % $\calP_{\text \no}$ is \emph{not} the uniform distribution over ``No''-instance -- do
% % you feel it should be put back?}
% % \rnote{No, I don't think we need it.}
% % }

\ignore{Picking a distribution $\D$ according to the process described
above is not equivalent to drawing uniformly at random a distribution in
$\calP_{\text \no}$. Indeed, while $\Delta$ (and thus $b$) is
chosen uniformly at random between all possible choices, there are far
more distributions for large values of $b$ than for small ones (as a
large value of $b$ implies the choice of a vector $\vartheta$ in a space
of dimension $2^b$.}

We note that while there is only one ``Yes''-distribution
$\calU$, it will sometimes be convenient for the analysis to think of
$\calU$  as resulting from the same initial process of picking a block size and
offset, but without the subsequent choice of a profile vector.
We sometimes refer to this as the ``fake construction'' of the
uniform distribution $\calU$ (the reason for this will be clear later).
%%%
% % \ignore{\mnote{Remove this?} This yields a random partition of $[N]$, on
% % which the distribution is exactly uniform (both between the blocks and
% % inside them).
% % }

The proof of \autoref{thm:intcond-test-uniform:lb} will be carried
out in two steps.  First we shall restrict the analysis to
\emph{non-adaptive algorithms}, and prove the lower bound for such algorithms.
This result will then be extended to the general setting by introducing
(similarly to \autoref{ssec:lb-Pcond-Dstar}) the notion of a
\emph{query-faking algorithm}, and reducing the behavior of adaptive
algorithms to non-adaptive ones through an appropriate sequence of
such query-faking algorithms.
\ignore{
\cnote{Is it worth emphasizing this query-faking idea here? It's already
been used in the same fashion in the LB for \PCOND known D*, but in this
section we specifically use it to deal with adaptiveness.}.
}

Before proceeding, we define the \emph{transcript} of the interaction
between an algorithm and a $\ICOND_\D$ oracle.
Informally, the transcript captures the entire history of
interaction between the algorithm and the $\ICOND_\D$ oracle during the
whole sequence of queries.

\BD\label{def:transcript}

  Fix any (possibly adaptive) testing algorithm $\Algo$ that  queries an
$\ICOND_\D$ oracle. The \emph{transcript} of \ignore{a sequence of requests
made by} $\Algo$ is a sequence $\mathcal{T}=(I_\newer{\ell}, s_\newer{\ell})_{\ell \in\N^\ast}$ of pairs,
where $I_\newer{\ell}$ is the $\newer{\ell}$-th interval provided by the
algorithm as input to $\ICOND_\D$, and $s_\newer{\ell} \in I_\newer{\ell}$
is the response that $\ICOND_\D$ provides to this query.
Given a transcript $\mathcal{T}$, we shall denote by
$\mathcal{T}|_k$ the partial transcript induced by the first $k$ queries, i.e.
$\mathcal{T}|_k=(I_\newer{\ell}, s_\newer{\ell})_{1\leq \newer{\ell} \leq k}$.

\ED

Equipped with these definitions, we now turn to proving the theorem in
the special case of non-adaptive testing algorithms. Observe that there
are three different sources of randomness in our arguments: (i)
the draw of the ``No''-instance from $\calP_{\text \no}$, (ii) the
internal randomness of the testing algorithm; and (iii) the random draws
from the oracle. Whenever there could be confusion we shall
explicitly state which probability space is under discussion.

\subsection{\new{A lower bound a}gainst non-adaptive algorithms}\label{sssec:unif-lb:intcond:non-adapt}

Throughout this subsection we assume that $\Algo$ is an arbitrary, fixed,
non-adaptive, randomized algorithm that makes exactly $q
\leq \tau \cdot {\frac {\log N}{\log \log N}}$ queries
to $\ICOND_\D$; here $\tau>0$ is some absolute constant
that will be determined in the course of the analysis.
(The assumption that $\Algo$ always makes exactly $q$ queries
is without loss of generality since if
in some execution the algorithm makes $q^\prime < q$ queries, it can
perform additional ``dummy'' queries).
In this setting algorithm $\Algo$ corresponds to a distribution $P_A$ over
$q$-tuples $\bar{I}= (I_1,\dots,I_q)$ of query intervals.\ignore{For
this reason, the transcript of any such algorithm running on a
distribution $\D$ can be described as only a
sequence $\bar{s}=(s_1,\dots,s_q)$ of samples.}
The following theorem will directly imply
\autoref{thm:intcond-test-uniform:lb} in the case of
non-adaptive algorithms:

\BT \label{theorem:non:adaptive:transcript:distribution:closeness}
%Fix any non-adaptive algorithm $\Algo$ that makes
%$q \leq \tau\cdot\frac{\log N}{\log\log N}$ calls to $\ICOND_\D$.
%There exists an absolute constant $\tau > 0$ such that
\BEQ \label{eq:intcond:unif:non-adapt}
\left|
\Pr_{\D \sim \calP_{\text \no}}[\Algo^{\ICOND_\D} \text{~outputs~} \accept] - \Pr[\Algo^{\ICOND_{\calU}} \text{~outputs~} \accept] \right| \leq 1/5.
\EEQ
\ET
Observe that in the first probability of Equation~\eqref{eq:intcond:unif:non-adapt}
the randomness is taken over the draw of $\D$ from $\calP_{\text \no}$,
the draw of $\bar{I} \sim P_A$ that $\Algo$ performs to select its sequence
of query intervals, and the randomness of the $\ICOND_\D$ oracle.
In the second one the randomness is just over the draw of $\bar{I}$ from
$P_A$ and the randomness of the $\ICOND_\calU$ oracle.

\medskip

\noindent
\textbf{Intuition for \autoref{theorem:non:adaptive:transcript:distribution:closeness}.}
The high-level idea is that the algorithm will
not be able to distinguish between the uniform distribution
and a ``No''-distribution unless it manages to learn something about the
``structure'' of the blocks in the ``No''-case, either by guessing
(roughly) the right block size, or by guessing (roughly) the
location of a block endpoint and querying a short interval containing such
an endpoint.

In more detail,
% for any ``No''-distribution $\D$ in the support of $\calP_{\text \no}$,
we define the following ``bad events'' (over the choice of $\D$ and the points $s_i$)
for a fixed sequence $\bar{I}= (I_1,\dots,I_q)$ of queries
(the dependence on $\bar{I}$ is omitted in the notation for
the sake of readability):
    \begin{align*}
    B^{\rm N}_{\rm size} &= \setOfSuchThat{ \exists \newer{\ell}
\in[q] }{ % \D \text{ has }
\Delta/\log N \leq \abs{I_\newer{\ell}} \leq \Delta \cdot (\log N)^2} &\\
    B^{\rm N}_{\rm boundary} &= \setOfSuchThat{ \exists \newer{\ell} \in[q]}
{ \abs{I_\newer{\ell}}
< \Delta/\log N \text{ and } I_\newer{\ell} \text{ intersects two blocks}} &\\
   \newer{ B^{\rm N}_{\rm middle} } &= \newer{ \setOfSuchThat{ \exists \newer{\ell}\in[q] }{ \abs{I_\newer{\ell}} < \Delta/\log N \text{ and } I_\newer{\ell} \text{ intersects both halves of the same block}} } &\\
    B^{\rm N}_{\newer{\ell},\rm outer} &= \{ \Delta \cdot
(\log N)^2 < \abs{I_\newer{\ell}} \text{
and } s_\newer{\ell} \text{ belongs to a block not contained entirely in }
 I_\newer{\ell} \} &\newer{\ell}\in[q] \\
    B^{\rm N}_{\newer{\ell},\rm collide} &= \{ \Delta \cdot (\log N)^2 <
\abs{I_\newer{\ell}} \text{ and } \exists j < \newer{\ell},\ s_\newer{\ell} \text{ and } s_j \text{ belong to the same block} \} &\newer{\ell}\in[q]
    \end{align*}

\ignore{

OLD DEFINITIONS:

    \begin{align*}
    B^{\rm N}_{\rm guess} &= \setOfSuchThat{ \exists i\in[q] }{ \D \text{ has } \abs{I_i}/\log N \leq \Delta \leq \abs{I_i} } &\\
    B^{\rm N}_{\rm boundary} &= \setOfSuchThat{ \exists i\in[q] }{ \abs{I_i} \leq \Delta \text{ and } I_i \text{ overlaps on two blocks}} &\\
    B^{\rm N}_{i,\rm draw} &= \{ \Delta\cdot\log N \leq \abs{I_i} \text{ and } s_i \text{ is drawn from a block } \subsetneq I_i \} &i\in[q] \\
    B^{\rm N}_{i,\rm coll} &= \{ \Delta\cdot\log N \leq \abs{I_i} \text{ and } \exists j < i,\ s_i \text{ and } s_j \text{ belong to the same block} \} &i\in[q]
    \end{align*}

END OF OLD DEFINITIONS}

The first \newer{three} events depend only on the draw of $\D$ from $\calP_{\text \no}$, which determines $\Delta$ and $y$,
while the last $2q$ events also depend on the random draws of $s_\newer{\ell}$
from the $\ICOND_\D$ oracle. We define in the same fashion the corresponding
bad events for the ``Yes''-instance (i.e. the uniform
distribution $\calU$) $B^{\rm Y}_{\rm size}$, $B^{\rm Y}_{\rm boundary}$, \newer{$B^{\rm Y}_{\rm middle}$,}
$B^{\rm Y}_{\newer{\ell},\rm outer}$ and $B^{\rm Y}_{\newer{\ell},\rm collide}$,
using the notion of the ``fake construction'' of $\calU$ mentioned above.

Events $B^{\rm N}_{\rm size}$ and $B^{\rm Y}_{\rm size}$
correspond to the possibility, mentioned above, that algorithm $\Algo$
``guesses'' essentially the right block size, and events
$B^{\rm N}_{\rm boundary}$, $B^{\rm Y}_{\rm boundary}$ \newer{and $B^{\rm N}_{\rm middle}$, $B^{\rm Y}_{\rm middle}$}
correspond to the possibility that algorithm $\Algo$ ``guesses'' a short
interval containing \newer{respectively} a block endpoint \newer{or a block midpoint}.
The final bad events correspond to $\Algo$ guessing a ``too-large'' block size
but ``getting lucky'' with the sample returned by $\ICOND$, either because the
sample belongs to one of the (at most two) outer blocks not entirely
contained in the query interval, or because $\Algo$ has already received a
sample from the same block as the current sample.

We can now describe the \emph{failure events} for both the uniform distribution
and for a ``No''-distribution as the union of the corresponding bad events:
  \begin{align*}
    B^{\rm N}_{(\bar{I})} &= B^{\rm N}_{\rm size}\cup B^{\rm N}_{\rm boundary}\cup \newer{ B^{\rm N}_{\rm middle}}\cup\left( \bigcup_{\newer{\ell}=1}^q B^{\rm N}_{\newer{\ell},\rm outer} \right)\cup\left( \bigcup_{\newer{\ell}=1}^q B^{\rm N}_{\newer{\ell},\rm collide} \right)\\
    B^{\rm Y}_{(\bar{I})} &= B^{\rm Y}_{\rm size}\cup B^{\rm Y}_{\rm boundary}\cup \newer{B^{\rm Y}_{\rm middle}}\cup\left( \bigcup_{\newer{\ell}=1}^q B^{\rm Y}_{\newer{\ell},\rm outer} \right)\cup\left( \bigcup_{\newer{\ell}=1}^q B^{\rm Y}_{\newer{\ell},\rm collide} \right)
  \end{align*}

These failure events can be interpreted, from the point of view of the
algorithm $\Algo$, as the ``opportunity to potentially learn
something;'' we shall argue below that if the failure events do not
occur then the algorithm gains no information about whether it is interacting
with the uniform distribution or with a ``No''-distribution.

\medskip
\noindent\textbf{Structure of the proof of \autoref{theorem:non:adaptive:transcript:distribution:closeness}.}
First, observe that since the transcript is the result of the interaction
of the algorithm and the oracle on a randomly chosen distribution, it is
itself a random variable; we will be interested in the distribution over
this random variable induced by the draws from the oracle and the choice
of $\D$.
More precisely, for a fixed sequence of query sets $\bar{I}$,
let $Z^{\rm N}_{\bar{I}}$
denote the random variable over
``No''-transcripts generated when $\D$ is drawn from $\calP_{\text \no}$.
Note that this is a random variable over the probability space defined by
the random draw of $\D$ and the draws of $s_i$ by $\ICOND_\D(I_\newer{\ell})$.
We define $\mathfrak{A}^{\rm N}_{\bar{I}}$ as the resulting distribution
over these ``No''-transcripts.
Similarly, $Z^{\rm Y}_{\bar{I}}$ will be the random variable
over ``Yes''-transcripts, with corresponding distribution
$\mathfrak{A}^{\rm Y}_{\bar{I}}$.
\ignore{
(in absence of ambiguity, we shall omit the subscript $\bar{I}$ from these
notations).
}

As noted earlier, the nonadaptive algorithm $\Algo$ corresponds to a distribution
$P_A$ over $q$-tuples $\bar{I}$ of query intervals.
We define $\mathfrak{A}^{\rm N}$ as the distribution over
transcripts corresponding to first drawing $\bar{I}$ from $P_A$ and
then making a draw from $\mathfrak{A}^{\rm N}_{\bar{I}}$.
Similarly, we define
$\mathfrak{A}^{\rm Y}$ as the distribution over
transcripts corresponding to first drawing $\bar{I}$ from $P_A$ and
then making a draw from $\mathfrak{A}^{\rm Y}_{\bar{I}}$.

To prove \autoref{theorem:non:adaptive:transcript:distribution:closeness}
it is sufficient to show that the two distributions over transcripts
described above are statistically close:

\BL\label{lemma:non:adaptive:transcript:distribution:closeness}
%There exists an absolute constant $\tau > 0$ such that
$\totalvardist{ \mathfrak{A}^{\rm Y} }{ \mathfrak{A}^{\rm N} } \leq 1/5$.
\EL
\ignore{
\cnote{I'm confused about the new formulation (with the remains of the old lemma (``There exists an absolute constant $\tau > 0$ such that''). Shouldn't we write somewhere the same thing as in \autoref{theorem:non:adaptive:transcript:distribution:closeness} (``if $q\leq \tau\cdot\log N/\log\log N$'')? Or write somewhere above that we assume once and for all  $q\leq \tau\cdot\log N/\log\log N$ for some constant $\tau$ to be determined?}
}

The proof of this lemma is structured as follows:
first, for any \emph{fixed} sequence of $q$ queries $\bar{I}$, we bound the probability of the failure events, both for the uniform and the
``No''-distributions:

\begin{clm}\label{claim:boundary:failure:proba}
For each fixed sequence $\bar{I}$ of $q$ query intervals, we have
\ignore{ if $q < \tau{\frac{\log N}{\log\log N}}$, then for $N$ large enough}
\[ \Pr\big[ B^{\rm Y}_{(\bar{I})} \big] \leq 1/10 \qquad\text{and}
\qquad \Pr_{\D\leftarrow\calP_{\text \no}}\big[ B^{\rm N}_{(\bar{I})} \big]
\leq 1/10. \]
\end{clm}
(Note that the first probability above is taken %only
over the randomness of
the $\ICOND_\calU$ responses \newer{and the choice of offset and size in the ``fake construction'' of \uniform}, while the second is over the random draw of $\D
\sim \calP_{\text \no}$ and over the $\ICOND_\D$ responses.)

Next we show that, provided the failure events do not occur,
the distribution over transcripts is exactly the same in both cases:

\begin{clm}\label{clm:cond:equality:in:law:ay:an}
Fix\ignore{any $q \geq 0$ and} any sequence $\bar{I}=(I_1,\dots,I_q)$ of $q$
queries.\ignore{
    \[ \shortexpect\!\left[\ \Algo^{\rm N}_{\bar{I}}\  \Big|\ \overline{ B^{\rm N}_{(\bar{I})} }\ \right] \operatorname*{=}^{\mathcal{L}}
      \shortexpect\!\left[\ \Algo^{\rm Y}_{\bar{I}}\  \Big|\ \overline{ B^{\rm Y}_{(\bar{I})} }\ \right]  \]
  that is, } Then, conditioned on their respective failure events not happening, $Z^{\rm N}_{\bar{I}}$ and $Z^{\rm Y}_{\bar{I}}$
are identically distributed:
\[\text{for every transcript~}\mathcal{T}=((I_1,s_1),\dots,(I_q,s_q)),
\quad \Pr\left[ Z^{\rm N}_{\bar{I}} = \mathcal{T}\ \Big|\
\overline{ B^{\rm N}_{(\bar{I})} } \right] = \Pr\left[ Z^{\rm Y}_{\bar{I}}
= \mathcal{T}\ \Big|\ \overline{ B^{\rm Y}_{(\bar{I})} } \right].  \]
\end{clm}
Finally we combine these two claims to show that the two overall distributions
of transcripts are statistically close:
\begin{clm}\label{clm:closeness:ay:an}
\ignore{Suppose
$q < \tau\frac{\log N}{\log\log N}$, where $\tau$ is as in
\autoref{claim:boundary:failure:proba}, and}
Fix any sequence of $q$ queries $\bar{I}=(I_1,\dots,I_q)$.
Then $\totalvardist{ \mathfrak{A}^{\rm N}_{\bar{I}} }
{ \mathfrak{A}^{\rm Y}_{\bar{I}} } \leq 1/5$.
\end{clm}
\autoref{lemma:non:adaptive:transcript:distribution:closeness}
(and thus
\autoref{theorem:non:adaptive:transcript:distribution:closeness})
directly follows from \autoref{clm:closeness:ay:an} since, using
the notation \mbox{$\bar{s} = (s_1,\dots,s_q)$} for a sequence of $q$ answers
to a sequence $\bar{I} = (I_1,\dots,I_q)$ of $q$ queries, which
together define a transcript $\mathcal{T}(\bar{I},\bar{s}) = ((I_1,s_1),\dots,(I_q,s_q))$,
%\dnote{Just spelled it out rather then referring to Yao}
\BEQN
  \dtv\left(\mathfrak{A}^{\rm Y},\mathfrak{A}^{\rm N}\right) &=& \frac{1}{2}  %\sum_{\mathcal{T}}
  \sum_{\bar{I}}\sum_{\bar{s}}
      \left| P_A(\bar{I}) \cdot \Pr\left[Z^Y_{\bar{I}}=\mathcal{T}(\bar{I},\bar{s})\right] -
            P_A(\bar{I}) \cdot \Pr\left[Z^N_{\bar{I}}=\mathcal{T}(\bar{I},\bar{s})\right]\right| \nonumber\\
  &=& \newer{\frac{1}{2}}\sum_{\bar{I}} P_A(\bar{I}) \cdot \sum_{\bar{s}}
     \left|  \Pr\left[Z^Y_{\bar{I}}=\mathcal{T}(\bar{I},\bar{s})\right]
           - \Pr\left[Z^N_{\bar{I}}=\mathcal{T}(\bar{I},\bar{s})\right]\right| \nonumber\\
  &\leq& \max_{\bar{I}}\left\{\dtv\left(\mathfrak{A}^{\rm Y}_{\bar{I}} , \mathfrak{A}^{\rm N}_{\bar{I}}\right) \right\}
    \leq 1/5\;.
\EEQN
%%%%
\ignore{
  To prove \autoref{lemma:non:adaptive:transcript:distribution:closeness}
  (and thus
  \autoref{theorem:non:adaptive:transcript:distribution:closeness}),
  it only remains to extend this variation distance upper bound
  to $\mathfrak{A}^{\rm Y},\mathfrak{A}^{\rm N}$;
  this is done by applying the easy direction of Yao's minimax principle to \autoref{clm:closeness:ay:an}.
  \rnote{It is the easy direction we're using here, right?}
  \ignore{It seems the situation is that
  \autoref{clm:closeness:ay:an} shows that for any fixed $\bar{I}$,
  the distributions
  $\mathfrak{A}^{\rm N}_{\bar{I}}$ and $\mathfrak{A}^{\rm Y}_{\bar{I}}$
  are statistically close.
  $\mathfrak{A}^{\rm N}$ is a mixture of the $\mathfrak{A}^{\rm N}_{\bar{I}}$'s,
  and
  $\mathfrak{A}^{\rm Y}$ is the same mixture of the
  $\mathfrak{A}^{\rm Y}_{\bar{I}}$'s.
  So really all we're using is that the average/mixture can't have
  worse variation distance than all of the components of the mixture.}
  \cnote{Yes, I think you're right -- if it seems too much to cite Yao's principle here, we could probably do without.}
}
%%%%

\noindent
This concludes the proof of
\autoref{lemma:non:adaptive:transcript:distribution:closeness}
modulo the proofs of the above claims; we give those proofs in
\autoref{sssec:unif-lb:intcond:claims} below.

%%%%%%%%%%%%%%%%%%%%%%%%%%%%%%%%%%%%%%%%%%%%%%%%%%%%%%%%%%%%%%%%
\subsubsection{Proof of Claims~\ref{claim:boundary:failure:proba} to \ref{clm:closeness:ay:an}}\label{sssec:unif-lb:intcond:claims}
%%%%%%%%%%%%%%%%%%%%%%%%%% Non-adaptive part%%%%%%%%%%%%%%%%%%%%%%%%%%

% % \ignore{
% % \rnote{I tried to make the analysis more concise throughout this
% % subsubsection.}
% % }

To prove \autoref{claim:boundary:failure:proba}
we bound the probability of each of the bad events separately, starting
with the ``No''-case.

%%%%%%%%%%%%%%%%%%%%%%%%%% Bad events
\begin{itemize}
%%%%% B_size
\item [(i)] \label{proofs:bad:events:guess:size}
\sloppy
Defining the event $B^{\rm N}_{\newer{\ell}, \rm size}$ as
\[
  B^{\rm N}_{\newer{\ell}, \rm size} =
     \{ \Delta/\log N \leq \abs{I_\newer{\ell}} \leq \Delta \cdot (\log N)^2 \}\;,
\]
we can use a union bound to get
$\Pr[ B^{\rm N}_{\rm size} ] \leq \sum_{\newer{\ell}=1}^q \Pr[ B^{\rm N}_{\newer{\ell}, \rm size} ].$
For any fixed setting of $I_\newer{\ell}$ there are $O(\log \log N)$ values of
$\Delta \in \{{\frac N {2^X}} \ | \ X \in \{{\frac 1 3} \log N, \dots,
{\frac 2 3} \log N\}\}$ for which
\mbox{$\Delta/\log N \leq I_\newer{\ell} \leq \Delta \cdot (\log N)^2$.}  Hence
we have
$\Pr[ B^{\rm N}_{\newer{\ell}, \rm size} ] = O((\log \log N)/\log N)$, and consequently
\mbox{$\Pr[ B^{\rm N}_{\rm size}]  = O(q(\log \log N)/\log N)$}.

% % \ignore{
% %
% % OLD ANALYSIS:
% %
% % Writing $L\eqdef\abs{I_1}$, and assuming wlog that $L \geq 1$, we have\footnote{Hereafter, we use $\indicSet{V}$ to denote the indicator function of the set (or event) $V$.}
% % \begin{align*}
% %   \Pr[ B^{\rm N}_{1, \rm size} ] &= \sum_{x} \indic{ L/\log N \leq x \leq L } \Pr[ \Delta = x ]
% %   =  \sum_{k=\frac{1}{3}\log N}^{\frac{2}{3}\log N} \frac{\indic{ L \leq N\cdot 2^{-k} \cdot\log N }\indic{ N\cdot2^{-k} \leq L}}{ \frac{1}{3}\log N + 1 } \\
% %   &\leq \frac{3}{ \log N + 3 } \sum_{k=\frac{1}{3}\log N}^{\frac{2}{3}\log N} \indic{ \log N - \log L \leq k \leq \log( N\log N ) - \log L } \\
% %   &\leq \frac{3}{ \log N + 3 } \left(\ \log( N\log N ) - \log N + 1\ \right) \leq 3\cdot\frac{\log \log N + 1}{ \log N }
% % \end{align*}
% % and therefore
% %   \begin{equation}\label{proba:bad:event:size}
% %     \Pr[ B^{\rm N}_{\rm size} ] \leq 3q\cdot\frac{\log \log N + 1}{ \log N }
% %   \end{equation}
% %
% % END OLD ANALYSIS
% %
% % }

%%%%% B_boundary
\item [(ii)] \label{proofs:bad:events:boundary:interval}
\sloppy
Similarly, defining  the event $B^{\rm N}_{\newer{\ell}, \rm boundary}$ as
\[
B^{\rm N}_{\newer{\ell}, \rm boundary} = \left\{ \abs{I_\newer{\ell}} < \Delta/\log N
\text{ and } I_\newer{\ell} \text{ intersects two blocks} \right\} \;,
\]
 we have
$\Pr[ B^{\rm N}_{\rm boundary} ] \leq
\sum_{\newer{\ell}=1}^q \Pr[ B^{\rm N}_{\newer{\ell}, \rm boundary} ] $.
For any fixed setting of $I_\newer{\ell}$, recalling the choice of a uniform
random offset $y \in [N]$ for the blocks, we have that
$\Pr[B^{\rm N}_{\newer{\ell}, \rm boundary}] \leq O(1/\log N)$, and consequently
$\Pr[ B^{\rm N}_{\rm boundary} ] = O(q/\log N)$.

\item [(iii)] \label{proofs:bad:events:middle:interval}
\sloppy\newer{
The analysis of $B^{\rm N}_{\rm middle}$ is identical (by considering the midpoint of a block instead of its endpoint), yielding directly
$\Pr[ B^{\rm N}_{\rm middle} ] = O(q/\log N)$.
}

\item [(iv)] Fix $\newer{\ell} \in [q]$ and recall that
$B^{\rm N}_{\newer{\ell},\rm outer} = \{ \Delta\cdot (\log N)^2 < \abs{I_\newer{\ell}}$
and $s_\newer{\ell}$ is drawn from a block $\subsetneq I_\newer{\ell} \}$.
Fix any outcome for $\Delta$ such that
$\Delta\cdot (\log N)^2 < \abs{I_\newer{\ell}}$
and let us consider only the randomness over the draw of
$s_\newer{\ell}$ from $I_\newer{\ell}$.
Since there are $\Omega((\log N)^2)$ blocks contained entirely in
$I_\newer{\ell}$, % it is easy to see that
the probability that $s_\newer{\ell}$ is drawn from a block not contained entirely in
$I_\newer{\ell}$ (there are at most two such blocks, one at each end of $I_\newer{\ell})$
is $O(1/(\log N)^2)$.
Hence we have
$\Pr[B^{\rm N}_{\newer{\ell}, \rm outer}] \leq O(1)/(\log N)^2.$

\ignore{

OLD ANALYSIS

Writing this time $L\eqdef\abs{I_i}$,
  \[
    \Pr[ B^{\rm N}_{i,\rm outer} ] = \Pr[ s_i \text{ is drawn from a block } \subsetneq I_i \ |\ \Delta\cdot\log N \leq L]\cdot\Pr[\Delta\cdot\log N \leq L]
  \]
  where the first factor can be upper bounded in a brute-force way, by taking the worst case -- which is having the two end blocks (the leftmost and rightmost) being respectively
  low-high and high-low, and optimizing according to the fractions $u,v$ of the half-blocks which fall within $I_i$:
  \begin{align*}
    \probaCond{ s_i\text{'s block } \subsetneq I_i }{\Delta\cdot\log N \leq L}
      &\leq \max\Big( \max_{u\in[0,\frac{1}{2}]}\frac{2u\cdot\Delta\frac{1+2\eps}{N}}{2u\cdot\Delta\frac{1+2\eps}{N} + \left( \left\lceil\frac{L}{\Delta}\right\rceil-2 \right)\frac{\Delta}{N}}, \\
      &\qquad\qquad \max_{v\in[0,\frac{1}{2}]} \frac{ \Delta\frac{1+2\eps}{N}+ 2v\Delta\frac{1-2\eps}{N} }{ \Delta\frac{1+2\eps}{N} + 2v\Delta\frac{1-2\eps}{N} + \left( \left\lceil\frac{L}{\Delta}\right\rceil-2 \right)\frac{\Delta}{N} }   \Big)\\
      &\leq \max\left( \frac{1+\eps}{1+\eps + \left( \left\lceil\frac{L}{\Delta}\right\rceil-2 \right)},\frac{2\Delta}{L} \right) = \frac{2\Delta}{L}
      \leq \frac{2N}{2^k L}
  \end{align*}
  From this,
  \[
    \Pr[ B^{\rm N}_{i,\rm outer} ] \leq \frac{6N}{ \log N }\sum_{k=\frac{1}{3}\log N}^{\frac{2}{3}\log N}\frac{\indic{ k \geq \log( N\log N ) -\log L}}{2^kL}
  \]
  It is easy to see that this quantity can only be maximized wrt $L$ if $\frac{L}{\log N}\in[n^{1/3}, n^{2/3}]$ (that is, the range for $\Delta$): if $L$ is too big, we have all terms possible in the sum, but the factor $\frac{1}{L}$ takes the whole quantity down; but if $L$ is too small, then we lose many possible terms from the sum; hence,
  \begin{align*}
    \Pr[ B^{\rm N}_{i,\rm outer} ] &\leq \frac{1}{L}\cdot\frac{6N}{ \log N }\sum_{k=\log\frac{N\log N}{L}}^{\frac{2}{3}\log N}\frac{1}{2^k} =
     \frac{1}{L}\cdot\frac{6N}{ \log N } \frac{2L}{N\log N}\left( 1 - \frac{1}{2^{\frac{2}{3}\log N - \log\frac{N\log N}{L}+1 }}\right)
  \end{align*}
  and finally, we obtain that
  \begin{equation}\label{proba:bad:event:outer}
    \Pr[ B^{\rm N}_{i,\rm outer} ] \leq \frac{12}{\log^2 N }
  \end{equation}

END OLD ANALYSIS

}

%%%%% B_{i,collide}
\item [(v)]
Finally, recall that
\[
B^{\rm N}_{\newer{\ell},\rm collide} =
\{ \Delta \cdot (\log N)^2 < \abs{I_\newer{\ell}} \mbox{ and }
\exists j < \newer{\ell} \mbox{ s.t. } s_\newer{\ell} \mbox{ and } s_j \mbox{ belong to the same block } \}\;.
\]
Fix $\newer{\ell} \in [q]$ and a query interval $I_\newer{\ell}$. Let $r_\newer{\ell}$ be the number of blocks in $I_\newer{\ell}$ within which resides
some previously sampled point $s_j$, $j \in [\newer{\ell}-1]$.
Since there are
$\Omega((\log N)^2)$ blocks in $I_\newer{\ell}$ and $r_\newer{\ell} \leq \newer{\ell}-1$,
the probability that $s_\newer{\ell}$ is drawn from
a block containing any $s_j$, $j < \newer{\ell}$, is $O(\newer{\ell}/(\log N)^2).$
%%%
\ignore{
    The worst possible case, when querying $I_i$, is that all $(i-1)$ previous
    samples $s_1,\dots,s_{i-1}$ are in $I_i$ and each of them occupies a
    different block.
    As in the preceding item, since there are
    $\Omega((\log N)^2)$ blocks in $I_i$, the probability that $s_i$ is drawn from
    a block containing any $s_j$, $j < i$, is $O(i/(\log N)^2).$
}
Hence we have
$\Pr[B^{\rm N}_{\newer{\ell},\rm collide}] =  O(\newer{\ell}/(\log N)^2).$

\ignore{

OLD ANALYSIS:

This is indeed the worst possible case, as it means that amongst all blocks for the new sample $s_i$ to fall, we have as little probability as possible to end up in a favorable one. Furthermore, and even this is already taken care of by another bad event, we will mark the two possibly incomplete ``end blocks'' as equally bad outcomes -- meaning that we now have $(i+1)$ bad blocks, all of them high, amongst $\frac{\abs{I_1}}{\Delta}$ possible blocks; the remaining ones being low.\\

After this discussion, it is not hard to see that this is actually \emph{exactly} the same analysis that for $\Pr[ B^{\rm N}_{i,\rm outer} ]$, except the number of bad blocks is now changed from $2$ to $(i+1)$. Hence,
  \begin{equation}\label{proba:bad:event:collision}
    \Pr[ B^{\rm N}_{i,\rm collide} ] \leq \frac{6(i+1)}{\log^2 N }
  \end{equation}

END OLD ANALYSIS
}

\end{itemize}

\noindent With these probability bounds for bad events in hand,
we can prove \autoref{claim:boundary:failure:proba}:

\begin{proofof}{\autoref{claim:boundary:failure:proba}}
Recall that $q \leq \tau \cdot \frac{\log N}{\log\log N}$.
Recalling the definition of $B^{\rm N}_{(\bar{I})}$,
a union bound yields
\begin{eqnarray*}
\Pr[ B^{\rm N}_{(\bar{I})} ]
    &\leq& \Pr[ B^{\rm N}_{\rm size} ] + \Pr[  B^{\rm N}_{\rm boundary} ] + \newer{\Pr[  B^{\rm N}_{\rm middle} ]} + \sum_{\newer{\ell}=1}^q \Pr[ B^{\rm N}_{\newer{\ell},\rm outer} ] + \sum_{\newer{\ell}=1}^q \Pr[ B^{\rm N}_{\newer{\ell},\rm collide} ] \\
&=& O\left( {\frac {q \cdot \log \log N}{\log N}} \right)
+ O\left( {\frac q {\log N}} \right)
+ \newer{O\left( {\frac q {\log N}} \right)}
+ \sum_{\newer{\ell}=1}^q O\left({\frac 1 {(\log N)^2}} \right)
+ \sum_{\newer{\ell}=1}^q O\left({\frac {\newer{\ell}} {(\log N)^2}} \right)\\
&\leq& {\frac 1 {10}}\;,
\end{eqnarray*}
where the last inequality holds for a sufficiently small choice of the
absolute constant $\tau.$

The same analysis applies unchanged for
$\Pr[ B^{\rm Y}_{\rm size} ]$, \newer{$\Pr[ B^{\rm Y}_{\rm middle} ]$} and $\Pr[  B^{\rm Y}_{\rm boundary} ]$,
using the ``fake construction'' view of $\calU$ as described earlier.
The arguments for $\Pr[B^{\rm Y}_{\newer{\ell}, \rm outer}]$ and
$\Pr[B^{\rm Y}_{\newer{\ell}, \rm collide}]$ go through unchanged as well,
and \autoref{claim:boundary:failure:proba} is proved.
\end{proofof}

\ignore{

by seeing the choice of the $Uniform$ distribution as a construction, as outlined in the first section. A similar argument also applies to $\Pr[ B^{\rm Y}_{i,\rm outer} ]$ and $\Pr[ B^{\rm N}_{i,\rm collide} ]$, resulting in the same upper bound. This yields the desired claim, both for $\Pr[ B^{\rm N}_{\rm size} ]$ and $\Pr[ B^{\rm Y}_{\rm size} ]$, by setting $\tau\eqdef\frac{1}{50}$.

}

\begin{proofof}{\autoref{clm:cond:equality:in:law:ay:an}}
Fix any $\bar{I}=(I_1,\dots,I_q)$ and any transcript
$\mathcal{T}=((I_1,s_1),\dots,(I_q,s_q))$.
\ignore{\footnote{Note that we do not have to worry about $q$ being too large, since we are conditioning on the ``bad events'' not happening - hence, as when $q$ grows too big, this events are bound to occur (w.p. 1), the probability space we are considering becomes empty and all properties trivially hold.\clement{ \bf{\color{black}[[C:} Remove this?{\color{black}]]}}}.
}
Recall that the length-$\ell$ partial transcript ${\cal T}|_\ell$
is defined to be $((I_1,s_1),\dots,(I_\ell,s_\ell))$.
We define the random variables
$Z^{\rm N}_{\bar{I},\ell}$
and
$Z^{\rm Y}_{\bar{I},\ell}$
to be the length-$\ell$ prefixes of
$Z^{\rm N}_{\bar{I}}$
and
$Z^{\rm Y}_{\bar{I}}$
respectively.
We prove \autoref{clm:cond:equality:in:law:ay:an} by
establishing the following, which we prove by induction on $\ell$:
\BEQ
\Pr\left[ Z^{\rm N}_{\bar{I}\newer{,\ell}} = \mathcal{T}\newer{|_\ell}\ \Big|\
\overline{ B^{\rm N}_{(\bar{I})} } \right] = \Pr\left[ Z^{\rm Y}_{\bar{I}\newer{,\ell}}
= \mathcal{T}\newer{|_\ell}\ \Big|\ \overline{ B^{\rm Y}_{(\bar{I})} } \right].
\label{eq:inductme}
\EEQ
For the base case, it is clear that (\ref{eq:inductme}) holds with
$\ell = 0.$
For the inductive step, suppose
(\ref{eq:inductme}) holds for all $k\in[\ell-1]$.
When querying $I_\ell$ at the $\ell$-th step,
one of the following cases must hold (since we conditioned on the
``bad events'' not happening):

\begin{itemize}

\item [(1)] $I_\ell$ is contained within a half-block
(more precisely, either entirely within the first half of a block
or entirely within the second half). In this case the
``{yes}'' and ``{no}'' distribution oracles behave {exactly} the same
since both generate $s_{\newer{\ell}}$ by sampling uniformly from $I_{\newer{\ell}}.$

\item [(2)]
The point $s_\ell$ belongs to a
block, contained entirely in $I_{\newer{\ell}}$, which is ``fresh'' in the sense
that it contains no $s_j$, $j < \newer{\ell}.$ In the ``No''-case
this block may either be high-low or low-high; but since both outcomes
have the same probability, there is another transcript with equal
probability in which the two profiles are switched.  Consequently
(over the randomness in the draw of $\D \sim \calP_{\no}$)
the probability of picking $s_\ell$ in the ``No''-distribution case is
the same as in the uniform distribution case (i.e., uniform on \newer{the fresh blocks contained in} $I_\ell$).

%     \begin{figure}[h!]\label{fig:proof:lemma:cond:equality:in:law:ay:an}
%       \begin{center}
%         \input{intcond_lb_05-30_drawing.tex}
%         \caption{Symmetric cases for $I_\ell$ ``big''}
%       \end{center}
%     \end{figure}

\ignore{ %%%%% WAS BUGGY -- FIXED BY INTRODUCING THE BAD EVENTS B^{\rm N}_{\rm middle}, B^{\rm Y}_{\rm middle}
\item [(3)]
$I_\ell$ is contained within one block, but not within one half-block
(i.e. $I_\ell$ intersects both the first and second halves of its block).
By the same symmetry argument as in (2), the profile of this block
could have been switched, and hence the distribution of $s_i$ is uniform
over $I_i$.
}

\end{itemize}

\noindent
This concludes the proof of \autoref{clm:cond:equality:in:law:ay:an}.
\end{proofof}

\ignore{
, for all transcript $\mathcal{T}$, we have $\Pr\left[ \Algo^{\rm N}_{\bar{I}} = \mathcal{T}\ \Big|\ \overline{ B^{\rm N}_{(\bar{I})} } \right] = \Pr\left[ \Algo^{\rm Y}_{\bar{I}} = \mathcal{T}\ \Big|\ \overline{ B^{\rm Y}_{(\bar{I})} } \right]$ (over the construction of the distribution and the sampling by $\ICOND$).
}

\begin{proofof}{\autoref{clm:closeness:ay:an}}
Given Claims~\ref{claim:boundary:failure:proba}
and \ref{clm:cond:equality:in:law:ay:an}, \autoref{clm:closeness:ay:an}
is an immediate consequence of the following basic fact:

\begin{fct} \label{fact:dtv-condit}
Let $\D_1$, $\D_2$ be two distributions over the same finite set $X$.  Let
$E_1,E_2,$ be two events such that
$\D_i[E_i] = \alpha_i \leq \alpha$ for $i=1,2$ and the conditional
distributions $(\D_i)_{\overline{E_i}}$ are identical, i.e.
$\dtv(
(\D_1)_{\overline{E_1}},
(\D_2)_{\overline{E_2}}
)=0.$
Then $\dtv(\D_1,\D_2) \leq \alpha.$
\end{fct}

\ignore{
\rnote{I was a little uncomfortable with some of the previous proof of
this -- there were references to things like ``$\Pr[F]$''
and ``$\Pr[G]$'' and it wasn't really clear to me
which distribution the probabilities were with respect to; I think this
may have been related to the ``independence'' discussion which also made me
a little uncomfortable. Let me know if the current argument seems OK.}
}

\BPF
We first observe that since $(\D_2)_{\overline{E_2}}(E_2)=0$ and
$(\D_1)_{\overline{E_1}}$ is identical to $(\D_2)_{\overline{E_2}}$, it must be
the case that $(\D_1)_{\overline{E_1}}(E_2)=0$, and likewise
$(\D_2)_{\overline{E_2}}(E_1)=0$.  This implies that
$\D_1(E_2 \setminus E_1) = \D_2(E_1 \setminus E_2)=0$.
Now let us write
\begin{eqnarray*}
2\dtv(\D_1,\D_2)
&=&\sum_{x \in X \setminus (E_1 \cup E_2)} |\D_1(x)-\D_2(x)| +
\sum_{x \in  E_1 \cap E_2} |\D_1(x)-\D_2(x)| + \\
&&\sum_{x \in E_1 \setminus E_2} |\D_1(x)-\D_2(x)| +
\sum_{x \in E_2 \setminus E_1} |\D_1(x)-\D_2(x)|.\\
\end{eqnarray*}
We may upper bound
$\sum_{x \in  E_1 \cap E_2} |\D_1(x)-\D_2(x)|$ by
$\sum_{x \in  E_1 \cap E_2} (\D_1(x) + \D_2(x)) =
\D_1(E_1 \cap E_2) + \D_2(E_1 \cap E_2)$,
and the above discussion gives
$\sum_{x \in E_1 \setminus E_2} |\D_1(x)-\D_2(x)| =
\D_1(E_1 \setminus E_2)$ and
$\sum_{x \in E_2 \setminus E_1} |\D_1(x)-\D_2(x)| = \D_2(E_2 \setminus E_1).$
We thus have
\begin{eqnarray*}
2\dtv(\D_1,\D_2)
&\leq&
\sum_{x \in X \setminus (E_1 \cup E_2)} |\D_1(x)-\D_2(x)| +
\D_1(E_1) + \D_2(E_2)\\
&\leq&\sum_{x \in X \setminus (E_1 \cup E_2)} |\D_1(x)-\D_2(x)| +
\alpha_1 + \alpha_2.
\end{eqnarray*}
Finally, since $\dtv((\D_1)_{\overline{E_1}},(\D_2)_{\overline{E_2}})=0$, we
have
\begin{eqnarray*}\sum_{x \in X \setminus (E_1 \cup E_2)} |\D_1(x)-\D_2(x)|
&=& \left|
\D_1(X \setminus (E_1 \cup E_2)) - \D_2(X \setminus (E_1 \cup E_2)\newer{)}
\right|\\
&=& \left|\D_1(\overline{E_1}) - \D_2(\overline{E_2}) \right| = |\alpha_1 - \alpha_2|.
\end{eqnarray*}
Thus $2 \dtv(\D_1,\D_2) \leq |\alpha_1 - \alpha_2| + \alpha_1 + \alpha_2 =
2 \max\{\alpha_1,\alpha_2\} \leq 2 \alpha$, and the fact is established.
\EPF

\noindent
This concludes the proof of \autoref{clm:closeness:ay:an}.
\end{proofof}

\ignore{

OLD ANALYSIS:

Fix an arbitrary non-adaptive, possibly randomized algorithm $\Algo$ making $q < \tau\frac{\log N}{\log\log N}$ queries. Write $G\eqdef\overline{ B^{\rm Y}_{(\bar{I})}}\cap\overline{ B^{\rm N}_{(\bar{I})}}$ and $F\eqdef B^{\rm Y}_{(\bar{I})}\cup B^{\rm N}_{(\bar{I})}$ (for ``Good'' and ``Fail''). Then,
\begin{align*}
 2\totalvardist{ \mathfrak{A}^{\rm N}_{\bar{I}} }{ \mathfrak{A}^{\rm Y}_{\bar{I}} }
  &= \sum_{\substack{T \text{ transcript} \\T=(\bar{I},\bar{s})\\ \abs{T}=\abs{\bar{I}}=q}} \abs{ \Pr[ \Algo^{\rm N}_{\bar{I}} = T ] - \Pr[ \Algo^{\rm Y}_{\bar{I}} = T ] } \\
  &= \sum_{T} \Big|\ \probaCond{ \Algo^{\rm N}_{\bar{I}} = T }{ G }\cdot\Pr[G]
    + \probaCond{ \Algo^{\rm N}_{\bar{I}} = T }{ F }\cdot\Pr[F] \\
    &\qquad\quad- \probaCond{ \Algo^{\rm Y}_{\bar{I}} = T }{ G }\cdot\Pr[G]
    - \probaCond{ \Algo^{\rm Y}_{\bar{I}} = T }{ F }\cdot\Pr[F]
    \ \Big| \\
  &\leq \sum_{T} \abs{\ \probaCond{ \Algo^{\rm N}_{\bar{I}} = T }{ G }
    - \probaCond{ \Algo^{\rm Y}_{\bar{I}} = T }{ G } \ }\cdot\Pr[G] \\
   &+ \sum_{T} \abs{\ \probaCond{ \Algo^{\rm N}_{\bar{I}} = T }{ F }
    - \probaCond{ \Algo^{\rm Y}_{\bar{I}} = T }{ F }
    \ }\cdot\Pr[F] \\
  &\leq \sum_{T} \abs{\ \probaCond{ \Algo^{\rm N}_{\bar{I}} = T }{ \overline{ B^{\rm N}_{(\bar{I})}} }
    - \probaCond{ \Algo^{\rm Y}_{\bar{I}} = T }{ \overline{ B^{\rm Y}_{(\bar{I})}} } \ }\cdot\Pr[G] \tag{by independence\footnotemark}\\
   &+ \sum_{T} \left( \probaCond{ \Algo^{\rm N}_{\bar{I}} = T }{ F }
    + \probaCond{ \Algo^{\rm Y}_{\bar{I}} = T }{ F }
    \right)\cdot\Pr[F] \\
    &\leq \sum_{T} \abs{\ \probaCond{ \Algo^{\rm N}_{\bar{I}} = T }{ \overline{ B^{\rm N}_{(\bar{I})}} }
    - \probaCond{ \Algo^{\rm Y}_{\bar{I}} = T }{ \overline{ B^{\rm Y}_{(\bar{I})}} } \ }\cdot 1 + 2\Pr[F]
\end{align*}
where the result for the second term follows from the fact that $\sum_{x} \probaCond{ X=x }{ E }\cdot\Pr[ E ] = \Pr[ E ]$. We know with \autoref{claim:boundary:failure:proba} and a union bound that the second term of the RHS is at most $2\cdot\frac{2}{10}=\frac{2}{5}$; and \autoref{clm:cond:equality:in:law:ay:an} ensures that the first term is zero. This concludes the proof that, for $N$ large enough, $\totalvardist{ \mathfrak{A}^{\rm N}_{\bar{I}} }{ \mathfrak{A}^{\rm Y}_{\bar{I}} }\leq 1/5$.
\footnotetext{To explain this argument, the best way is to consider the following thought experiment: we run in parallel two instances of the exact same algorithm $\Algo$, using the same source of randomness, on two independent oracles $\ICOND_\calU$ and $\ICOND_\D$ (where $\D$ is a ``No''-distribution); and we compare the distribution over the two kinds of transcripts we obtain. As this thought experiment shows, the bad events for the first algorithm have no influence on the transcript of the second, and vice-versa.}

END OLD ANALYSIS

}

%%%%%%%%%%%%%%%%%%%%%%%%%%%%%%%%%%%%%%%%%%%%%%%%%%%%%%%%%%%%%%%%

\subsection{A lower bound against adaptive algorithms:  Outline of the proof of
\autoref{thm:intcond-test-uniform:lb}}\label{sssec:unif-lb:intcond:adapt}
Throughout this subsection  $\Algo$ denotes a general
adaptive algorithm that makes $q \leq \tau \cdot {\frac {\log N}
{\log \log N}}$ queries, where as before $\tau>0$ is an absolute constant. \autoref{thm:intcond-test-uniform:lb} is a
consequence of the following theorem, which deals with adaptive algorithms:
\BT \label{theorem:adaptive:transcript:distribution:closeness}
%For any algorithm $\Algo$ making $q \leq \tau\cdot\frac{\log N}{\log\log N}$
%calls to $\ICOND_\D$,
\BEQ \label{eq:intcond:unif:adapt}
\left|
\Pr_{\D \sim \calP_{\text \no}}[\Algo^{\ICOND_\D} \text{~outputs~} \accept] - \Pr[\Algo^{\ICOND_{\calU}} \text{~outputs~} \accept] \right| \leq \newer{1/5}.
\EEQ
%where $\tau$ is as in \autoref{theorem:non:adaptive:transcript:distribution:closeness}.
\ET

The idea here is to \new{extend} the previous analysis for non-adaptive algorithms,
and argue that ``{adaptiveness does not really help}'' to distinguish between $\D = \calU$ and $\D \sim \calP_{\text \no}$
given access to $\ICOND_\D$.

%%% Clement: I actually removed stuff no longer relevant here -- it still can be found in the other uniformity_intcond*.tex; an important part is below:
%% Dana (Oct 25): I moved this closer to where we actually need it
% % As in \autoref{ssec:lb-Pcond-Dstar}, we will need
% % the notion of an algorithm \emph{faking queries}.
% %  Given an adaptive algorithm $\Algo$,
% % we define $\Algo^{(1)}$ as the algorithm that \emph{fakes} its first query,
% % in the following sense: If the first query made by $\Algo$ to the oracle is
% % some interval $I$, then the algorithm $\Algo^{(1)}$
% % does not call $\ICOND$ on $I$ but instead
% % chooses a point $s$ uniformly at random from $I$ and then behaves exactly as
% % $\Algo$ would behave if the $\ICOND$ oracle had returned $s$ in response
% % to the query $I$.
% % More generally, we define $\Algo^{(k)}$ for all $k \geq 0$
% % as the algorithm behaving like $\Algo$ but faking its first $k$
% % queries (note that $\Algo^{(0)}=\Algo$).

As in the non-adaptive case, in order to prove \autoref{theorem:adaptive:transcript:distribution:closeness},
it is sufficient to prove that the transcripts for uniform and ``No''-distributions are close in
total variation distance; i.e., that
%if $\Algo$ performs less than $\tau\frac{\log N}{\log\log N}$ queries, then
\begin{equation}\label{eqn:adaptive:transcript:distribution:closeness}
     \totalvardist{ \mathfrak{A}^{\rm Y} }{ \mathfrak{A}^{\rm N}} \leq \newer{1/5}.
\end{equation}
\newer{The key idea used to prove this will be to introduce a \emph{sequence} $\mathfrak{A}^{(k), \rm N}_{\rm otf}$ of distributions over transcripts (where ``otf'' stands for ``on the fly''), for $0\leq k \leq q$, such that (i) $\mathfrak{A}^{(0), \rm N}_{\rm otf}=\mathfrak{A}^{\rm Y}$ and $\mathfrak{A}^{(q), \rm N}_{\rm otf}=\mathfrak{A}^{\rm N}$, and (ii) the \new{distance} $\totalvardist{ \mathfrak{A}^{(k), \rm N}_{\rm otf} }{ \mathfrak{A}^{(k+1), \rm N}_{\rm otf} }$
% between any two adjacent terms
\new{for each $0 \leq k \leq q-1$}
is ``small''. This will enable us to conclude by the triangle inequality, as
\begin{equation}
%\begin{align*}
     \totalvardist{ \mathfrak{A}^{\rm N} }{ \mathfrak{A}^{\rm Y} }
     = \totalvardist{ \mathfrak{A}^{(0), \rm N}_{\rm otf} }{ \mathfrak{A}^{(q), \rm N}_{\rm otf} }
     \leq \sum_{k=0}^{q-1} \totalvardist{ \mathfrak{A}^{(k), \rm N}_{\rm otf} }{ \mathfrak{A}^{(k+1), \rm N}_{\rm otf} }\;.
%\end{align*}
\label{eq:tvd:otf:sum}
\end{equation}
To define this sequence,} in the next subsection we will introduce the notion of an \emph{extended transcript}, which in addition
to the queries and samples includes additional information about the ``local structure'' of the distribution
at the endpoints of the query intervals and the sample points. \newer{Intuitively, this} extra information will help us analyze the
interaction between the adaptive algorithm and the oracle.
\newer{We will then describe an alternative process according to which a
``faking algorithm'' (reminiscent of the similar notion from \autoref{ssec:lb-Pcond-Dstar}) can interact with an oracle to generate such an extended transcript.}
\newest{More precisely, we shall define a sequence of such faking algorithms, paramaterized
 by ``how much faking'' they perform. For both the original (``non-faking'')
algorithm $\Algo$ and for the faking algorithms, we will show how extended transcripts can be generated
``on the fly''. The aforementioned distributions $\mathfrak{A}^{(k), \rm N}_{\rm otf}$
over (regular) transcripts are obtained by \emph{truncating} the extended transcripts
that are generated on the fly (i.e., discarding the extra information), and we shall
argue that they satisfy requirements (i) and (ii) above.}
%%
% , while only defining the ``No''-distribution $\D$ ``on the fly''.
% Note that while this new process is \emph{not} equivalent to the one where $\D$ is
% drawn and fully committed to beforehand, we will argue that the distribution over (regular)
% transcripts obtained by \emph{truncating} the extended transcripts generated this way (i.e.,
% discarding the extra information) does satisfy requirements (i) and (ii) above.}
%%
%\clement{\par\textbf{C:} Not completely happy with the above attempt\dots }
%\dana{\par\textbf{Dana (Oct 25):} Modified a bit. Also the next sentence}

\newest{Before turning to the precise definitions and the analysis of
extended transcripts and faking algorithms, we provide the following variant of
 \autoref{fact:dtv-condit}, which will come in handy when we bound
 the right hand side of Equation~\eqref{eq:tvd:otf:sum}.}
% Before entering into the details of the argument, we start with the  following variant of
% \autoref{fact:dtv-condit}, that we shall require:

\begin{fct} \label{fact:dtv-condit2}
Let $\D_1$, $\D_2$ be two distributions over the same finite set $X$.  Let
$E$ be an event such that
$\D_i[E] = \alpha_i \leq \alpha$ for $i=1,2$ and the conditional
distributions $(\D_1)_{\overline{E}}$ and
$(\D_2)_{\overline{E}}$ are statistically close, i.e.
$\dtv(
(\D_1)_{\overline{E}}, (\D_2)_{\overline{E}})= \beta.$
Then $\dtv(\D_1,\D_2) \leq \alpha + \beta.$
\end{fct}

\BPF
As in the proof of \autoref{fact:dtv-condit}, let us write
\begin{align*}
2\dtv(\D_1,\D_2)
&= \sum_{x \in X \setminus E} |\D_1(x)-\D_2(x)| + \sum_{x \in  E} |\D_1(x)-\D_2(x)|.
\end{align*}
We may upper bound
$\sum_{x \in  E} |\D_1(x)-\D_2(x)|$ by
$\sum_{x \in  E} (\D_1(x) + \D_2(x)) =
\D_1(E) + \D_2(E) = \alpha_1+\alpha_2$; furthermore,
\begin{align*}
  \sum_{x \in \bar{E}} |\D_1(x)-\D_2(x)|
&= \sum_{x \in \bar{E}} \abs{ (\D_1)_{\bar{E}}(x)\cdot\D_1(\bar{E}) - (\D_2)_{\bar{E}}(x)\cdot\D_2(\bar{E}) }\\
&\leq \D_1(\bar{E})\cdot \sum_{x \in \bar{E}} \abs{ (\D_1)_{\bar{E}}(x) - (\D_2)_{\bar{E}}(x) } + \abs{\D_1(\bar{E})-\D_2(\bar{E})}\cdot(\D_2)_{\bar{E}}(\bar{E})\\
&\leq (1-\alpha_1)\cdot (2\beta) + \abs{\alpha_2-\alpha_1}\cdot 1 \leq 2\beta + \abs{\alpha_2-\alpha_1}
\end{align*}
Thus $2 \dtv(\D_1,\D_2) \leq 2\beta + |\alpha_1 - \alpha_2| + \alpha_1 + \alpha_2 =
2\beta + 2 \max\{\alpha_1,\alpha_2\} \leq 2 (\alpha + \beta)$, and the fact is established.
\EPF

% % \ignore{ %%% Start \ignore
% % Applying \newer{\autoref{fact:dtv-condit2}} to distributions
% % $\mathfrak{A}^{(k), \rm N}$ and ${\mathfrak{A}^{(k+1), \rm N}}$
% % and the event $E=\{(\Delta,y)$ is good$\}$, we get that
% % $\totalvardist{ \mathfrak{A}^{(k), \rm N} }{ \mathfrak{A}^{(k+1), \rm N} }
% % \leq \eta(N) + \beta(k,N)$, and consequently we have
% % $\totalvardist{ \mathfrak{A}^{\rm N} }{ \mathfrak{A}^{\rm Y} } \leq
% % \sum_{k=0}^{q} \left(\eta(N) + \beta(k,N)\right)$,
% % which is at most $1/4$ for a suitable choice of the absolute constant $\tau.$
% %
% % Thus it remains only to prove \autoref{lemma:faking:close:to:true}.  In the next subsection we describe
% % an alternative view of the random draw of $\D \sim \calP_{\text \no}$,
% % and in the following subsection we use this alternate view to prove the lemma.
% % } %%% End \ignore

\subsection{Extended transcripts and drawing \texorpdfstring{$\D \sim \calP_{\text \no}$}{a no-distribution D} on the fly.} \label{sssec:onthefly}
Observe that the testing algorithm, seeing only pairs of queries and answers,
does not have direct access to all the underlying information -- namely,
in the case of a ``No''-distribution,
whether the profile of the block that the sample point comes from
is $\downarrow \uparrow$ or $\uparrow \downarrow$.
\ignore{However, it may be
possible for $\Algo$ to infer some of this information, based on the
previous queries it made; to deal with this in the analysis,}
It will be
useful for us to consider an ``extended'' version of the transcripts,
which includes this information along with information about the
profile of the ``boundary'' blocks for each queried interval,
even though this information is not directly available to the
algorithm.

\BD\label{def:extended:transcript}  With the same notation
as in \autoref{def:transcript}, the \emph{extended transcript} of a
sequence of queries made by $\Algo$ and the corresponding responses
is a sequence
$\mathcal{E}=(I_\ell, s_\ell, b_\ell)_{\ell \in [q]}$ of triples,
where $I_\ell$ and $s_\ell$ are as before, and
$b_\ell = (b_\ell^L,b_\ell^{\rm samp},b_\ell^R) \in\{\downarrow\uparrow,
\uparrow\downarrow\}^3$ is a triple defined as follows:
Let $B_{i^L},\dots,B_{i^R}$ be the blocks that $I_\ell$ intersects, going from
left to right.  Then
\begin{enumerate}
\item $b_\ell^L$ is the profile of the block $B_{i^L}$;
\item $b_\ell^R$ is the profile of the block $B_{i^R}$;
\item $b_\ell^{\rm samp}$ is the profile of the block $B_\ell \in \{B_{i^L},\dots,B_{i^R}\}$
that $s_\ell$ belongs to.
\end{enumerate}
We define $\mathcal{E}|_k$ to be the length-$k$ prefix of an
extended transcript $\mathcal{E}$.
\ED

%\dana{ {\bf Dana (Oct 25):} Below is text that previously appeared earlier and was commented out . I think it fits here. I also moved some text about extended transcripts for faking algorithm from within Subsection~\ref{subsubsec:Ext:trans:Algo} to here, and elaborated.}

\newest{As was briefly discussed prior to the current subsection, we shall  be interested
in considering algorithms that \emph{fake} some answers to their queries.
Specifically,
given an adaptive algorithm $\Algo$,
we define $\Algo^{(1)}$ as the algorithm that \emph{fakes} its first query,
in the following sense: If the first query made by $\Algo$ to the oracle is
some interval $I$, then the algorithm $\Algo^{(1)}$
does not call $\ICOND$ on $I$ but instead
chooses a point $s$ uniformly at random from $I$ and then behaves exactly as
$\Algo$ would behave if the $\ICOND$ oracle had returned $s$ in response
to the query $I$.
More generally, we define $\Algo^{(k)}$ for all $0 \leq k \leq q$
as the algorithm behaving like $\Algo$ but faking its first $k$
queries (note that $\Algo^{(0)}=\Algo$).

In \autoref{subsubsec:Ext:trans:Algo} we explain how extended transcripts
can be generated for $\Algo^{(0)}=\Algo$ in an ``on the fly'' fashion so that the resulting distribution
over extended transcripts is the same as the one that would result from first
drawing $\D$ from $\calP_{\text \no}$ and then running algorithm $\Algo$ on it.
It follows that when we remove the extension to the transcript so as to obtain
a regular transcript, we get a distribution over transcripts that is identical to
$\mathfrak{A}^{\rm N}$.
In \autoref{subsubsec:Ext:trans:Algo:k} we explain how to generate extended
% Dana (Oct 25) I wrote $k\geq 0$ on purpose below (as indeed the description is well defined
% for $k=0$ and coincides with the first subsubsection.
transcripts for $\Algo^{(k)}$ where $0 \leq k \leq q$.
We note that for $k\geq 1$
the resulting distribution over extended transcripts is \emph{not} the same as
the one that would result from first
drawing $D$ from $\calP_{\text \no}$ and then running algorithm $\Algo^{(k)}$ on it.
However, this is not necessary for our purposes. For our purposes it is sufficient that
the distributions corresponding to pairs of consecutive indices $(k,k+1)$ are similar (including
the pair $(0,1)$), and that for $k=q$
the distribution over regular transcripts obtained by removing the extension to the transcript
is identical to $\mathfrak{A}^{\rm Y}$.
}

%\bigskip
\subsubsection{Extended transcripts for $\Algo \newest{ = \Algo^{(0)}}$}
\label{subsubsec:Ext:trans:Algo}
Our proof of Equation~\eqref{eqn:adaptive:transcript:distribution:closeness}
takes advantage of the fact that one can view the
 draw of a ``No''-distribution from $\calP_{\text \no}$ as being done ``on the fly''
 during the course of algorithm $\Algo$'s execution.
% (Actually, we will need such a view for an algorithm
% $\Algo^{(k)}$ which fakes the first $k\geq 1 $ queries,
% \newer{in which case the equivalence of the two views does no longer hold,}
% but first we consider
% the simpler case of  algorithm $\Algo = \Algo^{(0)}$ before turning to
% $\Algo^{(k)}$ for $k \geq 1$).
First, the size $\Delta$ and the offset $y$ are drawn at the very beginning, but we may view
the profile vector $\vartheta$ as having its components chosen
independently, coordinate by coordinate, only as $\Algo$ interacts with
\ICOND~--~each time an element $s_\ell$ is obtained in response to the
$\ell$-th query $I_\ell$, only then
are the elements of the profile vector $\vartheta$ corresponding to the
three coordinates of $b_\ell$ chosen
(if they were not already completely
determined by previous calls to $\ICOND$).
More precise details follow.
%  We now describe how the coordinates of the
% profile vector $\vartheta$ are generated sequentially as $\Algo$ interacts
% with $\ICOND$.

% \rnote{I thought it's clearer to explain the non-faking case first.}
% \dnote{I agree}

Consider the $\ell$-th query $I_\ell$ that $\Algo$ makes to $\ICOND_{\D}$.  Inductively some coordinates of $\vartheta$ may have been already set by previous queries.
Let $B_{i^L},\dots,B_{i^R}$ be the blocks that $I_\ell$ intersects.
First, if the coordinate of $\vartheta$ corresponding to block $B_{i^L}$ was
not already set by a previous query, a fair
coin is tossed to choose a
setting from $\{\downarrow\uparrow,\uparrow\downarrow\}$ for this coordinate.
Likewise, if the coordinate of $\vartheta$ corresponding to block $B_{i^R}$
was not already set (either by a previous query or because $i^R = i^L$), a fair
coin is tossed to choose a
setting from $\{\downarrow\uparrow,\uparrow\downarrow\}$ for this coordinate.
\ignore{
\dnote{I merged the two cases, since there was a lot of overlap (in particular since the second case also dealt with having the profile of one of the extreme blocks set). I also elaborated a bit more on what is important to know about the distribution over blocks.}\cnote{Read it -- I am happy with the new discussion.}
}

At this point, the values of $b^L_\ell$ and $b^R_\ell$ have been set.
A simple but important observation is that
these outcomes of $b^L_\ell$ and $b^R_\ell$ completely determine the
probabilities (call them $\alpha^L$ and $\alpha^R$ respectively)
that the block $B_\ell$ from which $s_\ell$ will be chosen is $B_{i^L}$ (is $B_{i^R}$
respectively), as we explain in more detail next. If $i^R=i^L$ then there is no
choice to be made, and so assume that $i^R>i^L$. For $K\in \{L,R\}$ let
$\rho^K_1\cdot \Delta$ be the size of the intersection of $I_\ell$ with the
first (left) half of $B_{i^K}$ and let $\rho^K_2\cdot \Delta$ be the size of the intersection of $I_\ell$ with the second (right) half of $B_{i^K}$. Note that $0 < \rho^K_1+\rho^K_2 \leq 1$
and that $\rho^L_1=0$ when $\rho^L_2 \leq 1/2$ and similarly $\rho^R_2=0$ when $\rho^R_1 \leq 1/2$. If $b^K_\ell = \uparrow\downarrow$
then let $w^K = \rho^K_1\cdot(1+2\eps) + \rho^K_2\cdot(1-2\eps)
  = \rho^K_1+\rho^K_2 + 2\eps(\rho^K_1-\rho^K_2)$, and if
  $b^K_\ell = \downarrow\uparrow$
then let $w^K =  \rho^K_1+\rho^K_2 - 2\eps(\rho^K_1-\rho^K_2)$.
We now set $\alpha^K = \frac{w^K}{w^L+w^R + (i^L-i^R-1)}$.
The block $B_{i^L}$ is selected with probability $\alpha^L$, the block
$B_{i^R}$ is selected with probability $\alpha^R$, and for $i^R \geq i^L+2$, each
of the other blocks is selected with equal probability, $\frac{1}{w^L+w^R + (i^L-i^R-1)}$.

% \ignore{
% To see this, observe
% that every outcome of $D \sim \calP_{\text No}$
% must put the exact same amount of total weight on blocks $B_2,\dots,B_{j-1}$.
% Thus the values of $\alpha_L$ and $\alpha_R$ are completely determined by the
% number of blocks, $j$, the relative sizes of the intersection of $I_\ell$
% with $B_1$ and with $B_j$, and the settings $b^L_\ell$ and $b^R_\ell$
% of the profiles of $B_1$ and $B_j$.
% \footnote{If $j=1$ then we take $\alpha_L=1$ and $\alpha_R=0.$}
% Hence at this point a well-defined random choice can be made for which
% block $B$ the point
% $s_\ell$ belongs to:  with probability $\alpha_L$ the block $B$ is chosen to be
% $B_1$, with probability $\alpha_R$ it is chosen to be $B_j$,
% and with probability $1-\alpha_L - \alpha_R$ it is chosen to be a
% uniformly selected block from $B_2,\dots,B_{j-1}.$  (This last choice is
% uniform because each block $B_i \in \{B_2,\dots,B_{j-1}\}$ has the same
% probability under any outcome of $D \sim \calP_{\text \no}.$)
% We note that it is not important for our purposes to compute the
% exact expressions for these probabilities.  What is important is that they
% are well defined, that the blocks $B_2,\dots,B_{j-1}$ are equally distributed,
% and that the probability \red{$\alpha_L$} of selecting $B_1$ (similarly,
% the probability \red{$\alpha_R$ of selecting} $B_j$) deviates
% from $\abs{B_1\cap I_\ell}/\abs{I_\ell}$
% (similarly, $\abs{B_j\cap I_\ell}/\abs{I_\ell}$) by at most a factor of
% $(1\pm \red{\bigO{\eps}})$.
% }

Given the selection of
the block $B_\ell$ as described above, the element $s_\ell$ and the profile
$b^{\rm samp}_\ell$ of the block to which it belongs are selected as follows.
If the coordinate of $\vartheta$ corresponding to $B_{\ell}$ has already been
determined, then $b^{\rm samp}_\ell$ is set to this value
and $s_{\ell}$ is drawn from $B_\ell$
as determined by the $\downarrow \uparrow$ or $\uparrow \downarrow$ setting
of $b^{\rm samp}_\ell$.  Otherwise, a fair coin is tossed, $b^{\rm samp}_\ell$
is set either to $\downarrow \uparrow$ or to $\uparrow \downarrow$ depending
on the outcome, and $s_\ell$ is drawn from $B_\ell$ as in the previous case
(as determined by the setting of $b^{\rm samp}_\ell$).
Now all of $I_\ell$, $s_\ell$, and $b_\ell=(b^L_\ell,b^{\rm samp}_\ell,b^R_\ell)$
have been determined and
the triple $(I_\ell,s_\ell,b_\ell)$ is taken as the
$\ell$-th element of the extended transcript.

\newest{We now define $\mathfrak{A}^{(0),\rm N}_{\rm otf}$ as follows.
A draw from this distribution over
(non-extended) transcripts is obtained by first drawing an
extended transcript $(I_1,s_1,b_1),\dots,(I_q,s_q,b_q)$ from the
on-the-fly process described above, and then removing the third element
of each triple to yield $(I_1,s_1),\dots,(I_q,s_q).$  This is exactly
the distribution over transcripts that is obtained by first drawing $\D$
from $\calP_{\text No}$ and then running $\Algo$ on it.}

% By the above discussion,
% the distribution over extended transcripts resulting from this
% ``on the fly'' process is exactly the same as the distribution
% over extended transcripts that results from first drawing $D$ from
% $\calP_{\text \no}$ and then running algorithm $\Algo$ on it.

\subsubsection{Extended transcripts for $\Algo^{(k)}$, \newest{$k\geq 0$}}
\label{subsubsec:Ext:trans:Algo:k}
% \newer{We are now in position of defining $\mathfrak{A}^{(k),\rm N}_{\rm otf}$, for $0\leq k \leq q$. % Recalling that $\Algo = \Algo^{(0)}$, we turn to} extended transcripts for $\Algo^{(k)}$ where
% $k > 0$.
\newest{In this subsection we define the
distribution $\mathfrak{A}^{(k),N}_{\rm otf}$ for $0 \leq k \leq q$ (the
definition we give below will coincide with our definition from the
previous subsection for $k=0$).}
Here too the size $\Delta$ and the offset $y$ are drawn at the very beginning, and
the coordinates of the profile vector $\vartheta$ are chosen on the fly, together with the
sample points.
For each $\ell > k$, the pair $(s_\ell,b_\ell)$ is selected exactly as was described for $\Algo$, conditioned on the length-$k$ prefix of the extended transcript and the new
query $I_\ell$ \newest{(as well as the choice of $(\Delta,y)$)}.
It remains to explain how the selection is made for $1 \leq \ell \leq k$.

Consider a value $1 \leq \ell \leq k$ and the $\ell$-th
query interval $I_\ell$.
As in our description of the ``on-the-fly'' process
for $\Algo$, inductively some coordinates of $\vartheta$ may have been already
set by previous queries.
Let $B_{i^L},\dots,B_{i^R}$ be the blocks that $I_\ell$ intersects.
As in the process  for $\Algo$,
if the coordinate of $\vartheta$ corresponding to block $B_{i^L}$ was
not already set by a previous query, a fair
coin is tossed to choose a
setting from $\{\downarrow\uparrow,\uparrow\downarrow\}$ for this coordinate.
Likewise, if the coordinate of $\vartheta$ corresponding to block $B_{i^R}$
was not already set (either by a previous query or because $i^L=i^R$), a fair
coin is tossed to choose a
setting from $\{\downarrow\uparrow,\uparrow\downarrow\}$ for this coordinate.
Hence, $b^L_\ell$ and $b^R_\ell$ are set exactly the same as described for
$\Algo$.

\ignore{
\new{By the definition of $\Algo^{(k)}$, the point
$s_\ell$ should be selected uniformly from $I_\ell$. As for the setting of
$b_\ell^{\rm samp}$, if the coordinate of $\vartheta$ corresponding to the
block $B_\ell$ that $s_\ell$ belongs to is  already set
(either by a previous query or because $\ell \in \{i^L,i^R\}$), then
$b_\ell^{\rm samp}$ is determined. Otherwise, we would like it to be set so that
the joint distribution of $s_\ell$ and $b_\ell^{\rm samp}$ is the
same as described for $\Algo$. We next describe how this is done,
where in order to aid in the comparison between the two distributions
\newer{$\mathfrak{A}^{(k),\rm N}_{\rm otf}$ and $\mathfrak{A}^{(k+1),\rm N}_{\rm otf}$},\footnote{\newer{Recall that our goal is to show the quantity $\totalvardist{ \mathfrak{A}^{(k), \rm N}_{\rm otf} }{ \mathfrak{A}^{(k+1), \rm N}_{\rm otf} }$ is small.}}
%\rnote{\green{I am not sure it is clear what are the two distributions
% we are referring to here --- also not sure how to describe them}}
the (uniform) selection of $s_\ell$ in $I_\ell$ is done by first selecting
the block to which $s_\ell$ belongs, and then selecting $s_\ell$ within
this block.
}
% it should be set
% with equal probability to be either  $\downarrow\uparrow$ or $\uparrow\downarrow$,
% independently from all coordinates of  $\vartheta$ that were already set.
% This will ensure that it is also set the same as described for $\Algo$.
% We next describe how to set $s_\ell$ and $b_\ell^{\rm samp}$ so that each
% is distributed as defined above, and furthermore, the {\em joint\/}
% distribution
}

We now explain how to set the probabilities $\alpha^L$ and $\alpha^R$
of selecting the block $B_\ell$ (from which $s_\ell$ is chosen) to be $B_{i^L}$ and
$B_{i^R}$, respectively.
% Similarly to the process described for $\Algo$, we first
% explain how to set the probabilities $\alpha^L$ and $\alpha^R$
% of selecting the block $B_\ell$ (from which $s_\ell$ is chosen) to be $B_{i^L}$ and
% $B_{i^R}$, respectively.
%
Since the ``faking'' process  should choose $s_\ell$ to be
a uniform point from $I_\ell$, the probability $\alpha^L$ is simply
$|B_{i^L} \cap I_\ell|/|I_\ell|$, and similarly for $\alpha^R$.
(If $i^L=i^R$ we take $\alpha^L=1$ and $\alpha^R=0.$)
Thus the
values of $\alpha^L$ and $\alpha^R$ are completely determined by the
number of blocks $j$ and the relative sizes of the intersection of $I_\ell$
with $B_{i^L}$ and with $B_{i^R}$.
Now, with probability $\alpha^L$ the block $B_\ell$ is chosen to be $B_{i^L}$,
with probability $\alpha^R$ it is chosen to be $B_{i^R}$ and
with probability $1 - \alpha^L - \alpha^R$
it is chosen uniformly among $\{B_{i^L+1},\dots,B_{i^R-1}\}$.

%%%
%% Dana (Oct 27) Was a bit lengthy. Made more similar to previous subsection
%%
% It remains only to select the element $s_\ell$ and the profile
% $b^{\rm samp}_\ell$ of the block to which $s_\ell$ belongs; these are chosen
% as follows.  First, the block $B_\ell \in \{B_{i^L},\dots,B_{i^R}\}$ to which
% $s_\ell$ belongs is selected as described above, and $s_\ell$ is chosen
% to be a uniform random element of $B_\ell \cap I_\ell$.
%%
\newest{Given the selection of the block $B_\ell$ as described above, $s_\ell$ is chosen
to be a uniform random element of $B_\ell \cap I_\ell$.}
The profile $b^{\rm samp}_\ell$ of $B_\ell$ is selected as follows:
\begin{enumerate}\label{discussion:uniform:lb:drawing:on:the:fly}

\item If the coordinate of $\vartheta$ corresponding to $B_\ell$
has already been determined (either by a previous query or because
$B_\ell \in \{B_{i^L},B_{i^R}\}$), then $b^{\rm samp}_\ell$ is set accordingly.

\item Otherwise, the profile of $B_\ell$ was not already set;
note that in this case it must hold that $B_\ell \notin \{B_{i^L},B_{i^R}\}$.
 We look at the half of $B_\ell$ that $s_\ell$ belongs to, and
toss a biased coin to set its profile $b^{\rm samp}_\ell\in\{\downarrow\uparrow,
\uparrow\downarrow\}$:  If $s_\ell$ belongs to the first half, then
the coin
toss's probabilities are $((1-{2}\eps)/2, (1+{2}\eps)/2)$; otherwise, they
are $((1+{2}\eps)/2, (1-{2}\eps)/2)$.
  \end{enumerate}
  \newest{
Let $\mathfrak{E}^{(k), \rm N}_{\rm otf}$ denote the distribution induced by the above process
over extended transcripts, and let $\mathfrak{A}^{(k), \rm N}_{\rm otf}$ be the
corresponding distribution over regular transcripts (that is, when removing the profiles
from the transcript). As noted in \autoref{subsubsec:Ext:trans:Algo}, for
$k=0$ we have that $\mathfrak{A}^{(0), \rm N}_{\rm otf} = \mathfrak{A}^{\vphantom{(0),}\rm N}_{\vphantom{otf}}$.
In the other extreme, for $k=q$, since each point $s_\ell$ is selected uniformly
in $I_\ell$ (with no dependence on the selected profiles) we have that
$\mathfrak{A}^{(q), \rm N}_{\rm otf} = \mathfrak{A}^{\vphantom{(0),}\rm Y}_{\vphantom{otf}}$.
In the next subsection we bound the total variation distance between
$\mathfrak{A}^{(k), \rm N}_{\rm otf}$ and $\mathfrak{A}^{(k+1), \rm N}_{\rm otf}$
for every $0 \leq k \leq q-1$
by bounding the distance between the corresponding distributions
$\mathfrak{E}^{(k), \rm N}_{\rm otf}$ and $\mathfrak{E}^{(k+1), \rm N}_{\rm otf}$.
Roughly speaking, the only difference between the two (for each $0 \leq k\leq q-1$) is in the distribution over $(s_{k+1},b^{\rm samp}_{k+1})$. As we argue in more detail and
formally in the next subsection, conditioned on certain events (determined, among other
things, by the choice of $(\Delta,y)$), we have that $(s_{k+1},b^{\rm samp}_{k+1})$
are distributed the same under
$\mathfrak{E}^{(k), \rm N}_{\rm otf}$ and $\mathfrak{E}^{(k+1), \rm N}_{\rm otf}$.
}

%\dana{{\bf Dana (Oct 27):} The above is of course not precise (e.g., what do we mean ``distributed the same''? There is the issue of conditioning. But I think that at this point it is intuitive, and it is Ok since in the next subsection we formalize things.}

\ignore{
\new {We make the following observations, which will play an important role in the next subsection.
\begin{itemize}
\item If $i^L = i^R$, that is, $I_\ell$ intersects a single block, and
furthermore, it intersects a single half block, then $(s_\ell,b_\ell^{\rm samp})$
is distributed the same as for $\Algo$.
\item If $i^L < i^R$, then
conditioned on $B_\ell \notin \{B_{i^L},B_{i^R}\}$, the block
$B_\ell$ is chosen uniformly among the remaining blocks, which corresponds
precisely to its distribution in the ``on-the-fly'' process for $\Algo$.
Moreover, conditioned on $B_\ell \notin \{B_{i^L},B_{i^R}\}$ and furthermore
on
the profile of $B_\ell$ not having been already set,
the pair $(s_\ell,b^{\rm samp}_\ell)$ is
distributed the same as for $\Algo$.
\end{itemize}
}
}

\ignore{
\dana{\textbf{D:} Done or need to elaborate more?}
\clement{\textbf{C:} I think it is fine (this and the new red paragraph from the previous page give together a good idea of what matters and why).}
}
\ignore{
\dnote{I hope this is clear and does not require any more elaboration. Also, I think that we state here the important property of what the process gives us. In the next subsection we shall need to deal with those cases in which $(s_{k+1},b_{k+1})$ is distributed differently under the two distributions.}
}

\ignore{
Let $\mathfrak{E}^{(k), \rm N}$ denote the distribution induced by the above process
over extended transcripts, and $\mathcal{E}^{(k)}$ the corresponding random variable. We observe that
the profile vector generated is in accordance with a draw from
$\calP_{\text \no}$, i.e., every component of it is independently
uniform random from $\{\downarrow\uparrow, \uparrow\downarrow\}$.
This, together with the way the sample points $s_\ell$ are selected implies that
if we discard the third component $b_\ell$ from each triple $(I_\ell,s_\ell,b_\ell)$, then from $\mathfrak{E}^{(k), \rm N}$
we get the \newer{truncated distribution $\mathfrak{A}^{(k),\rm N}_{\rm otf}$ as referred to} on \autopageref{eqn:adaptive:transcript:distribution:closeness}. \newer{Note that by this definition it is not hard to see that indeed $\mathfrak{A}^{(0),\rm N}_{\rm otf}=\mathfrak{A}^{\vphantom{(0),}\rm N}_{\vphantom{otf}}$
and $\mathfrak{A}^{(q),\rm N}_{\rm otf}=\mathfrak{A}^{\vphantom{(q),}\rm Y}_{\vphantom{otf}}$.}
}

\subsection{Bounding $\totalvardist{ \mathfrak{A}^{(k), \rm N}_{\rm otf} }{ \mathfrak{A}^{(k+1), \rm N}_{\rm otf} }$}\label{sssec:unif-lb:intcond:adapt:proof}

As per the foregoing discussion, we can focus on bounding the total variation distance between extended transcripts
\[
\totalvardist{ \mathfrak{E}^{(k), \rm N}_{\newest{\rm otf}} }{ \mathfrak{E}^{(k+1), \rm N}_{\newest{\rm otf}} }
\]
for arbitrary fixed $k\in\{0,\dots, q-1\}$.
Before diving into the proof, we start by defining the probability space we shall be working in, as well as explaining the different sources of randomness
that are in play and how they fit into the random processes we end up analyzing.

\paragraph{The probability space.} Recall the definition of an extended transcript: for notational convenience, we reserve the notation $\mathcal{E}=(I_\ell, s_\ell, b_\ell)_{\ell \in [q]}$ for extended transcript valued random variables, and will write $E=(\iota_\ell, \sigma_\ell, \pi_\ell)_{\ell \in [q]}$ for a fixed outcome. We denote by $\Sigma$ the space of all such tuples $E$, and by $\Lambda$ the set of all possible outcomes for $(\Delta,y)$. The sample space we are considering is now defined as $X\eqdef \Sigma\times\Lambda$: that is, an extended transcript along with the underlying choice of block size and offset\footnote{We emphasize the fact that the algorithm, whether faking or not, has access neither to the ``extended'' part of the transcript nor to the choice of $(\Delta,y)$; however, these elements are part of the events we analyze.}. The two probability measures on $X$ we shall consider will be induced by the execution of $\Algo^{(k)}$ and $\Algo^{(k+1)}$, as per the process detailed below.\medskip

A key thing to observe is that, as we focus on two ``adjacent'' faking algorithms $\Algo^{(k)}$ and $\Algo^{(k+1)}$, it will be sufficient to consider the following equivalent view of the way an extended transcript is generated:
\begin{enumerate}
  \item\label{item:gen:ext:transcript:1} up to (and including) stage $k$, the faking algorithm generates on its own both the queries $\iota_\ell$ and the uniformly distributed samples $\sigma_\ell\in\iota_\ell$; it also chooses its $(k+1)$-st query $\iota_{k+1}$;
  \item\label{item:gen:ext:transcript:2} then, at that point only is the choice of $(\Delta,y)$ made; and the profiles $\pi_\ell$ ($1\leq \ell\leq k$) of the \emph{previous} blocks decided upon, \new{as described in \autoref{sssec:onthefly}};
  \item\label{item:gen:ext:transcript:3} after this, the sampling and block profile selection is made exactly according to the previous ``on-the-fly process'' description.
\end{enumerate}
% Dana (Oct 27): moved the discussion outside. Was previously inside the second item.
      \newest{The reason that we can defer the choice of $(\Delta,y)$ and the setting of
      the profiles in the manner described above is the following: For both $\Algo^{(k)}$ and $\Algo^{(k+1)}$, the choice of each $\sigma_\ell$
       for $1 \leq \ell \leq k$ depends only on $\iota_\ell$ and the choice of each
       $\iota_\ell$ for
      $1 \leq \ell \leq k+1$ depends only on $(\iota_1,\sigma_1),\dots,(\iota_{\ell-1},\sigma_{\ell-1})$.
      That is, there is no dependence on $(\Delta,y)$ nor on any $\pi_{\ell'}$ for $\ell' \leq \ell$. }
% This distinction allows us to defer the choice of the pair $(\Delta,y)$, and thus to
\newest{By deferring the choice of the pair $(\Delta,y)$ we may consider the randomness coming in its draw only at the $(k+1)$-st stage (which is the pivotal stage here).}
Note that, both for $\Algo^{(k)}$ and $\Algo^{(k+1)}$, the resulting distribution over $X$ induced by the description above exactly matches the one from the ``on-the-fly'' process. In the next paragraph, we go into more detail, and break down further the randomness and choices happening in this new view.

\paragraph{Sources of randomness.}
To define the probability measure on this space, we describe the process that, up to stage $k+1$, generates
the corresponding part of the extended transcript and the $(\Delta,y)$ for $\Algo^{(m)}$ (where $m\in\{k,k+1\}$)
(see the previous subsections for precise descriptions of how the following random choices are made):
\begin{enumerate}
  \renewcommand{\theenumi}{R\arabic{enumi}}
  \renewcommand{\labelenumi}{(\theenumi)}
  \item\label{item:process:iotak1} $\Algo^{(m)}$ draws $\iota_1,\sigma_1,\dots, \iota_k,\sigma_k$ and finally $\iota_{k+1}$ by itself;  % \Gamma_1 "is decided" here
  \item\label{item:process:deltay} the outcome of $(\Delta,y)$ is chosen: this ``retroactively''  fixes the partition of the $\iota_\ell$'s ($1\leq \ell \leq k+1$) into blocks $B^{(\ell)}_{i_L},\dots,B^{(\ell)}_{i_R}$;
  \item\label{item:process:profiles:prev} the profiles of $B^{(\ell)}_{i_L}$, $B^{(\ell)}_{i_R}$ and $B_{\ell}$ (i.e., the values of the triples $\pi_\ell$, for $1\leq \ell \leq k$) are drawn;
  \item\label{item:process:profiles:endblocks} the profiles of $B^{(k+1)}_{i_L}$, $B^{(k+1)}_{i_R}$ are chosen;
  \item\label{item:process:block:selection} the block selection (choice of the block $B_{k+1}$ to which $\sigma_{k+1}$ will belong to) is made:
    \begin{enumerate}
      \item\label{item:process:block:inner:outer:selection} whether it will be one of the two end blocks, or one of the inner ones
\new{(for $\Algo^{(k+1)}$ this is
based on the respective sizes of the end blocks, and for $\Algo^{(k)}$ this
is based on the weights of the end blocks, using the profiles of the end
blocks)};
  \item the choice itself:
        \begin{itemize}
          \item\label{item:process:block:outer:selection} if one of the outer ones, draw it based on either the respective sizes (for $\Algo^{(k+1)}$) or the respective weights (for $\Algo^{(k)}$, using the profiles of the end blocks)
          \item\label{item:process:block:inner:selection} if one of the inner ones, uniformly at random \new{among} all inner blocks;
        \end{itemize}
    \end{enumerate}
 %   \item\label{item:process:profiles:block} the profile $\pi_{k+1}$ of $B_{k+1}$
 % is chosen;
 %   \item\label{item:process:sample:selection} the sample $\sigma_{k+1}$ is
 % drawn (uniformly in $B_{k+1}$ for $\Algo^{(k+1)}$, based on $\pi_{k+1}$
 % for $\Algo^{(k)}$);
    \item \label{item:process:profile-sample:selection} the sample $\sigma_{k+1}$ and the profile $\pi_{k+1}^{\rm samp}$ are chosen;
         \ignore{
         (for $\Algo^{(k)}$, if the profile was not already set, then it is selected with equal probability to be either
         $\downarrow\uparrow$ or $\uparrow\downarrow$, and $\sigma_{k+1}$ is drawn accordingly,
         while for $\Algo^{(k+1)}$, the point $\sigma_{k+1}$ is drawn uniformly in $\iota_{k+1}$
         and the profile is set accordingly);
         }
    \item the rest of the transcript, for $k+1,\dots,q$, is iteratively chosen (in the same way for $\Algo^{(k)}$ and $\Algo^{(k+1)}$) according to the on-the-fly process discussed before.
\end{enumerate}
Note that the only differences between the processes for $\Algo^{(k)}$ and $\Algo^{(k+1)}$ lie in steps \eqref{item:process:block:inner:outer:selection}, \eqref{item:process:block:outer:selection} and \eqref{item:process:profile-sample:selection} of the $(k+1)$-st stage.

\paragraph{Bad events and outline of the argument}

Let $G(\iota_{k+1})$  (where `$G$' stands for `Good') denote the
settings of $(\Delta,y)$ that satisfy the following: Either
\new{(i)} $|\iota_{k+1}| > \Delta\cdot(\log N)^2$ or \new{(ii)} $|\iota_{k+1}| < \Delta/\log N$ and $\iota_{k+1}$
is contained entirely within a single half block.
 We next define three indicator random variables for a given
\new{element $\omega=(E,(\Delta,y))$ of the sample space $X$}, where
 $E = ((\iota_1, \sigma_1, \pi_1),\dots,(\iota_q, \sigma_q, \pi_q))$.
The first, $\Gamma_1$, is zero when $(\Delta,y) \notin G(\iota_{k+1})$.
Note that the randomness for $\Gamma_1$ is over
the choice of $(\Delta,y)$ \newest{and the choice of $\iota_{k+1}.$}
 The second, $\Gamma_2$, is zero when 
 % Dana Oct 30: just changed the order so that will be more similar to the modified $\Gamma_3$
 \newest{$\iota_{k+1}$ intersects at least two blocks and  the block
 $B_{k+1}$ is one of the two extreme blocks intersected by $\iota_{k+1}$.}
\ignore{Note that $\Gamma_2$ depends on the sources
of randomness \eqref{item:process:iotak1}, \eqref{item:process:deltay}, \eqref{item:process:profiles:prev}, \eqref{item:process:profiles:endblocks} and \eqref{item:process:block:inner:outer:selection}.}
The third, $\Gamma_3$, is zero when 
\newest{$\iota_{k+1}$ is not contained entirely within a single half block and
$B_{k+1}$  is a block whose
profile had already been set (either because it contains a selected
point $\sigma_\ell$ for $\ell \leq k$ or because it belongs to one of the
two extreme blocks for some queried interval $\iota_\ell$ for $\ell \leq k$).}
For notational ease we write $\overline{\Gamma}(E)$ to denote the triple
$(\Gamma_1,\Gamma_2,\Gamma_3)$.
Observe that these indicator variables are well defined, and correspond to events that are indeed subsets of our space $X$:
 given any element $\omega\in X$, whether $\Gamma_i(\omega)=1$ (for $i\in\{1,2,3\}$) is fully determined.\medskip

\newer{Define $\D_1$, $\D_2$ as the two distributions over $X$ induced by the executions of respectively $\Algo^{(k)}$ and $\Algo^{(k+1)}$ (in particular, by only keeping the first marginal of $\D_1$ we get back $\mathfrak{E}^{(k),\rm N}$). Applying \autoref{fact:dtv-condit2} to  $\D_1$ and $\D_2$, we obtain that}
\begin{align}\label{eq:everything:is:going:to:be:alright} % http://www.oldielyrics.com/lyrics/johnny_nash/i_can_see_clearly_now.html
  %\totalvardist{ \mathfrak{E}^{(k),\rm N} }{ \mathfrak{E}^{(k+1),\rm N} }
  \newer{\totalvardist{ \D_1 }{ \D_2 } }
  &\leq \probaOf{ \overline{\Gamma} \neq (1,1,1) }
      + \totalvardist{ \newer{\D_1} \;\vert\;  \overline{\Gamma} = (1,1,1) }{ \newer{\D_2} \;\vert\; \overline{\Gamma} = (1,1,1) } \notag\\
  &\leq \probaOf{ \Gamma_1 = 0 } + \probaCond{ \Gamma_2 = 0 }{ \Gamma_1 = 1 }  + \probaCond{ \Gamma_3 = 0 }{ \Gamma_1 = \Gamma_2 = 1 }  \notag\\
    &\quad+ \totalvardist{ \newer{\D_1} \;\vert\;  \overline{\Gamma} = (1,1,1) }{ \newer{\D_2} \;\vert\; \overline{\Gamma} = (1,1,1) }\;.
\end{align}

\noindent To conclude, we can now deal which each of these 4 summands separately:
\begin{clm}\label{claim:adapt:proba:gamma:1}
We have that $\probaOf{ \Gamma_1 = 0 } \leq \eta(N)$, where 
$\eta(N) = O\left({\frac {\log \log N}{\log N}}\right)$.
\end{clm}
\begin{proof}
% \clement{\sc Todo.}
\newest{Similarly to the proof of \autoref{claim:boundary:failure:proba}, 
for any fixed
setting of $\iota_{k+1}$, there are $O(\log \log N)$ values of
$\Delta \in \setOfSuchThat{{\frac N {2^j}} }{ j \in \{{\frac 1 3} \log N, \dots,
{\frac 2 3} \log N\} }$ for which
\mbox{$\Delta/\log N \leq \iota_{k+1} \leq \Delta \cdot (\log N)^2$.} 
Therefore, the probability that one of these (``bad'') values of $\Delta$ is selected is
$O\left({\frac {\log \log N}{\log N}}\right)$. If the choice of $\Delta$ is such that
 $|\iota_{k+1}| < \Delta/\log N$, then, by the choice of the random offset $y$,
 the probability that $\iota_{k+1}$ is not entirely contained within a single
 half block is $O(1/\log N)$. The claim follows.
}
\end{proof}

\begin{clm}\label{claim:adapt:proba:gamma:2}
We have that $\probaCond{ \Gamma_2 = 0 }{ \Gamma_1 = 1 } \leq \eta(N)$.
\end{clm}
\begin{proof}
% \clement{\sc Todo.}
\newest{If $\Gamma_1=1$ because $|\iota_{k+1}| < \Delta/(\log N)^2$ and 
$\iota_{k+1}$ is  entirely contained within a single
 half block, then $\Gamma_2 = 1$ (with probability 1). 
Otherwise, $|\iota_{k+1}| > \Delta \cdot (\log N)^2$, so that 
$\iota_{k+1}$ intersects at least $(\log N)^2$ blocks. The probability
that one of the two extreme blocks is selected is hence
$O(1/(\log N)^2)$, and the claim follows.}
\end{proof}

\begin{clm}\label{claim:adapt:proba:gamma:3}
We have that $\probaCond{ \Gamma_3 = 0 }{ \Gamma_1 = \Gamma_2 = 1 } \leq \eta(N)$.
\end{clm}
\begin{proof}
% \clement{\sc Todo.}
\newest{If $\Gamma_1=1$ because $|\iota_{k+1}| < \Delta/(\log N)^2$ and
$\iota_{k+1}$ is  entirely contained within a single
 half block, then $\Gamma_3 = 1$ (with probability 1). 
 Otherwise, $|\iota_{k+1}| > \Delta \cdot (\log N)^2$, so that
$\iota_{k+1}$ intersects at least $(\log N)^2$ blocks. Since $\Gamma_2=1$,
the block $B_{k+1}$ is uniformly selected from $(\log N)^2-2$ non-extreme blocks.
Among them there are at most $3k = \bigO{ {\frac {\log N}{\log\log  N}} }$ blocks
whose profiles were already set. The probability that one of them is selected
(so that $\Gamma_3=1$) is $\bigO{ {\frac {1}{\log N\log \log N}} } = \bigO{ {\frac {\log \log N}{\log N}} }$, and the claim follows.}
\end{proof}

We are left with only the last term, $\totalvardist{ \newer{\D_1} \;\vert\;  \overline{\Gamma} = (1,1,1) }{ \newer{\D_2} \;\vert\; \overline{\Gamma} = (1,1,1) }$. But as we are now ruling out all the ``bad events'' that would induce a difference between the distributions of the extended transcripts under $\Algo^{(k)}$ and $\Algo^{(k+1)}$, it becomes possible to argue that this distance is actually zero:
\begin{clm}\label{claim:adapt:tv:all:is:good}
  $\totalvardist{ \newer{\D_1} \;\vert\;  \overline{\Gamma} = (1,1,1) }{ \newer{\D_2} \;\vert\; \overline{\Gamma} = (1,1,1) } = 0$.
\end{clm}
\begin{proof}
Unrolling the definition, we can write  $\totalvardist{ \newer{\D_1} \;\vert\;  \overline{\Gamma} = (1,1,1) }{ \newer{\D_2} \;\vert\; \overline{\Gamma} = (1,1,1) }$ as
\[
\sum_{E,(\Delta,y)} \abs{ \probaCond{ \mathcal{E}^{(k)}=E, \mathcal{Y}^{(m)}=(\Delta,y) }{ \overline{\Gamma} = (1,1,1) } - \probaCond{ \mathcal{E}^{(k+1)}=E,\mathcal{Y}^{(m)}=(\Delta,y) }{ \overline{\Gamma} = (1,1,1) } }\;.
\]
where $\mathcal{Y}^{(m)}$ denotes the \newer{$\Lambda$}-valued random variable corresponding to $\Algo^{(m)}$. In order to bound this sum, we will show that each of its terms is zero: i.e., that for
any fixed $(E,(\Delta,y))\in\Sigma\times\Lambda$ we have
\[
\probaCond{ \mathcal{E}^{(k)}=E,\mathcal{Y}^{(k)}=(\Delta,y) }{ \overline{\Gamma} = (1,1,1) }=\probaCond{ \mathcal{E}^{(k+1)}=E,\mathcal{Y}^{(k+1)}=(\Delta,y) }{ \overline{\Gamma} = (1,1,1) } \;.
\]
We start by observing that, for $m\in\{k,k+1\}$,
\begin{align*}
  &\probaCond{ \mathcal{E}^{(m)}=E, \mathcal{Y}^{(m)}=(\Delta,y) }{ \overline{\Gamma} = (1,1,1) } \\
  &\qquad\quad=
  \probaCond{ \mathcal{E}^{(m)}=E }{ \overline{\Gamma} = (1,1,1), \mathcal{Y}^{(m)}=(\Delta,y) }\probaCond{\mathcal{Y}^{(m)}=(\Delta,y)}{\overline{\Gamma} = (1,1,1)}
\end{align*}
and that the term $\probaCond{\mathcal{Y}^{(m)}=(\Delta,y)}{\overline{\Gamma} = (1,1,1)} = \probaOf{\mathcal{Y}^{(m)}=(\Delta,y)}$ is identical for $m=k$ and $m=k+1$. Therefore, it is sufficient to show that
\[
   \probaCond{ \mathcal{E}^{(k)}=E }{ \overline{\Gamma} = (1,1,1), \mathcal{Y}^{(k)}=(\Delta,y) } = \probaCond{ \mathcal{E}^{(k+1)}=E }{ \overline{\Gamma} = (1,1,1), \mathcal{Y}^{(k+1)}=(\Delta,y) }\;.
\]
\noindent Let $\omega=(E,\newer{(\Delta,y)})\in X$ be arbitrary, with $E=((\iota_1, \sigma_1, \pi_1),\dots,(\iota_q, \sigma_q, \pi_q))\in\Sigma$, and let $m\in\{k,k+1\}$. We can express $\Phi^{(m)}(\omega)\eqdef\probaCond{ \mathcal{E}^{(m)}=E }{ \overline{\Gamma} = (1,1,1), \mathcal{Y}^{(m)}=(\Delta,y)  }$ as
\newer{the product of the following 5 terms:
\begin{enumerate}
\renewcommand{\theenumi}{T\arabic{enumi}}
\renewcommand{\labelenumi}{(\theenumi)}
  \item\label{item:T1} $p^{(m),\rm int, samp}_k(\omega)$, defined as
    \begin{align*}
    p^{(m),\rm int, samp}_k(\omega)
      &\eqdef \probaCond{\mathcal{E}^{(m), \rm int, samp}|_k = E^{\rm int, samp}|_k}{ \overline{\Gamma} = (1,1,1), \mathcal{Y}^{(m)}=(\Delta,y) } \\
      &= \probaOf{\mathcal{E}^{(m), \rm int, samp}|_k = E^{\rm int, samp}|_k} \;,
  \end{align*}
  where $E^{\rm int, samp}_\ell$ denotes $(\iota_\ell,\sigma_\ell)$
  and $E^{\rm int, samp}|_k$ denotes $(E^{\rm int, samp}_1,\dots,E^{\rm int, samp}_k)$;
\item\label{item:T2} $p^{(m), \rm prof}_k(\omega)$, defined as
  \begin{align*}
   p^{(m),\rm prof}_k(\omega)
      &\eqdef \probaCond{\mathcal{E}^{(m),\rm prof}|_k = E^{\rm prof}|_k}{\mathcal{E}^{(m),\rm int, samp}|_k = E^{\rm int, samp}|_k, \overline{\Gamma} = (1,1,1), \mathcal{Y}^{(m)}=(\Delta,y) } \\
      &= \probaCond{\mathcal{E}^{(m),\rm prof}|_k = E^{\rm prof}|_k}{\mathcal{E}^{(m),\rm int, samp}|_k = E^{\rm int, samp}|_k, \mathcal{Y}^{(m)}=(\Delta,y) }\;,
  \end{align*}
  where $E^{\rm prof}|_k$ denotes $(\pi_1,\dots,\pi_k)$;
\item\label{item:T3} $p^{(m), \rm int}_{k+1}(\omega)$, defined as
  \begin{align*}
    p^{(m),\rm int}_{k+1}(\omega)
      &\eqdef \probaCond{I_{k+1} = \iota_{k+1}}{\mathcal{E}^{(m),\rm int, samp}|_k = E^{\rm int, samp}|_k, \overline{\Gamma} = (1,1,1), \mathcal{Y}^{(m)}=(\Delta,y) } \\
      &= \probaCond{I_{k+1} = \iota_{k+1}}{\mathcal{E}^{(m),\rm int, samp}|_k = E^{\rm int, samp}|_k }\;;
  \end{align*}
\item\label{item:T4}
\newest{$p^{(m), \rm samp,prof}_{k+1}(\omega)$, defined as
\begin{align*}
&p^{(m), \rm samp,prof}_{k+1}(\omega) \\
&\quad\eqdef \probaCond{(s_{k+1},b_{k+1}) = (\sigma_{k+1}, \pi_{k+1}) }{ I_{k+1}=\iota_{k+1}, \mathcal{E}|_{k}^{(m)} = E|_{k}, \overline{\Gamma} = (1,1,1), \mathcal{Y}^{(m)}=(\Delta,y) };
  \end{align*}
  }
\item\label{item:T5} and the last term $p^{(m)}_{k+2}(\omega)$, defined as
  \begin{align*}
    p^{(m)}_{k+2}(\omega)
      &\eqdef \probaCond{ \mathcal{E}^{(m)}|_{k+2,\dots, q} = E|_{k+2,\dots, q} }{ \mathcal{E}^{(m)}|_{k+1} = E|_{k+1}, \overline{\Gamma} = (1,1,1), \mathcal{Y}^{(m)}=(\Delta,y) },
  \end{align*}
  where $E|_{k+1}=((\iota_{k+1},\sigma_{k+1},\pi_{k+1}),\dots,(\iota_{q},\sigma_{q},\pi_{q}))$.
\end{enumerate}
}
Note that we could remove the conditioning on $\bar{\Gamma}$ for the first three 
% quantities, 
\newest{terms},
as they only depend on the length-$k$ prefix of the (extended) transcript and the choice of $\iota_{k+1}$, that is, on the randomness from \eqref{item:process:iotak1}.
\new{The important observation is that the above probabilities are
independent of whether $m=k$ or $m=k+1$.}
%\rnote{The wording ``the above four probabilities $p^{\ast}_{\ast}$ are the same for $m= k$ and and $m=k+1$.'' seemed clunky, but I'm not sure this is any better.}
 \newest{We first verify this for \eqref{item:T1}, \eqref{item:T2}, \eqref{item:T3} and \eqref{item:T5}, and
then turn to the slightly less straightforward term \eqref{item:T4}.}
This is true for $p^{(m),\rm int, samp}_k(E)$  because $\Algo^{(k)}$ and $\Algo^{(k+1)}$ select their interval queries in exactly the same manner, and for $1 \leq \ell \leq k$, the $\ell$-th sample point is uniformly selected in the $\ell$-th queried interval. Similarly we get that $p^{(k),\rm int}_{k+1}(E) = p^{(k+1),\rm int}_{k+1}(E)$. The probabilities $p^{(k),\rm prof}_k(E)$ and $p^{(k+1),\rm prof}_k(E)$ are induced in the same manner by \eqref{item:process:deltay} and \eqref{item:process:profiles:prev}, \newest{and}
 % Finally, 
 $p^{(k)}_{k+2}(E)=p^{(k+1)}_{k+2}(E)$ since for both $\Algo^{(k)}$ and $\Algo^{(k+1)}$, the pair $(s_\ell,b_\ell)$ is distributed the same for every $\ell \geq k+2$
 \newest{(conditioned on any length-$(k+1)$ prefix of the (extended) transcript and the 
 choice of $(\Delta,y)$).}
 
% But now, 
\newest{Turning to \eqref{item:T4},}
observe that $\Gamma_1=\Gamma_2=\Gamma_3 =1$ (by conditioning). 
% the term in \eqref{eq:key:conditioned:expression} is identical under both distributions. 
% To verify this, 
Consider first the case that
$\Gamma_1=1$ because $|\iota_{k+1}| < \Delta/\log N$ and $\iota_{k+1}$
is contained entirely within a single half block. 
\newest{For this case there are two subcases. In the first subcase, the profile
of the block that contains $\iota_{k+1}$ was already set. This implies that $b_{k+1}$ is fully
determined (in the same manner) for both $m=k$ and $m=k+1$. In the second subcase, the profile 
of the block that contains $\iota_{k+1}$ (which is an extreme block) is set independently and with equal probability to either $\downarrow\uparrow$ or $\uparrow\downarrow$ for both $m=k$ and $m=k+1$.
In either subcase, $s_{k+1}$ is uniformly distributed in $\iota_{k+1}$ for both $m=k$ and $m=k+1$.
}
% since the profile of this block was not yet set (as $\Gamma_3 = 1$), 
% for both $m=k$ and $m=k+1$ the $(k+1)$-th sampled point is uniformly distributed 
% in $\iota_{k+1}$, and the profile of the block
% is set independently and with equal probability to either $\downarrow\uparrow$ or 
% $\uparrow\downarrow$.

Next, consider the remaining case that
$\Gamma_1=1$ because $|\iota_{k+1}| > \Delta \cdot (\log N)^2$.
In this case, since $\Gamma_2=1$, the block $B_{k+1}$ is not
an extreme block, and since $\Gamma_3=1$, the profile of the block
$B_{k+1}$ was not previously set.  Given this, it follows from the
discussion at the end of \autoref{sssec:onthefly} that the distribution of
$(s_{k+1},b_{k+1})$ is identical whether $m=k$ (and $\Algo^{(m)}$ does not fake
the $(k+1)$-th query) or $m=k+1$ (and $\Algo^{(m)}$ fakes the $(k+1)$-th query).
\end{proof}

\newer{Assembling the pieces, the 4 claims above together with Equation~\eqref{eq:everything:is:going:to:be:alright} yield $\totalvardist{ \mathfrak{E}^{(k),\rm N} }{ \mathfrak{E}^{(k+1),\rm N} } \leq \totalvardist{\D_1}{\D_2} \leq 3\eta(N)$, and finally
\begin{align*}
     \totalvardist{ \mathfrak{A}^{\rm N} }{ \mathfrak{A}^{\rm Y} } = \totalvardist{ \mathfrak{A}^{(0), \rm N}_{\rm otf} }{ \mathfrak{A}^{(q), \rm N}_{\rm otf} }
      &\leq \sum_{k=0}^{q-1} \totalvardist{ \mathfrak{A}^{(k), \rm N}_{\rm otf} }{ \mathfrak{A}^{(k+1), \rm N}_{\rm otf} } \\
     &\leq \sum_{k=0}^{q-1} \totalvardist{ \mathfrak{E}^{(k),\rm N} }{ \mathfrak{E}^{(k+1),\rm N} }
     \leq 3q\cdot \eta(N) \\
     &\leq \newer{1/5}
\end{align*}
}for a suitable choice of the absolute constant $\tau$.
\end{proofof}

\section{Conclusion}  \label{sec:conclusion}

We have introduced a new conditional sampling framework for testing probability
distributions and shown that it allows significantly more query-efficient
algorithms than the standard framework for a range of problems.
This new framework presents many potential directions for future work.

One specific goal is to strengthen the upper and lower bounds for problems
studied in this paper. As a concrete question along these lines,
we conjecture that $\COND$ algorithms for testing equality of two unknown
distributions $\D_1$ and $\D_2$ over $[N]$ require $(\log N)^{\Omega(1)}$
queries.  A broader goal is to study more properties of distributions
beyond those considered in this paper; natural candidates here,
which have been well-studied in the standard model, are monotonicity
{(for which we have preliminary results)},
independence between marginals of a joint distribution, and entropy.%\rnote{add cites?}
Yet another goal is to study distributions over other structured
domains such as the Boolean hypercube $\{0,1\}^n$~--~here it would seem
natural to consider ``subcube'' queries, analogous to the $\ICOND$ queries
we considered when the structured domain is the linearly ordered set $[N]$.
A final broad goal is to study distribution \emph{learning}
(rather than testing) problems in the conditional sampling framework.

\newest{
\section*{Acknowledgements}

We are sincerely grateful to the anonymous referees for their close reading of this paper
and for their many helpful suggestions, which significantly improved the exposition of the
final version.
}

%%%%%%%%%%%%%%%%%%%%%%%%%%%%%%%%%%%%%%%%%%%%%%%%%%%%%%%%%%%%%%%%%%%%%%%%%%%%%%%%%%%%%%

\ignore{
  While we have shown that our new framework allow property testers to
  perform significantly better in a wide range of natural problems, and
  provided both upper and lower bounds for those, many directions remain
  for future work. One of them is to investigate more properties of
  distributions, such as monotonicity\footnote{For which we already have a
  {\bf BLAH}-query \PCOND algorithm.} or {\bf ???}, under the various
  restrictions of our \COND oracle. We also believe that testing whether
  $\D_1 = \D_2$ given $\COND$ access requires $(\log N)^{\Omega(1)}$
  queries: proving such a lower bound would give some insight on the
  limitations of our model, and put our current results in perspective.
  Another goal is to study learning problems in this new setting, and see
  whether the additional power it brings to testing distributions
  generalizes to learning -- and, if so, to what extent.

  As our results show that the variants we studied, \ICOND and \PCOND, are
  anything but equivalent, it would also be of interest to define some
  notion of a ``query complexity measure'', which could capture their
  respective power, and help characterize and compare \COND variants.

  Finally, similarly to the \ICOND variant we introduced, which naturally
  comes up when considering totally ordered domains, one can also
  transpose our conditional query model to other structured domains (e.g,
  ``subcube'' queries for distributions over the hypercube $\{0,1\}^n$),
  and see what results these variants would enable.

  %%%%%%%%%%%%%%%%%%%%%%%%%%%%%%%%%%%%%%%%%%%%%%%%%%%%%%%%%%%%%%%%%%%%%

  Discuss directions for future work.  Some things we might want to touch
  on:

  \BI
    \item Specific problems:  we could (if we want) conjecture that
    testing whether $\calD \equiv \calD'$ given $\COND_{\calD}$
    and $\COND_{\calD'}$ requires $(\log N)^{\Omega(1)}$ queries.
    Any other conjectures we want to formulate?

    \item We've considered $\ICOND$ queries which make sense when the
    domain is a totally ordered set.  Analogue of $\ICOND$ queries for
    other structured domains, and what results can be obtained.
    Example -- could consider ``subcube'' queries for distributions
    over $\{0,1\}^n.$

    \item Talk a bit about learning (or no)?

    \item Mention the monotonicity testing here?

    \item Since we've seen that \ICOND and \PCOND are anything but equivalent, idea of
     a ``query complexity measure'' which could help rank and compare \COND variants?
   \EI
    \rnote{Maybe leave out the ``query complexity measure'' direction, I   think it's not as crisp as the others.}
}

% %%%%%%%%%%%%%%%%%%%%%%%%%%%%%%%%%%%%%%%%%%%%%%%

% \bibliography{allrefs-dana}

\bibliography{cond-refs}
\bibliographystyle{alpha}

% %%%%%%%%%%%%%%%%%%%%%%%%%%%%%%%%%%%%%%%%%%%%%%%

\end{document}